\documentclass[prb,twocolumn,showpacs,preprintnumbers,amsmath,amssymb,superscriptaddress]{revtex4}

\usepackage{dcolumn}
\usepackage{bm}
\usepackage{wrapfig}
\usepackage{subfig}

\def\SetFigFont#1#2#3#4#5{%
  \fontsize{#1}{#2pt}%
  \fontfamily{#3}\fontseries{#4}\fontshape{#5}%
  \selectfont}

\usepackage[pdftex]{graphicx,color}

\newcommand\Dfrtl[1]{\ensuremath{\,\mathrm{d}#1\,}}
\newcommand\imag{\ensuremath{\mathrm{i}}}
\renewcommand\Re{\ensuremath{\mathrm{Re}\,}}
\renewcommand\Im{\ensuremath{\mathrm{Im}\,}}
\newcommand\euler[1]{\ensuremath{\mathrm{e}^{#1}}}
\newcommand\sign{\ensuremath{\mathrm{sgn}}}
\newcommand\diag{\ensuremath{\mathrm{diag}\,}}

\begin{document}

\title{
Imaginary-time quantum many-body theory out of equilibrium II: Analytic
continuation of dynamic observables and transport properties
}

\date{\today}
\author{Andreas Dirks}
\affiliation{Institut f\"ur Theoretische Physik, Universit\"at G\"ottingen, D-37077
G\"ottingen, Germany}
\author{Jong E.~Han}
\affiliation{Department of Physics, State University of New York at
Buffalo, Buffalo, NY 14260, USA}
\author{Mark Jarrell}
\affiliation{Department of Physics and Astronomy, Louisiana State University,
Baton Rouge, LA 70803, USA}
\author{Thomas Pruschke}
\affiliation{Institut f\"ur Theoretische Physik, Universit\"at G\"ottingen, D-37077
G\"ottingen, Germany}


\begin{abstract}
Within the imaginary-time theory for nonequilibrium in quantum dot systems the calculation
of dynamical quantities like Green's functions is possible via a suitable quantum Monte-Carlo
algorithm. The challenging task is to analytically continue the imaginary-time data for both
complex voltage and complex frequency onto the real variables. To this end 
a function-theoretical description of dynamical observables is introduced
and discussed within the framework of the mathematical theory of several complex 
variables. We construct a feasible maximum-entropy algorithm for the analytical continuation
by imposing a continuity assumption on the analytic structure and provide results for spectral functions
in stationary non-equilibrium and current-voltage characteristics for different values of the dot charging
energy.
\end{abstract}

\maketitle

\section{Introduction}

Dynamic observables play a crucial role in the Matsubara voltage approach introduced by Han and
Heary to address steady-state non-equilibrium properties of models for quantum dots.\cite{prl07} 
In a previous publication (hereafter referred to as I)\cite{dirks_I} we showed under what conditions the
real-time Keldysh and imaginary-time Matsubara-voltage approaches are formally equivalent, and 
how a proper analytical continuation must be performed. 
In I, this scheme was applied to static quantities obtained from a continuous-time quantum Monte-Carlo
(CT-QMC) algorithm\cite{gull_review,dirks} combined with a standard maximum-entropy (MaxEnt) approach\cite{mem} 
to obtain results for  steady-state expectation values of quantum dot
models at finite bias.

We consider a single-impurity Anderson
model for the quantum dot system\cite{dirks_I},
{with the Hamiltonian
\begin{eqnarray}
\hat{H} & = & \sum_{\alpha k\sigma}\epsilon_{\alpha
k\sigma}c^\dagger_{\alpha k\sigma}
c_{\alpha k\sigma}+\epsilon_d\sum_\sigma d^\dagger_\sigma d_\sigma
\nonumber \\
& & -\sum_{\alpha k\sigma}\frac{t_\alpha}{\sqrt\Omega}(d^\dagger_\sigma
c_{\alpha k\sigma}+\text{h.c.}) \\
& & + U \cdot n_{d,\uparrow} n_{d,\downarrow}\;\;.
\nonumber
\end{eqnarray}
}
{This Hamiltonian describes a quantum dot device
which consists of the quantum dot orbital operator $d^\dagger_\sigma$ of spin
$\sigma$ and
with source and drain leads, represented by conduction electron
operators $c^\dagger_{\alpha k\sigma}$ with the continuum index $k$,
the spin index $\sigma$ and the reservoir index $\alpha=L,R$ for
the source (left) and drain (right), respectively.}
The model is characterized by few parameters:\cite{pustilnik} The local energy or gate voltage $\epsilon_d$, which 
controls the number of electrons on the dot; the charging energy $U>0$ due to the small 
capacitance; and finally the coupling of the dot to the leads, which can in many cases be collected in
two quantities, $\Gamma_L$ and $\Gamma_R$.

{ Although some thermodynamic
observables can be calculated directly without analytic continuation of
imaginary-frequency~\cite{dirks_I},} only a restricted set of
observables can be handled in this manner. Unfortunately,
the current operator $\langle I\rangle$ is not suitable, and the rather important question about
the transport through a quantum dot, both electrically and thermally driven, must be addressed
in a different manner. For simple quantum dot geometries one can employ the result by Meir and
Wingreen,\cite{wingreen} who showed that for single quantum dots and not too different properties in
the left and right leads, one can express the current through the dot due to a finite external bias
$eV_B\equiv\Phi$  via the density-of-states (DOS) 
${\cal N}_\sigma(\epsilon;\Phi)$ on the dot as
\begin{equation}
\label{eq:meir_wingreen}
I(\Phi)=I_0\sum_\sigma\int d\epsilon\left[f_L(\epsilon)-f_R(\epsilon)\right]\,{\cal N}_\sigma(\epsilon;\Phi)
\end{equation}
where $f_\alpha(\epsilon)$ denotes Fermi's function for the left respectively right lead. Note that the
bias enters in two distinct ways: First, in the Fermi functions through the chemical potential of the leads
as\footnote{We use $L=+$ and $R=-$ in mathematical expressions.} $\mu\pm\Phi/2$, and second,
through the DOS. Usually, for $\Phi\to0$, one ignores the latter dependency and can thereby recover
the results from linear-response theory.

Besides its relevance for calculating the current, the DOS is an interesting quantity in its own right, and
{its} dependency as function of frequency and bias
{is} still a matter of debate. In equilibrium, it is well-known
that the DOS develops a very sharp resonance, the so-called Kondo resonance, pinned at $\epsilon=0$,
which in linear response leads to the pinning of the conductance to the
{unitary} limit. The precise way {that}
this resonance dies under the influence of finite bias is actually unknown, and different techniques provide
different answers.

Within the Matsubara-voltage imaginary-time approach by Han and Heary\
the nonequilibrium steady state is mapped to an infinite set of effective equilibrium systems
by introducing a bosonic Matsubara voltage $\varphi_m=4\pi m/\beta$.\cite{prl07,prb10,dirks,dirks_I} It has to be 
analytically continued to a variable $z_\varphi\in \mathbb{C}$ in order to
compute the limit $z_\varphi\to\Phi \pm\imag \delta$ to obtain the physical quantity at the chosen bias $\Phi$.

Considering dynamic observables, the simultaneous presence of the fermionic Matsubara
frequency $\imag\omega_n$, which must be analytically 
continued to $z_\omega \in \mathbb{C}$ to eventually obtain the
corresponding real-frequency quantity, implies that a
{double-complex-variable} function $G(\underline
z)$, with $\underline z := (z_\varphi, z_\omega)^T \in \mathbb{C}^2$, must be
studied.
Without the presence of $\imag\varphi_m$, i.e. within conventional Matsubara
theory, Green's functions are analytic on the upper and lower half-planes,
$\mathbb{H}$ and $\mathbb{H}^*$. Due to the rapid decay of $G(z_\omega)$ as
$z_\omega\to\infty$, a Lehmann spectral representation with respect to the
real axis is used.\cite{negele} Conversely, from a knowledge of all Matsubara-frequency data
one can in principle infer the spectral function. Numerically, this is a known to be an ill-conditioned
problem. An approach, particularly suited for QMC data, is the MaxEnt technique.\cite{mem} 

From a mathematical point of view, the branch cut on the real axis represents 
the only set of points $z_0\in \mathbb{C}$ for which the Green's function is
not holomorphic. 
The very location of the branch cut gives rise to the spectral representation 
in Matsubara theory, i.e.\ it yields, due the nonsingular structure at
$\infty$, a unique characterization of the analytic structure.
In order to perform an analytic continuation for the two-variable 
function $G(\underline z)$ the natural question arises
{as to} which minimal set of 
quantities characterizes its analytic structure in a unique fashion.
{Therefore} the referral to mathematical results in the analysis of
several complex variables is necessary {
(see in particular the
discussion following section \ref{subsubsec:Shilov})}.\cite{dirks}

{In the original implementation of Han and his
co-worker~\cite{prl07,prb10}, analytic continuation on $(z_\varphi,
z_\omega)$ produced rather smooth spectra in good agreement with other
numerical results~\cite{prb10,han10}. However, the employed methods were
crude fit with ad-hoc smoothening and annealing to the numerical data
without regards to statistical analysis. In this work we provide
mathematical foundation to multi-variable analytic continuation and
develop systematic numerical implementation.}

\begin{figure*}
\resizebox{\linewidth}{!}{
\begin{picture}(0,0)%
\includegraphics{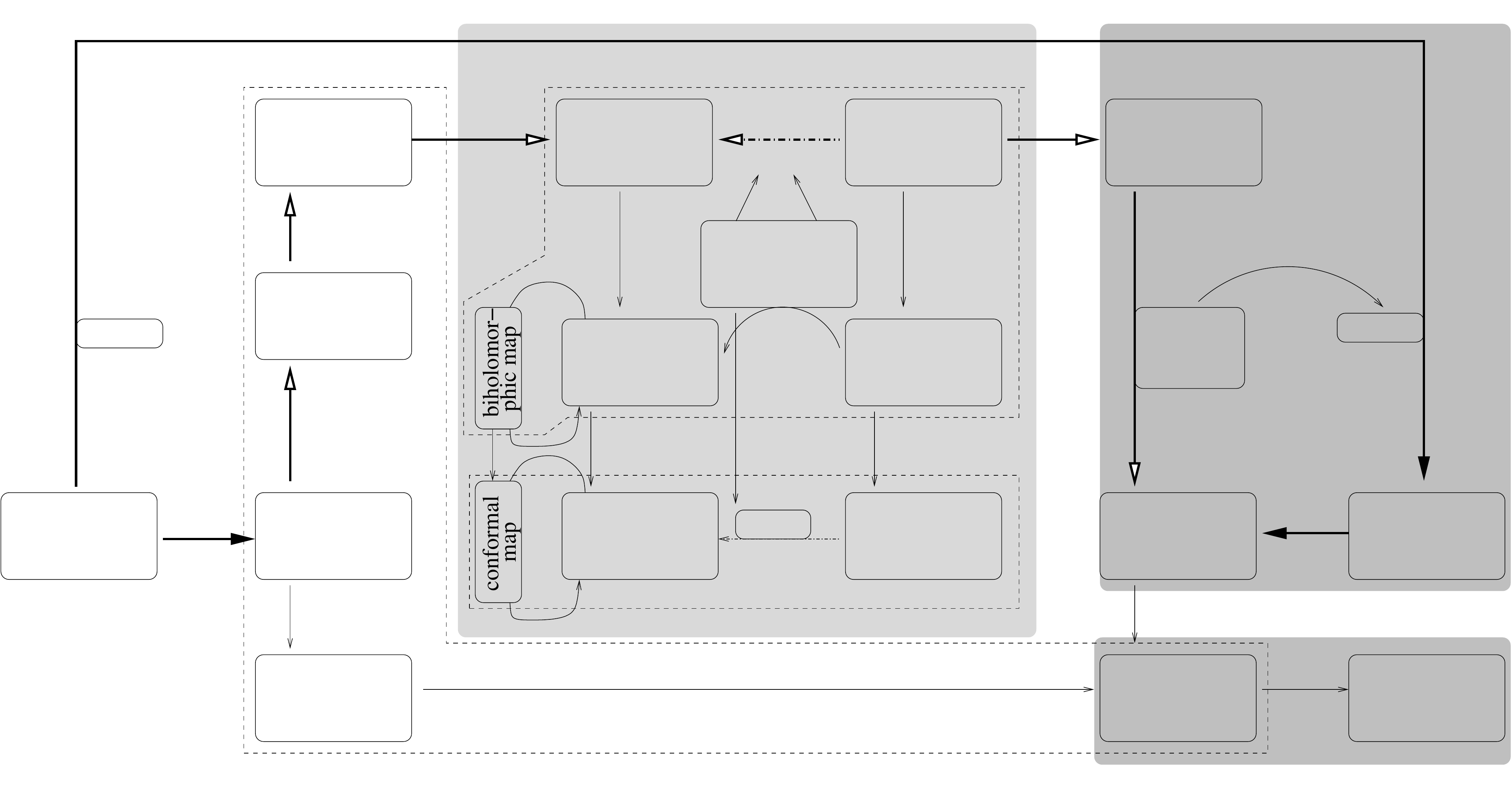}%
\end{picture}%
\setlength{\unitlength}{3947sp}%
\begingroup\makeatletter\ifx\SetFigFont\undefined%
\gdef\SetFigFont#1#2#3#4#5{%
  \reset@font\fontsize{#1}{#2pt}%
  \fontfamily{#3}\fontseries{#4}\fontshape{#5}%
  \selectfont}%
\fi\endgroup%
\begin{picture}(19588,10285)(-3236,-8777)
\put(8401,-361){\makebox(0,0)[lb]{\smash{{\SetFigFont{17}{20.4}{\familydefault}{\mddefault}{\updefault}{\color[rgb]{0,0,0}edges}%
}}}}
\put(8026,-5386){\makebox(0,0)[lb]{\smash{{\SetFigFont{17}{20.4}{\familydefault}{\mddefault}{\updefault}{\color[rgb]{0,0,0}(topological)}%
}}}}
\put(8176,-5686){\makebox(0,0)[lb]{\smash{{\SetFigFont{17}{20.4}{\familydefault}{\mddefault}{\updefault}{\color[rgb]{0,0,0}boundary}%
}}}}
\put(2401,-211){\makebox(0,0)[lb]{\smash{{\SetFigFont{17}{20.4}{\familydefault}{\mddefault}{\updefault}{\color[rgb]{0,0,0}comprised}%
}}}}
\put(676,-541){\makebox(0,0)[lb]{\smash{{\SetFigFont{17}{20.4}{\familydefault}{\mddefault}{\updefault}{\color[rgb]{0,0,0}wedges}%
}}}}
\put(751,-211){\makebox(0,0)[lb]{\smash{{\SetFigFont{17}{20.4}{\familydefault}{\mddefault}{\updefault}{\color[rgb]{0,0,0}set of}%
}}}}
\put(-2774,-5236){\makebox(0,0)[lb]{\smash{{\SetFigFont{17}{20.4}{\familydefault}{\mddefault}{\updefault}{\color[rgb]{0,0,0}data from}%
}}}}
\put(-2549,-5536){\makebox(0,0)[lb]{\smash{{\SetFigFont{17}{20.4}{\familydefault}{\mddefault}{\updefault}{\color[rgb]{0,0,0}QMC}%
}}}}
\put(-2849,-5836){\makebox(0,0)[lb]{\smash{{\SetFigFont{17}{20.4}{\familydefault}{\mddefault}{\updefault}{\color[rgb]{0,0,0}simulations}%
}}}}
\put(4351,-3061){\makebox(0,0)[lb]{\smash{{\SetFigFont{17}{20.4}{\familydefault}{\mddefault}{\updefault}{\color[rgb]{0,0,0}domains of}%
}}}}
\put(4351,-3391){\makebox(0,0)[lb]{\smash{{\SetFigFont{17}{20.4}{\familydefault}{\mddefault}{\updefault}{\color[rgb]{0,0,0}holomorphy}%
}}}}
\put(7801,-3061){\makebox(0,0)[lb]{\smash{{\SetFigFont{17}{20.4}{\familydefault}{\mddefault}{\updefault}{\color[rgb]{0,0,0}Bergman-Shilov}%
}}}}
\put(4876,-1786){\makebox(0,0)[lb]{\smash{{\SetFigFont{17}{20.4}{\familydefault}{\mddefault}{\updefault}{\color[rgb]{0,0,0}are}%
}}}}
\put(2851,-511){\makebox(0,0)[lb]{\smash{{\SetFigFont{17}{20.4}{\familydefault}{\mddefault}{\updefault}{\color[rgb]{0,0,0}of}%
}}}}
\put(676,-1486){\makebox(0,0)[lb]{\smash{{\SetFigFont{17}{20.4}{\familydefault}{\mddefault}{\updefault}{\color[rgb]{0,0,0}give rise to}%
}}}}
\put(6151,-136){\makebox(0,0)[lb]{\smash{{\SetFigFont{17}{20.4}{\familydefault}{\mddefault}{\updefault}{\color[rgb]{0,0,0}Vladimirov's}%
}}}}
\put(8626,-1486){\makebox(0,0)[lb]{\smash{{\SetFigFont{17}{20.4}{\familydefault}{\mddefault}{\updefault}{\color[rgb]{0,0,0}can be}%
}}}}
\put(8626,-1786){\makebox(0,0)[lb]{\smash{{\SetFigFont{17}{20.4}{\familydefault}{\mddefault}{\updefault}{\color[rgb]{0,0,0}thought}%
}}}}
\put(8626,-2086){\makebox(0,0)[lb]{\smash{{\SetFigFont{17}{20.4}{\familydefault}{\mddefault}{\updefault}{\color[rgb]{0,0,0}of as}%
}}}}
\put(10201,-211){\makebox(0,0)[lb]{\smash{{\SetFigFont{17}{20.4}{\familydefault}{\mddefault}{\updefault}{\color[rgb]{0,0,0}yield}%
}}}}
\put(11101,-7411){\makebox(0,0)[lb]{\smash{{\SetFigFont{17}{20.4}{\familydefault}{\mddefault}{\updefault}{\color[rgb]{0,0,0}spectral function}%
}}}}
\put(11626,-7786){\makebox(0,0)[lb]{\smash{{\SetFigFont{17}{20.4}{\familydefault}{\mddefault}{\updefault}{\color[rgb]{0,0,0}$A(\omega)$}%
}}}}
\put(11626,-61){\makebox(0,0)[lb]{\smash{{\SetFigFont{17}{20.4}{\familydefault}{\mddefault}{\updefault}{\color[rgb]{0,0,0}real-time}%
}}}}
\put(11626,-391){\makebox(0,0)[lb]{\smash{{\SetFigFont{17}{20.4}{\familydefault}{\mddefault}{\updefault}{\color[rgb]{0,0,0}structure}%
}}}}
\put(11401,-736){\makebox(0,0)[lb]{\smash{{\SetFigFont{17}{20.4}{\familydefault}{\mddefault}{\updefault}{\color[rgb]{0,0,0}(set of edges)}%
}}}}
\put(11551,-6511){\makebox(0,0)[lb]{\smash{{\SetFigFont{17}{20.4}{\familydefault}{\mddefault}{\updefault}{\color[rgb]{0,0,0}contains}%
}}}}
\put(14626,-5836){\makebox(0,0)[lb]{\smash{{\SetFigFont{17}{20.4}{\familydefault}{\mddefault}{\updefault}{\color[rgb]{0,0,0}distribution}%
}}}}
\put(14701,-5536){\makebox(0,0)[lb]{\smash{{\SetFigFont{17}{20.4}{\familydefault}{\mddefault}{\updefault}{\color[rgb]{0,0,0}probability}%
}}}}
\put(14851,-5236){\makebox(0,0)[lb]{\smash{{\SetFigFont{17}{20.4}{\familydefault}{\mddefault}{\updefault}{\color[rgb]{0,0,0}inferred}%
}}}}
\put(6376,-5761){\makebox(0,0)[lb]{\smash{{\SetFigFont{17}{20.4}{\familydefault}{\mddefault}{\updefault}{\color[rgb]{0,0,0}represents}%
}}}}
\put(676,-3886){\makebox(0,0)[lb]{\smash{{\SetFigFont{17}{20.4}{\familydefault}{\mddefault}{\updefault}{\color[rgb]{0,0,0}separated}%
}}}}
\put(676,-4216){\makebox(0,0)[lb]{\smash{{\SetFigFont{17}{20.4}{\familydefault}{\mddefault}{\updefault}{\color[rgb]{0,0,0}by (Fig. \ref{fig:branchcuts})}%
}}}}
\put(5026,-6661){\makebox(0,0)[lb]{\smash{{\SetFigFont{17}{20.4}{\familydefault}{\mddefault}{\updefault}{\color[rgb]{0,0,0}\emph{conventional function theory}}%
}}}}
\put(4201,464){\makebox(0,0)[lb]{\smash{{\SetFigFont{17}{20.4}{\familydefault}{\mddefault}{\updefault}{\color[rgb]{0,0,0}\emph{function theory of several complex variables}}%
}}}}
\put(5251,-8536){\makebox(0,0)[lb]{\smash{{\SetFigFont{17}{20.4}{\familydefault}{\mddefault}{\updefault}{\color[rgb]{0,0,0}\emph{Matsubara voltage theory}}%
}}}}
\put(5551,-7711){\makebox(0,0)[lb]{\smash{{\SetFigFont{17}{20.4}{\familydefault}{\mddefault}{\updefault}{\color[rgb]{0,0,0}analytic continuation}%
}}}}
\put(12976,-1861){\makebox(0,0)[lb]{\smash{{\SetFigFont{17}{20.4}{\familydefault}{\mddefault}{\updefault}{\color[rgb]{0,0,0}improves}%
}}}}
\put(-2174,-2911){\makebox(0,0)[lb]{\smash{{\SetFigFont{17}{20.4}{\familydefault}{\mddefault}{\updefault}{\color[rgb]{0,0,0}\emph{MaxEnt}}%
}}}}
\put(14176,-2836){\makebox(0,0)[lb]{\smash{{\SetFigFont{17}{20.4}{\familydefault}{\mddefault}{\updefault}{\color[rgb]{0,0,0}\emph{MaxEnt}}%
}}}}
\put(226,-5311){\makebox(0,0)[lb]{\smash{{\SetFigFont{17}{20.4}{\familydefault}{\mddefault}{\updefault}{\color[rgb]{0,0,0}discrete data}%
}}}}
\put(226,-5686){\makebox(0,0)[lb]{\smash{{\SetFigFont{17}{20.4}{\familydefault}{\mddefault}{\updefault}{\color[rgb]{0,0,0}set $G(\imag\varphi_m,\imag\omega_n)$}%
}}}}
\put(601,-7411){\makebox(0,0)[lb]{\smash{{\SetFigFont{17}{20.4}{\familydefault}{\mddefault}{\updefault}{\color[rgb]{0,0,0}function}%
}}}}
\put(526,-7786){\makebox(0,0)[lb]{\smash{{\SetFigFont{17}{20.4}{\familydefault}{\mddefault}{\updefault}{\color[rgb]{0,0,0}$G(z_\varphi,z_\omega)$}%
}}}}
\put(6376,-1711){\makebox(0,0)[lb]{\smash{{\SetFigFont{17}{20.4}{\familydefault}{\mddefault}{\updefault}{\color[rgb]{0,0,0}theory of}%
}}}}
\put(6451,-2011){\makebox(0,0)[lb]{\smash{{\SetFigFont{17}{20.4}{\familydefault}{\mddefault}{\updefault}{\color[rgb]{0,0,0}integral}%
}}}}
\put(6001,-2311){\makebox(0,0)[lb]{\smash{{\SetFigFont{17}{20.4}{\familydefault}{\mddefault}{\updefault}{\color[rgb]{0,0,0}representations}%
}}}}
\put(13276,-5236){\makebox(0,0)[lb]{\smash{{\SetFigFont{17}{20.4}{\familydefault}{\mddefault}{\updefault}{\color[rgb]{0,0,0}approx.}%
}}}}
\put(13126,1289){\makebox(0,0)[lb]{\smash{{\SetFigFont{17}{20.4}{\familydefault}{\mddefault}{\updefault}{\color[rgb]{0,0,0}\textbf{section \ref{sec:introQ}}}%
}}}}
\put(12901,-8686){\makebox(0,0)[lb]{\smash{{\SetFigFont{17}{20.4}{\familydefault}{\mddefault}{\updefault}{\color[rgb]{0,0,0}\textbf{section \ref{sec:results}}}%
}}}}
\put(11176,-5311){\makebox(0,0)[lb]{\smash{{\SetFigFont{17}{20.4}{\familydefault}{\mddefault}{\updefault}{\color[rgb]{0,0,0}positive definite}%
}}}}
\put(11476,-5686){\makebox(0,0)[lb]{\smash{{\SetFigFont{17}{20.4}{\familydefault}{\mddefault}{\updefault}{\color[rgb]{0,0,0}$\tilde A(x_\varphi,x_\omega)$}%
}}}}
\put(826,-6361){\makebox(0,0)[lb]{\smash{{\SetFigFont{17}{20.4}{\familydefault}{\mddefault}{\updefault}{\color[rgb]{0,0,0}defines}%
}}}}
\put(6376,-2911){\makebox(0,0)[lb]{\smash{{\SetFigFont{17}{20.4}{\familydefault}{\mddefault}{\updefault}{\color[rgb]{0,0,0}represents}%
}}}}
\put(6376,-5386){\makebox(0,0)[lb]{\smash{{\SetFigFont{17}{20.4}{\familydefault}{\mddefault}{\updefault}{\color[rgb]{0,0,0}Cauchy}%
}}}}
\put(601,-6661){\makebox(0,0)[lb]{\smash{{\SetFigFont{17}{20.4}{\familydefault}{\mddefault}{\updefault}{\color[rgb]{0,0,0}(appendix \ref{apx:uniqueness})}%
}}}}
\put(4576,-5461){\makebox(0,0)[lb]{\smash{{\SetFigFont{17}{20.4}{\familydefault}{\mddefault}{\updefault}{\color[rgb]{0,0,0}domains}%
}}}}
\put(226,-2836){\makebox(0,0)[lb]{\smash{{\SetFigFont{17}{20.4}{\familydefault}{\mddefault}{\updefault}{\color[rgb]{0,0,0}$G$ in $\mathbb{C}^2$}%
}}}}
\put(226,-2461){\makebox(0,0)[lb]{\smash{{\SetFigFont{17}{20.4}{\familydefault}{\mddefault}{\updefault}{\color[rgb]{0,0,0}branch cuts of}%
}}}}
\put(8176,-3361){\makebox(0,0)[lb]{\smash{{\SetFigFont{17}{20.4}{\familydefault}{\mddefault}{\updefault}{\color[rgb]{0,0,0}boundary}%
}}}}
\put(5776,1289){\makebox(0,0)[lb]{\smash{{\SetFigFont{17}{20.4}{\familydefault}{\mddefault}{\updefault}{\color[rgb]{0,0,0}\textbf{section \ref{sec:GFincontextofsevcomplexvars}}}%
}}}}
\put(-1049,-5311){\makebox(0,0)[lb]{\smash{{\SetFigFont{17}{20.4}{\familydefault}{\mddefault}{\updefault}{\color[rgb]{0,0,0}approx.}%
}}}}
\put(6376,-4336){\makebox(0,0)[lb]{\smash{{\SetFigFont{17}{20.4}{\familydefault}{\mddefault}{\updefault}{\color[rgb]{0,0,0}$\widehat{=}$}%
}}}}
\put(8176,-4336){\makebox(0,0)[lb]{\smash{{\SetFigFont{17}{20.4}{\familydefault}{\mddefault}{\updefault}{\color[rgb]{0,0,0}$\widehat{=}$}%
}}}}
\put(4501,-4336){\makebox(0,0)[lb]{\smash{{\SetFigFont{17}{20.4}{\familydefault}{\mddefault}{\updefault}{\color[rgb]{0,0,0}$\widehat{=}$}%
}}}}
\put(2926,-4411){\makebox(0,0)[lb]{\smash{{\SetFigFont{17}{20.4}{\familydefault}{\mddefault}{\updefault}{\color[rgb]{0,0,0}$\widehat{=}$}%
}}}}
\put(11551,-3061){\makebox(0,0)[lb]{\smash{{\SetFigFont{17}{20.4}{\familydefault}{\mddefault}{\updefault}{\color[rgb]{0,0,0}assumption}%
}}}}
\put(11701,-3361){\makebox(0,0)[lb]{\smash{{\SetFigFont{17}{20.4}{\familydefault}{\mddefault}{\updefault}{\color[rgb]{0,0,0}Eq.~$\eqref{eq:sharedRPassumption}$}%
}}}}
\put(11626,-2761){\makebox(0,0)[lb]{\smash{{\SetFigFont{17}{20.4}{\familydefault}{\mddefault}{\updefault}{\color[rgb]{0,0,0}continuity}%
}}}}
\put(14776,-7411){\makebox(0,0)[lb]{\smash{{\SetFigFont{17}{20.4}{\familydefault}{\mddefault}{\updefault}{\color[rgb]{0,0,0}transport}%
}}}}
\put(14776,-7786){\makebox(0,0)[lb]{\smash{{\SetFigFont{17}{20.4}{\familydefault}{\mddefault}{\updefault}{\color[rgb]{0,0,0}properties}%
}}}}
\put(6151,-586){\makebox(0,0)[lb]{\smash{{\SetFigFont{17}{20.4}{\familydefault}{\mddefault}{\updefault}{\color[rgb]{0,0,0}formula \eqref{eq:poissonkernelepsdomain}}%
}}}}
\put(13276,-7336){\makebox(0,0)[lb]{\smash{{\SetFigFont{17}{20.4}{\familydefault}{\mddefault}{\updefault}{\color[rgb]{0,0,0}Eq.~\eqref{eq:meir_wingreen}}%
}}}}
\put(4501,-361){\makebox(0,0)[lb]{\smash{{\SetFigFont{17}{20.4}{\familydefault}{\mddefault}{\updefault}{\color[rgb]{0,0,0}wedges}%
}}}}
\end{picture}%
}
\caption{{
Contents of the paper as a flowchart. Full arrows represent
numerical procedures within our approach. Empty arrows 
represent the formal steps necessary for a derivation of the MaxEnt kernel used for numerical
results. Shaded areas are discussed in the corresponding sections. 
The rather unconventional two-dimensional function theory with respect to the
simultaneously variables $(z_\varphi,z_\omega)$ is explained by
analogies ($\widehat{=}$) with the conventional theory in the light-shaded
area. It may be skipped at first reading. Concepts such as the biholomorphic
maps are used to obtain explicit equations. Dash-dotted lines denoting either Vladimirov's
formula or Cauchy's integral equation put an emphasize on the fact that the
content of a domain (of holomorphy) is parametrized by its (Bergman-Shilov)
boundary. In the numerical implementation it is thus an inverse problem to reconstruct function values 
on the (Bergman-Shilov) boundary which requires a MaxEnt approach.
}
}
\label{fig:overview}
\end{figure*}
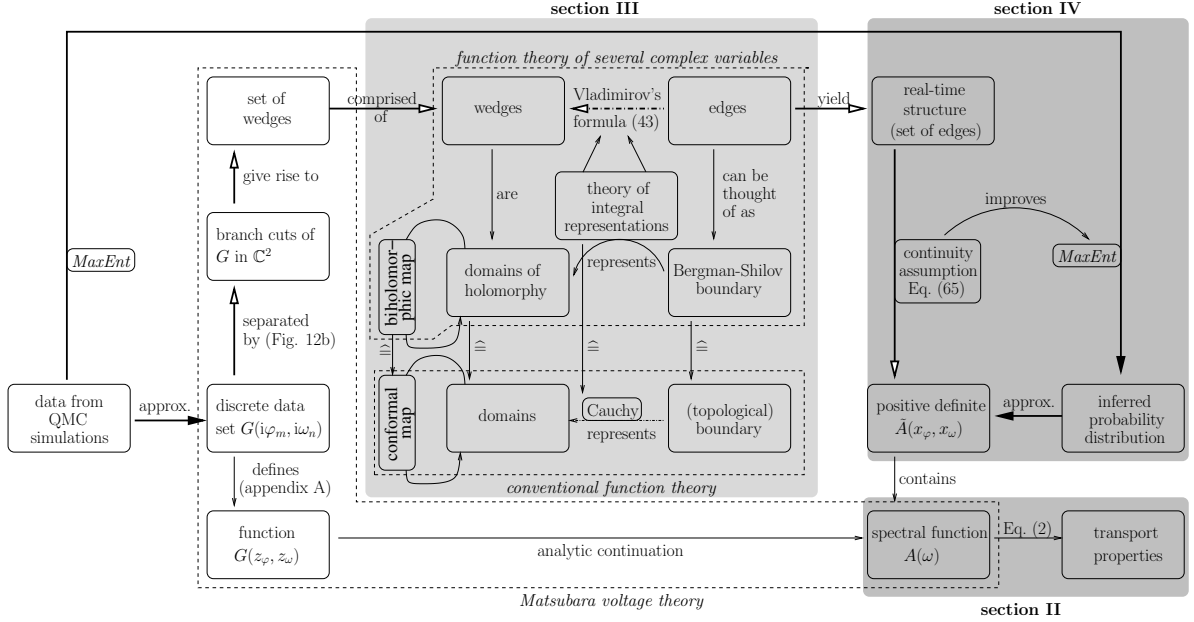

The paper is structured as follows.
Since the mathematical structure dealing with functions of several complex variables is 
probably very alien to the reader, we start with a presentation of the central results for spectral
functions and transport properties of the single-impurity Anderson model in steady-state non-equilibrium in 
section \ref{sec:results}. The underlying mathematical framework is developed in the succeeding sections.
Starting from conventional one-dimensional complex analysis, 
section \ref{sec:GFincontextofsevcomplexvars} provides an introduction to
the basic concepts of theory of several complex variables. We then use this theory to 
systematically analyze the analytical structure of
the Matsubara-voltage Green's function and provide an
axiomatic description of it. 
In section \ref{sec:introQ}, an 
asymptotically exact continuity assumption 
is employed to construct an integral representation for Matsubara Green's function
$G(\imag\varphi_m,\imag\omega_n)$.
It allows {us} to include more information  within the
Bayesian inference process of the maximum entropy method (MaxEnt) as compared to
our previous approach in Ref.\ \onlinecite{dirks}. 
The resulting structure connects 
to the earlier suggestions by Han and Heary.\cite{prl07, prb10}
For future applications, an unbiased extension of the integral representation beyond the 
continuity assumption is proposed in section
\ref{sec:unbiasedQ}.

{
Figure \ref{fig:overview} provides an overview of the paper as a flowchart.
It may serve as a guide. Mathematically less inclined readers may
skip the lightly shaded section \ref{sec:GFincontextofsevcomplexvars} and study the numerical
results in section \ref{sec:results}, which are based on the continuity
ansatz and MaxEnt procedure described in section \ref{sec:introQ} and
corresponding appendices. In the chart full arrows 
denote numerical steps in the computation. Empty arrows denote formal
analytical steps required to derive the relations involved into the MaxEnt
procedure for analytic continuation. 

All numerical results provided in the paper rely on a highly precise continuous-time quantum
Monte-Carlo (QMC) implementation which was introduced in
Ref.~\onlinecite{dirks}. The data provided by QMC simulations, see left box in figure 
\ref{fig:overview}, give rise to a discrete grid
of well-estimated imaginary-time Green's function values. Due to the
structure of the Matsubara-voltage formalism,
however, these data in general belong to different analytic sheets of the two-variable
Green's function (Fig.~\ref{fig:branchcuts}), due to the presence of an
infinite set of branch cuts. The analytic sheets are defined on so-called
wedges in the complex variable space $\mathbb{C}^2$. 
Depending on the considered wedge, an analytic continuation within the sheet to real frequencies 
and voltages may not have a direct physical interpretation. However, QMC data
from these sheets should be used to reconstruct the physical real-time limit.
For this sake, the mathematical structure of the Green's function's
sheets is systematically investigated in section \ref{sec:GFincontextofsevcomplexvars}. 
Since it may require a lot of effort to study the details of the latter, the reader is recommended to first
study the results section and possibly skip section
\ref{sec:GFincontextofsevcomplexvars} at first reading.

The results section \ref{sec:results} discusses numerical results for the
dot electron spectral function out of equilibrium and consequent transport
characteristics. It relies on a MaxEnt procedure which
infers a probability distribution based on a linear relation
\eqref{eq:inverseProblemQ} which is derived
in section \ref{sec:introQ} but has to be inverted. Since as in the
conventional Wick rotation of imaginary-time data,\cite{mem} the inversion is an
ill-posed problem, the MaxEnt provides a most probable solution by means of
Bayes' theorem. Central to the function-theoretical derivation of the
relation \eqref{eq:inverseProblemQ} is a continuity
assumption to the real-time structure of the theory, Eq.~\eqref{eq:sharedRPassumption}. 
It gives rise to the kernel operator \eqref{eq:QoperatorDefinition} which
defines the inverse problem \eqref{eq:inverseProblemQ}. It is
shown in section \ref{sec:introQ} that the continuity assumption improves the MaxEnt 
algorithm for the determination of spectral functions dramatically, as compared 
to the earlier approach introduced in Ref.~\onlinecite{dirks}, such that
nontrivial nonequilibrium spectra could be obtained.

Let us now briefly discuss the more mathematically involved section
\ref{sec:GFincontextofsevcomplexvars} by means of the flowchart in figure
\ref{fig:overview}.
Since the analytical derivation makes use of
two-dimensional complex analysis, it shall provide an introduction to that field,
comparing its fundamental notions to those of conventional function theory.
In particular, the theory of integral representations of functions on domains
of holomorphy is discussed. Such functions may often be parametrized by their values
on the so-called Bergman-Shilov boundary. In an analogous way, conventional
complex analysis parametrizes functions on domains by their boundary values
using, e.g., Cauchy's integral equation. Also the concept of the conformal map has
the analogon of a biholomorphic map which is widely used for formal
derivations in the present work. Section
\ref{sec:GFincontextofsevcomplexvars} also systematically points out that wedges are 
the domains of holomorphy for the two-variable Green's function and provides constraints to
the Green's function which give rise to Vladimirov's integral
representation \eqref{eq:poissonkernelepsdomain} which is central to the
constructed MaxEnt algorithm. The representation links function values on the real-time
boundary (the edge) of a considered wedge to data within the wedge.
The dash-dotted lines in figure \ref{fig:overview} represent such linear relations which are practically used in the
reverse direction and thus bring along an inverse problem. 
}

\section{Results for spectral functions and transport}
\label{sec:results}
\begin{figure}
\includegraphics[width=\linewidth]{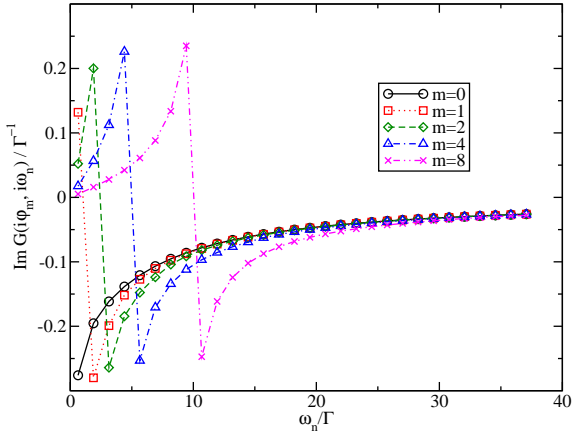}
\caption{
{
Effective-equilibrium data as obtained from CT-QMC simulations\cite{dirks} for
$U=8\Gamma$, $e\Phi=0.1\Gamma$, $\beta=5\Gamma^{-1}$. The integer number $m$ specifies
the respective index of the Matsubara voltage $\varphi_m = 4\pi m / \beta$.
The discontinuity at $\omega_n = \varphi_m/2$ is the principal branch cut
which will in particular be discussed in section \ref{subsec:holostructGF}.
}
}
\label{fig:matsubaradata}
\end{figure}

In the following, we 
present results obtained from CT-QMC data using the algorithm described in Ref.\ \onlinecite{dirks}.
{
Figure \ref{fig:matsubaradata} provides an example for the raw simulation
output. It has to be analytically continued with respect to both, Matsubara voltage and frequency.
}
{
This was accomplished using   MaxEnt applied to the 
functional relation between Matsubara-domain data and spectral function
developed in section \ref{sec:introQ}.
}
As already discussed in Ref.~\onlinecite{dirks}, the Green's function is analytical in certain
cones in the four-dimensional variable space. Previously, we used only the
cone closest to $i\varphi_m=0$ for providing information to the MaxEnt. This
turned out to be not sufficient to generate reliable and reproducable spectra. 
As will be discussed in detail in section \ref{sec:introQ} and appendix
\ref{app:maxentmultiwedge}, we here assume a certain property of the Green's
function, namely that its real part at the meeting point of the cones is
independent of the cone it was approached from. This allows to map data in
different cones by means of linear transformations into the actual data space
and hence improve the accuracy of the MaxEnt tremendously. We must
emphasize that the validity of this crucial property cannot proven
rigorously; however, the results obtained can be
taken as evidence that its violation does not influence the physical
structures too much. Furthermore, in section \ref{sec:unbiasedQ} we provide a route to
improve on this approximation systematically, at the expense of a more
complex algorithm.

We concentrate on the evolution of the spectral functions as function of Coulomb parameter $U$ and
bias voltage $\Phi$.
Using the relation \eqref{eq:meir_wingreen} we also calculate $I(V)$ characteristics and compare them
to results obtained with other techniques. 

\subsection{Weak-Coupling Regime}

\begin{figure}[htb]
\centering
\includegraphics[width=0.47\textwidth]{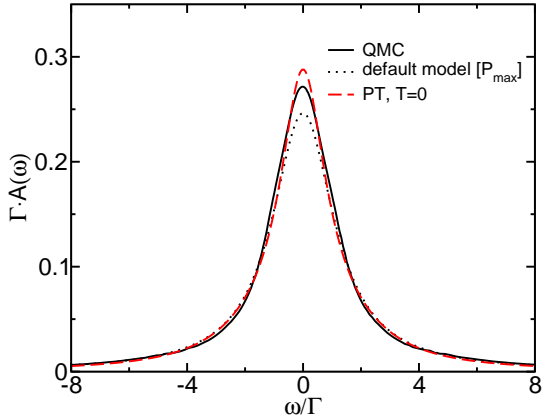}
\caption{(color online) Spectral function of the dot electrons as inferred for the 
nonequilibrium weak-coupling case 
$U=2\Gamma$, $e\Phi=\Gamma$, $\beta=5\Gamma^{-1}$, compared to
zero-temperature
second-order perturbation theory.}
\label{fig:AU2V1b5GenApproach}
\end{figure}

The resulting nonequilibrium spectral function for $U=2\Gamma$,
$\beta=5\Gamma^{-1}, e\Phi=\Gamma$, obtained by evaluating 
$A(\omega) = \tilde A(\Phi, \omega)$, where $\tilde A$ is some
two-dimensional MaxEnt-inferred quantity, is displayed in figure
\ref{fig:AU2V1b5GenApproach}. A good agreement with the zero-temperature
second-order perturbation theory provided by Ref.~\onlinecite{muehlbacher} is observed.
Presumably due to the finite
temperature, the quasi-particle weight is slightly smaller than the
perturbative one. 
\begin{figure*}[htb]
\subfloat[bias voltage $e\Phi=0.125\Gamma$]{\includegraphics[width=0.48\linewidth]{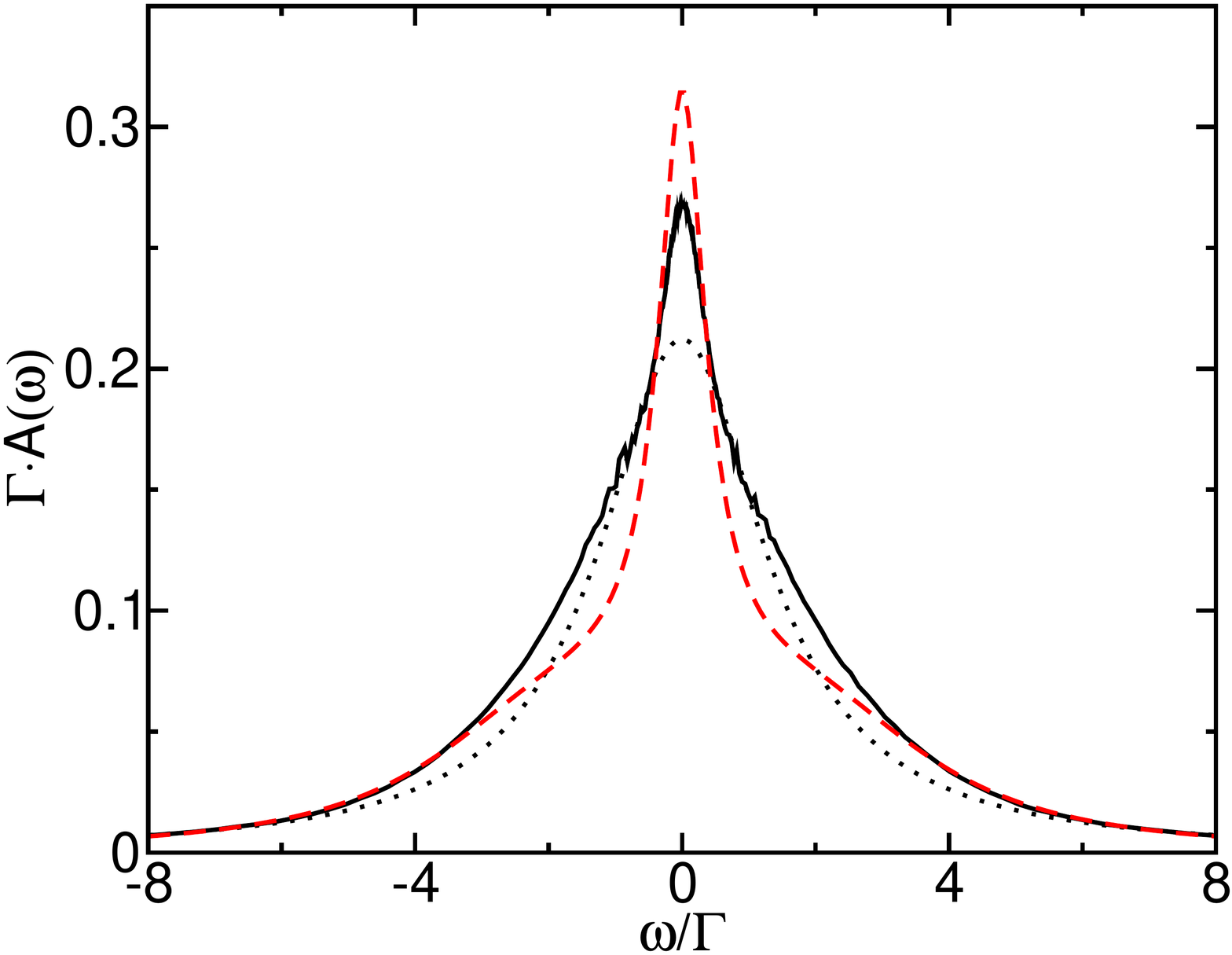}}
\subfloat[bias voltage $e\Phi=0.25\Gamma$]{\includegraphics[width=0.48\linewidth]{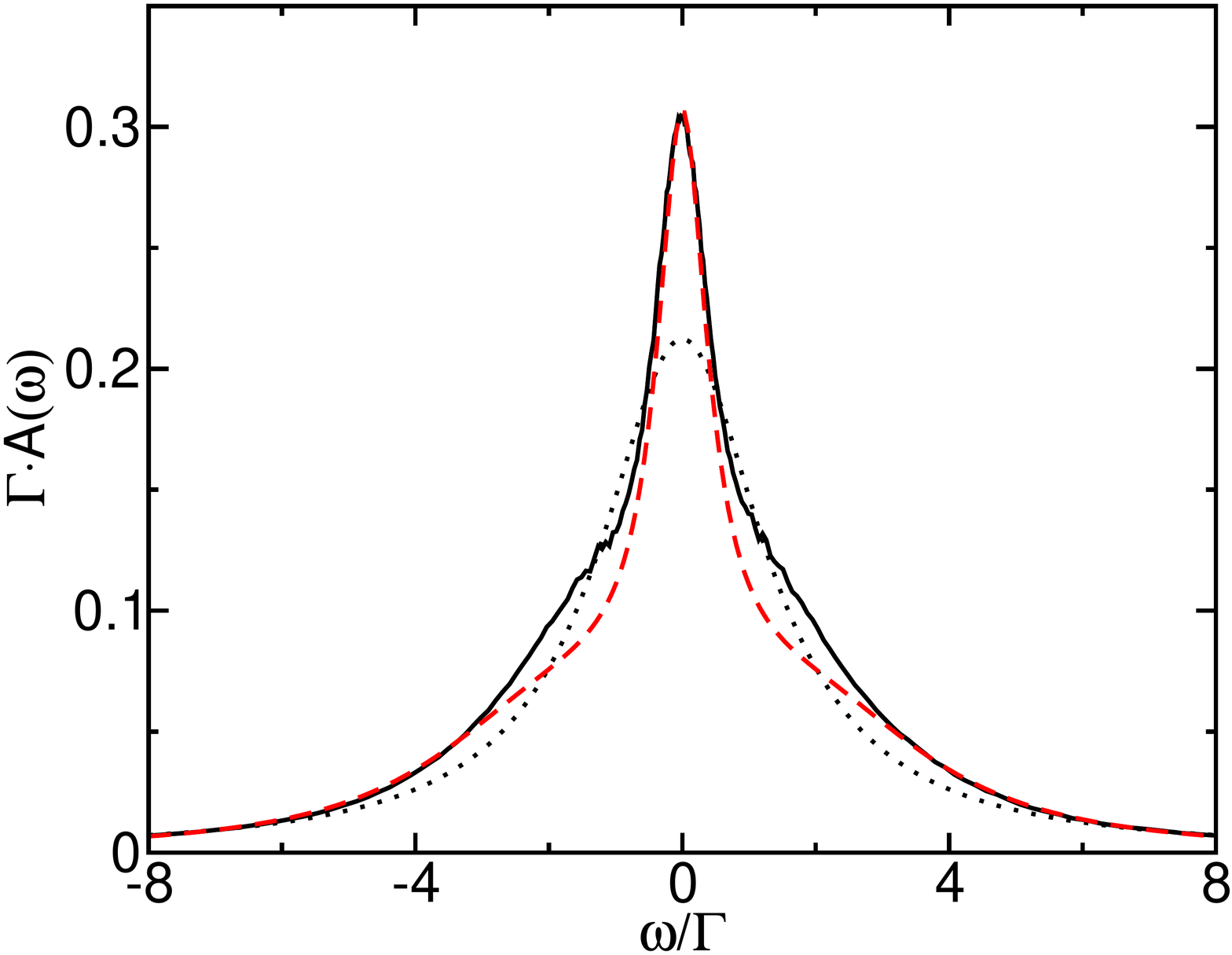}}\\
\subfloat[bias voltage $e\Phi=0.5\Gamma$]{\includegraphics[width=0.48\linewidth]{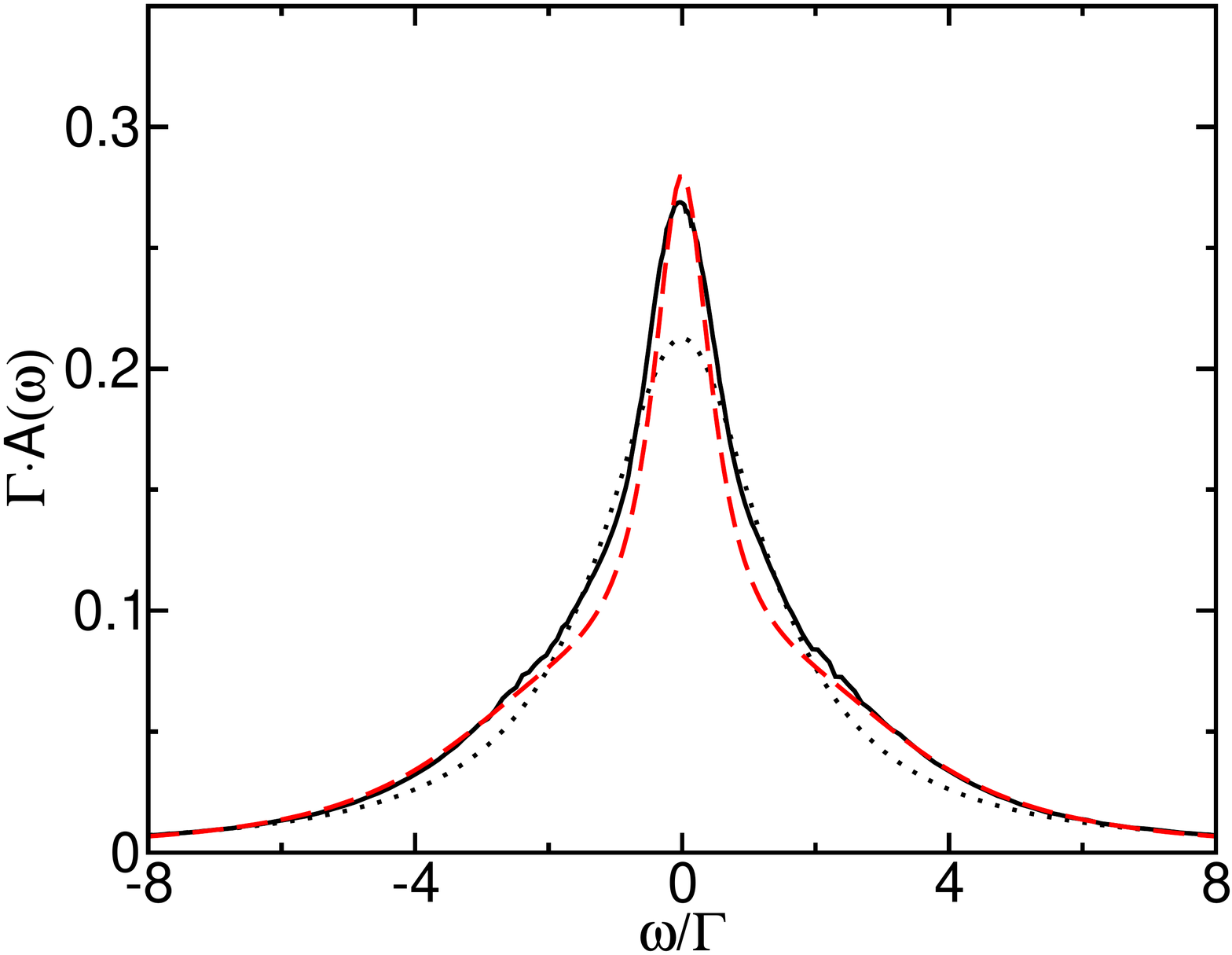}}
\subfloat[bias voltage $e\Phi=1.0\Gamma$]{\includegraphics[width=0.48\linewidth]{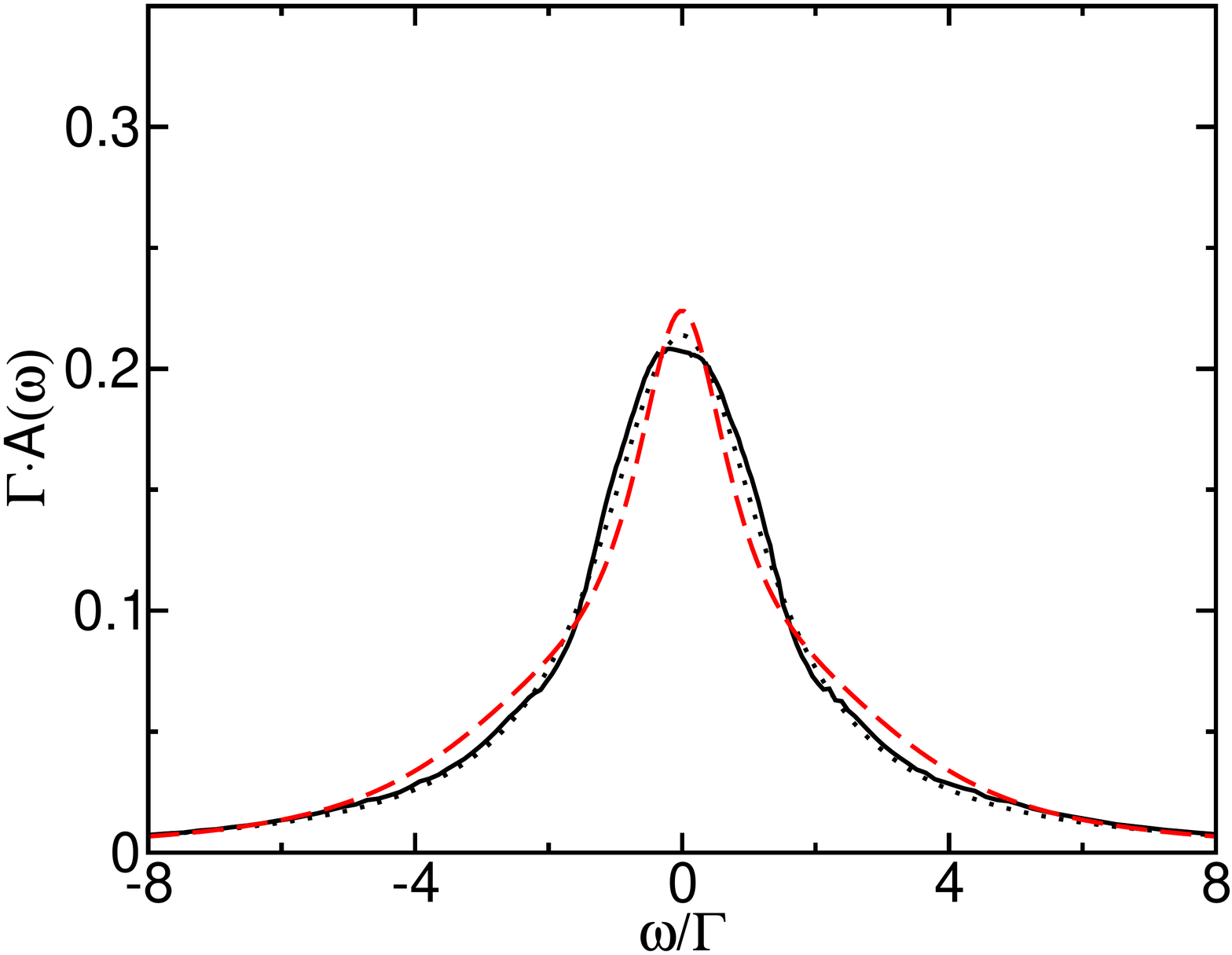}}
\caption{(color online) Nonequilibrium spectral functions for $U=4\Gamma$ and inverse
temperature $\beta=5\Gamma^{-1}$ of the leads, as compared to zero-temperature second-order
perturbation theory. Line legends are the same as in
Fig.~\ref{fig:AU2V1b5GenApproach}.}
\label{fig:intermediatecouplingspectraU4}
\end{figure*}
Also included is the default model supplied to the MaxEnt, a Lorentzian whose
width is determined by the Bayesian procedure outlined in appendix
\ref{app:maxentmultiwedge}. For the weak coupling case, the width is essentially the one for the true spectrum,
as expected.

\subsection{Intermediate-Coupling Regime}

Figure \ref{fig:intermediatecouplingspectraU4} shows spectral functions computed at
$U=4\Gamma$ and inverse temperature $\beta=5\Gamma^{-1}$ for different bias
voltages $\Phi$. An excellent agreement with second-order perturbation theory is
obtained for the cases $e\Phi = 0.25\Gamma$, $e\Phi = 0.5 \Gamma$, $e\Phi =
\Gamma$. Although one should not expect it to be very different, the 
quasiparticle resonance  in the MaxEnt result for
$e\Phi = 0.125\Gamma$
is significantly reduced as compared to $e\Phi = 0.25\Gamma$ and 
disagrees with the almost unchanged second-order perturbation theory. This is a 
systematic MaxEnt artifact which prefers to reproduce the default model in case of missing
Bayesian evidence. In our case, Bayesian evidence is indeed decreased for very small voltages,
because the simulation data at $\varphi_m=0$ cannot be taken into account as the
Matsubara voltage Green's function has a branch cut there (see Ref.\ \onlinecite{prb10} for details).

The normalization of the MaxEnt spectra is reasonably close to one.
In particular at the small voltages,
and for larger  frequencies some 
side-bands form which tend to increase the total spectral weight unphysically. However,
the good description of the low-energy physics seems to be unaffected by this type of artifact. Note that again
the optimal width of the default model is the same as the one for the final spectrum.

\begin{figure*}[htb]
\subfloat[bias voltage $e\Phi=0.125\Gamma$]{\includegraphics[width=0.49\linewidth]{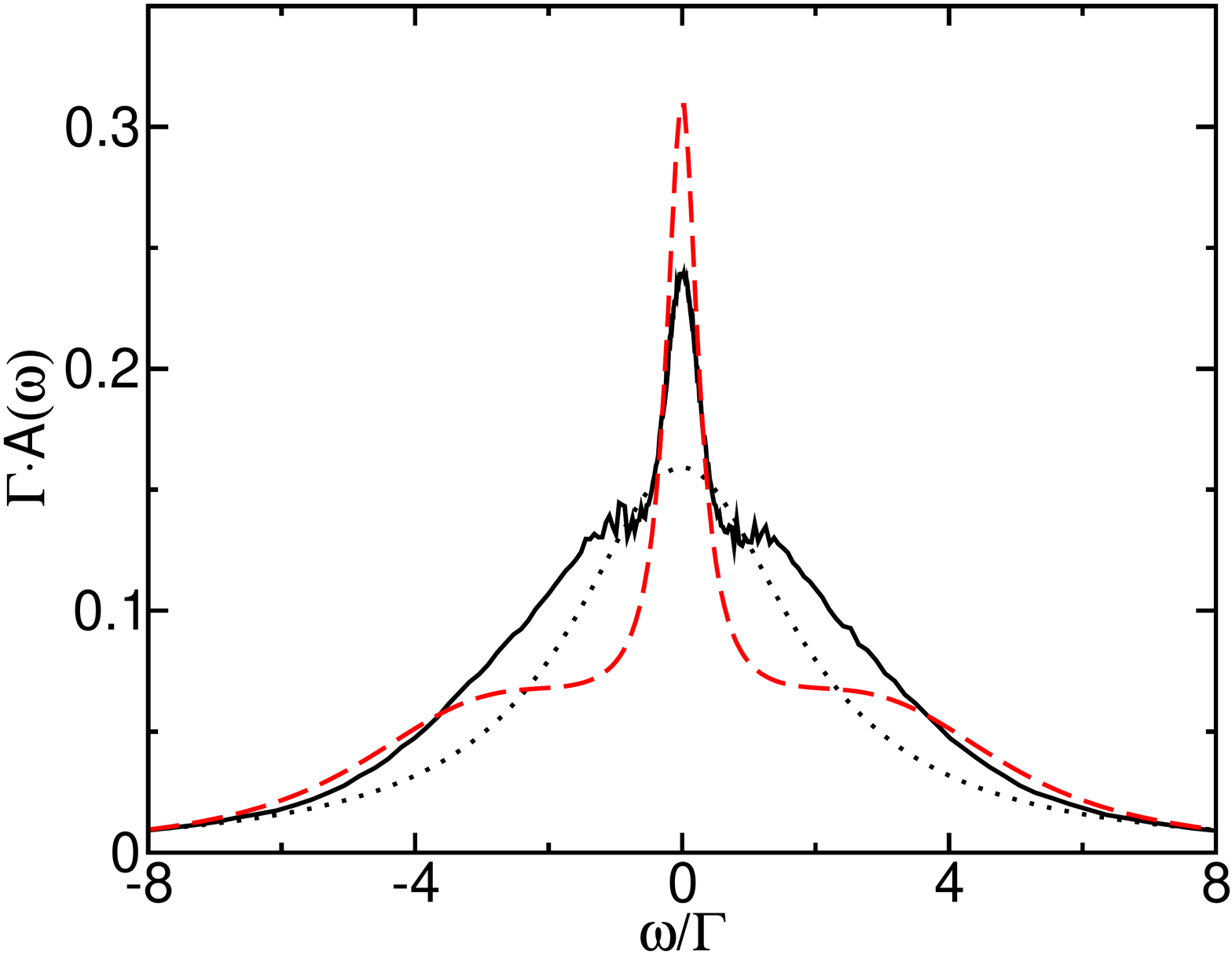}}
\subfloat[bias voltage $e\Phi=0.25\Gamma$]{\includegraphics[width=0.49\linewidth]{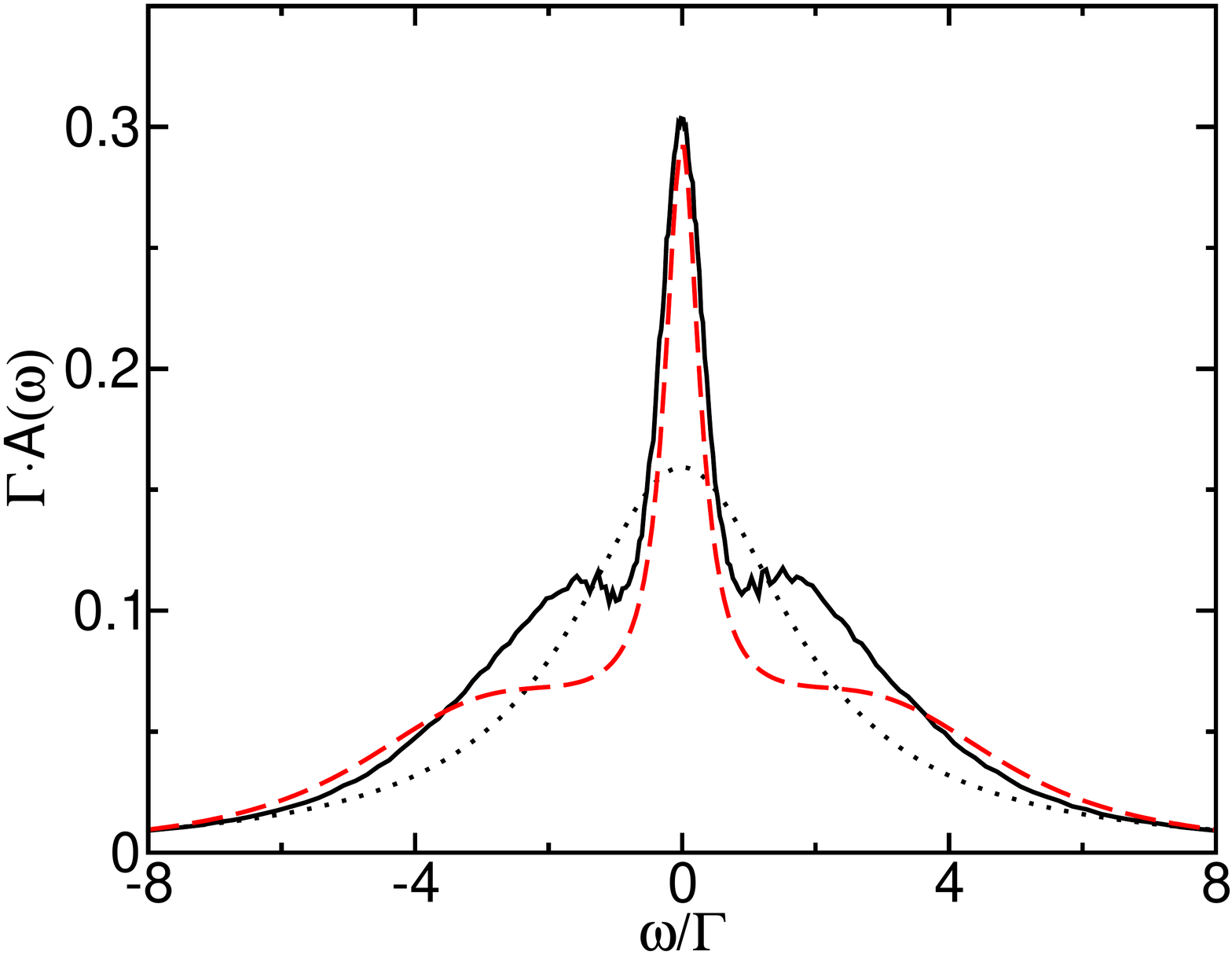}}\\
\subfloat[bias voltage $e\Phi=0.5\Gamma$]{\includegraphics[width=0.49\linewidth]{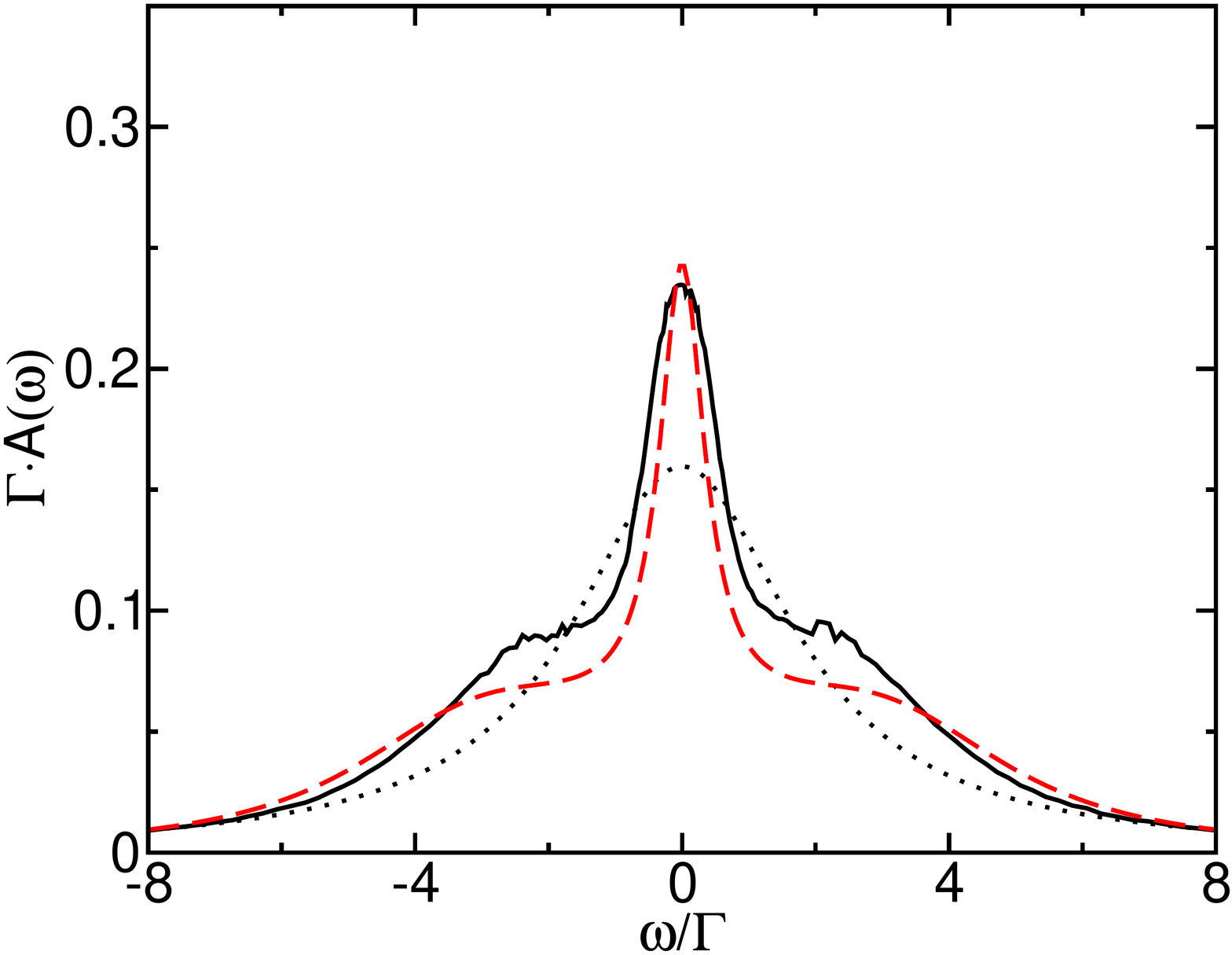}}
\subfloat[bias voltage $e\Phi=1.0\Gamma$]{\includegraphics[width=0.49\linewidth]{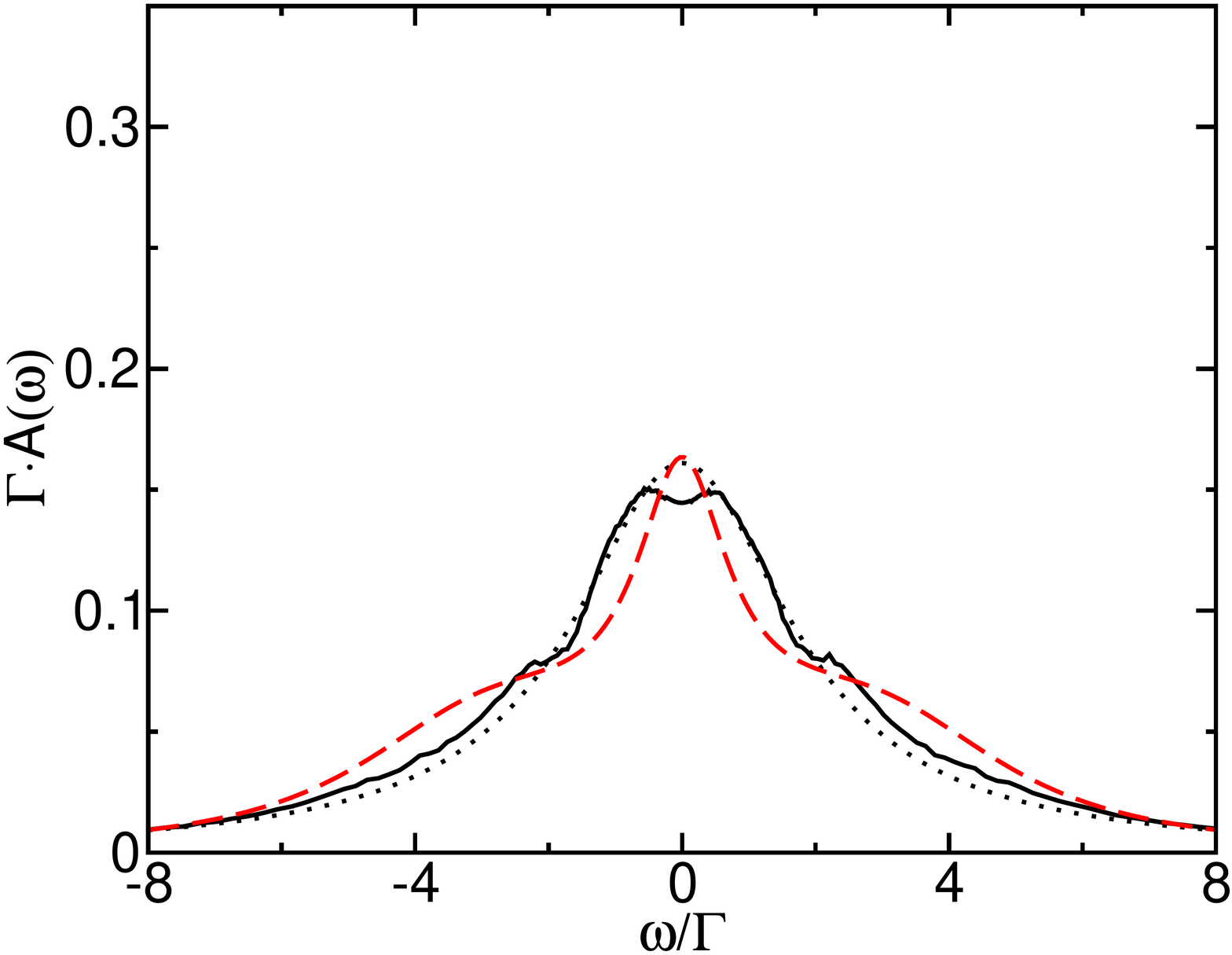}}
\caption{(color online) Nonequilibrium spectral functions for
$U=6\Gamma$ and inverse temperature $\beta=5\Gamma^{-1}$, as compared to zero-temperature second-order
perturbation theory. Line legends are the same as in Fig.~\ref{fig:AU2V1b5GenApproach}.}
\label{fig:intermediatecouplingspectraU6}
\end{figure*}

Inferred spectral functions for an even stronger interaction $U=6\Gamma$ are displayed in 
Fig.~\ref{fig:intermediatecouplingspectraU6} for intermediate to
large bias voltage. For the equilibrium situation we already are in a regime with a distinct 
three-peak structure with an Abrikosov-Suhl resonance (ASR) at $\omega=0$ characteristic for the Kondo regime. An estimate for the equilibrium Kondo scale gives
$T_\text{K}\approx0.1\Gamma$. 
Again, the general low-energy behavior agrees very well with 
the results from perturbation theory -- which should still be valid for this value of $U$ --
and the weight at $e\Phi=0.125\Gamma$ is again
underestimated. At $e\Phi=0.5\Gamma$ one now observes a distinctly larger broadening of the ASR as
compared to perturbation theory, and at $\Phi = \Gamma$ a clear double-peak structure  is visible.
This structure is compatible with a Kondo peak splitting. However, due to the
approximations involved, we feel unable to decide at present
whether this feature is actually a prediction of the Matsubara-voltage theory 
itself. It is interesting to note that here the default model is strongly renormalized for small bias, while at large bias
the default model, apart from the double peak structure around $\omega=0$, again is already a reasonable
estimate for the full spectrum. 

As before, the spectral weight in the now developing Hubbard bands is strongly enhanced as compared to
perturbation theory, pointing towards an overestimation of the integral weight by MaxEnt. On the other hand, the
position is in good agreement with perturbation theory. 

\begin{figure*}[htb]
\subfloat[bias voltage
$e\Phi=0.1\Gamma$]{\includegraphics[width=0.49\linewidth]{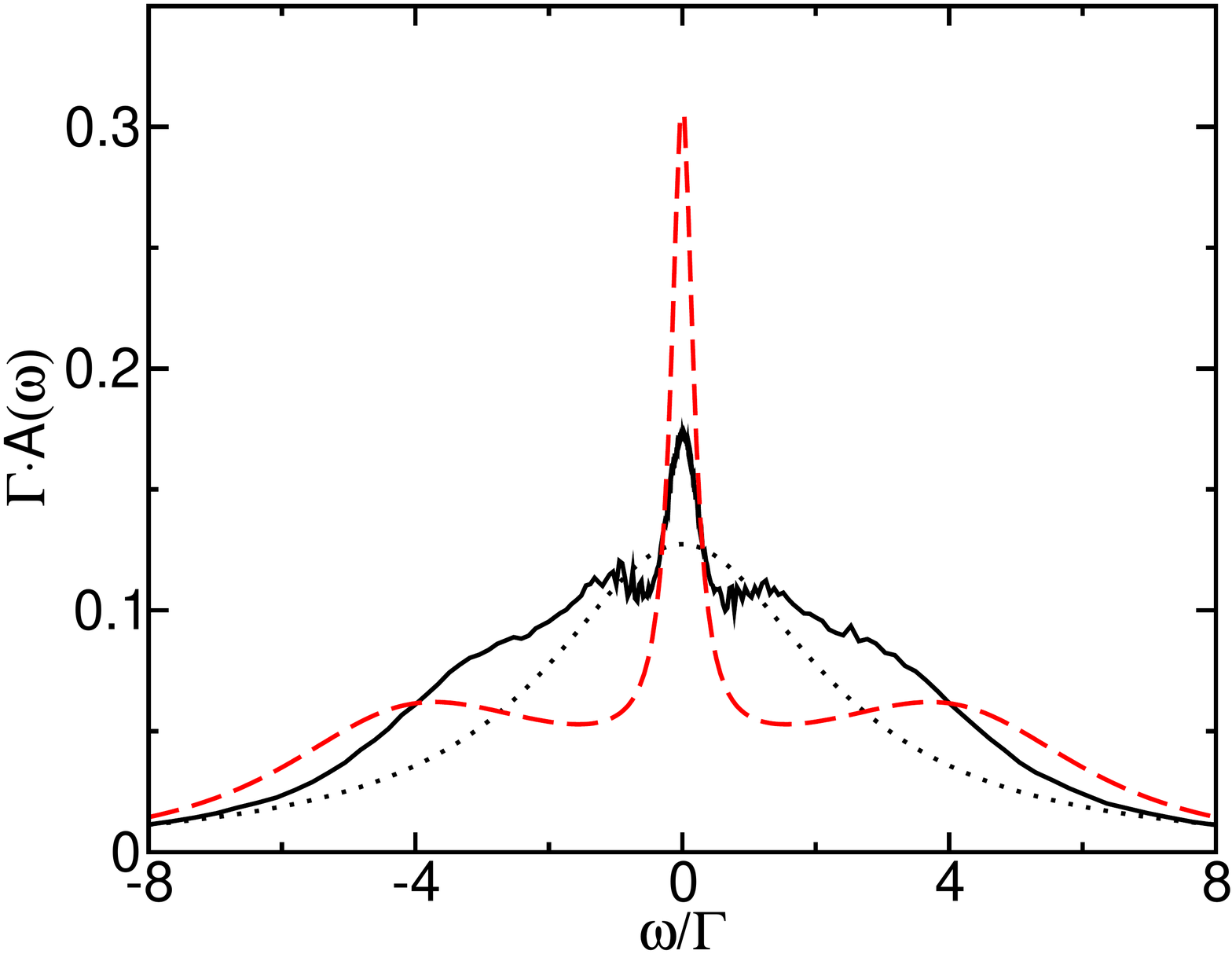}}
\subfloat[bias voltage
$e\Phi=0.2\Gamma$]{\includegraphics[width=0.49\linewidth]{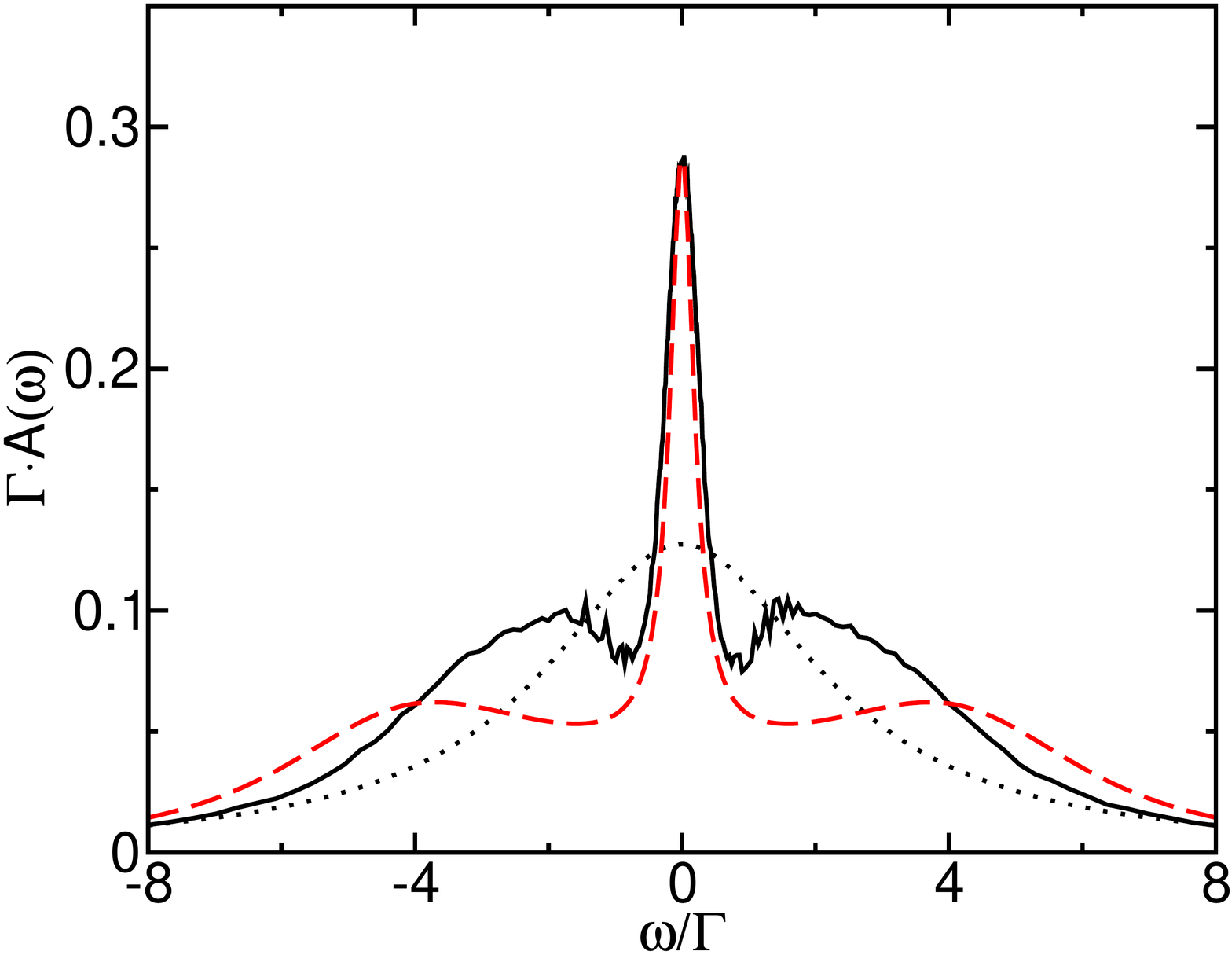}}\\
\subfloat[bias voltage
$e\Phi=0.5\Gamma$]{\includegraphics[width=0.49\linewidth]{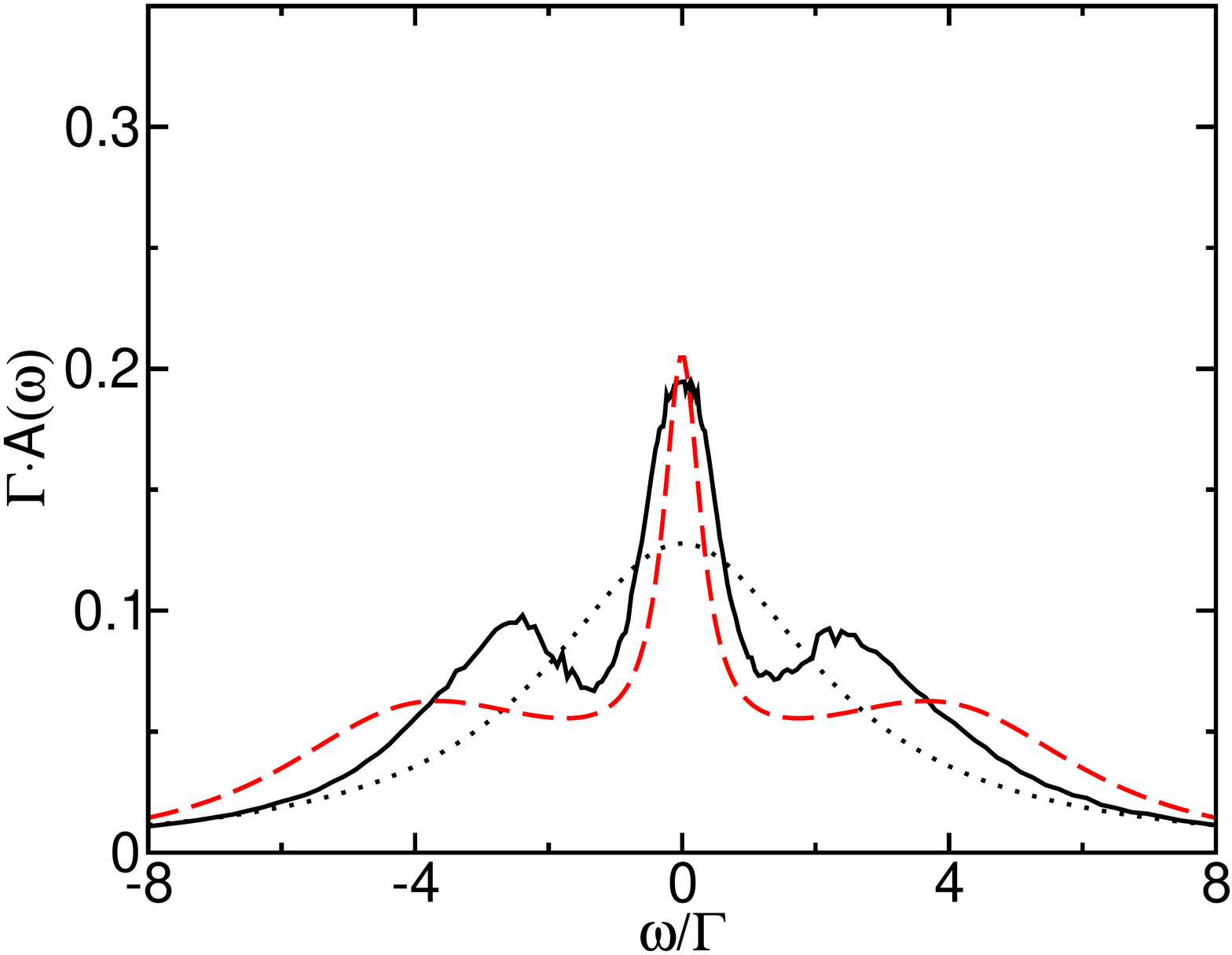}}
\subfloat[bias voltage
$e\Phi=1.0\Gamma$]{\includegraphics[width=0.49\linewidth]{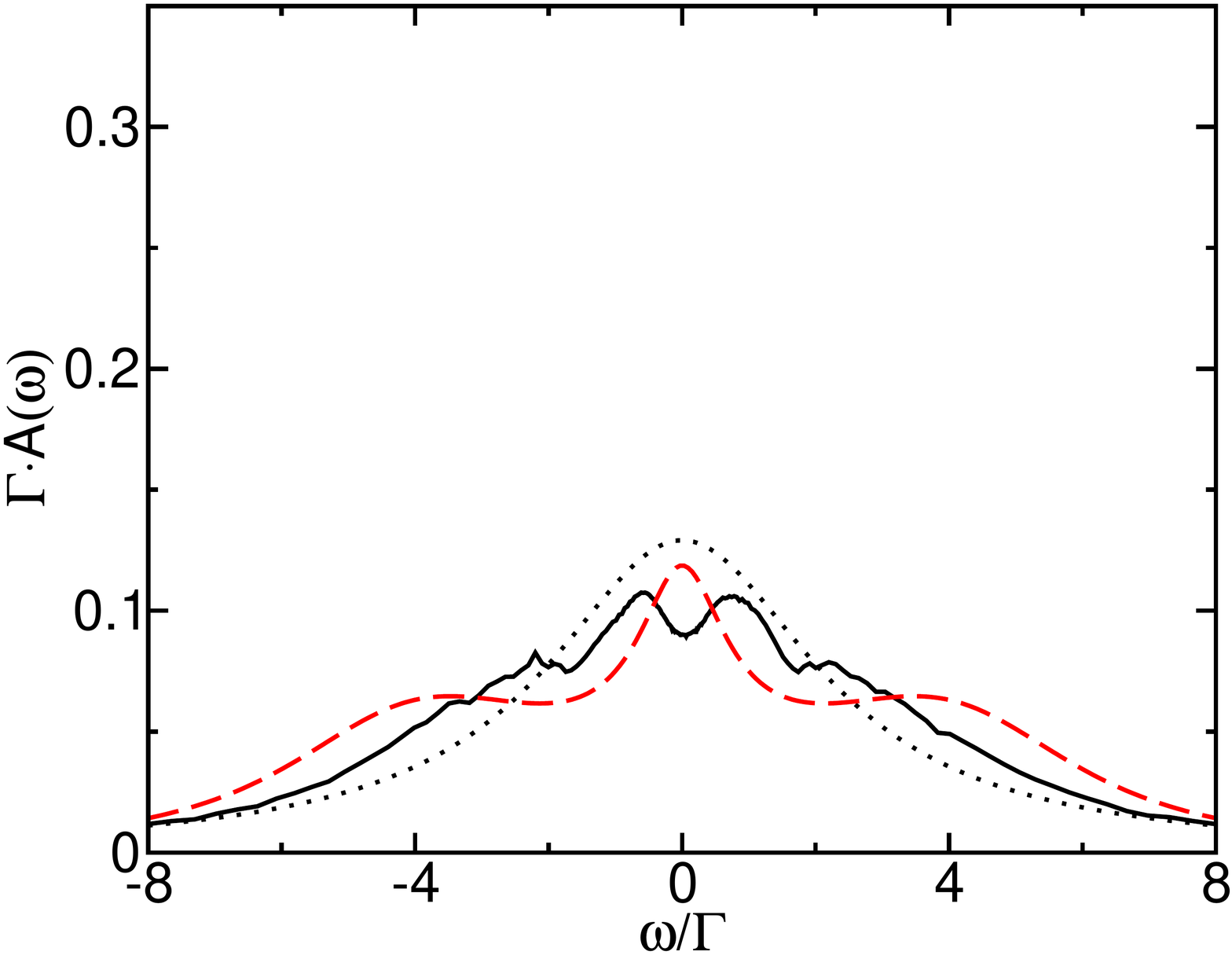}}
\caption{(color online) Nonequilibrium spectral functions for $U=8\Gamma$ at inverse temperature
$\beta=5\Gamma^{-1}$, as compared to zero-temperature second-order perturbation theory. 
Line legends are the same as in
Fig.~\ref{fig:AU2V1b5GenApproach}.}
\label{fig:intermediatecouplingspectraU8}
\end{figure*}

Figure \ref{fig:intermediatecouplingspectraU8} shows a similar set of curves
for interaction strength $U=8\Gamma$. As compared to the lower values of $U$,
a similar behaviour of the algorithm is observed. The perturbative prediction for the ASR is 
again essentially reproduced. The ASR is however again broadened as a function of the bias voltage
and appears to split eventually at $e\Phi=\Gamma$. Again, at the small bias
voltage $e\Phi=0.1\Gamma$, lack of Bayesian evidence causes an underestimation of the ASR.
The increased interaction again broadens the spectra as compared to
smaller values of $U$. However, the MaxEnt procedure does not clearly predict
the correct Hubbard peak positions at $\pm U/2$. A possible reason for this
is the partially rather slow decay behaviour of the kernel function which was
derived for the MaxEnt. It may result in decreased resolution in the high-frequency range, as
compared to a conventional MaxEnt procedure for the Wick rotation. 

\begin{figure*}[htb]
\subfloat[bias voltage
$e\Phi=0.5\Gamma$]{\includegraphics[width=0.49\linewidth]{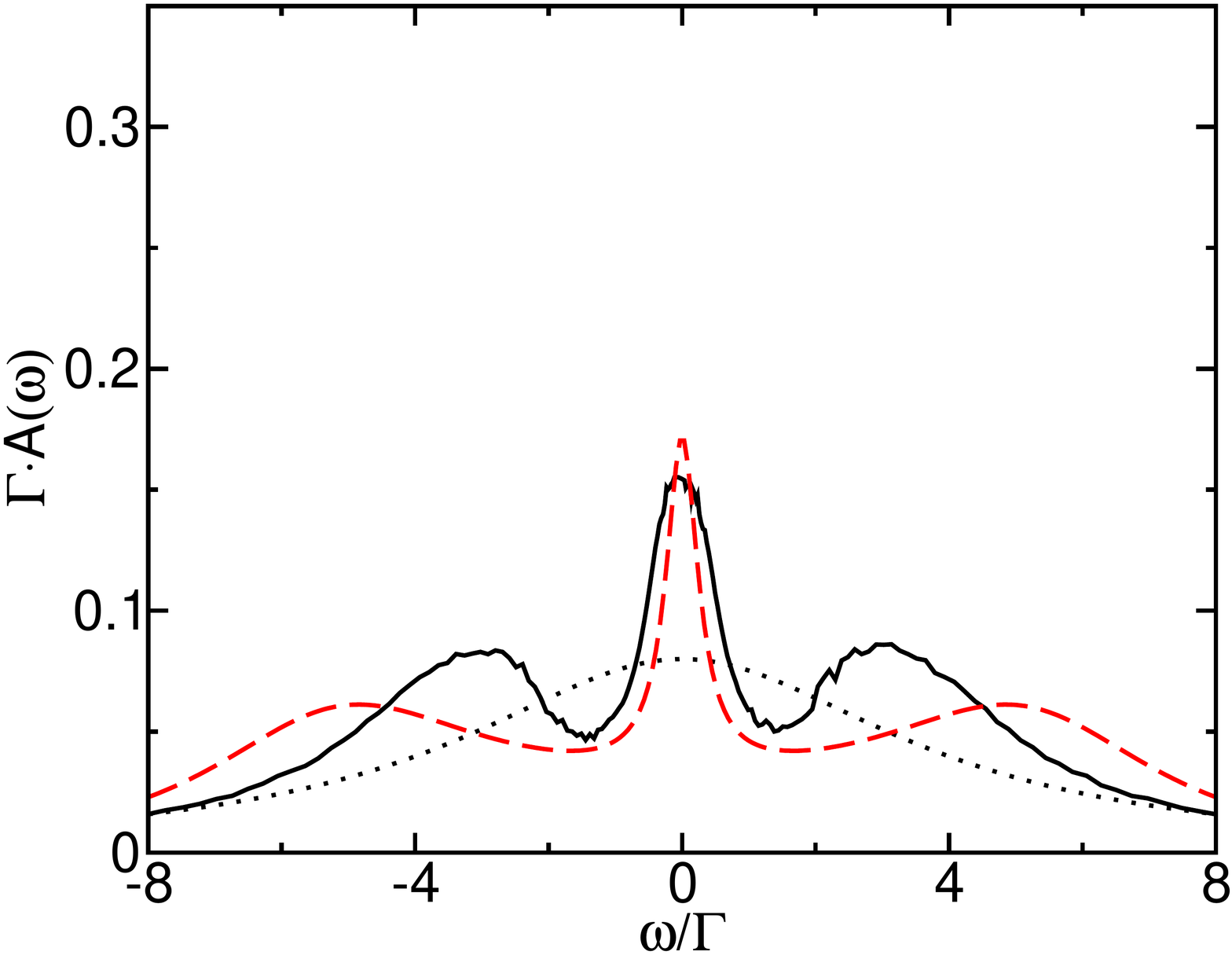}} 
\subfloat[bias voltage
$e\Phi=1.0\Gamma$]{\includegraphics[width=0.49\linewidth]{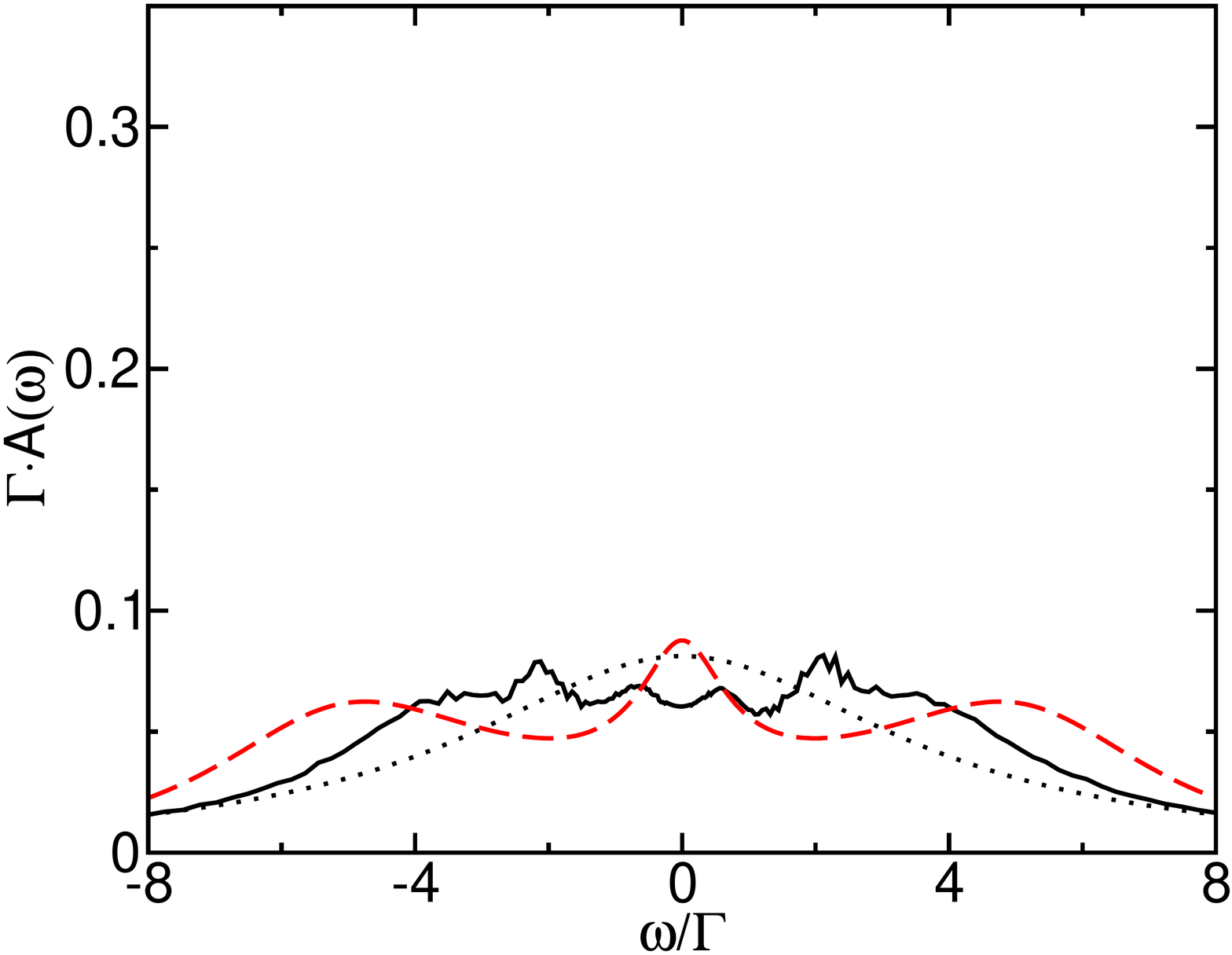}}
\caption{(color online) Nonequilibrium spectral functions for $U=10\Gamma$ at inverse temperature
$\beta=5\Gamma^{-1}$, as compared to zero-temperature second-order perturbation theory.  Line legends are the same as in
Fig.~\ref{fig:AU2V1b5GenApproach}.}
\label{fig:intermediatecouplingspectraU10}
\end{figure*}

Spectra for interaction strength $U=10\Gamma$ are displayed in figure
\ref{fig:intermediatecouplingspectraU10}.
Again, the solution is very similar to the perturbative prediction, and still
a splitting of the ASR at larger bias voltages is observed. The MaxEnt resolution issue for 
the Hubbard bands is again observed. 

\subsection{Approaching Lower Temperatures}
For data at lower temperature, namely $\beta = 10 \Gamma^{-1}$, the behaviour
of MaxEnt solutions is similar to the one described above. Nevertheless, 
sharper structures, such as the Kondo peak, which emerge at lower
temperatures, make the MaxEnt procedure more challenging. A common way to deal with 
this problem is the so-called ``annealing procedure''.\cite{jarrell, aryanpour} Here, a fine 
temperature grid is imposed in order to freeze out low-energy features step by step. The
procedure starts with a featureless default model at very high temperatures.
At lower temperature, the MaxEnt result of the next higher temperature is
used as default model, and so forth, until the target temperature is reached.

\begin{figure*}[htb]
\subfloat[bias voltage
$e\Phi=0.5\Gamma$]{\includegraphics[width=0.49\linewidth]{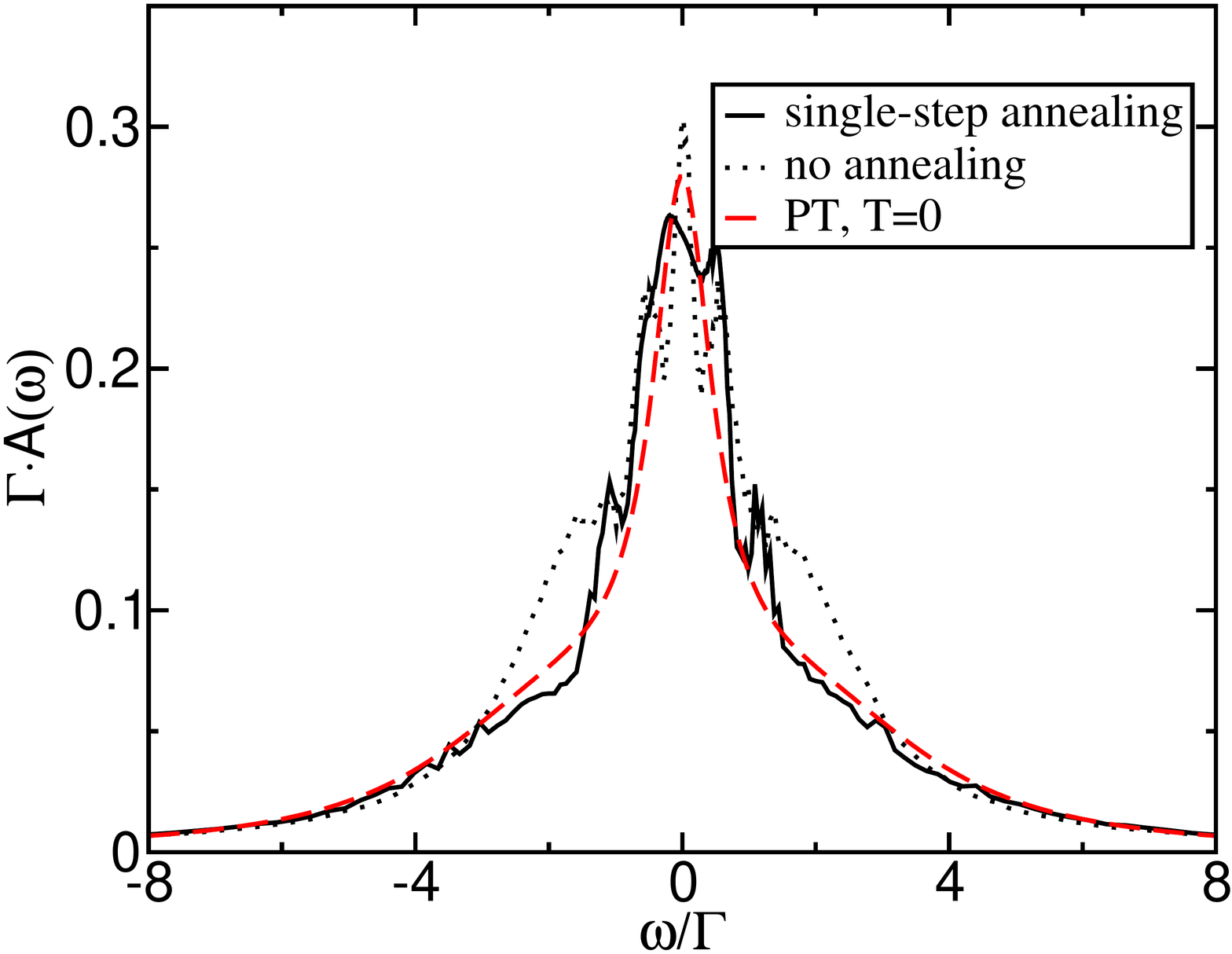}}
\subfloat[bias voltage
$e\Phi=1.0\Gamma$]{\includegraphics[width=0.49\linewidth]{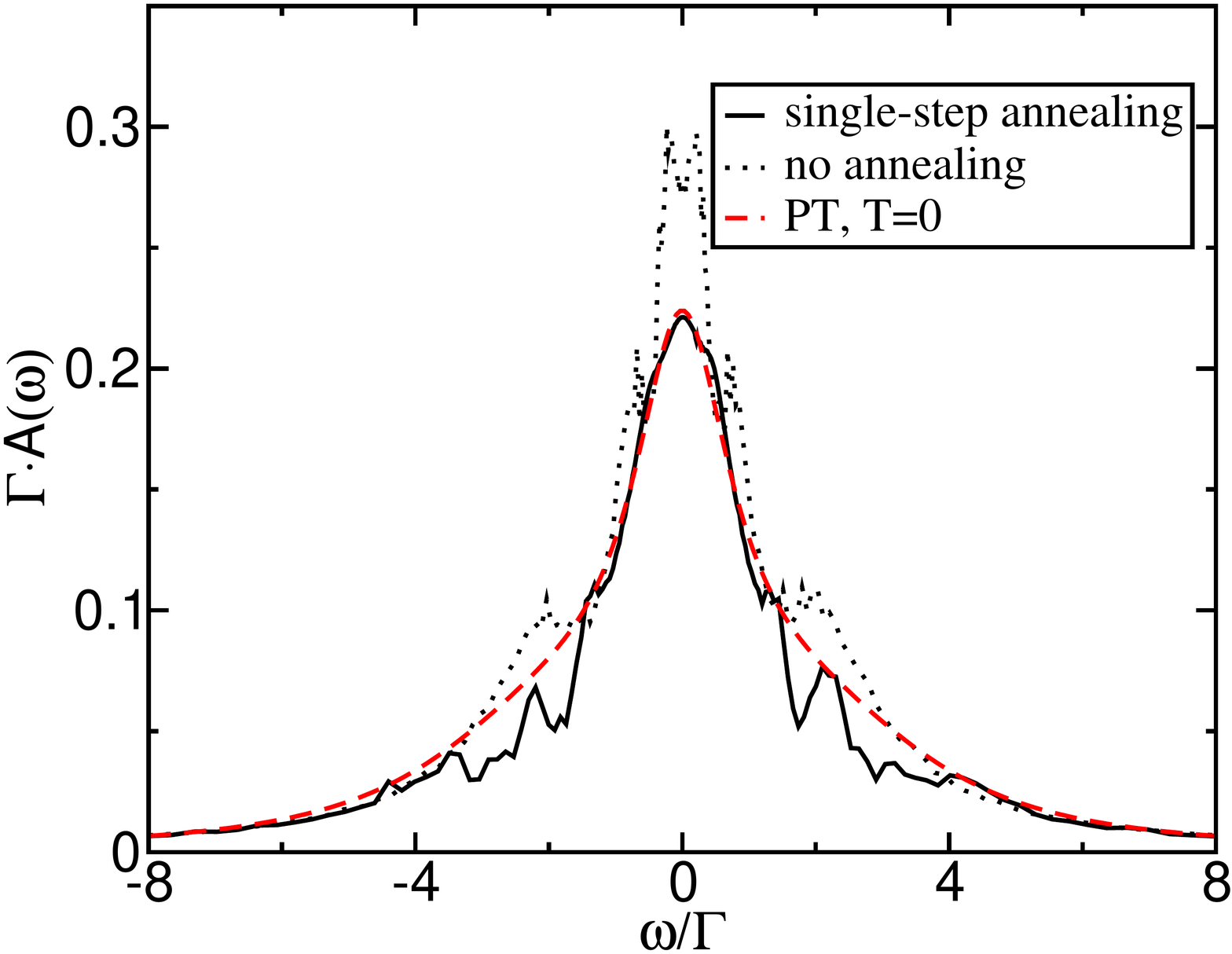}}
\caption{(color online) Lower-temperature spectral functions as inferred for $U=4\Gamma$, $\beta=10\Gamma^{-1}$ with or without an 
annealing step. The default model for the single-step procedure is taken from temperature $\beta=5\Gamma^{-1}$.}
\label{fig:annealingexample}
\end{figure*}

We found earlier \cite{dirks} that it is of great use
also within the two-dimensional analytic continuation problem. Also in the
present extension of the approach in Ref.~\onlinecite{dirks}, the MaxEnt yields more 
well-behaved solutions if a higher temperature is used as default model. This was investigated by a
simple single-step annealing procedure, using results from a
$\beta=5\Gamma^{-1}$ run. In fact, the occurance of unphysical
normalization-violating sidebands may already be avoided in some cases
for this rather rough temperature grid. 
Figure \ref{fig:annealingexample} shows two examples in which the one-step annealing procedure
was able to improve the results significantly. 

\subsection{Transport Properties}
\begin{figure}[htb]
\includegraphics[width=0.99\linewidth]{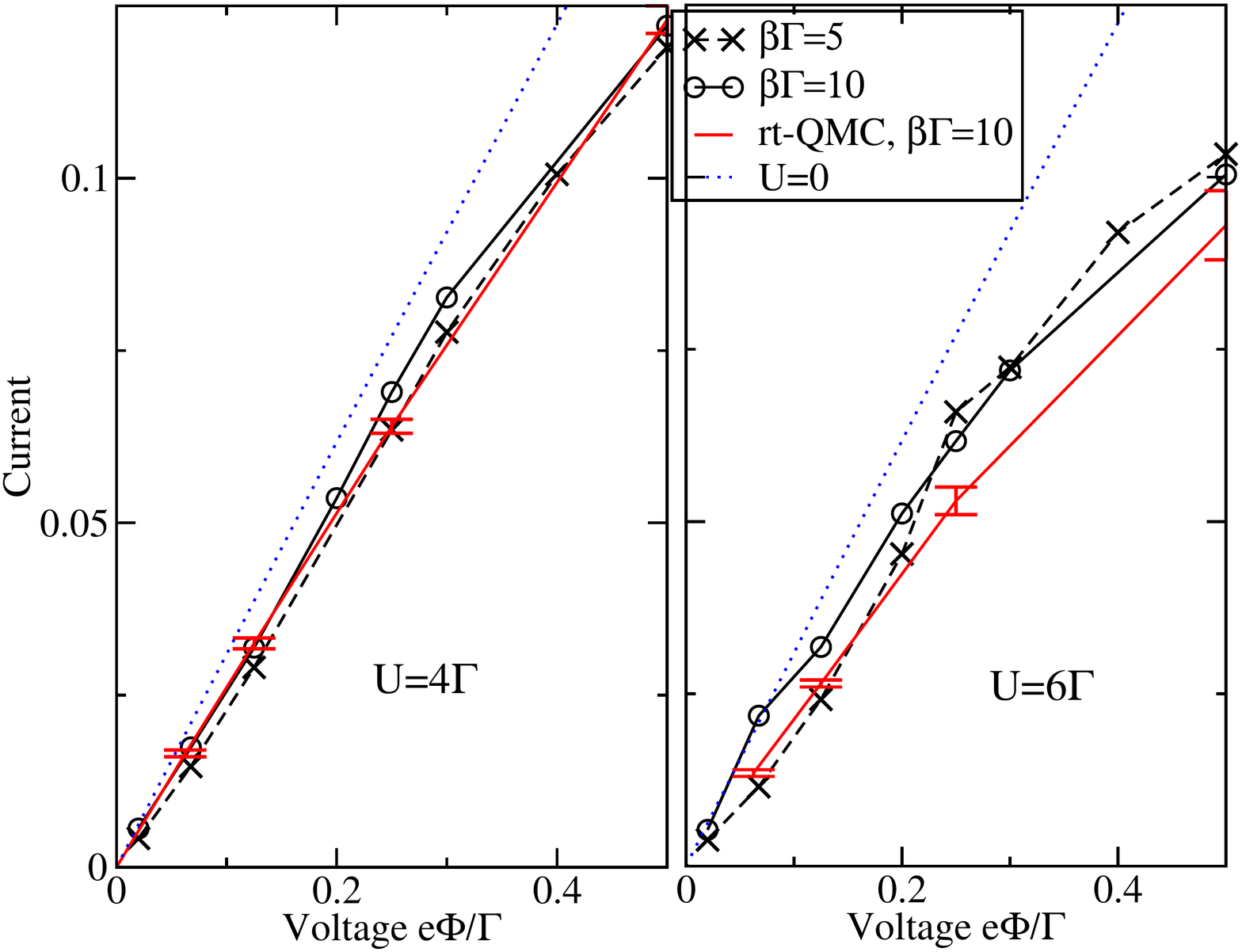}
\caption{(color online) Transport characteristics as compared to real-time QMC data from
Ref.~\onlinecite{werner10}.}
\label{fig:comparisonIVrtqmc}
\end{figure}

Using the Meir-Wingreen equation \eqref{eq:meir_wingreen}, we are able to compute transport 
properties based on spectral functions resulting from the MaxEnt analytic continuation
procedure.
Figure \ref{fig:comparisonIVrtqmc} compares results obtained at Coulomb interaction
strengths $U=4\Gamma$ and $U=6\Gamma$ to real-time quantum Monte-Carlo data from
Ref.~\onlinecite{werner10}.

The low-temperature data were obtained without the employment of an expensive
annealing procedure:
the most distinct feature of the low-temperature data is an increase in current at low-voltages 
for $\beta=10\Gamma^{-1}$ and $U=6\Gamma$. It is obtained no matter whether
the single-step annealing procedure described above is employed or not. 

As compared to the real-time QMC calculations, we obtain a good
agreement at interaction strength $U=4\Gamma$. 
Since the only significant deviation is at $e\Phi=0.25$, the discrepancy
may still be due to statistical errors within the Monte-Carlo procedures.
At higher interaction $U=6\Gamma$, the current predicted by our method appears to be 
systematically higher at voltages $e\Phi > 0.2\Gamma$, for both temperatures.

{
In the following, we will discuss the low-voltage region of both cases,
$U=4\Gamma$ and $U=6\Gamma$.
}
As shown in the previous section, in the case $\beta\Gamma=5$, the quasi-particle
weight is underestimated in the voltage range $e\Phi \leq 0.125\Gamma$ for the reason
that $\imag\varphi_m=0$ data cannot be taken into account for the analytic
continuation. This is compatible with the fact that the current is
underestimated as compared to the $\beta\Gamma=10$ data, as well as the
real-time QMC data.

\begin{figure}[htb]
\includegraphics[width=0.99\linewidth]{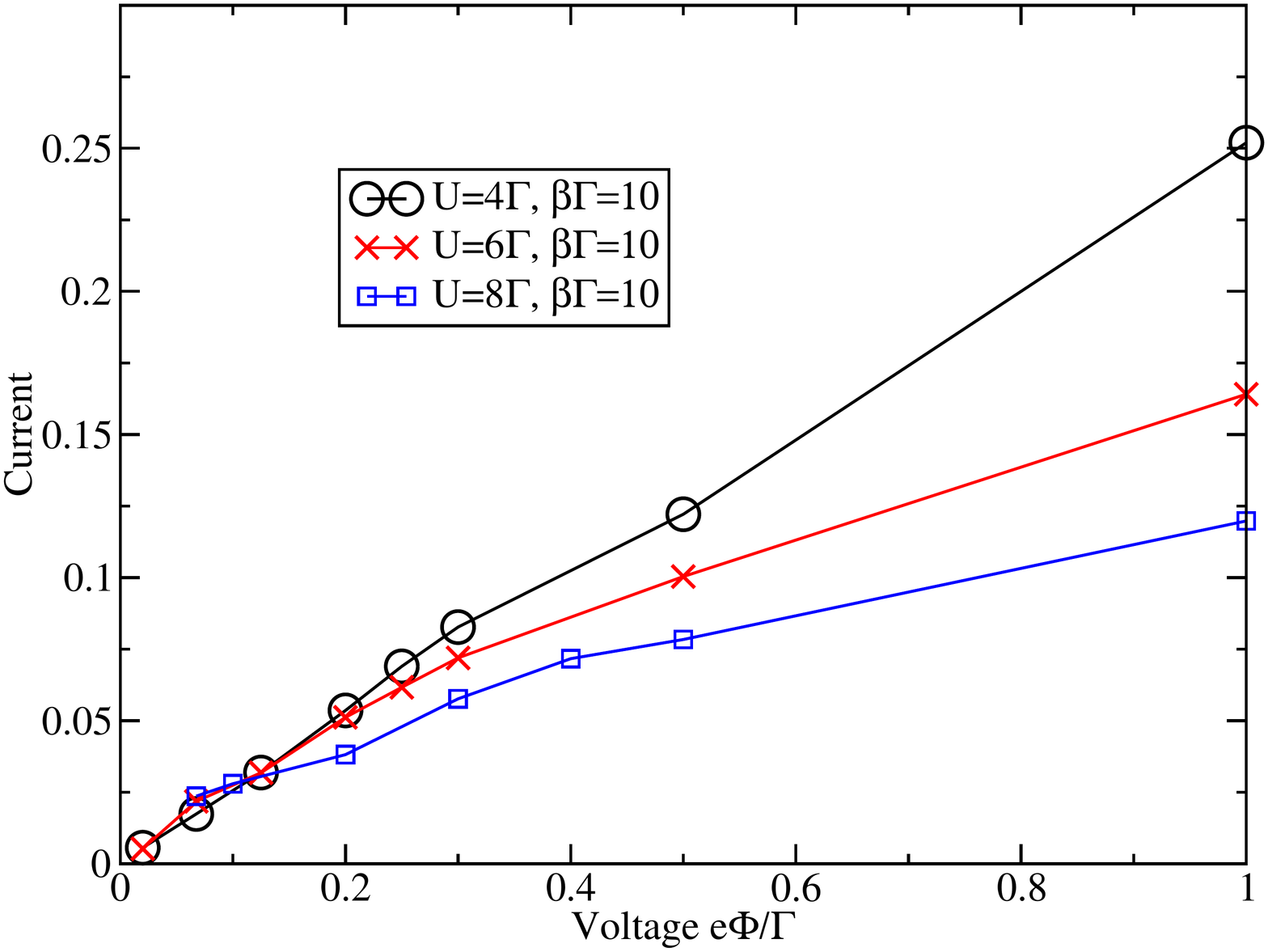}
\caption{(color online) Current as a function of the voltage for different interaction
strengths.}
\label{fig:currentsbeta10noanneal}
\end{figure}

Figure \ref{fig:currentsbeta10noanneal} shows current-voltage curves for
different values of $U$ at $\beta\Gamma=10$.
In any case, an S-shaped current-voltage characteristics is obtained, the
first increase of which is due to the ASR, and the second of which is due to
the Hubbard bands. The reduced weight of the ASR at increased interaction 
brings about an earlier departure of the $U=8\Gamma$ curve as compared to the
curves at $U=4\Gamma$ and $U=6\Gamma$. Same is true for the $U=6\Gamma$ curve
as compared to the case $U=4\Gamma$.

{
In the next section of the paper, mathematically involved details of the structure 
of the Green's function will be analyzed systematically. It comes along with
an introduction of the relevant concepts of the theory of several complex
variables. Those are in particular needed for the derivation of the MaxEnt
inverse problem, equation \eqref{eq:inverseProblemQ}, which leads to the results discussed in 
the present section.
The latter mathematical derivation is provided in section \ref{sec:introQ}.
It may be convenient to skip section \ref{sec:GFincontextofsevcomplexvars} on
first reading and directly go to the MaxEnt derivations and discussions of
section \ref{sec:introQ}.
}

\section{Several complex variables and the Green's function}
\label{sec:GFincontextofsevcomplexvars}

We want to provide a detailed physicist's introduction to the basic
mathematical structures of functions with several complex variables,
essential
for a full understanding of the Green's function within the Matsubara-voltage formalism. 
In particular,  we will attempt to answer the question by what means intuition
from conventional function theory is appropriate or misguiding in the
context of dynamic expectation values.

As next step we will describe the analytic structure of dynamic correlation functions 
$G(z_\varphi, z_\omega)$ appearing as a fundamental object in the Matsubara-voltage 
formalism (see I for  details) by means of the function theory for multiple complex variables. 
Uniqueness of the
mathematical procedure involved in the analytical continuations is proven and
the connection to maximum entropy approaches for the inference of spectral
functions is made.

\subsection{Holomorphy of Functions of Several Complex Variables}
In order to discuss the notion of holomorphy in the context of functions with several complex variables 
we will partially follow the book by Vladimirov.\cite{vladimirovbook}

A function $G$ of one complex variable is \emph{holomorphic} at a point $z_0$, if and 
only if the Cauchy-Riemann differential equation
\begin{equation}
\left.\frac{\partial G}{\partial \bar z_\omega}\right|_{z_0} = 0
\label{eq:CRDEQ}
\end{equation}
is satisfied.
The notion for holomorphy of functions of several complex variables is a
natural extension of this definition: 
A function $f(\underline z)$, with $\underline z \in \mathbb{C}^d$, is holomorphic
with respect to $\underline z$ at the point $\underline z^{(0)}$
if and only if it is holomorphic with respect to each individual variable,
\begin{equation}
\left.\frac{\partial f}{\partial
\bar z_i}\right|_{\underline z^{(0)}} = 0,\quad i = 1, \dots, d.
\label{eq:CRDGLmanyvars}
\end{equation}
{
Note that in the following, we will always denote vectors in
$\mathbb{C}^d (d>1)$
by an underlined symbol such as $\underline z$.
\emph{Hartogs' fundamental theorem} asserts 
that definition \eqref{eq:CRDGLmanyvars} is also equivalent to the Weierstra\ss~definition of holomorphy for several variables.
}
The latter calls a function holomorphic at $\underline z_0$ if and only if
there exists an open neighbourhood $M$ of $\underline z^{(0)}$, such that for all  
$\underline z \in M$ the function $f$ may be written as an absolutely convergent power 
series $f(\underline z) = \sum_{|\alpha|=0}^\infty a_\alpha (\underline z - \underline
z^{(0)})^\alpha$.
$\alpha$ denotes the multi-index for the monomial 
$\underline z^\alpha := \prod_{n=0}^d z_n^{\alpha_n}$, and 
$|\alpha| :=\sum_{n=0}^d \alpha_n$. An analytic complex function of
several variables is holomorphic.

The notion of holomorphy implied by Eqs.~\eqref{eq:CRDGLmanyvars} is thus as
natural and intuitive as in the one-dimensional case.

\subsection{Domains of Holomorphy and \\ Biholomorphic Transformations}
The major qualitative difference between single- and multi-variable complex analysis 
is contained in  the notion of a domain, and the geometric
{equivalence}
 among holomorphic functions arising from classes of domains. This has far-reaching
consequences to the theory itself, such as the construction of integral representations. 
In the context of our formalism we aim at integral representations. 
{
We will
thus first comment on the basic structures which integral
representations operate on. Furthermore, we will point out the most prominent
differences to conventional function theory.
}
The notion of a domain in single-variable complex analysis is replaced by the
notion of a \emph{domain of holomorphy} in multi-variable complex analysis -- the
notion of a conformal map is replaced by the notion of a \emph{biholomorphic
map}.
We will address the most prominent differences by first reminding the reader of basic
structural properties of one-dimensional complex analysis and 
then introducing the corresponding terminologies of the more-dimensional theory.

\subsubsection{One-dimensional function theory}

In one-dimensional complex analysis, \emph{domains} are defined as open 
connected subsets of $\mathbb{C}$. For the time being, we will restrict the discussion to 
simply connected open sets, i.e.~open connected sets with
no holes.

Conformal maps between domains $U$, $V$, namely functions
\begin{equation}
m : U \to V, \quad z \mapsto m(z)
\end{equation}
which are holomorphic and invertible (one-to-one),
provide links between certain classes of domains. 
The \emph{Riemann mapping theorem} states that conformal maps between any 
simply connected $U \neq \mathbb{C}, \emptyset$
and the unit disk exists. I.e.~all simply connected domains $\neq \mathbb{C},\emptyset$ are conformally
equivalent:
their structures of holomorphic functions map one-to-one to each other and
are conformally diffeomorphic.
Generalizing the concept of holomorphy to Riemannian surfaces,
conformal maps exist only for surfaces of the same topological genus. 
The \emph{uniformization theorem} finds that
for
simply connected Riemann surfaces (topological genus 0) up to conformal equivalence three 
classes exist:
\begin{itemize} 
\item the unit disk $D_1 = \{z\in\mathbb{C} :|z| < 1\}$, 
\item the complex plane $\mathbb{C}$, 
\item the
Riemann sphere $\mathbb{C}\cup \{\infty\}$.
\end{itemize}
{
These three surfaces form the so-called
moduli space of genus 0, defined as the space of conformally inequivalent Riemann
surfaces of genus 0.
}

In general, the size of the moduli space of a Riemann surface of genus $g$
grows as a function of $g$. Each modulus represents an equivalence class of holomorphic
functions.

Conformal equivalence plays an important role in physical applications
such as two-dimensional potential flows around airfoils or conformal quantum
field theory.

One natural consequence of the conformal equivalence of all non-empty simply connected domains
$U\subsetneq \mathbb{C}^1$ is that there always exists a function which is not
analytically continuable beyond the domain: the function 
\begin{equation}
f_0(z) = \sum_{\alpha = 0}^\infty z^{\alpha!}
\end{equation}
is holomorphic on the unit disk but may not be analytically continued
to larger domains. \cite{vladimirovbook}
Using a conformal map $m$ from the unit disk to $U$, which exists due to the Riemann
mapping theorem, one finds the function $f_0 \circ m^{-1}$ which cannot be
analytically continued beyond $U$.
One also calls the unit disk the \emph{domain of holomorphy of $f_0$},
i.e.~the largest domain for which $f_0$ is holomorphic. $U$ is the domain
of holomorphy of $f_0\circ m^{-1}$. For a given domain $G$, \emph{if}
there exists any function $f$ for which $G$ is the largest possible 
domain in which $f$ is holomorphic,
the domain is called a \emph{domain of holomorphy}.

In general, any domain is also a domain of holomorphy in conventional
complex analysis. Due to the simple structures arising from these
far-reaching
equivalences, the conventional function theory of one complex variable is 
widely considered to be a finalized field of mathematical research.

\subsubsection{Multi-dimensional function theory}
The two final statements of the last section are completely incorrect for several complex
variables $\mathbb{C}^d$, $d \geq 2$.

In several complex variables,
\begin{itemize}
\item
a domain is \emph{not necessarily} a domain of holomorphy;
\item
domains of holomorphy are \emph{usually not} biholomorphically equivalent.
\end{itemize}
{
An example of a domain which is no domain of holomorphy is the hollow sphere
(see paragraph b below), because \emph{all} analytic functions in the hollow
sphere can be analytically continued to the sphere.
}
As compared to one-dimensional function theory, the emerging structures are 
thus very rich. 
{
Depending on the domain geometry, very different \emph{sheaves of holomorphic
functions}\footnote{Sheaves are very general mathematical concepts, but here one may just
imagine the set of holomorphic functions $U\to \mathbb{C}$ carried by a given
domain $U\subset \mathbb{C}^d$.} will arise.
}

\paragraph{Biholomorphic maps.}
The tool of a biholomorphic map, as mentioned in the second point, is the
generalization of a conformal map to several complex variables. If a
holomorphic mapping $m : U \to U'$, with $U,U'\subset \mathbb{C}^d$ is
invertible,  it is called a \emph{biholomorphic map}.
Two domains $U$, $U'$ are \emph{biholomorphically equivalent}, if and only if such a
map exists. They will have an equivalent sheaf of holomorphic
functions. 
Biholomorphic maps do not necessarily preserve angles. Therefore, they are
usually not conformal. Nevertheless, with respect to the holomorphic
structure, they are the natural generalization of conformal maps on
$\mathbb{C}$, because 
they are complex 
diffeomorphisms: the inverse of a biholomorphic map is also holomorphic.\cite{vladimirovbook}

\paragraph{Domains of holomorphy and holomorphic envelopes.}
A striking example of a domain $\subset \mathbb{C}^2$ which is not a domain of
holomorphy is given by the hollow sphere
$M:= \{\underline z = (z_1, z_2)^T \in \mathbb{C}^2 : \frac{1}{2} < \sqrt{|z_1|^2 + |z_2|^2} < 1\}$. In sharp
contrast to the single-variable case, one can show that
\emph{any} holomorphic function $f:M\to\mathbb{C}$ may be analytically continued to
the unit sphere. The unit sphere is, in fact, a domain of holomorphy and is
thus named the \emph{holomorphic envelope} of $M$.\cite{vladimirovbook} 
This extends to the famous result by Friedrich Hartogs that isolated
singularities are always removable for analytic functions $\mathbb{C}^d\to\mathbb{C}$,
$d\geq 2$ (\emph{Hartogs' lemma}). While isolated singularities play an essential role in the residue
calculus in the $d=1$ case, the $d\geq 2$ case is, due to Hartogs' result,
of an entirely different nature. As we will see in section \ref{sec:intreps}
the theory of integral representations of complex functions for $d\geq 2$ has
consequently a very different character as compared to the $d=1$ case.

Since ordinary domains such as the hollow sphere are rather
friendly as far as
analytic continuation is concerned, mathematicians restrict to the systematic study of 
corresponding envelopes of holomorphy, i.e.~the \emph{domains of holomorphy}.
One can show that a domain is a domain of holomorphy if and only if it is a
so-called \emph{pseudoconvex} domain. Pseudoconvexity is a certain generalization of
convexity from $\mathbb{R}^d$ to $\mathbb{C}^d$.\cite{vladimirovbook}
{
For this reason, the fundamental domains of holomorphy in our application, the
wedges, are indeed convex (cf.~section \ref{subsec:holostructGF}).
}

\paragraph{Biholomorphic equivalence.}
In order to provide a classification of domains of holomorphy in
$\mathbb{C}^d$, $d\geq 2$, the fruitful strategy from $d=1$, namely using
biholomorphic equivalences, is adopted.
Non-empty simply connected domains of holomorphy $\subsetneq \mathbb{C}^d$
are usually not biholomorphically equivalent. There are many different types
of holomorphic structures, depending on the domain geometry.

A prominent example of biholomorphically inequivalent domains of 
holomorphy is given by the unit ball $|\underline z| < 1$
 and the bicylinder $D_1\times D_1$ in $\mathbb{C}^2$.
The unit ball in higher dimensions is with regard to the holomorphic
structure in no way related to the unit disks and carries in fact a rather
singular holomorphic structure.

\label{sec:bihomequivalence}

\subsection{Integral Representations}
\label{sec:intreps}
As one might have already guessed from Hartogs' lemma and the domain
dependence of the mathematical structures, finding an
{analogy} to Cauchy-like integral 
representations, yielding, for example, the spectral representation of our
two-variable
Green's function, turned out to be a cumbersome task. 

At present, one knows several, more or less general, possibilities to construct such representations, 
even with different dimensions of the integration manifolds.
The Bochner-Martinelli representation \cite{vladimirovbook} is, for example, probably the most general 
integral representation, but the integration manifold is $2d-1$ dimensional.

For our practical purposes, a
minimal integration space dimension is, of course, most desirable to
reduce the number of fitting variables 
when reconstructing the function by a Maximum-Entropy-like Bayesian inference
technique.

\subsubsection{The Shilov Boundary}
\label{subsubsec:Shilov}

In $d=1$ complex analysis, Cauchy's integral representation reconstructs all
values of a simply connected open domain of finite radius using the limit values on the 
topological boundary, \emph{as long as the boundary values are
continuous}. I.e.~the structure on the topological boundary
determines the structure of the interior.

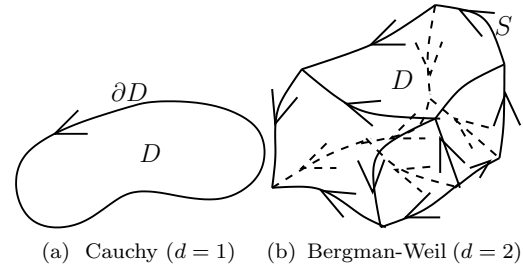
\begin{figure}
\subfloat[~Cauchy ($d=1$)]{\resizebox{0.45\linewidth}{!}{
\begin{picture}(0,0)%
\includegraphics{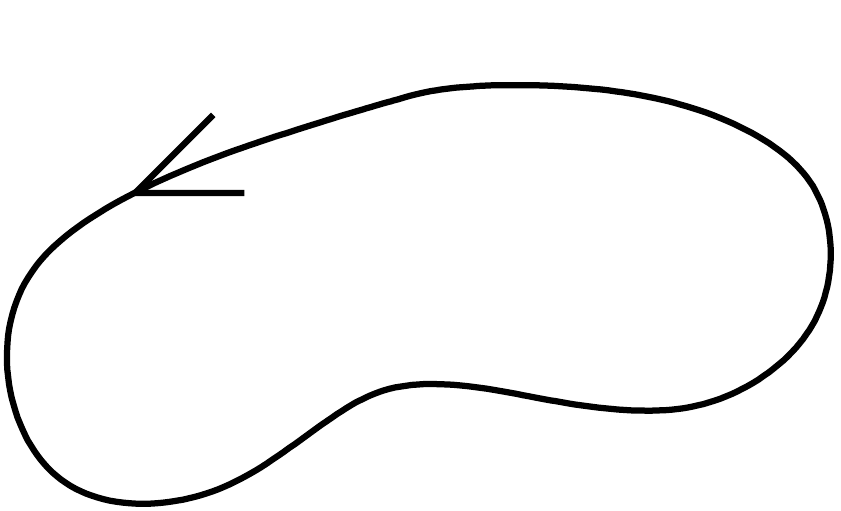}%
\end{picture}%
\setlength{\unitlength}{3947sp}%
\begingroup\makeatletter\ifx\SetFigFont\undefined%
\gdef\SetFigFont#1#2#3#4#5{%
  \reset@font\fontsize{#1}{#2pt}%
  \fontfamily{#3}\fontseries{#4}\fontshape{#5}%
  \selectfont}%
\fi\endgroup%
\begin{picture}(4020,2448)(2803,-2935)
\put(4351,-886){\makebox(0,0)[lb]{\smash{{\SetFigFont{29}{34.8}{\familydefault}{\mddefault}{\updefault}{\color[rgb]{0,0,0}$\partial D$}%
}}}}
\put(4801,-1861){\makebox(0,0)[lb]{\smash{{\SetFigFont{29}{34.8}{\familydefault}{\mddefault}{\updefault}{\color[rgb]{0,0,0}$D$}%
}}}}
\end{picture}%
 }}
\subfloat[~Bergman-Weil
($d=2$)]{\resizebox{0.45\linewidth}{!}{
\begin{picture}(0,0)%
\includegraphics{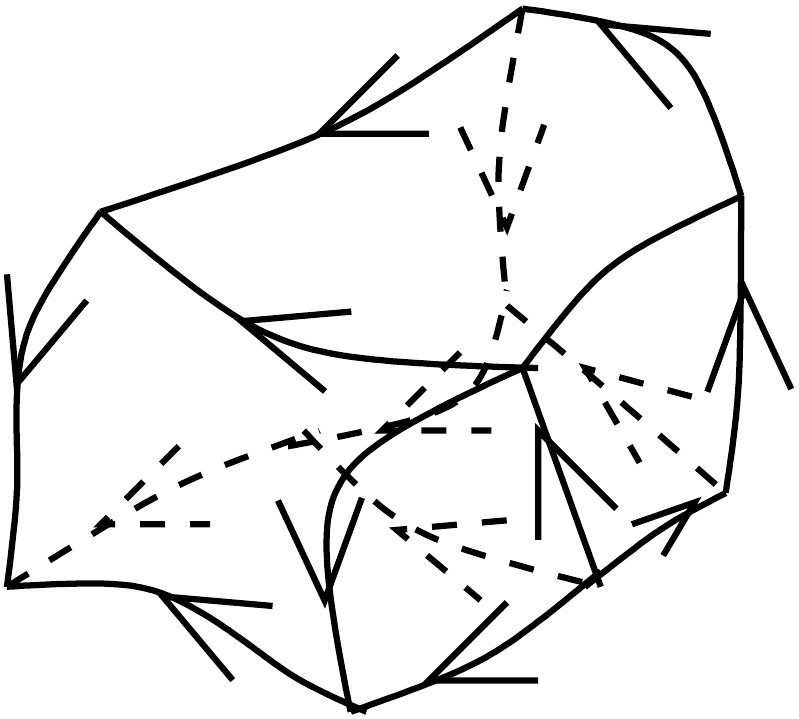}%
\end{picture}%
\setlength{\unitlength}{3947sp}%
\begingroup\makeatletter\ifx\SetFigFont\undefined%
\gdef\SetFigFont#1#2#3#4#5{%
  \reset@font\fontsize{#1}{#2pt}%
  \fontfamily{#3}\fontseries{#4}\fontshape{#5}%
  \selectfont}%
\fi\endgroup%
\begin{picture}(3831,3441)(3718,-7744)
\put(7126,-4786){\makebox(0,0)[lb]{\smash{{\SetFigFont{29}{34.8}{\familydefault}{\mddefault}{\updefault}{\color[rgb]{0,0,0}$S$}%
}}}}
\put(5551,-5611){\makebox(0,0)[lb]{\smash{{\SetFigFont{29}{34.8}{\familydefault}{\mddefault}{\updefault}{\color[rgb]{0,0,0}$D$}%
}}}}
\end{picture}%
}\label{subfig:bergman}}
\caption{Comparison of Cauchy and Bergman-Weil integral representation theories.
Integrations run over the (a) full topological boundary $\partial D$ and (b) 
the Bergman-Shilov boundary $S\subset \partial D$ of an analytic polyhedron, respectively.
}
\label{fig:BergmanvsCauchy}
\end{figure}

For $d>1$, the topological boundary can in this sense be reduced to an often
much smaller set, the 
so-called \emph{distinguished} or \emph{Bergman-Shilov
boundary}.\cite{EOM}
In the 1930s, Stefan Bergman discovered the distinguished boundary in the context of
$\mathbb{C}^2$ for bounded domains with piecewise smooth boundaries. He found that
under certain regularity conditions, an integral representation with respect
to an only two-dimensional manifold $S$ of all intersections of the smooth boundaries was possible, using the
so-called Bergman kernel function. A precise geometric and analytic description of these 
results is provided in great detail in chapter XI of Ref.~\onlinecite{bergmanbook}.
A pictorial comparison of the resulting \emph{Bergman-Weil representation} to the
Cauchy formula is provided in figure \ref{fig:BergmanvsCauchy}.
These concepts where independently discovered by Shilov in the 1940s
in the rather different context of commutative Banach algebras.
\cite{EOM}

\paragraph{Generalization.}
In the modern terminology, the Shilov boundary, as a generalization of
the Bergman-Shilov boundary, may be defined for any
compact space \emph{with respect to} an algebra of continuous complex-valued
functions on the space.\cite{EOM}
If we for example find physical constraints imposing certain conditions to
the set of holomorphic functions, the Shilov boundary with respect to these
functions may be reduced to a smaller set. 
If a sufficiently elaborate kind of integral representation is used, this may
enable us to again reduce the number of linear fit parameters significantly
in the Bayesian inference problem.

\paragraph{Examples.} For the domains with piecewise smooth boundaries which
Bergman investigated, $S$ is given by the unification of all possible
intersections between the smooth boundary hypersurfaces, as long as certain
regularity conditions hold. 
We return to our two examples from part \ref{sec:bihomequivalence}, the \emph{bicylinder} and the \emph{unit
sphere} and comment on their Bergman-Shilov boundaries.

The bicylinder,  $D_1 \times D_1$, is one of the most easily
accessible domains of holomorphy, because it simply factorizes into two
$D_1$-disks in $\mathbb{C}$. A minimal integral 
representation of a holomorphic function which is continuous on the closure
$\overline{D_1\times D_1}$, 
is simply given by the product of two conventional Cauchy representations
(see Theorem 2.2.1 in Ref.~\onlinecite{hoermanderbook}), 
\begin{equation}
f(\underline z) = \frac{1}{(2\pi \imag)^2} 
\iint_{S_1\times S_1}\-\-\mathrm{d}^2 \zeta \,
\frac{f(\underline \zeta)}{\prod_{k}(z_k-\zeta_k)}.
\label{eq:CauchyRepBicyl}
\end{equation}
Therefore, the Bergman-Shilov boundary of $D_1 \times D_1$ is given by
the only two-dimensional toroidal subset $S_1\times S_1 = \partial D_1\times \partial
D_1$ of the three-dimensional
topological boundary $\partial(D_1\times D_1) = \partial D_1\times D_1 \cup D_1 \times \partial D_1$.
Similarly, integral representations of domains which are direct products of $\mathbb{C}^1$ 
domains can be constructed from the conventional Cauchy integral formula.

It was already mentioned in part \ref{sec:bihomequivalence} that the 
\emph{unit sphere} in $\mathbb{C}^2$ is not
biholomorphically equivalent to the bicylinder. In contrast to the two 
connected smooth boundary hypersurfaces of the bicylinder, 
the boundary hypersurface of the unit sphere
is not even a smooth hypersurface in Bergman's notion. 
One can show that here the Bergman-Shilov boundary is, in fact, identical to the
topological boundary. 
{Thus, any integral representation for holomorphic
functions in domains such as the unit sphere must invoke at least
\emph{three} real integrals. It is an example of the strong distinguishments
which have to be made between certain classes of domains of holomorphy.
}

\subsection{Holomorphic Structure of the Green's Functions}

\label{subsec:holostructGF}
As a next step, we will systematically analyze the mathematical structure of
the Green's function arising in the Matsubara-voltage approach.
The bare Green's function with respect to the two variables $\imag\varphi_m$,
$\imag \omega_n$ reads \cite{prl07}
\begin{eqnarray}
\label{eq:G0}
&&G_0(\imag\varphi_m, \imag\omega_n) =  \\
&&\sum_{\alpha=\pm1}
\frac{\Gamma_\alpha/\Gamma}
{
\imag\omega_n -\alpha (\imag \varphi_m - \Phi) / 2 - \varepsilon_d + 
\imag \Gamma \mathrm{sgn}_{nm}
}. \nonumber
\end{eqnarray}
Here, $\mathrm{sgn}_{nm} := \sign (\omega_n -\alpha \varphi_m/2)$.
Performing the analytic continuations $\imag\omega_n \to z_\omega$, 
$\imag \varphi_m \to z_\varphi$, the sign function in the denominator 
results in an ambiguity, as far as the definition of domains, for which $G_0$
is holomorphic, is concerned.

\begin{figure*}
\subfloat[branch cut structure of $G_0(z_\varphi, z_\omega)$]{
\includegraphics[width=.47\textwidth]{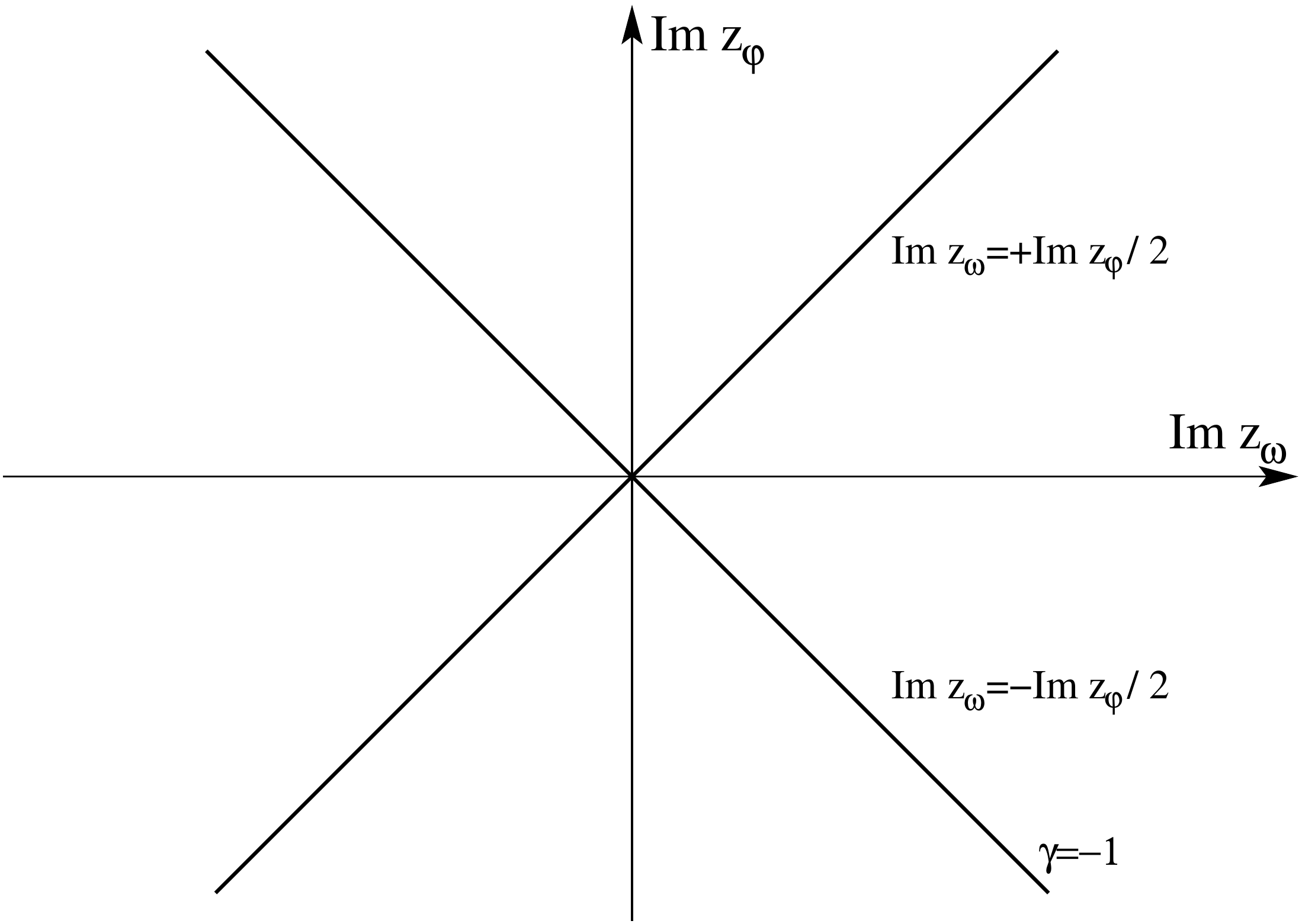}
\label{fig:branchcuts0}
}
\subfloat[branch cut structure of $G(z_\varphi, z_\omega)$]{
\includegraphics[width=.47\textwidth]{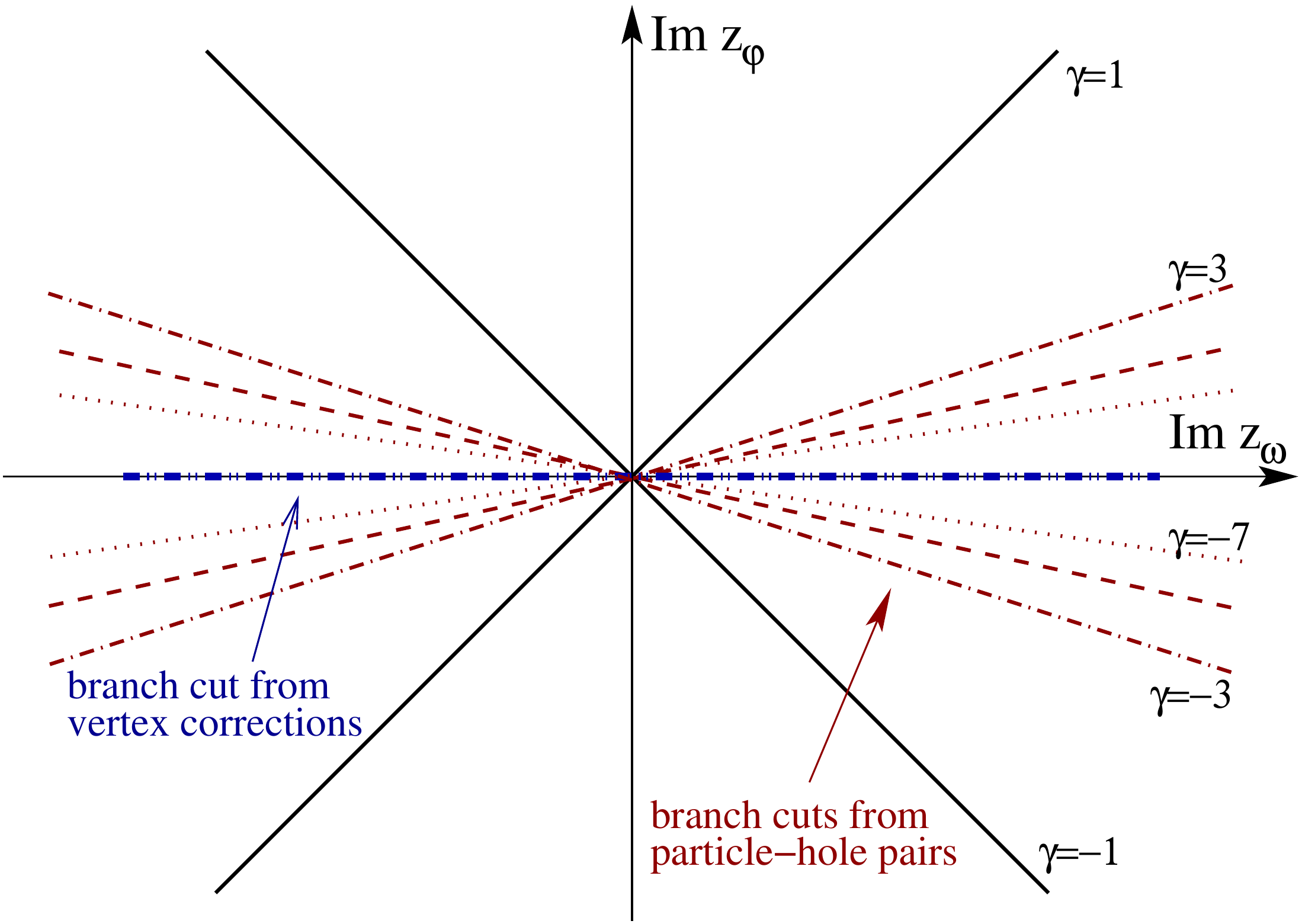}
\label{fig:branchcuts}
}
\caption{(color online) {
Branch cut structure of the Green's
function. The fully interacting structure is obtained from the perturbative expansion in $U/\Gamma$.}}
\end{figure*}

Choosing a branch cut structure which corresponds to the continuation
\begin{equation}
\sign (\omega_n \pm \varphi_m /2) \to 
\sign (\Im z_\omega \pm \Im z_\varphi / 2)
\end{equation}
appears to be most sensible from both, a mathematic and a physical point of view: 

From the former perspective, in contrast to other choices the resulting domains are 
also \emph{domains of holomorphy} and are thus ``maximal'' with respect to the
holomorphic structure. The four domains of holomorphy are given by
$\mathbb{C}^2$ separated into wedges by the
two branch-cut hyperplanes $\Im z_\omega \pm \Im z_\varphi / 2 = 0$ (see
Fig.~\ref{fig:branchcuts0}).
From the latter, it is just 
the imaginary part of the linear combinations of
$z_\omega$ and $z_\varphi$, appearing in the denominators of perturbative
expressions in $U$,
which yields the crucial sign-switching delta functions generating the
non-analytic terms separating the domains for which $G$ is holomorphic.
See, e.g., appendix B of paper I.

Consequently, for a finite interaction $U\neq 0$, the appearance of a branch cut must be expected for any
new kind of linear combination of $z_\omega$ and $z_\varphi$ in the denominators
of the integrands in the perturbation expansion.  For example, second-order perturbation theory from Ref.~\onlinecite{prl07} indicated
that particle-hole bubbles create higher-order branch cuts
$\Im z_\omega - \frac{\gamma}{2} \Im z_\varphi = 0$, with $\gamma\in\mathbb{Z}$ odd, due to the
structure of convolutions which are involved. In a later publication\cite{han10} it was 
pointed out that vertex corrections seem to introduce yet another branch cut for $\Im z_\varphi=0$. 
The physical retarded Green's function is then given by 
\begin{widetext}
\begin{equation}
G_\text{ret}(\omega) =
\lim_{z_\omega \to \omega + \imag 0^+}
\frac{\lim_{\delta_\varphi\to 0^+} + 
      \lim_{\delta_\varphi\to 0^-} }
     {2}
G(\Phi + \imag \delta_\varphi, z_\omega),
\label{eq:DefPhysicalLimit}
\end{equation}
\end{widetext}
i.e.~one has to average over the two possible limits with respect to
$\delta_\varphi$.

Note that the latter subtlety was not taken into account in the direct
continuation using the cone $C_\varepsilon$ in Ref.~\onlinecite{dirks}, and also
not in the initial approach in Ref.~\onlinecite{prl07}.

\subsubsection{Resulting Mathematical Assumptions}

The following assumptions are being made for the mathematical structure of
the Green's function:

\begin{enumerate}
\item
By means of holomorphy, we obtain cone-like constraints for the combinations of imaginary parts,
as depicted in Fig.~\ref{fig:branchcuts}. More precisely, we require $G$ to
be a solution of the Cauchy-Riemann equations \eqref{eq:CRDGLmanyvars} for
any $\underline{z}^{(0)}$ \emph{except
for those $\underline{z}^{(0)}$ for which}
\begin{equation}
\begin{split}
&\Im z^{(0)}_\omega = \frac{\gamma}{2}\Im z^{(0)}_\varphi 
\,\,\text{or}\\ 
\,\,&
\Im
z_\varphi^{(0)}= 0,
\end{split}
\label{eq:AssumptionI}
\end{equation}
with some $\gamma \in 2\mathbb{Z}+1$. Those $\underline{z}^{(0)}$ define the
branch cut hyperplanes and delimit the wedges for which $G$ is holomorphic.
\item
We will require the interacting Green's function $G(\underline
z)$ to be bounded, i.e.
\begin{equation}
\sup_{\underline z \in \mathbb{C}^2} |G(\underline z)| < \infty. 
\label{eq:AssumptionII}
\end{equation}
\item
We will assume that the Green's function $G(\underline z)$ is \emph{uniquely} defined by the
discrete function values $G(\imag\varphi_m, \imag \omega_n)$ which are obtained from
the effective-equilibrium computations. I.e.~we require that the continuation
to a multisheeted holomorphic function
\begin{equation}
G(\imag \varphi_m,\imag\omega_n) \to 
G(z_\varphi,z_\omega)
\label{eq:AssumptionIII}
\end{equation}
is unique.
\end{enumerate}
The second assumption is justified by the structure of the convolution equations in
perturbation theory and the boundary conditions that terms $\euler{\imag
\varphi_m \beta/2}$ and $\euler{\imag \omega_n \beta}$, evaluate to 1
and -1 before the analytic continuations are carried out.

{
A proof of the third statement, which is of course crucial for the physical
theory itself, will be provided in appendix \ref{apx:uniqueness}.
}
It is based on assumption 1 and 2 and the assumption 3' which sharpens the
requirements on the $\underline z \to \infty$-asymptotics:

\begin{itemize}
\item[3'.] Given arbitrary $\underline{x}^{(0)}\in
\mathbb{R}^2\setminus\{0\}$ and $\zeta\in \mathbb{C}$, we have 

\begin{equation}
\begin{split}
&\lim_{\zeta\to\infty} \left|\zeta \cdot G(\zeta\underline x^{(0)})\right| < \infty \\
&\,\Leftrightarrow\,
x^{(0)}_\omega \neq \pm x^{(0)}_\varphi / 2.
\end{split}
\label{eq:AssumptionIIIPrime}
\end{equation}
In other words, $G(\underline z)$ is required to decay like a usual Green's function
as a function of $\zeta$, where $\underline z = \zeta \cdot
\underline{x}^{(0)}$, if and only if $\underline{x}^{(0)}$
satisfies the \emph{regularity condition} $x^{(0)}_\omega \neq \pm x^{(0)}_\varphi / 2$.
\end{itemize}

\subsubsection{Justification of assumption 3'}

The assumption 3' may be justified as follows.

Consider the absolute value of the free Green's function \eqref{eq:G0},
\begin{widetext}
$$
|G_0(z_\varphi, z_\omega)| \leq
\sum_{\alpha =\pm1} \frac{\Gamma_\alpha}{\Gamma}
\frac{
1
}
{
|
(z_\omega - \alpha (z_\varphi - \Phi)/2 - \epsilon_d + \imag \Gamma \sign\Im
(z_\omega - \alpha z_\varphi / 2)
|
}
$$
\end{widetext}
 It is obvious that it decays $\propto\frac{1}{\zeta}$ when
$z_\varphi = \zeta \cdot x^{(0)}_\varphi$ and
$z_\omega = \zeta \cdot x^{(0)}_\omega$
 for the nonsingular combinations of $x^{(0)}_\varphi$ 
and $x^{(0)}_\omega$. 
It does not decay at all in the singular cases $x^{(0)}_\omega =
\pm x^{(0)}_\varphi / 2$.

It is easy to check that interaction $U>0$ does not change this \emph{high-energy
structure}. 
Let us examine the second-order self-energy expression (Eq.~(15) in
Ref.~\onlinecite{prl07}):
\begin{widetext}
\begin{eqnarray}
\label{eq:Sigma2ndOrder}
\Sigma^{(2)}(\underline z) = 
U^2 \sum_{\alpha_i}\left[\prod_{i=1}^3
\int\Dfrtl{\epsilon_i}\frac{\Gamma_{\alpha_i}}{\Gamma}\right] \cdot 
\frac{
f_{\alpha_1}(1-f_{\alpha_2})f_{\alpha_3} +
(1-f_{\alpha_1})f_{\alpha_2}(1-f_{\alpha_3})
}
{
z_\omega - (\alpha_1 -\alpha_2 + \alpha_3)\frac{z_\varphi-\Phi}{2} 
-\epsilon_1 + \epsilon_2 - \epsilon_3
}\;\;,
\nonumber
\end{eqnarray}
\end{widetext}
with $f_{\alpha_i} = f(\epsilon_i-\alpha_i\Phi/2)$.

Due to the structure of the denominator, we see that on top of the singular 
directions of the bare Green's function,
${x}^{(0)}_\omega = \pm {x}^{(0)}_\varphi / 2$, we also have the singular
directions ${x}^{(0)}_\omega = \pm \frac{3}{2} {x}^{(0)}_\varphi$.

Consequently, assumption 3' is incorrect for the second-order \emph{self-energy}.
Nevertheless, when inserted into Dyson's equation,
\begin{equation}
G^{(2)}(\underline z) =
\frac{G_0(\underline z)} 
{
1-G_0(\underline z) \Sigma^{(2)}(\underline z)
},
\end{equation}
we see that for the directions ${x}^{(0)}_\omega = \pm \frac{3}{2}
{x}^{(0)}_\varphi$ the limiting behaviour of $G_0$ is adopted, i.e.~the
behaviour \eqref{eq:AssumptionIIIPrime}.

Note that the uniqueness proof of the appendix also holds when directly applied
to the self-energy, because the singular directions of the second-order
perturbation theory, ${x}^{(0)}_\omega = \pm
\frac{3}{2} {x}^{(0)}_\varphi$, are not required to be regular in the proof.
This is because the direction ${x}^{(0)}_\omega = \pm
\frac{3}{2} {x}^{(0)}_\varphi$ also defines a branch cut (assumption 1).

\subsection{Tubular Cone Domains (``Wedges'')}
As we have seen, the structure of $G_0$ combined with the structure of
convolutions in the perturbation theory with respect to $U$, indicates that the numerous 
branch cut hyperplanes divide $\mathbb{C}^2$ into several, in fact infinitely many, 
wedges of the form $T^C = \mathbb{R}^2 + \imag C$. $C$ is by definition a convex cone 
with its vertex at zero. See also the pictorial discussion in Ref.~\onlinecite{dirks}.
Due to the convexity of $C$, $T^C$ is pseudoconvex and thus a domain of holomorphy.
\cite{vladimirovbook}
In the mathematical classification scheme, domains like these are called 
tubular cone domains. 

\subsubsection{Geometry of the Cones}

We briefly introduce certain notions of the description of the analytic geometry of 
cones in $\mathbb{R}^d$. This is necessary to thoroughly follow the
mathematical formulae which are involved
in the description of the analytic structure of $T^C$.
The cone $C$ with vertex at zero is formally defined by the scaling property 
$\underline x \in C \Rightarrow \forall \lambda >0 : \lambda \underline x \in C$.

Its \emph{dual cone} $C^*$ is defined via the standard scalar product
\begin{equation}
(\underline \xi, \underline x) := \sum_{k=1}^d \xi_k x_k;\quad \underline \xi, \underline x \in\mathbb{R}^d,
\label{eq:scalarproductedge}
\end{equation}
by 
\begin{equation}
C^* := \{\underline \xi \in \mathbb{R}^d\,|\,\forall \underline x \in C :
(\underline \xi, \underline x) \geq 0\}.
\end{equation}
$C^*$ represents the space of positive semi-definite linear functionals on $C$ when
the functional form \eqref{eq:scalarproductedge} is considered.
The dual cone is important, because the construction of kernel functions
often involves Fourier transforms. 

A \emph{convex cone} is a cone for which the straight line between any pair
of points within the cone is also contained by the cone.
We will also use the analytic continuation of the scalar product 
\eqref{eq:scalarproductedge} with respect to $\underline x$. We continue 
$\underline x \to \underline z$ holomorphically in \eqref{eq:scalarproductedge}:
\begin{equation}
(\underline \xi, \underline z) := \sum_{k=1}^d \xi_k z_k;\quad \underline \xi
\in \mathbb{R}^d, \underline z \in \mathbb{C}^d.
\label{eq:scalarproductwedge}
\end{equation}

\subsubsection{Analytic Structure and Biholomorphic Equivalence Classes}
\label{subsubsec:anastructbiholomorphiceq}

Tubular cones and domains are well-known objects in the theory of several complex
variables, because they naturally arise in certain fields of mathematics. 
As a consequence, many efforts were put in {for} a detailed understanding of their
structure.
In the prominent physical example, axiomatic quantum field theory, the cones
represent forward and backward light cones, in four-dimensional spacetime,
$d=4$. A celebrated result was Bogolyubov's edge-of-the-wedge theorem.
\footnote{See section \ref{subsec:edgeofwedgetheorems}.}
Mathematical examples include 
Fourier analysis, functional analysis, the theory of hyperfunctions,
and the theory of partial differential equations. 

A key component of the wedge is given by its edge, namely the real
subspace associated with the vertex of the cone.
Because in our case the vertex is located at zero, the edge of the wedge $T^C$ can be
formally identified with an oriented copy of the real subspace, 
\begin{equation}
\text{Edge}_{T^C} := \mathbb{R}^d + \imag 0^C, 
\label{eq:DefEdge}
\end{equation}
where $0^C$ is an infinitesimal vector within the cone $C$.
Although there exist, depending on the direction
approaching the origin within $C$, several infinitesimals $0^C$, the Edge is well-defined through \eqref{eq:DefEdge}, 
because all infinitesimals in $C$ are obviously equivalent with respect to holomorphic continuation
in $T^C$.
$T^C$ may be regarded as a generalization of the upper half plane.

Let $C,C'$ be arbitrary convex cones in $\mathbb{R}^d$, $C\neq C'$.
$T^C$ and $T^{C'}$ are in general \emph{not} biholomorphically
equivalent. This means that the sets of holomorphic functions living on 
them are structured differently. Wedges within the space $\mathbb{C}^2$
are fortunately an exception to this rule: \emph{$T^C$ and $T^{C'}$ are biholomorphically 
equivalent for any combination $C, C'$.}
See also the introductory notes in the corresponding part of the second volume of 
reference \onlinecite{encyclMathSciences}. 
In $\mathbb{C}^2$,
biholomorphisms between $T^C$ and $T^{C'}$ may be constructed easily using
complexified rotations and dilations of $\mathbb{R}^2$: Consider that the
real (non-singular) matrix 
$M : \mathbb{R}^2 \to \mathbb{R}^2,\, \underline x \mapsto
\begin{pmatrix}
a & b \\
c & d 
\end{pmatrix} \underline x
$
maps $C$ to $C'$, i.e.~ $C' = M C$. The corresponding biholomorphism between
$T^C$ and $T^{C'}$ is obtained from the complexified map 
$\tilde M : \mathbb{C}^2 \to \mathbb{C}^2,\, \underline z \mapsto 
\begin{pmatrix}
a & b \\
c & d 
\end{pmatrix} \underline z$. It is easy to see that $T^{C'} = M T^C$, that
$M$ is holomorphic, and invertible.

If the complexified linear map is a rotation, we will also call it a
\emph{biholomorphic rotation} when we want to emphasize the biholomorphic
character of the mapping.

Note that the helpful notion of the Bergman-Shilov boundary is \emph{not} directly
applicable to $T^C$, because $T^C$ is \emph{unbounded}. However, as will be
discussed next, sequences of bounded
domains approaching $T^C$ from its interior may be used to understand the holomorphic structure on $T^C$. 

\subsubsection{Bergman-Weil Representations}
\label{subsubsec:BergmanRepresentations}
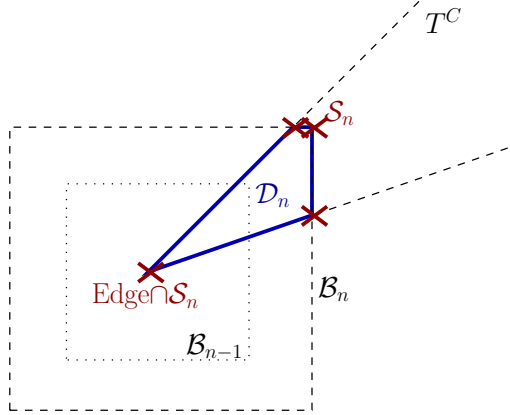
\begin{figure}
\resizebox{0.9\linewidth}{!}{
\begin{picture}(0,0)%
\includegraphics{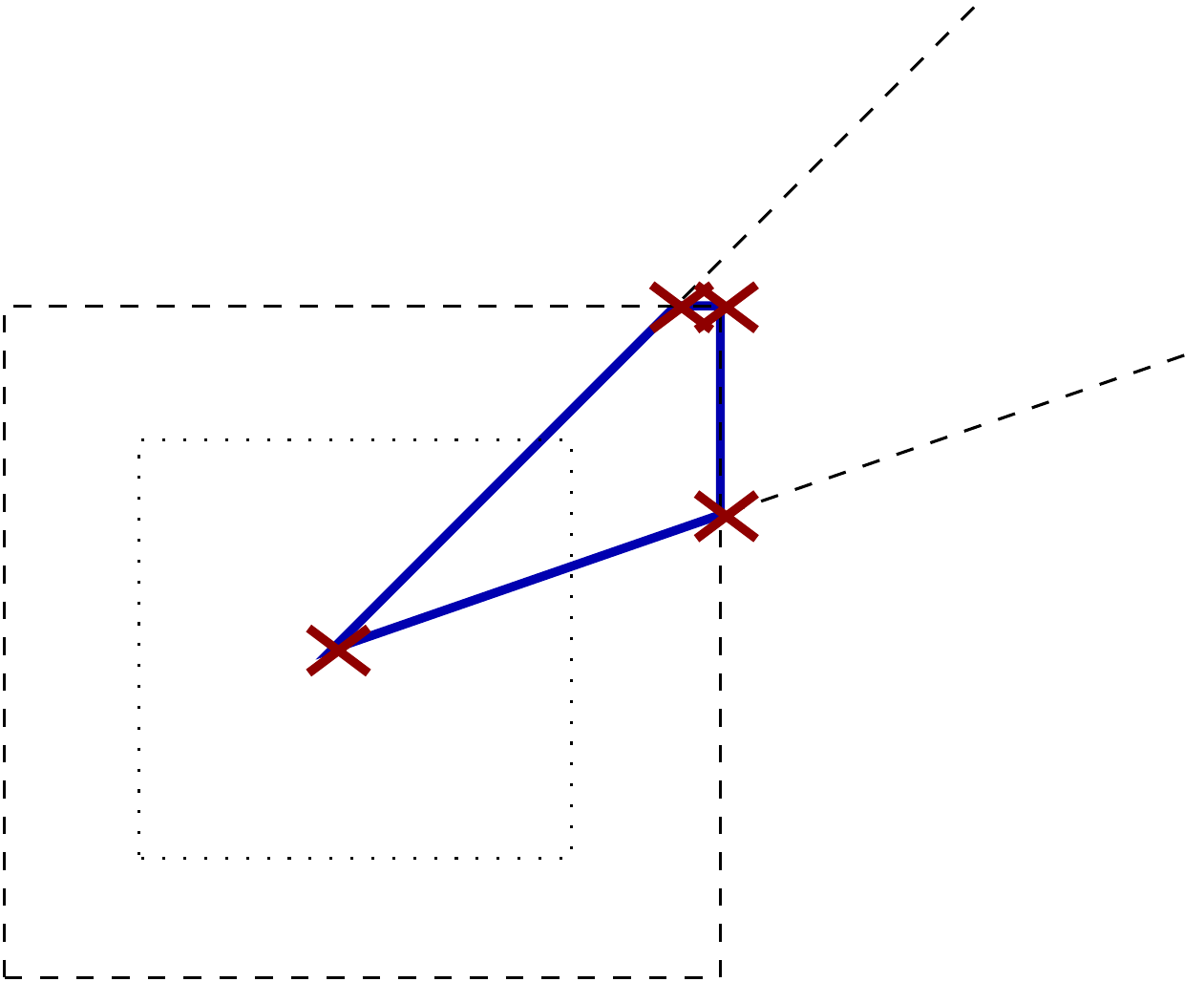}%
\end{picture}%
\setlength{\unitlength}{3947sp}%
\begingroup\makeatletter\ifx\SetFigFont\undefined%
\gdef\SetFigFont#1#2#3#4#5{%
  \reset@font\fontsize{#1}{#2pt}%
  \fontfamily{#3}\fontseries{#4}\fontshape{#5}%
  \selectfont}%
\fi\endgroup%
\begin{picture}(6044,4919)(1779,-5558)
\put(5476,-4186){\makebox(0,0)[lb]{\smash{{\SetFigFont{20}{24.0}{\familydefault}{\mddefault}{\updefault}{\color[rgb]{0,0,0}$\mathcal{B}_n$}%
}}}}
\put(6751,-1036){\makebox(0,0)[lb]{\smash{{\SetFigFont{20}{24.0}{\familydefault}{\mddefault}{\updefault}{\color[rgb]{0,0,0}$T^C$}%
}}}}
\put(5551,-2086){\makebox(0,0)[lb]{\smash{{\SetFigFont{20}{24.0}{\familydefault}{\mddefault}{\updefault}{\color[rgb]{.56,0,0}$\mathcal{S}_n$}%
}}}}
\put(4726,-3061){\makebox(0,0)[lb]{\smash{{\SetFigFont{20}{24.0}{\familydefault}{\mddefault}{\updefault}{\color[rgb]{0,0,.56}$\mathcal{D}_n$}%
}}}}
\put(2776,-4261){\makebox(0,0)[lb]{\smash{{\SetFigFont{20}{24.0}{\familydefault}{\mddefault}{\updefault}{\color[rgb]{.56,0,0}Edge$\cap \mathcal{S}_n$}%
}}}}
\put(3901,-4861){\makebox(0,0)[lb]{\smash{{\SetFigFont{20}{24.0}{\familydefault}{\mddefault}{\updefault}{\color[rgb]{0,0,0}$\mathcal{B}_{n-1}$}%
}}}}
\end{picture}%
}
\caption{(color online) An artist's impression of asymptotically filling the wedge $T^C$
with the sequence $\mathcal{D}_n$.
$\mathcal{D}_n$ is a domain with piecewise
smooth boundaries created by intersecting with a growing bicylinder $\mathcal{B}_n$,
creating the analytic polyhedra $\mathcal{D}_n$. 
The corresponding sequence of Bergman-Shilov boundaries
improperly converges to the set $\text{Edge} \cup \{\infty\}$.
Therefore, the sheaf of holomorphic functions on $T^C$ is solely characterized by
its edge values and its asymptotic behaviour $\underline z \to \infty$.
}
\label{fig:fillupcone}
\end{figure}

A sequence of bounded domains $\mathcal{D}_n\subset T^C$,
$\lim_{n\to\infty}\mathcal{D}_n = T^C$, 
with piecewise smooth boundaries like in
Fig.~\ref{subfig:bergman} may be easily constructed, such that the edge of the wedge
contains a part of the Bergman-Shilov boundary $\mathcal{S}_n$ of
$\mathcal{D}_n$ and such that the
other subsets of $\mathcal{S}_n$ disappear to $\infty$ as $n\to \infty$.
In our case, $d=2$, an explicit construction of such a sequence may be obtained by intersecting 
$T^C$ with a growing bicylinder $\mathcal{B}_n := R_n\cdot(D_1\times D_1)$,
$\mathcal{D}_n := \mathcal{B}_n \cap T^C$, with 
{
radius 
}
$R_n\propto n$.
The procedure is sketched in Fig.~\ref{fig:fillupcone}.
One finds that the sequence of Bergman-Shilov boundaries improperly converges
to
\begin{equation}
\mathcal{S_\infty} := \mathrm{Edge}_{T^C} \cup \{\infty\},
\end{equation}
where ``$\infty$'' shall informally denote the point or merely a list of
points which emerge when $T^C$ is, depending on the holomorphic structure, 
compactified suitably. At first glance, each direction for approaching
$\infty$ might yield a different point in $\{\infty\}$. 
\emph{The points ``$\infty$''
carry the additional information which is necessary to turn the structure on
$\mathrm{Edge}_{T_C}$ into a unique description of the holomorphic structure on $T^C$.
}

{
The Bergman representation for an analytic polyhedron in $\mathbb{C}^2$ may
be obtained by the following, rather technical, procedure, whose details are
not particularly relevant but enable us to investigate rather explicitly the
structure of a suitably large class of holomorphic functions on a wedge. An excellent pedestrian's
introduction to it is provided by Bergman's original monograph, Ref.~\onlinecite{bergmanbook}.
The book also provides a comprehensive introduction to
the Bergman-Shilov boundary and biholomorphic maps, using the example of
analytic polyhedra.
}
Equations of the
form
\begin{equation}
\zeta^{(k)} = \mathfrak{f}^{(k)}(\underline z, \lambda_k);
\end{equation}
with $\lambda_k \in \Lambda_k\subset \mathbb{R}, k = 1, \dots, K$,
where $\mathfrak{f}^{(k)}(\underline z, \lambda_k)$ are 
$\lambda$-parametrized families of analytic functions of
$\underline z$, shall define the analytic polyhedron.
Each equation yields a surface in $\underline z$-space for a given $\lambda$ and 
a hypersurface $h_k$ in $\underline z$-space as $\lambda$ is varied continuously.
The mutual intersections $S_{kl}=h_k \cap h_l$ yield the Bergman-Shilov boundary
surface $S=\bigcup_{k,l}S_{kl}$. $S_{kl}$ is then parametrized by a function $\underline z =
\underline{\mathfrak{g}}^{(kl)} (\lambda_k, \lambda_l)$.
A holomorphic function $f$ on the analytic polyhedron may then be written
with respect to the Bergman-Shilov boundary of the latter
using the \emph{Bergman kernel function}
\begin{equation}
\begin{split}
B_{kl} = & \left|\frac{\partial (\mathfrak{g}_1^{(kl)},\mathfrak{g}_2^{(kl)})}
{\partial(\lambda_k,\lambda_l)}\right| 
\cdot  \\
&
\quad \cdot\Bigg(
\frac{ \mathfrak{f}_l(z_1, z_2, \lambda_l)
       \mathfrak{f}_k(z_1, \mathfrak{g}_2^{(kl)},\lambda_k)
}
{
(\mathfrak{g}_1^{(kl)} - z_1) (\mathfrak{g}_2^{(kl)} - z_2)
}\,\,- \\
&
\qquad\quad
\frac{
      \mathfrak{f}_k(z_1, z_2, \lambda_k)
       \mathfrak{f}_l(z_1, \mathfrak{g}_2^{(kl)},\lambda_l)
}
{
(\mathfrak{g}_1^{(kl)} - z_1) (\mathfrak{g}_2^{(kl)} - z_2)
}\Bigg).
\end{split}
\label{eq:bergmankernel}
\end{equation}
The integral representation with respect to $\left.f\right|_S$ then reads
\begin{equation}
f(\underline z) = -\frac{1}{8\pi^2} \sum_{k\neq l} \int_{\lambda_k,\lambda_l} 
\frac{f(\mathfrak{g}_1^{(kl)}, \mathfrak{g}_2^{(kl)}) B_{kl}}
{\mathfrak{f}^{(k)}(\underline z, \lambda_k)
\mathfrak{f}^{(l)}(\underline z, \lambda_l)
}
\end{equation}
and can be applied directly to our $\mathcal{D}_n$ domains.

An explicit test on whether a Bergman integral representation for $G_0$ on the
$\mathcal{D}_n=\mathcal{B}_n \cap T^C$ domains is feasible yields that the subsets 
of $\mathcal{S}_n$ which go to $\infty$ may \emph{not} be neglected for 
$G(\underline z)$ functions.
This is because $G_0(\underline z)$ 
has a nonzero limit as $\underline z\to \infty$ if one goes along the cross-shaped submanifold $\Re
z_\omega = \pm \Re z_\varphi / 2$ and keeps $\Im \underline z$ constant. Due
to the independence of $\Im \underline z$ this problem
occurs for each of the wedges.
Hence, the Bergman kernel function \eqref{eq:bergmankernel} is only of
limited use for us. We will thus not go into further details of this rather
clumsy computation here.

{
\emph{
The formal use of the sequence $\mathcal{D}_n$ enables us to see
very explicitly that the Edge of $T^C$ 
is with respect to the representation \eqref{eq:bergmankernel} the only carrier of 
structural information which involves finite values of
$\underline z$. The rest of the information,
then uniquely defining the holomorphic structure on $T^C$, is encoded in the several possible
classes of limiting behaviour as $\underline z$ approaches infinity.
}
}

\subsubsection{Cauchy-Bochner Integral Representation}

\label{subsubsec:CauchyBochner}

As a straightforward consequence of this, assuming a certain limiting behaviour of the 
considered set of functions on $T^C$, integral representations with respect
to the Edge $\mathbb{R}^d + \imag 0^C$ may be derived.
Even more generally, a constraint on the function class which also limits the
Edge behaviour, can be imposed in such a way that the Edge function yields a
unique description. The several possible $\underline z \to \infty$ behaviours are
then, using the information from the constraint, encoded in the Edge.
This appears to be linked deeply to the extension of the notion of the Bergman-Shilov
boundary to the notion of the Shilov boundary mentioned in the course of section
\ref{subsubsec:Shilov}:
Considering a subset of the sheaf of holomorphic functions on a certain
(compactified) domain may cause the Shilov boundary to ``shrink''.

{
We are now going to discuss one of the earliest developments going beyond
simply restricting the considered function set such that a na\"ive extrapolation of Bergman representations, 
like in Fig.~\ref{fig:fillupcone}, holds. 
It was another extension of the Cauchy integral formula with respect to tubular
cones by Salomon Bochner.
He considers a function $f$ which is holomorphic on $T^C$ and satisfies the
constraint 
}
\begin{equation}
\|f(\underline x + \imag \underline y)\| \leq
M_{\epsilon,f}(C')\euler{\epsilon |y|}
\label{eq:bochnercondition}
\end{equation}
which has to hold for any compact cone $C' \subset C$ and for any $\epsilon >
0$, where $M_{\epsilon,f}(C')$ is a suitably chosen real number. $\|\cdot\|$ is a norm
which integrates out the $\underline x$ variable, $\|f(\underline x
+\imag\underline y)\|^2 := \int \mathrm{d}^dx \, |f(\underline x +\imag \underline y)|^2$.\cite{vladimirovbook}

Inequality \eqref{eq:bochnercondition} constrains the limiting behaviour 
$\underline z \to \infty$ in a sufficiently strong way such that an integral
representation with respect to the Edge may be constructed.

Namely, the Cauchy-Bochner representation then allows a function $f$
satisfying \eqref{eq:bochnercondition} to be written as 
\begin{equation}
f(\underline z) = \frac{1}{(2\pi)^d} \int \mathrm{d}^d{x'}\,
\mathcal{K}_C(\underline z - \underline{x}')f(\underline{x}' + \imag 0^C);
\label{eq:bochnerrep}
\end{equation}
with the Edge values $f(\underline{x}' + \imag 0^C)$.
Here, the so-called \emph{Cauchy kernel}  \cite{vladimirovbook}
of the cone $C$, {defined as}
\begin{equation}
\mathcal{K}_C(\underline z) := \int_{C^*}\mathrm{d}^d \xi\, \euler{\imag
(\underline \xi, \underline z)},
\label{eq:CauchyKernelBochner}
\end{equation}
was introduced. 
It is straightforward to compute the Cauchy kernel for our wedges with this
formula: see the Appendix of Ref.~\onlinecite{dirks}.
For our purposes, we will provide a general but 
easily applicable expression for further numerical and analytical computations in 
section \ref{sec:genintrepwedge}.

Unfortunately, as in the Bergman approach, a numerical test of 
Eq. \eqref{eq:bochnerrep} for $f=G_0$ using
an arbitrary wedge for which $G_0$ is holomorphic shows that the Cauchy-Bochner 
representation \eqref{eq:bochnerrep} is also incorrect for $G_0$. As a consequence, we find 
that the Green's functions does not satisfy \eqref{eq:bochnercondition}.
This is compatible with the fact that the left hand side of
\eqref{eq:bochnercondition} diverges in the case $f=G_0$, no matter which $C'$ is
considered.

Nevertheless, as we will see, the Cauchy-Bochner kernel
{ 
  \eqref{eq:CauchyKernelBochner}  
}
will serve as a building block for the construction of an exact integral
representation for a different class of
holomorphic functions which in fact contains the holomorphic branches of our
Green's function $G(\underline z)$ on the respective wedges.
{
As such it is essential as a connection of real-time and imaginary-time
structure of the Green's function.
}

\subsubsection{The tubular octant $\mathbb{H}\times\mathbb{H}$ and
\\ Biholomorphic Equivalence to the Bicylinder}

\label{subsubsec:tuboctbiheqbicyl}

Due to the biholomorphic equivalence of all $T^C$ in $\mathbb{C}^2$ 
any of our wedges which arise for the
Green's function may be mapped biholomorphically 
to the tubular octant $\mathbb{H}\times \mathbb{H} =\mathbb{R}^2 +\imag\,(\mathbb{R}^+\times
\mathbb{R}^+)$, where $\mathbb{H}$ is the upper half plane of $\mathbb{C}$. This domain may itself be
mapped biholomorphically to the bicylinder $D_1\times D_1$ via a piecewise
M\"obius transformation of the coordinates. 
\emph{Hence, all wedges of the Green's function are biholomorphically equivalent to
the bicylinder.} Let us denote a corresponding biholomorphism by $\mathfrak{m}_C : T^C \to
D_1\times D_1$.
We would like to comment on this due to the striking 
simplicity of the bicylinder and of domains which are direct products of $\mathbb{C}^1$
domains with respect to the construction of integral representations.

From the point of view of this construction, the tubular octant may be regarded as the simplest representant of the
biholomorphic equivalence class of all wedges in $\mathbb{C}^2$.

Due to the premises of the Cauchy integral formula, a usage of the representation 
\eqref{eq:CauchyRepBicyl} for the biholomorphically transformed sheaf of holomorphic 
functions is feasible in case the transformed Green's function
$\left.G\right|_{T^C} \circ \mathfrak{m}_C^{-1}$
is continuous on its topological boundary $\partial (D_1 \times D_1)$. Note that under the biholomorphic
transformations, $S_1\times S_1$ is mapped to the edges of the $T^C$ wedges.
The ``$\infty$'' in $T^C$ maps to ``$\infty$'' in $\mathbb{H}\times \mathbb{H}$ under
biholomorphic transformation, and that again maps biholomorphically, using the component-wise
M\"obius transformation $\frac{z_k-\imag}{z_k+\imag}$ to the points
$\mathfrak{p}_\infty := \{1\}\times S_1 \cup S_1 \times \{1\}$. 
Note that the points $\mathfrak{p}_\infty \subset \partial (D_1 \times D_1)$ are also part of the 
distinguished boundary torus $S_1 \times S_1$. 

\emph{The boundary behaviour of the transformed Green's function
$\left.G\right|_{T^C} \circ \mathfrak{m}_C^{-1}$ is not continuous at the
intersection point $(1,1)$ of the two circles $\mathfrak{p}_\infty$, due to the 
properties of $\left.G_0\right|_{T^C}$ 
at $\infty$ leading to singular directions, as summarized in assumption
3' of section \ref{subsec:holostructGF}.}

\begin{figure}
\resizebox{0.9\linewidth}{!}{
\begin{picture}(0,0)%
\includegraphics{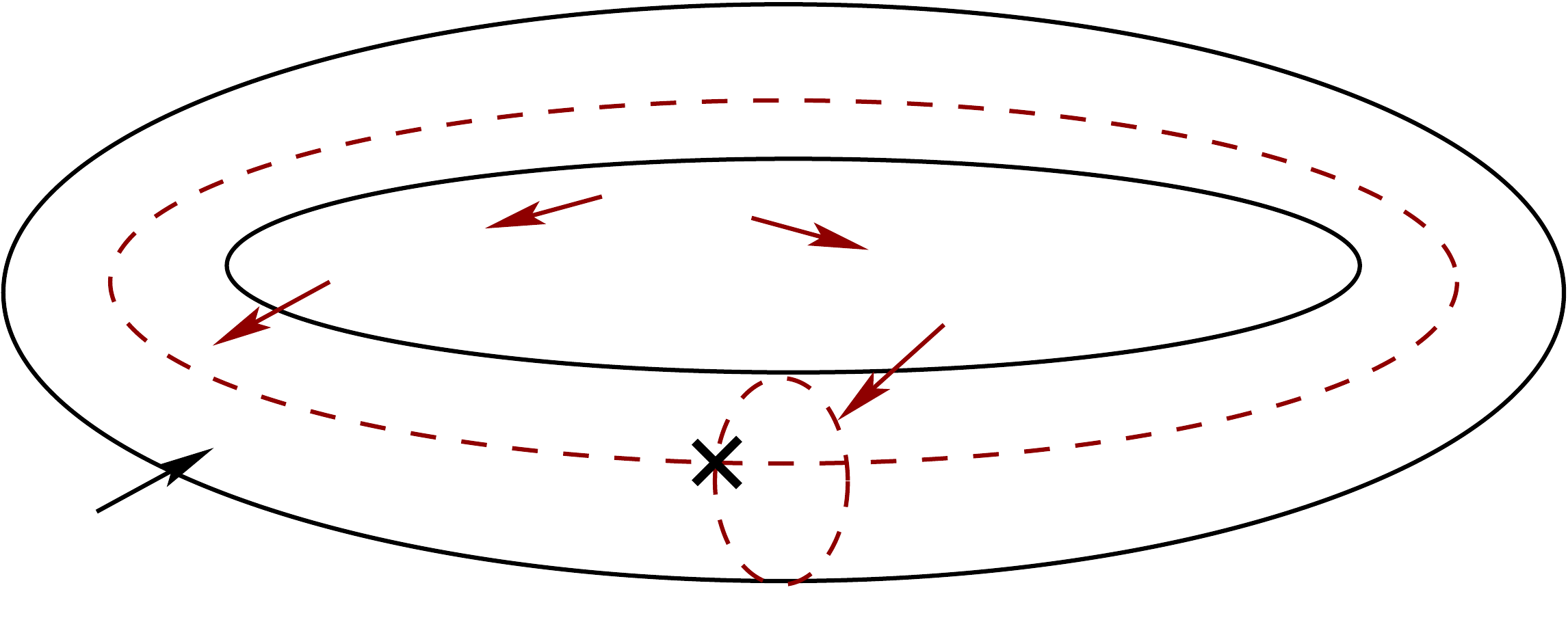}%
\end{picture}%
\setlength{\unitlength}{3947sp}%
\begingroup\makeatletter\ifx\SetFigFont\undefined%
\gdef\SetFigFont#1#2#3#4#5{%
  \reset@font\fontsize{#1}{#2pt}%
  \fontfamily{#3}\fontseries{#4}\fontshape{#5}%
  \selectfont}%
\fi\endgroup%
\begin{picture}(10996,4343)(728,-4831)
\put(4741,-3601){\makebox(0,0)[lb]{\smash{{\SetFigFont{34}{40.8}{\familydefault}{\mddefault}{\updefault}{\color[rgb]{0,0,0}$(1,1)$}%
}}}}
\put(751,-4486){\makebox(0,0)[lb]{\smash{{\SetFigFont{34}{40.8}{\familydefault}{\mddefault}{\updefault}{\color[rgb]{0,0,0}$\mathfrak{m}_C(\text{Edge})$}%
}}}}
\put(3001,-2386){\makebox(0,0)[lb]{\smash{{\SetFigFont{34}{40.8}{\familydefault}{\mddefault}{\updefault}{\color[rgb]{.56,0,0}$S_1\times 1$}%
}}}}
\put(9601,-4636){\makebox(0,0)[lb]{\smash{{\SetFigFont{34}{40.8}{\familydefault}{\mddefault}{\updefault}{\color[rgb]{0,0,0}$S_1\times S_1$}%
}}}}
\put(5176,-1936){\makebox(0,0)[lb]{\smash{{\SetFigFont{34}{40.8}{\familydefault}{\mddefault}{\updefault}{\color[rgb]{.56,0,0}$\mathfrak{p}_\infty$}%
}}}}
\put(6976,-2611){\makebox(0,0)[lb]{\smash{{\SetFigFont{34}{40.8}{\familydefault}{\mddefault}{\updefault}{\color[rgb]{.56,0,0}$1\times S_1$}%
}}}}
\end{picture}%
}
\caption{On the distinguished boundary surface of the bicylinder, 
the picture of $\text{Edge}_{T^C}$ under the component-wise M\"obius
transformation $\mathfrak{m}_C$
is delimited by the dash-dotted lines $\mathfrak{p}_\infty$.
A discontinuity of $\left.G\right|_{T^C}\circ \mathfrak{m}_C^{-1}$ 
occurs at the intersection point $(1,1)$ of the two circles and
prevents a Cauchy representation from being applicable.
The violation of the Cauchy-Bochner condition \eqref{eq:bochnercondition} appears to be
related to the occurance of the discontinuity.
}
\label{fig:torusDiscont}
\end{figure}
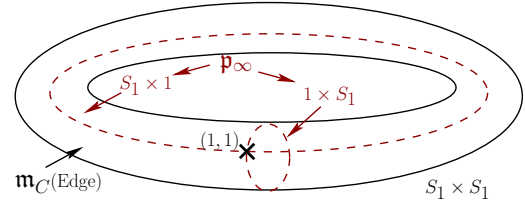

An illustration may be found in Fig.~\ref{fig:torusDiscont}.

Therefore, using the biholomorphic equivalence to the bicylinder is
not immediately
helpful for the construction of an integral representation of the Green's
function $G$. Nevertheless, it
is essential in the application of Vladimirov's approach which will be subject of
the next section.

\subsection{Vladimirov's Integral Formula}
Vladimirov provided a generalization 
of the so-called Herglotz-Nevanlinna 
representation for the upper half plane to tubular cone domains. 
His investigations were motivated by applications in the
field of linear passive systems in mathematical physics
\cite{encyclMathSciences,
VladimirovMathPhysRev, dirks}.
Due to its generality, the approach is
applicable to the analytic wedges of the interacting Green's function $G$.

\subsubsection{Herglotz-Nevanlinna Representation ($d=1$)}
We will discuss the conventional Herglotz formulae.
Herglotz' representation theorem considers holomorphic functions in the open unit 
disk $D_1$ which have a positive real part, the so-called Carath\'eodory functions
\cite{alpay,herglotzcaratheo,Nevanlinna}.
 By separating a phase factor out of the
function one can also consider functions with positive or negative imaginary
part, and so on.
 Since the open unit disk can
be conformally mapped to the upper half plane $\mathbb{H}$, using the
M\"obius transformation $\frac{z-\imag}{z+\imag}$, as mentioned above, the representation 
can under certain circumstances be also used for $\mathbb{H}$. 

By considering Carath\'eodory functions, Herglotz' theorem only imposes assumptions 
on the positivity of the real (imaginary) part of the function. In contrast to 
Cauchy's integral formula, no assumptions about the behaviour of the
Carath\'eodory functions on the boundary of the
disk are made, such as the continuity.

The theorem states that every Carath\'eodory function $f$ can be represented by 
\begin{equation}
f(z) = \imag \cdot \Im f(0) + \int_{0}^{2\pi} \frac{\euler{\imag t}+z}{\euler{\imag t}-z}
\Dfrtl{\sigma (t)},
\label{eq:HerglotzFormulaUnitDisk}
\end{equation}
where $\Dfrtl\sigma$ is a nonnegative finite measure. \cite{alpay}

\emph{Regarding a different set of functions}, a formally very similar representation is the 
so-called \emph{Poisson formula},
which is the analog of the Cauchy formula to the real analysis of harmonic
functions 
{(solutions of Laplace's equation)}, and can in fact be derived from it. 
It provides an integral kernel for the solution to the Dirichlet 
problem for the Laplace equation on the unit disk in $\mathbb{R}^2$.
For a \emph{continuous} function $\mathfrak{f}:\partial D_1 \to \mathbb{R}$ it
allows to construct a harmonic function $u : D_1 \to \mathbb{R}$ as follows
(see pp.~169ff.~in Ref.~\onlinecite{BehnkeSommer}):
\begin{equation}
u(z) = \frac{1}{2\pi}
\Re \int_{0}^{2\pi} \frac{\euler{\imag t} + z}{\euler{\imag t} - z}
\mathfrak{f}(\euler{\imag t}) \Dfrtl t
\label{eq:PoissonFormulaUnitDisk}
\end{equation}

A comparison of Eqs.~\eqref{eq:HerglotzFormulaUnitDisk} and
\eqref{eq:PoissonFormulaUnitDisk} shows that the measure $\Dfrtl{\sigma}$ 
of the Herglotz formula is in fact defined by the (possibly singular) boundary 
limit of the holomorphic function.

As a natural extension of the Poisson formula, the \emph{Schwarz integral formula} 
reconstructs a holomorphic function $f$ on the \emph{closed} unit disk from the
real part of its boundary values, up to a constant imaginary offset.
It reads (p.~171~in Ref.~\onlinecite{BehnkeSommer})
\begin{equation}
f(z) = 
 \imag \cdot \Im f(0)
+
\frac{1}{2\pi}
 \int_{0}^{2\pi} \frac{\euler{\imag t} + z}{\euler{\imag t} - z}
f(\euler{\imag t}) \Dfrtl t.
\label{eq:SchwarzFormulaUnitDisk}
\end{equation}
Apparently, the only formal difference between \eqref{eq:PoissonFormulaUnitDisk} 
and \eqref{eq:SchwarzFormulaUnitDisk} is the different measure.

Due to the conformal equivalence, for a holomorphic function $f$ on the \emph{closed} 
upper half plane $\Im z \geq 0$, 
under the assumption that there is an $\alpha>0$ for which $|z^\alpha f(z)|$ is bounded, 
one has the Schwarz representation in the following form:

\begin{equation}
f(z) = \frac{1}{\pi \imag} 
\int_{-\infty}^\infty \frac{\Re f(x+\imag 0^+)}{x-z} \Dfrtl x.
\label{eq:SchwarzRepUpHalf}
\end{equation}
Note the formal equivalence to the spectral representation of a conventional
Matsubara Green's function, $G(z) = \int\Dfrtl x\frac{1}{-\pi} \frac{\Im G(x+\imag
0^+)}{z-x}$.

Similarly, the Poisson kernel for the closed upper half plane is
\begin{equation}
P_y(x) = \frac{y}{x^2 + y^2},
\end{equation}
yielding the representation
\begin{equation}
u(x+\imag y) = \frac{1}{\pi} \int_{-\infty}^\infty P_y (x-t) \mathfrak{f}(t)
\Dfrtl{t},
\label{eq:PoissonRepUpHalf}
\end{equation}
with $\mathfrak{f} \in L^p(\mathbb{R})$.

The full Herglotz-Nevanlinna representation of arbitrary analytic functions with positive real
part for the \emph{open} upper half-plane reads \cite{VladimirovMathPhysRev}
\begin{equation}
\begin{split}
f(z) = &\frac{z+\imag}{\pi\imag} \int \frac{\Dfrtl{\mu}(x')}{(x'-\imag)(x'-z)} \\
       & - \frac{1}{\pi} \int \frac{\Dfrtl{\mu}(x')}{1+x'^2} -\imag a z + b.
\end{split}
\label{eq:HerglotzNevanlinna1D}
\end{equation}
Here, $\mu$ is given by the boundary-value distribution of $\Re f$:
\begin{equation}
\mu = \Re \text{bv}\,f,
\end{equation}
i.e.~$\mu(x) \,\text{``$=$''}\,\Re f(x+\imag 0^+)$.
The linear coefficient 
\begin{equation}
a = \Re f(\imag) - \frac{1}{\pi} \int \frac{\Dfrtl{\mu} (x')}{1+x'^2},
\end{equation}
and the constant term $b=\Im f(\imag)$. 

For example, in the case of the function $f(z) = \imag / \pi z$, $\mu$ is the Dirac
measure $\mu(x) = \Re \text{bv} f = \delta(x)$ and the coefficients $a=b=0$.
The case $\mu(x) = -a\cdot \delta(x)$ is not permitted by construction.

At first glance, Eqs. \eqref{eq:SchwarzRepUpHalf} and \eqref{eq:PoissonRepUpHalf}
and the connection to the Herglotz-Nevanlinna representation
\eqref{eq:HerglotzNevanlinna1D}
seem to be rather straightforward applications of the Cauchy integral formula. 
However, attempting the multidimensional generalization, we found that in our case, 
$d>1$, the Cauchy-Bochner way of invoking $\mathcal{K}_C$ for a
representation, equation \eqref{eq:CauchyKernelBochner}, 
is not valid for the noninteracting Green's function $\left.G_0\right|_{T^C}$:
see Section \ref{subsubsec:CauchyBochner}. 
Remarkably, as found in section \ref{subsubsec:tuboctbiheqbicyl}, taking
assumption 2 from section \ref{subsec:holostructGF} in to account, we find that when transformed to the bicylinder, the
Green's function $\left.G\right|_{T^C}\circ \mathfrak{m}_C^{-1}$ is
{closely} related 
to a Carath\'eodory function, but the Cauchy-Bochner representation is invalid.

These subtleties are apparently reflected by the central assumption
\eqref{eq:bochnercondition} of Cauchy-Bochner representations.

\subsubsection{Functions with positive real or \\ imaginary part in $T^C$}
Note that the representation \eqref{eq:HerglotzNevanlinna1D} can be also used
for bounded functions on the upper half plane. This can be seen by formally
introducing a shift in $f$ which makes the real part of the function of
consideration positive definite.

While the signs of its real and imaginary parts will vary, the Green's function 
$\left.G\right|_{T^C}$ is in fact a bounded function, assumption
\eqref{eq:AssumptionII}. This is why Vladimirov's
integral representation for functions with positive imaginary part turned out
to be applicable. \cite{dirks}
Let us denote the set of holomorphic functions with positive imaginary parts on
$T^C$ by $H_+(T^C)$.
Due to the biholomorphic equivalence of the Green's function's wedges to the 
bicylinder, one may think of $H_+(T^C)$ as a generalization of the Carath\'eodory 
functions.
Note that in the literature, sometimes functions with positive real and
sometimes functions with positive imaginary parts are considered, resulting in
marginal differences in the equations. 

\subsubsection{Vladimirov's Kernel Functions for $T^C$}
We will now study the generalization of the Herglotz-Nevanlinna representation to 
$d$-dimensional wedges.
\cite{encyclMathSciences, VladimirovMathPhysRev}
Vladimirov's approach may be found for positive \emph{real} parts in
Ref.~\onlinecite{VladimirovMathPhysRev} 
and for positive \emph{imaginary} parts in Ref.~\onlinecite{encyclMathSciences}.

Because our original work \cite{dirks} referred to Ref. \onlinecite{encyclMathSciences} we would like
to switch to considering the class of functions with positive \emph{imaginary} part, 
$H_+(T^C)$, in the following.

Let us first introduce Vladimirov's
generalizations of the Poisson and
Schwarz kernels, using the Cauchy kernel $\mathcal{K}_C$ from
Eq.~\eqref{eq:CauchyKernelBochner} as a starting point.

The (generalized) \emph{Poisson kernel} for the wedge $T^C$ is defined
by\cite{encyclMathSciences}
\begin{equation}
\mathcal{P}_C(\underline z) := 
\frac{|\mathcal{K}_C(\underline z)|^2}
{
(2\pi)^d \mathcal{K}_C(2\imag \underline y)
};
\quad \underline z = \underline x + \imag \underline y.
\label{eq:PoissonKernelVladimirov}
\end{equation}
In case of the tubular octant, it is the product of usual Poisson kernels,
however it is no longer simply proportional to the imaginary part of the 
Cauchy kernel.
The (generalized) \emph{Schwarz kernel with respect to a point
$\underline{z}^{(0)} = \underline{x}^{(0)} + \imag \underline{y}^{(0)} \in T^C$}
is given by
\begin{equation}
\begin{split}
\mathcal{S}_C(\underline z, \underline{z}^{(0)})
:= &
\frac{
  2\mathcal{K}_C(\underline z)
  \overline{\mathcal{K}_C(\underline{z}^{(0)})}
}
{
  (2\pi)^d \mathcal{K}_C\left(\underline z - \overline{\underline{z}^{(0)}}\right)
} \\
&\quad-
\mathcal{P}_C (\underline{x}^{(0)}, \underline{y}^{(0)}).
\end{split}
\label{eq:SchwarzKernelVladimirov}
\end{equation}

For a measure $\mu(\underline x)$ we call
\begin{equation}
P_C[\Dfrtl{\mu}](\underline z) :=
\int \mathrm{d}\mu(\underline x)\, \mathcal{P}_C(\underline z - \underline x)
\end{equation}
the \emph{Poisson integral with respect to $\mu$}.

\subsubsection{Vladimirov's Theorem}

We already pointed out that in the case $d\geq 3$ two arbitrary different wedges, $T^C$ and
$T^{C'}$, are usually \emph{not} biholomorphically equivalent. In particular,
the wedge $T^C$ is not necessarily biholomorphically equivalent to
$\mathbb{H}^d$ when $d\geq 3$. Hence, one may expect the structural similarity 
to the Carath\'eodory functions to break down more easily in higher dimensions. 
{
This is the reason why Vladimirov's $d$-dimensional generalization\cite{encyclMathSciences} of Herglotz' theorem is stated in
a comparably cryptic way which will simplify considerably in our case $d=2$, for the
reasons above: 
}

\textbf{Theorem.} (Vladimirov, 1978/79) The following conditions for a
function $f\in H_+(T^{C})$ are equivalent for a cone $C \subset \mathbb{R}^d$
and $\mu(\underline x) := \Im f(\underline x + \imag 0^C)$:
\begin{enumerate}
\item The Poisson integral $P_C[\Dfrtl \mu]$ is pluriharmonic in $T^C$;
\item the function $\Im f(\underline z)$, $\underline z=\underline x + \imag
\underline y\in T^C$, 
is represented by the Poisson formula
\begin{equation}
\Im f(\underline z) = P_C[\Dfrtl \mu](\underline z) + (\underline a,
\underline y),
\label{eq:VladimirovPoisson}
\end{equation}
for some $\underline{a} \in C^*$, where $C^*$ is the dual cone of $C$;
\item for all $\underline z^0\in T^C$, under the assumption that $C$ is
regular, the Schwarz representation
\begin{equation}
\begin{split}
f(\underline z) = &\imag \int_{\mathbb{R}^d} \mathcal{S}_C (\underline z -
\underline t, \underline z^0 - \underline t) \Dfrtl \mu (t) \\ & 
+ (\underline a, \underline z) + \underline b
\end{split}
\label{eq:VladimirovSchwarz}
\end{equation}
holds, with $b = b(\underline z^0) = \Re f(\underline z^0) - (\underline a,
\underline x^0)$.\hfill$\Box$
\end{enumerate}
Note that pluriharmonic functions are the natural multidimensional generalization of
harmonic functions. A regular cone $C$ in our context is a cone for which $1/\mathcal{K}_C$ is
non-singular in $T^C$. In the cases $d=1,2,3$ all pointed cones are regular 
\cite{VladimirovMathPhysRev}.

{Using the equivalence of all $T^C$ in the case $d=2$,} in Ref.~\onlinecite{dirks} we verified that the first statement of the theorem is true for
$\left.G\right|_{T^C}$. This is so because it is known from the literature
(see p.~134 in Ref.~\onlinecite{VladimirovMathPhysRev}) that 
the Poisson integral is pluriharmonic for any function $H_+(\mathbb{H}^d)$.
Due to the biholomorphic equivalence of all $T^C$ to $\mathbb{H}^2$ in $\mathbb{C}^2$, the two
integral representations provide exact relations for all holomorphic sheets of the interacting
Green's function. A parametrization of the Green's function with respect to
their Edge values is gained by this.
In our case, the validity of the
representation has due to the biholomorphic equivalences first been shown by Kor\'anyi and 
Puk\'ansky's work on the polycylinder.\cite{koranyi}

Note that the linear growth term $\underline{a}$ is zero for the Green's
function, because it is bounded, as required by assumption \eqref{eq:AssumptionII}.

\subsection{Application to the Green's function}
\label{sec:genintrepwedge}
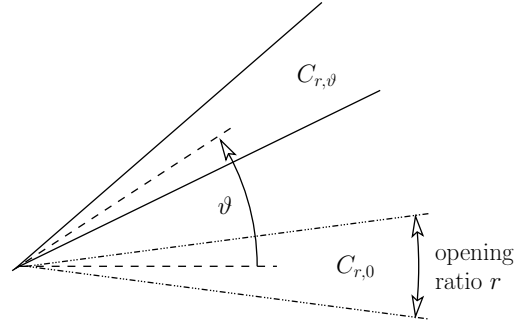
\begin{figure}
\resizebox{0.9\linewidth}{!}{
\begin{picture}(0,0)%
\includegraphics{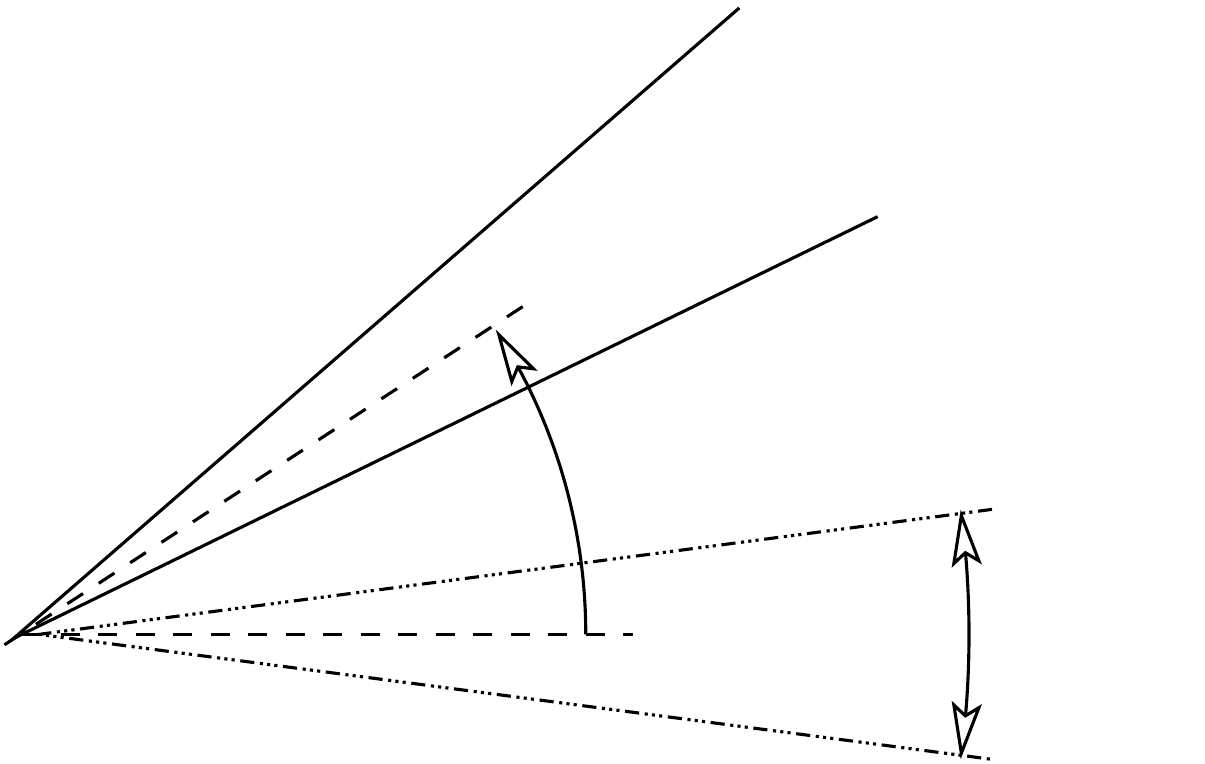}%
\end{picture}%
\setlength{\unitlength}{3947sp}%
\begingroup\makeatletter\ifx\SetFigFont\undefined%
\gdef\SetFigFont#1#2#3#4#5{%
  \reset@font\fontsize{#1}{#2pt}%
  \fontfamily{#3}\fontseries{#4}\fontshape{#5}%
  \selectfont}%
\fi\endgroup%
\begin{picture}(5775,3651)(2439,-3383)
\put(4801,-2161){\makebox(0,0)[lb]{\smash{{\SetFigFont{17}{20.4}{\familydefault}{\mddefault}{\updefault}{\color[rgb]{0,0,0}$\vartheta$}%
}}}}
\put(7276,-3016){\makebox(0,0)[lb]{\smash{{\SetFigFont{17}{20.4}{\familydefault}{\mddefault}{\updefault}{\color[rgb]{0,0,0}ratio $r$}%
}}}}
\put(5701,-661){\makebox(0,0)[lb]{\smash{{\SetFigFont{17}{20.4}{\familydefault}{\mddefault}{\updefault}{\color[rgb]{0,0,0}$C_{r,\vartheta}$}%
}}}}
\put(6151,-2836){\makebox(0,0)[lb]{\smash{{\SetFigFont{17}{20.4}{\familydefault}{\mddefault}{\updefault}{\color[rgb]{0,0,0}$C_{r,0}$}%
}}}}
\put(7276,-2686){\makebox(0,0)[lb]{\smash{{\SetFigFont{17}{20.4}{\familydefault}{\mddefault}{\updefault}{\color[rgb]{0,0,0}opening}%
}}}}
\end{picture}%
}
\caption{An arbitrary cone $C$ with 0 as vertex can be parametrized by an
opening ratio $r$ and an orientation angle $\vartheta$.
The rotation $R_{\vartheta}$ which maps $C_{r,0}$ to $C_{r,\vartheta}$
induces a biholomorphism between $T^{C_{r,0}}$ and 
$T^{C_{r,\vartheta}}$.
}
\label{fig:wedgethetaparam}
\end{figure}

It turns out to be reasonable to specify a given cone domain arising from the
branch cut structure by an angle $\vartheta$ and an opening ratio $r$.
See figure \ref{fig:wedgethetaparam}.
It is sufficient to consider the case $\vartheta=0$ first, because relations
for finite $\vartheta$ may be reconstructed from biholomorphic rotations, as explained
in section \ref{subsubsec:anastructbiholomorphiceq}.

\subsubsection{Kernels for $\vartheta=0$}

For the case $\vartheta =0$, a computation of the kernels 
$\mathcal{K}_{C_{r,\vartheta}}$ and $\mathcal{P}_{C_{r,\vartheta}}$ 
has been provided in
the appendix of Ref.~\onlinecite{dirks} already, where the opening ratio $r$ was
named $\varepsilon$ for technical reasons. \footnote{Because in our application $\varepsilon$ will not
necessarily be small, we changed to the notation $r$.}

We used the definition
\begin{equation}
C_{r,\vartheta} := 
\bigcup_{\lambda \in (-r,r)}
\{
(x_1,x_2)\in \mathbb{R}^2\,|\,x_2 >0 \wedge x_1 =
\lambda x_2
\}
\end{equation}
and computed the Cauchy and Poisson kernels via equations
\eqref{eq:CauchyKernelBochner} and \eqref{eq:PoissonKernelVladimirov}.
The resulting 
Cauchy kernel is
\begin{equation}
\mathcal{K}_{C_{r,0}}(\underline z) = 
-2r \prod_{\mu=\pm 1} \frac{1}{r z_2 - \mu z_1 },
\end{equation}
and the resulting Poisson kernel is
\begin{equation}
\mathcal{P}_{C_{r,0}} (\underline z) = 
\frac{r}{\pi^2} \prod_{\mu=\pm 1} 
\frac{ry_2 - \mu y_1}
{
(rx_2 - \mu x_1)^2 + (ry_2 - \mu y_1)^2
}.
\label{eq:poissonkernelepsdomain}
\end{equation}
We have not used an explicit formula for the Schwarz kernel
\eqref{eq:SchwarzKernelVladimirov} yet, because the
occurance of the reference point $\underline{z}^{(0)}$ appears to introduce additional
technical complications. The shape of equation
\eqref{eq:poissonkernelepsdomain} is so simple because it can be computed
from the tubular octant, whose Poisson kernel is the product of usual Poisson
kernels. A simple real-valued $2\times 2$ matrix acts as the biholomorphism which
converts the two representations. 

\subsubsection{Operator Notation}

In order to put a stronger emphasize on the functional-analytic nature of the 
integral representations which interrelate edge and wedge values of the Green's
function, let us introduce an operator 
notation for the Poisson integral and also for the biholomorphic rotations.

Let us denote the set of all a-priori admitted Green's functions on $T^C$
by $\mathcal{G}_C$. By ``a-priori admitted'' we mean those analytic functions
$\left.G\right|_{T^C}: T^C \to \mathbb{C}$ for which
axiom 2 from section \ref{subsec:holostructGF} holds ($T^C$ has to comply
with axiom 1). Furthermore, let us denote the corresponding space of boundary value
distributions 
{(edge functions)} 
$\left.G\right|_{T^C}(\underline x + \imag 0^C)$ by $\mathcal{E}_C$.

In order to focus on the Poisson kernel, we introduce the
corresponding spaces of imaginary parts, $\mathcal{G}_C^{(I)}$ and
$\mathcal{E}_C^{(I)}$.

We denote by the operator $\mathcal{P}_C$ the linear map
\begin{eqnarray}
\mathcal{P}_C: \mathcal{E}_C^{(I)} &\to& \mathcal{G}_C^{(I)}, \\
                \Im G(\underline x + \imag 0^C) &\mapsto&
                   \int \mathrm{d}x^2\, \mathcal{P}_C (\underline z - \underline x)  \Im G(\underline x + \imag 0^C).
\nonumber
\end{eqnarray}
Note that the Schwarz kernel \eqref{eq:SchwarzKernelVladimirov} does not directly yield a comparable map from
$\mathcal{E}_C$ to $\mathcal{G}_C$, due to the occurrence of the
$\underline{z}^{(0)}$ reference point.

Furthermore, the rotation $R_\vartheta$, which maps the cone $C_{r,0}$ to the
cone $C_{r,\vartheta}$, induces a biholomorphic map 
$\tilde{R}_\vartheta : T^{C_{r,0}} \to T^{C_{r,\vartheta}}$ (see section
\ref{subsubsec:anastructbiholomorphiceq}). 
At this point we would again like to emphasize that the biholomorphism does
not connect the different branches of the Green's function on the
wedges. It merely yields a counterpart of a given holomorphic branch on a
biholomorphically equivalent wedge which can be formally operated with. It is
in that sense 
{that is}
 analogous to the concept of a conformal map.
This biholomorphism maps functions
$f\in \mathcal{G}_{C_{r,0}}$ to functions $f\in
\mathcal{G}_{C_{r,\vartheta}}$.
This can be similarly represented by the linear operator
\begin{eqnarray}
\mathcal{R}_\vartheta : \mathcal{G}_{C_{r,0}} & \to & \mathcal{G}_{C_{r,\vartheta}}, \\
                        f(\underline z)       & \mapsto & f(\tilde{R}^{-1}_\vartheta(\underline z)).
\end{eqnarray}

The operator $\mathcal{R}_\vartheta$ also naturally extends to a linear map from
$\mathcal{E}_{C_{r,0}}$ to $\mathcal{E}_{C_{r,\vartheta}}$ which we will
denote by the same symbol $\mathcal{R}_\vartheta$.

\subsubsection{Kernel functions at finite $\vartheta$}

Consequently, for finite $\vartheta$, the Poisson kernel operator of
$T^{C_{r,\vartheta}}$ is
\begin{equation}
\mathcal{P}_{C_{r,\vartheta}}
=
\mathcal{R}_\vartheta \mathcal{P}_{C_{r,0}}
\mathcal{R}^{-1}_\vartheta.
\label{eq:PoissonOpFiniteTheta}
\end{equation}
Equivalently, the Poisson kernel function of $T^{C_{r,\vartheta}}$ is given
by
\begin{equation}
\mathcal{P}_{C_{r,\vartheta}} (\underline z) = 
\mathcal{P}_{C_{r,0}} (R_{\vartheta}^{-1} \cdot \underline z).
\label{eq:poissonkernelrotatedepsdomain}
\end{equation}
In practical computations, the function can be evaluated combining equation 
\eqref{eq:poissonkernelepsdomain} 
and the rotation matrix
\begin{equation}
R_\vartheta^{-1} = 
\begin{pmatrix}
\cos\vartheta & -\sin \vartheta \\
\sin\vartheta & \cos\vartheta
\end{pmatrix}.
\end{equation}

\subsubsection{Edge Properties of $G_0$}

Since we essentially reduced the structure of the Green's function to the
edge values of their holomorphic branches, it seems worthwhile to investigate
the edge structure of $G_0$,
and later also the perturbative structure of the theory in $U$,
more carefully. See section \ref{subsec:edgeofwedgetheorems} for the deeper
analysis.

The edge limit of the bare Green's function \eqref{eq:G0}, as a function of the edge
orientation $\vartheta$, is given by
\begin{equation}
G_0^\text{(edge)}(\vartheta;\underline x) =
\sum_{\alpha=\pm1} G_0^{\text{(edge)},\alpha}(\vartheta;\underline x),
\label{eq:edgelimitG0}
\end{equation}
where
\begin{equation}
\begin{split}
&G_0^{\text{(edge)},\alpha}(\vartheta;\underline x) =  \\
&\qquad
\frac{\Gamma_\alpha/\Gamma}
{
x_\omega -\alpha (x_\varphi - \Phi) / 2 - \varepsilon_d + 
\imag \Gamma \mathrm{sgn}_\vartheta
}
\end{split}
\label{eq:edgelimitG0Addends}
\end{equation}
and
\begin{equation}
\mathrm{sgn}_\vartheta:= \sign \left(\cos \vartheta - \frac{\alpha}{2} \sin
\vartheta\right).
\label{eq:edgelimitG0Switch}
\end{equation}

Apparently, the edge function only changes as a function of $\vartheta$ whenever 
$\cos \vartheta \pm \sin \vartheta / 2$ crosses zero. This reflects the
equivalence of all directions 
\begin{equation}
\underline x + \imag 0^\vartheta
:= \underline x + \imag \cdot
\begin{pmatrix}
\sin \vartheta \\
\cos \vartheta
\end{pmatrix} 0^+
\end{equation}
when approaching the edge
within a holomorphic branch $T^C$, as discussed in section
\ref{subsubsec:anastructbiholomorphiceq}.
The edge function changes whenever $\vartheta$ crosses a branch cut, namely
for the following singular orientations in the interval $[0,2\pi)$:
\begin{eqnarray}
\label{eq:singularorientations}
\vartheta^\text{(sing)}_1 &=& \arctan 2; \\
\nonumber
\vartheta^\text{(sing)}_2 &=& \pi - \arctan 2; \\
\nonumber
\vartheta^\text{(sing)}_3 &=& \pi + \arctan 2; \\
\nonumber
\vartheta^\text{(sing)}_4 &=& 2\pi - \arctan 2. 
\end{eqnarray}
{ These are the angles corresponding to the four half-lines emerging from the
origin in figure \ref{fig:branchcuts0}.} The orientations $\vartheta^\text{(sing)}_i$ are also identical to the
singular directions of assumption 3' in section \ref{subsec:holostructGF}.

There is another subtle feature of the edge behaviour of the bare Green's function.
The real part
\begin{equation}
\begin{split}
&
\Re G_0^\text{(edge)}(\vartheta;\underline x) =  \\
& \qquad
\sum_{\alpha=\pm1}
\frac{\frac{\Gamma_\alpha}{\Gamma} (x_\omega -\alpha (x_\varphi - \Phi) / 2 -
\varepsilon_d) }
{
(x_\omega -\alpha (x_\varphi - \Phi) / 2 - \varepsilon_d)^2 + 
\Gamma^2
}
\end{split}
\label{eq:ReG0Edge}
\end{equation}
is completely $\vartheta$-independent. As a consequence, for any branch of
$G_0$, the 
edge limit $\Re G_0 (\underline x + \imag 0^C)$ is identical.

Another property is that, following the instructions
\eqref{eq:DefPhysicalLimit} to obtain the physical limit as far as the
orientation of the limiting procedure is concerned, the function
\begin{equation}
\tilde A_0 (\underline x) :=
- \frac{1}{\pi}
\Im G_0\left( \underline x + \imag  \cdot
0^{\vartheta=0} 
\right)
\label{eq:defAtilde0}
\end{equation}
is positive definite:

\begin{equation}
\tilde A_0 (\underline x) 
= 
\sum_{\alpha=\pm1}
\frac{\Gamma_\alpha / \pi}
{
(x_\omega -\alpha (x_\varphi - \Phi) / 2 - \varepsilon_d)^2 + 
\Gamma^2
}.
\label{eq:nonintAtilde}
\end{equation}

 In particular, the non-interacting spectral function
\begin{equation}
A_0(\omega) = \tilde A_0(\Phi, \omega).
\end{equation}

Again one can see in equation \eqref{eq:nonintAtilde} that $\tilde
A_0$ does not decay to zero as a function of $\underline x
\to\infty$ along the singular directions $\vartheta^\text{(sing)}_i$
($\underline{x}^{(0)}$ in axiom 3', section
\ref{subsubsec:anastructbiholomorphiceq}).
This is because the singular directions are an essential feature of the edges and lead to a
discontinuity at $\infty$ when one compactifies the edge as shown in 
figure \ref{fig:torusDiscont}. 

\subsection{Bayesian Inference of Spectral Functions}
\label{sec:bayesianinfspecfuncs}

In Ref.~\onlinecite{dirks} we used Vladimirov's integral representation in order to
reconstruct a function $\tilde A$ which was defined by equation
\eqref{eq:defAtilde0}
for the interacting system.
We chose a cone domain with orientation zero, $T^{C_{\varepsilon,0}}$, and 
assumed the constrained Green's function
$G|_{T^{C_{\varepsilon,0}}}$ to
be analytic for sufficiently small cone opening ratios $\varepsilon$.
This was justified, because the higher-order branch cuts of particle-hole character
(see figure \ref{fig:branchcuts}) occur only in high-order terms in $U$. 
The vertex-correction type of branch cut pointed out in Ref.~\onlinecite{han10} was ignored. 

Then the standard maximum entropy procedure\cite{mem} for inferring spectral functions 
from quantum Monte-Carlo data could be adopted to the inference of $\tilde A$
and therefore the spectral function.

The procedure was found to work well in the equilibrium limit, $\Phi=0$.
However, entering the nonequilibrium regime, the
{ill-posedness} 
of the inverse
problem increased. Similar to the intertwined geometric dependencies between
the function structures on edge and
wedge coming to the surface in the appendix' uniqueness proof, a geometric
dependency of the quality of Bayesian inference was found.

Decreasing the parameter $\varepsilon$ provided a limit to a holomorphic
function (leaving aside the vertex-correction branch cut) on the one hand, 
but on the other hand increased the ill-posedness of the inverse problem for a
finite-$\Phi$ spectral function. A discussion of how this is reflected by the
structure of the Poisson kernel function may be found in Ref.~\onlinecite{dirks}.

Apparently, the problem is very much related to restricting to the sheet
$T^{C_{\varepsilon,0}}$ only taking $G(\imag\varphi_m,\imag\omega_n)$ data from the
sheet into account, \emph{discarding the others}.

The only possible way to alleviate the increasing ill-posedness is to provide
a link between the holomorphic branches of the Green's function, being able
to take into account data from not one but several wedges in order to perform the
analytic continuation \eqref{eq:DefPhysicalLimit}.

\subsection{Bogolyubov's edge-of-the-wedge theorem}
\label{subsec:edgeofwedgetheorems}
A candidate of such a link was provided by Bogolyubov's famous edge-of-the-wedge theorem
in the context of axiomatic quantum field theory. It considered the analytic continuation of
Wightman functions\cite{vladimirovbook} in order to establish certain dispersion relations.
From a mathematical point of view, it also introduced a generalization of the
very notion of analytic continuation.\cite{vladimirovbook}

There are several versions of the theorem. A simple version which captures
the essential idea may be found in
the book by H\"ormander on partial differential operators
\cite{hoermanderPDE}.
It roughly considers two functions $f^\pm$ which are holomorphic on the tube cones
$T^{\pm C}$, where $C$ is a convex open cone with vertex at zero.
Consequently, the edges of $T^C$ and $T^{-C}$ are  ``infinitesimal neighbours''.
If the functions have the same boundary value distributions, $f^+(\underline
x + \imag 0^C) = f^-(\underline x - \imag 0^C)=:f_0$, then $f_0$ is an
analytic function. $f_0$ provides an analytic continuation of both, $f^+$ and
$f^-$.

In its more general formulations, the theorem actually demands the functions $f^\pm$ to be holomorphic only
locally at the edge and establishes certain facts about the domain in which
$f_0$ is analytic (global edge-of-the-wedge theorem).

An extension to several cones whose edges meet in a single point is Martineau's theorem.
As in Bogolyubov's theorem, locally, holomorphic functions may be found which 
constitute analytic continuations of pairs of functions living on wedges.
Again, the edge values of the considered set of functions have to be
interrelated in a more or less direct way.

\section{Systematic extension of the continuation procedure by use of edge relations}
\label{sec:introQ}

{
In the previous section, we systematically analyzed the function-theoretical
structure of the Green's function with regard to the two complex variables
$z_\omega$ and $z_\varphi$. The former comes along with the analytic
continuation of the fermionic Matsubara frequency $\imag\omega_n$ associated to the dynamical
properties of the effective-equilibrium systems. The latter comes along with
the analytic continuation with respect to the Matsubara voltage. A fundamental property of the Green's function with regard to
the two variables is the branch cut structure shown in figure
\ref{fig:branchcuts}. It separates the holomorphic sheets of the Green's
function which live on wedges. Their edges meet in a branch point. 
For a holomorphic sheet, we were able to derive 
an integral representation of the Green's function with regard to two real
variables, using kernel functions such the one in Eq.~\eqref{eq:poissonkernelepsdomain}.
By this, the functions values on the wedge are represented linearly by boundary values
on its edge and vice versa. The physical limit \eqref{eq:DefPhysicalLimit} of the theory
corresponds to approaching the branch point in figure \ref{fig:branchcuts} along a certain
direction. In order to use data from several wedges for
physical results, it is thus necessary to find more or less explicit relations between 
function values on edges of different wedges.
The so-called edge-of-the-wedge theorem (section
\ref{subsec:edgeofwedgetheorems}) provides some insight along this line.
In order to construct an explicit functional-analytic approach to the
analytic continuation which would enable us to extend the numerical
implementation of the maximum entropy 
approach, it is only of indirect use, however.
}

{
It is clear that any simple relation between edges of the different
branches of the Green's function provides a rather direct link between
integral representations of the respective wedges. Based on a continuity
approximation to function values at the branch point around which the edges are aligned, 
the present section derives the MaxEnt procedure which was used to infer
the numerical results of section \ref{sec:results}.
}

\subsection{Continuous real part at branch point}

{
The relation is an exact identity of the bare Green's function. Namely $\Re G_0(\underline x +
\imag 0^\vartheta)$ is identical for any
edge orientation $\vartheta$ of the bare Green's function, see equation \eqref{eq:ReG0Edge}.
It is easy to verify that same is true for the second-order self-energy
\eqref{eq:Sigma2ndOrder} and also for the functions which are parametrized by Han and
Heary's original fitting approach in reference \onlinecite{prl07}.
}

We have
\begin{eqnarray}
\Re G_0(\underline x + \imag 0^\vartheta) &=&
\Re G_0(\underline x + \imag 0^{\vartheta'}),  
\label{eq:sharedRealPartG0}
\\
\Re \Sigma^{(2)}(\underline x + \imag 0^\vartheta) &=&
\Re \Sigma^{(2)}(\underline x + \imag 0^{\vartheta'}),
\label{eq:sharedRealPartSigma2}
\end{eqnarray}
for all $\vartheta,\vartheta'\in [0,2\pi)$.
This structure is similar to the conventional Green's function causality
relation 
\begin{equation}
G(z^*) = G(z)^*.
\label{eq:CausalityConventionalG}
\end{equation}
 There, we consequently have 
\begin{equation}
\Re G(\omega+\imag 0^+) = \Re G(\omega-\imag 0^+).
\label{eq:CausalityConventionalRealPart}
\end{equation}
However, in our case we only know for sure the symmetry 
\begin{equation}
G(z_\varphi^*, z_\omega^*) = G(z_\varphi, z_\omega)^*.
\label{eq:symmrelG}
\end{equation}
From equation \eqref{eq:symmrelG} only the edge relation
\begin{equation}
\Re G(\underline x + \imag 0^{\vartheta + \pi}) = 
\Re G(\underline x + \imag 0^{\vartheta}),\, \vartheta \in [0,\pi)
\label{eq:exactEdgeSymmetryRelation}
\end{equation}
can be derived. I.e.~conjugate wedges $T^C$, $T^{-C}$ carry the same real
parts of $G(\underline z)$ on their edges.

\subsection{Range of the continuity assumption}
\label{subsec:rangesharedrp}
We will now investigate to 
{what} extent the relations 
\eqref{eq:sharedRealPartG0} and \eqref{eq:sharedRealPartSigma2} 
also hold for higher-order contributions to the fully interacting Green's 
function $G(\underline z)$, i.e.~to {what} extent we a-priori expect the
approximation
\begin{equation}
\Re G(\underline x + \imag 0^{\vartheta}) \approx\Re G(\underline x + \imag 0^{\vartheta=0})
\label{eq:sharedRPassumption}
\end{equation}
to hold.
It is insightful to study the algebraic properties of a conventional Green's 
function $G(z)$ first. Subsequently, the two-variable function
$G(\underline z)$ is discussed with respect to its formal structure and 
regarding empirical findings from the continuous-time QMC simulation data.

\paragraph{Conventional Green's function.}
As a simple example, let us consider the summation of the Dyson series\cite{negele}
\begin{eqnarray}
\label{eq:DysonSeries}
G &=& G_0 - G_0 \Sigma G \\ 
  &=& G_0 - G_0\Sigma G_0 + G_0 \Sigma G_0 \Sigma G_0  \cdots.
\nonumber
\end{eqnarray}
The entities $G(z)$, $G_0(z)$, and $\Sigma(z)$ satisfy the causality relation
\eqref{eq:CausalityConventionalG}. For the equation to hold, the
product (and the sum) of two \eqref{eq:CausalityConventionalG}-satisfying
quantities $A(z)$, $B(z)$ 
shall also satisfy \eqref{eq:CausalityConventionalG}.
This is obviously the case, because 
\begin{eqnarray*}
\Re (AB)(\omega + \imag 0^+) &=& 
\Re A(\omega + \imag 0^+) \Re B(\omega + \imag 0^+)  \\
&& - \Im A(\omega + \imag 0^+) \Im B(\omega + \imag 0^+) \\
&=& \Re A(\omega - \imag 0^+) \Re B(\omega - \imag 0^+) \\
&&  - (-\Im A(\omega - \imag 0^+)) \\
&&   (-\Im B(\omega - \imag 0^+)) \\
&=& \Re (AB)(\omega - \imag 0^+).
\end{eqnarray*}
Same can be shown for the imaginary part.
{
It is crucial to recall that the mutual conjugation of \emph{imaginary parts}
of upper and lower functions has been used explicitly for closedness of
\eqref{eq:CausalityConventionalG} under multiplication.
}
In other words: the set of functions with \emph{only} the property
\eqref{eq:CausalityConventionalRealPart} is not closed under multiplication.

\paragraph{Two-variable Green's function.} 
The closedness under multiplication is in general violated for functions with
solely a continuous real part on the branch point.
This is due to the fact that no statement about the imaginary part is
made, and case of the causality relation \eqref{eq:CausalityConventionalG},
conjugateness of the imaginary parts is needed for closedness of the real
part's continuity under multiplication.
 For instance, one can
easily verify that $G_0(\underline z) \cdot
G_0(\underline z)$ yields different real parts on the edges. Same can be
shown for $G_0(\underline z)^{-1}$.

\subsection{Structure of the residual term}

\label{subsec:strucresterm}
{
Nevertheless, as discussed in this paragraph, we are able to show that the continuity assumption
\eqref{eq:sharedRPassumption} is recovered for a certain energy range. 
Additionally, empirical findings for the structure
of CT-QMC data, as discussed in appendix \ref{app:empiricalrelations},
partially support the assumption by observing continuity relations between
edge functions. Last but not least, the assumption is justified a posteriori
for a rather large collection of wedges via the obtained numerical results (see
section \ref{subsec:numericalImplementationQ} for further discussion).
It is found in section \ref{subsec:numericalImplementationQ}
that including upper (and lower) wedges, $|\omega_n|>|\varphi_m/2|$
(and $|\omega_n|<-|\varphi_m/2|$) into the considered collection of wedges
causes the numerical procedure to fail. Otherwise, it converges. This
observation is coherent with a strong violation of the continuity assumption
at the principal branch cuts $\Im z_\omega = \pm \Im z_\varphi/2$.
}

It is insightful to study how the resulting difference of two given edge
functions is structured, namely to study the local residue
\begin{equation}
R^{(\vartheta,\vartheta')}(\underline x)
:=
\Re G(\underline x + \imag 0^\vartheta) - \Re G(\underline x + \imag
0^{\vartheta'})
\end{equation}
for arbitrary values $\vartheta,\vartheta'\in[0,2\pi)$ as a function of $\underline x \in
\mathbb{R}^2$.

\paragraph{Angular structure.}

Obviously, due to Eq.~\eqref{eq:exactEdgeSymmetryRelation} we have
\begin{equation}
R^{(\vartheta,\vartheta)} \equiv R^{(\vartheta,\vartheta+\pi)} \equiv 0,
\end{equation}
for all $\vartheta$.
$R^{(\vartheta,\vartheta)} \equiv 0$ if $\vartheta$ and $\vartheta'(+\pi)$ belong
to the same wedge.

\paragraph{Structure due to continuity of imaginary-time data.}
Using the empirical fact that the continuous-time quantum Monte Carlo data 
$\Sigma(\imag \varphi_m, \imag \omega_n)$ are continuous as a function of
$\varphi_m$ and $\omega_n$, we can derive certain continuity relations for
$R^{(\vartheta,\vartheta')}$. They are provided in appendix
\ref{app:empiricalrelations}.

\paragraph{High-energy structure.}
Let us also consider the high-energy limit 
\begin{equation}
|\underline x| \gg 
\max\{\Gamma, |U|, |\Phi|, |\epsilon_d|\}.
\end{equation}
For this, $|G(\underline x + \imag 0^\vartheta)|$ is 
significantly larger 
than zero {only} if $x_\omega\approx \pm x_\varphi/2$, according to point 3',
section \ref{subsubsec:anastructbiholomorphiceq}. The ``$+$'' and ``$-$''
cases imply a separation of energy scales.
\emph{For both of these two energy scales, the closedness under multiplication is
recovered.}

One can easily see the recovery of the multiplicative structure in the
high-energy limit by investigating
the bare Green's function \eqref{eq:edgelimitG0}. The ``$+$'' and ``$-$''
energy scales are given by the $\alpha = -$ and $\alpha =+$ addends
\eqref{eq:edgelimitG0Addends} in
Eq.~\eqref{eq:edgelimitG0}, respectively. Each of the addends satisfies the
multiplication rule, because the absolute value of their imaginary part
remains the same for all $\vartheta$ at fixed $\underline x$. 
Same is true for the sum of the corresponding $\sum_i\alpha_i = \mp 1$ addendends in the second-order self-energy 
\eqref{eq:Sigma2ndOrder}.

Therefore, we conclude that
\begin{equation}
\lim_{\underline x \to \infty}R^{(\vartheta,\vartheta')} = 0 \text{ for all
$\vartheta,\vartheta'$},
\end{equation}
in contrast to the limiting behaviour of $G$ itself.

\paragraph{Consequences for the analytic structure.}

Hence, although the assumption \eqref{eq:sharedRPassumption}
is apparently only approximate, the error $R^{(\vartheta,\vartheta')}(\underline x)$ which is being done by assuming the relation
is localized around $0$ in the $\underline x$-space.
It is remarkable that the intermediate-coupling numerical data presented in 
section \ref{sec:results} appear to be rather precise in the low-energy
region, 
although the violation terms are a-priori expected to be strong at low energies.

From the a-priori perspective, the assumption \eqref{eq:sharedRPassumption} gives a correct picture of how
the wedges are related in the high-energy range. When assuming the relation,
additional low-energy degrees of freedom have to be introduced in order to
reobtain an exact continuation theory (cf.~section \ref{sec:unbiasedQ}). Empirical data discussed in appendix
\ref{app:empiricalrelations} indicate that these
degrees of freedom are comparably well-behaved.

\subsection{Functional-analytic consequences of the shared-real-part assumption}

We will see that the continuity assumption \eqref{eq:sharedRPassumption}
leads to a complete description of the entire function $G(\underline z)$ on
\emph{all} wedges \emph{only} as a function of the single edge $\Im G(\underline x + \imag 0^{\vartheta =0})$. 
This is extraordinarily attractive from a numerical point of view, because by this, the number of degrees of freedom when doing the
maximum entropy inference is not increased, but \emph{all} imaginary-time
theory data $G(\imag\varphi_m,\imag\omega_n)$ may be taken into account,
without any a-priori constraint. As in the single-wedge approach of
Ref.~\onlinecite{dirks}, 
the spectral function can still be directly extracted from the MaxEnt result.
We will see that for functions which comply with
the assumption, it in fact alleviates the ill-posedness of the inverse problem, as desired.

\paragraph{Construction of the kernel.}
{
Starting from equation \eqref{eq:sharedRPassumption} we can derive a
representation of the Green's function
with respect to $\Im G(\underline x + \imag 0^{\vartheta =0})$ in the following way.
}
First, we introduce the Hilbert transform operator $\mathcal{H}$ as
\begin{equation}
(\mathcal{H}f)(\underline x) := \frac{1}{\pi} \mathcal{P}\!\!\!\!\!\!\int
\!\!\Dfrtl{x_2'}
\frac{f(x_1, x_2')}{x_2-x_2'}.
\end{equation}
Then we can use condition 3' (section \ref{subsec:holostructGF}) in order to
apply the Hilbert transform for computing real and imaginary part from each
other on the boundary of certain $\mathbb{H}$-isomorphic complex lines.

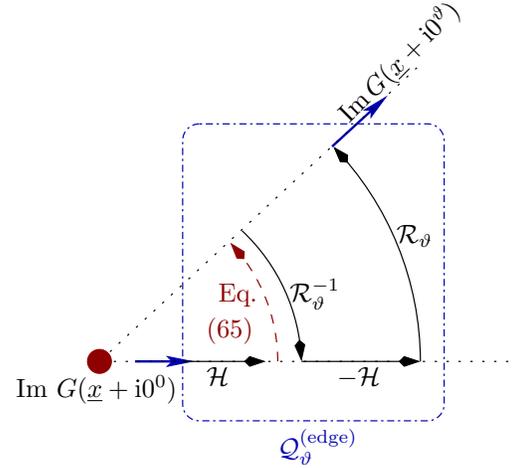
\begin{figure}
\resizebox{0.9\linewidth}{!}{
\begin{picture}(0,0)%
\includegraphics{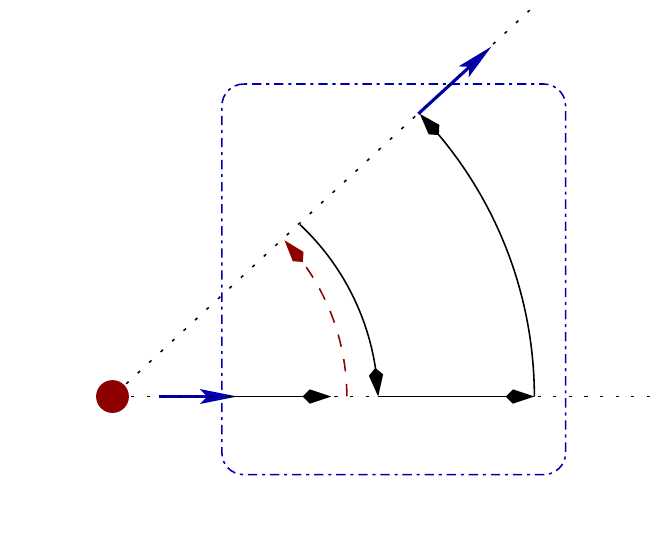}%
\end{picture}%
\setlength{\unitlength}{3947sp}%
\begingroup\makeatletter\ifx\SetFigFont\undefined%
\gdef\SetFigFont#1#2#3#4#5{%
  \reset@font\fontsize{#1}{#2pt}%
  \fontfamily{#3}\fontseries{#4}\fontshape{#5}%
  \selectfont}%
\fi\endgroup%
\begin{picture}(3177,2551)(2236,-4325)
\put(3901,-4261){\makebox(0,0)[lb]{\smash{{\SetFigFont{10}{12.0}{\familydefault}{\mddefault}{\updefault}{\color[rgb]{0,0,.69}$\mathcal{Q}^\text{(edge)}_\vartheta$}%
}}}}
\put(4651,-2911){\makebox(0,0)[lb]{\smash{{\SetFigFont{10}{12.0}{\familydefault}{\mddefault}{\updefault}{\color[rgb]{0,0,0}$\mathcal{R}_\vartheta$}%
}}}}
\put(4351,-2161){\rotatebox{45.0}{\makebox(0,0)[lb]{\smash{{\SetFigFont{10}{12.0}{\familydefault}{\mddefault}{\updefault}{\color[rgb]{0,0,0}$\Im G(\underline x+\imag0^\vartheta)$}%
}}}}}
\put(3976,-3286){\makebox(0,0)[lb]{\smash{{\SetFigFont{10}{12.0}{\familydefault}{\mddefault}{\updefault}{\color[rgb]{0,0,0}$\mathcal{R}_\vartheta^{-1}$}%
}}}}
\put(3451,-3511){\makebox(0,0)[lb]{\smash{{\SetFigFont{10}{12.0}{\familydefault}{\mddefault}{\updefault}{\color[rgb]{.56,0,0}\eqref{eq:sharedRPassumption}}%
}}}}
\put(3526,-3286){\makebox(0,0)[lb]{\smash{{\SetFigFont{10}{12.0}{\familydefault}{\mddefault}{\updefault}{\color[rgb]{.56,0,0}Eq.}%
}}}}
\put(2251,-3886){\makebox(0,0)[lb]{\smash{{\SetFigFont{10}{12.0}{\familydefault}{\mddefault}{\updefault}{\color[rgb]{0,0,0}$\Im\,G(\underline x + \imag 0^{0})$}%
}}}}
\put(4276,-3811){\makebox(0,0)[lb]{\smash{{\SetFigFont{10}{12.0}{\familydefault}{\mddefault}{\updefault}{\color[rgb]{0,0,0}$-\mathcal{H}$}%
}}}}
\put(3451,-3811){\makebox(0,0)[lb]{\smash{{\SetFigFont{10}{12.0}{\familydefault}{\mddefault}{\updefault}{\color[rgb]{0,0,0}$\mathcal{H}$}%
}}}}
\end{picture}%
}
\caption{(color online) Action of the operator
$-\mathcal{Q}_{\vartheta}^\text{(edge)}$. 
{
It translates between functions living on edges of wedges with two different angular orientations. The
orientation $\vartheta$ is the one of the considered
data wedge, and $0$ is the orientation of the physical limiting procedure
\eqref{eq:DefPhysicalLimit}. Initially acting on the physical edge function, the 
consecutive formal operations which comprise
$-\mathcal{Q}_{\vartheta}^\text{(edge)}$ either change the angular rotation
or leave it invariant, as indicated by the respective arrows.
}
}
\label{fig:pictrepQedge}
\end{figure}

One can show that 
\begin{equation}
\Im G|_{T^{C_{r,\vartheta}}} = -\frac{1}{\pi}\mathcal{Q}_{r,\vartheta}
\cdot \Im G(\underline x + \imag 0^{\vartheta =0}),
\label{eq:inverseProblemQ}
\end{equation}
where we introduced the operator
\begin{equation}
\mathcal{Q}_{r,\vartheta} :=
 \mathcal{P}_{r,\vartheta}
\mathcal{R}_\vartheta \mathcal{H} \mathcal{R}^{-1}_\vartheta
\mathcal{H}.
\label{eq:QoperatorDefinition}
\end{equation}
Let us also introduce a symbol for the right part of the operator sequence,
\begin{equation}
\mathcal{Q}_{\vartheta}^\text{(edge)}
:= \mathcal{R}_\vartheta \mathcal{H} \mathcal{R}^{-1}_\vartheta
\mathcal{H}.
\end{equation}

The action of $-\mathcal{Q}_{\vartheta}^\text{(edge)}$ on $\Im G(\underline x
+ \imag 0^{\vartheta=0})$ is depicted in figure
\ref{fig:pictrepQedge}: First, the Hilbert transform $\mathcal{H}$ with respect to the
$x_\omega$ variable yields $\Re G(\underline x + \imag 0^{\vartheta =0})$.
Then, via equation \eqref{eq:sharedRPassumption} it is identified with $\Re
G(\underline x + \imag 0^{\vartheta})$. In order to obtain $\Im
G(\underline x + \imag 0^{\vartheta})$ one formally has to transform to
the biholomorphic equivalent of $G|_{T^{C_{r,\vartheta}}}$ in the domain $T^{C_{r,0}}$ via
the operator $\mathcal{R}_\vartheta^{-1} = \mathcal{R}_{-\vartheta}$. The inverse Hilbert transform $-\mathcal{H}$ yields 
the imaginary part of the edge value of the function $T^{C_{r,0}} \to \mathbb{C}$, 
$\underline z \mapsto (\mathcal{R}^{-1}_\vartheta (G|_{T^{C_{r,\vartheta}}}))(\underline z)$.
Transforming the function back to the edge of $T^{C_{r,\vartheta}}$ using
 $\mathcal{R}_\vartheta$ yields
the result $\Im G (\underline x + \imag 0^\vartheta)$.

Using the Poisson kernel $\mathcal{P}_{r,\vartheta}$,
Eq.~\eqref{eq:poissonkernelrotatedepsdomain}, the Green's function is
obtained in the desired wedge $T^{C_{r,\vartheta}}$. The entire procedure is contained in 
$\mathcal{Q}_{r,\vartheta}$.

\paragraph{Feature of $\mathcal{Q}_{\vartheta}^\text{(edge)}$:
decoding branch cut geometry from single edge function.}
\label{paragraph:propQedge}
The operator $\mathcal{Q}_{\vartheta}^\text{(edge)}$ is a map of pure 
``edge'' character. Therefore, it is worthwhile to study it separately.
$\mathcal{Q}_{\vartheta}^\text{(edge)}$ is well-defined for any
square-integrable function $f(\underline x)$, no matter which orientation $\vartheta$
is considered.

Considering an edge function which is compatible with the Green's function properties 1,2,3' 
(section \ref{subsec:holostructGF}), $\mathcal{Q}_{\vartheta}^\text{(edge)}$
is not defined for the singular orientations \eqref{eq:singularorientations}.
For example, a straightforward calculation shows that applying $\mathcal{Q}_{\vartheta}^\text{(edge)}$ step by step to
$G_0(\underline x + \imag 0^{\vartheta=0})$ yields exactly the formula
\eqref{eq:edgelimitG0}, with the switching behaviour
\eqref{eq:edgelimitG0Switch} whose value is undefined for the
orientations \eqref{eq:singularorientations}. The missing square-integrability of the edge
functions along these directions is the corresponding mathematical reason. In particular, 
whenever $\mathcal{Q}_{\vartheta}^\text{(edge)}$ crosses a
singular orientation of $\Im G_0(\underline x + \imag 0^{\vartheta=0})$, it
exactly generates the jump 
in $\Im G_0(\underline x + \imag 0^{\vartheta})$ as a function of $\vartheta$.

Consequently, assuming \eqref{eq:sharedRPassumption} is correct, both, not
only the holomorphic structure, but also the complete information about the entire \emph{branch
cut structure}, namely the exact geometry of the branch cuts, are encoded in 
the single edge function $\Im G(\underline x + \imag 0^0)$. 
Same is true for the $(\vartheta=0)$-edge limit of the second-order
self-energy, $\Im \Sigma^{(2)}(\underline x + \imag 0^0)$, due to equation
\eqref{eq:sharedRealPartSigma2}. It is always square-integrable, except for
 the directions $x_\omega = -3/2 x_\varphi$, $x_\omega = -
x_\varphi /2$, $x_\omega = x_\varphi / 2$, $x_\omega = 3/2 x_\varphi$, namely
for the geometry of the 2nd-order branch cuts.

{
For practical computations, we find that an exploitation of symmetries of $\mathcal{Q}_{\vartheta}^\text{(edge)}$
is mandatory. Those are \emph{translational invariance} and \emph{scale
invariance}, but \emph{no rotational invariance}: 
}
For the translation operator
\begin{equation}(\mathcal{T}_{\underline{x}'}f)(\underline{x}) := f(\underline{x} - \underline{x}'),
\label{eq:definitionTranslationOp}
\end{equation} and 
for the homogenous scaling operator 
\begin{equation}(\Lambda_{\lambda}f)(\underline{x}) := \lambda^2
f(\lambda\underline{x}),\  \lambda > 0,
\label{eq:definitionScalingOp}
\end{equation}
we have
\begin{eqnarray}
\left[\mathcal{Q}_{\vartheta}^\text{(edge)},\mathcal{T}_{\underline x} \right] &=& 0, 
\label{eq:commutatorQthetaTranslation}
\\
\left[\mathcal{Q}_{\vartheta}^\text{(edge)},\Lambda_{\lambda} \right] &=& 0, 
\label{eq:commutatorQthetaScaling}
\\
\left[\mathcal{Q}_{\vartheta}^\text{(edge)},\mathcal{R}_{\vartheta'} \right]
&\neq& 0.
\label{eq:commutatorQthetaRotation}
\end{eqnarray}
The proof of these commutator relations is provided in appendix
\ref{sec:commrelQedge}. Note that because \emph{directional} scaling implies a
shear and therefore nonconserved angles in the shapes of $\Im G(\underline x +
\imag 0^0)$, it is no symmetry of the operator,
in contrast to uniform scaling.

\subsection{Numerical implementation of $\mathcal{Q}_{r,\vartheta}$}
\label{subsec:numericalImplementationQ}

The numerical implementation of the kernel $\mathcal{Q}_{r,\vartheta}$ is
nontrivial. Assuming that $G(\underline x + \imag 0^0)$ is sufficiently
smooth, we can represent it by superimposing localized test functions which
span the space of edge functions.

\paragraph{Integral structure of the mapping.}
$\mathcal{Q}_{r,\vartheta}$ introduces a quadruple integral. The first two
integrals are the two principal value integrals which come with the Hilbert transforms.
The second ones are included by the Poisson kernel $\mathcal{P}_{r,\vartheta}$.
The integrations are formally very similar to a sequence of convolutions
$A*(B*(C*e))$, where $e$ is an edge function.
A crucial point is that due to the distributional nature of both, the
principal values and the edge functions, the associativity rule 
cannot be expected to hold for these convolutions (see section 4.2 in
Ref.~\onlinecite{hoermanderPDE}): Principal value and our type of edge functions
(functions with singular directions, Eqs.~\eqref{eq:singularorientations}) are no
distributions with compact support.
 Therefore, it is impossible to simply
contract some ``inner integrals'' within $\mathcal{Q}_{r,\vartheta}$
analytically in order to obtain a simple kernel function for
$\mathcal{Q}_{r,\vartheta}$.
The use of a set of test functions which spans the space of edge functions
is mandatory.

\paragraph{Construction of the test functions.}
The test functions would preferably be structured in a way which allows the
quadruple integral in the operator $\mathcal{Q}_{r,\vartheta}$ to be solved
essentially analytically. 
{
Using the translation operators
$\mathcal{T}_{\underline X}$, Eq.~\eqref{eq:definitionTranslationOp}, and scaling operators
$\Lambda_{1/\varepsilon}$, Eq.~\eqref{eq:definitionScalingOp}, we define the functions
\begin{equation}
f_{\underline{X},\varepsilon} := \mathcal{T}_{\underline X} \Lambda_{1/\varepsilon} f,
\label{eq:numtestfuncQgeneral}
\end{equation}
with
\begin{equation}
f(\underline x) := \frac{1}{\pi^2} \prod_{\alpha = \pm 1} \frac{1}{(x_\omega - \alpha x_\varphi / 2)^2 + 1}.
\label{eq:numtestfuncQspecial}
\end{equation}
They turn out to be a good choice as test functions:
}
First, we have the Dirac delta distribution
\begin{equation}
\lim_{\varepsilon \to 0} f_{\underline{X},\varepsilon} = \delta (\underline x
- \underline X)
\label{eq:numtestfuncQgeneralconvdelta}
\end{equation}
as a limit.
{
Second, due to the symmetries \eqref{eq:commutatorQthetaTranslation} and
\eqref{eq:commutatorQthetaScaling}, the use of scaling and translation operators yields 
-- regarding the action of the integrals in $\mathcal{Q}^\text{(edge)}_{\vartheta}$ -- the much more simple expression
\begin{equation}
(\mathcal{Q}_{r,\vartheta}f_{\underline{X},\varepsilon})(\underline x) = 
(\mathcal{P}_{r,\vartheta}\mathcal{T}_{\underline X}\Lambda_{1/\varepsilon}
(\mathcal{Q}^\text{(edge)}_{\vartheta}f))(\underline x)
\label{eq:Qsimplified}
\end{equation}
rather than
$(\mathcal{P}_{r,\vartheta}(\mathcal{Q}^\text{(edge)}_{\vartheta}\mathcal{T}_{\underline X}\Lambda_{1/\varepsilon}
f))(\underline x)$ as a matrix element of $\mathcal{Q}_{r,\vartheta}$.
}
Third, the simple pole structure of \eqref{eq:numtestfuncQspecial} allows us
to compute most of the integrals analytically.
Note that the simpler looking symmetric Lorentzian function $\frac{1}{x_\omega^2 + x_\varphi^2 +
1}$ is, in fact, no good alternative to $f$, because the poles
with respect to $z_\omega$ or $z_\varphi$ contain square roots of
$z_\varphi$ or $z_\omega$, respectively. Similar problems arise for localized
Gaussians. 

The directional arbitrariness $x_\omega \pm x_\varphi / 2$ arising in
Eq.~\eqref{eq:numtestfuncQspecial} from
choosing a product of 1-dimensional Lorentzians in
\eqref{eq:numtestfuncQspecial} is still to be discussed.
For example, one could also have chosen it to be $x_\omega \pm x_\varphi$, adjusting the
normalization factor from $\frac{1}{\pi^2}$ to $\frac{2}{\pi^2}$ in order to assert
\eqref{eq:numtestfuncQgeneralconvdelta}. A conceptional advantage of our choice of $f$
is, however, that for any domain $T^{C_{r,\vartheta}}$, for which $G$ is
holomorphic, we have $r \leq 2$. Consequently, due to the pole structure of
\eqref{eq:numtestfuncQspecial}, $f$ is holomorphic in the domain
$T^{C_{r,0}} \subset T^{C_{2,0}}$, whose edge is the starting point of the
$\mathcal{Q}_{r,\vartheta}$ transform. Nevertheless, a certain ambiguity
remains which could be technically useful.

\paragraph{Computation of the matrix elements.} 
We found it feasible to calculate at least the first three integrals of the
right-hand side 
of expression \eqref{eq:Qsimplified} analytically, using a computer algebra system.
In order to compute the fourth integral, an adaptive numerical quadrature can
be used.

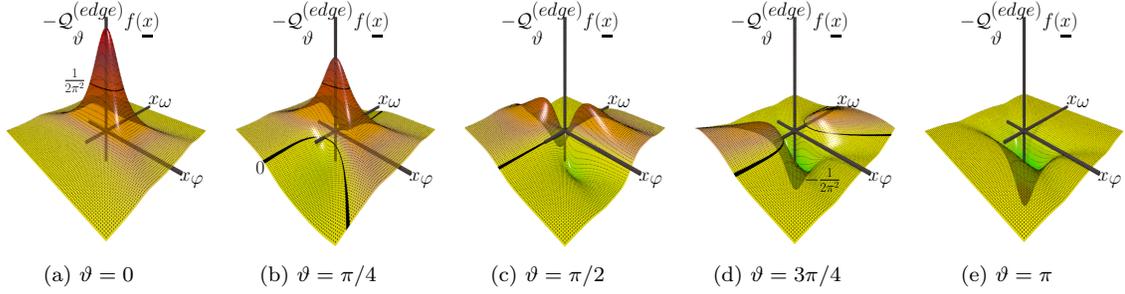
\begin{figure*} 
\subfloat[$\vartheta=0$]{
     \resizebox{0.18\linewidth}{!}{
\begin{picture}(0,0)%
\includegraphics[width=0.7\textwidth]{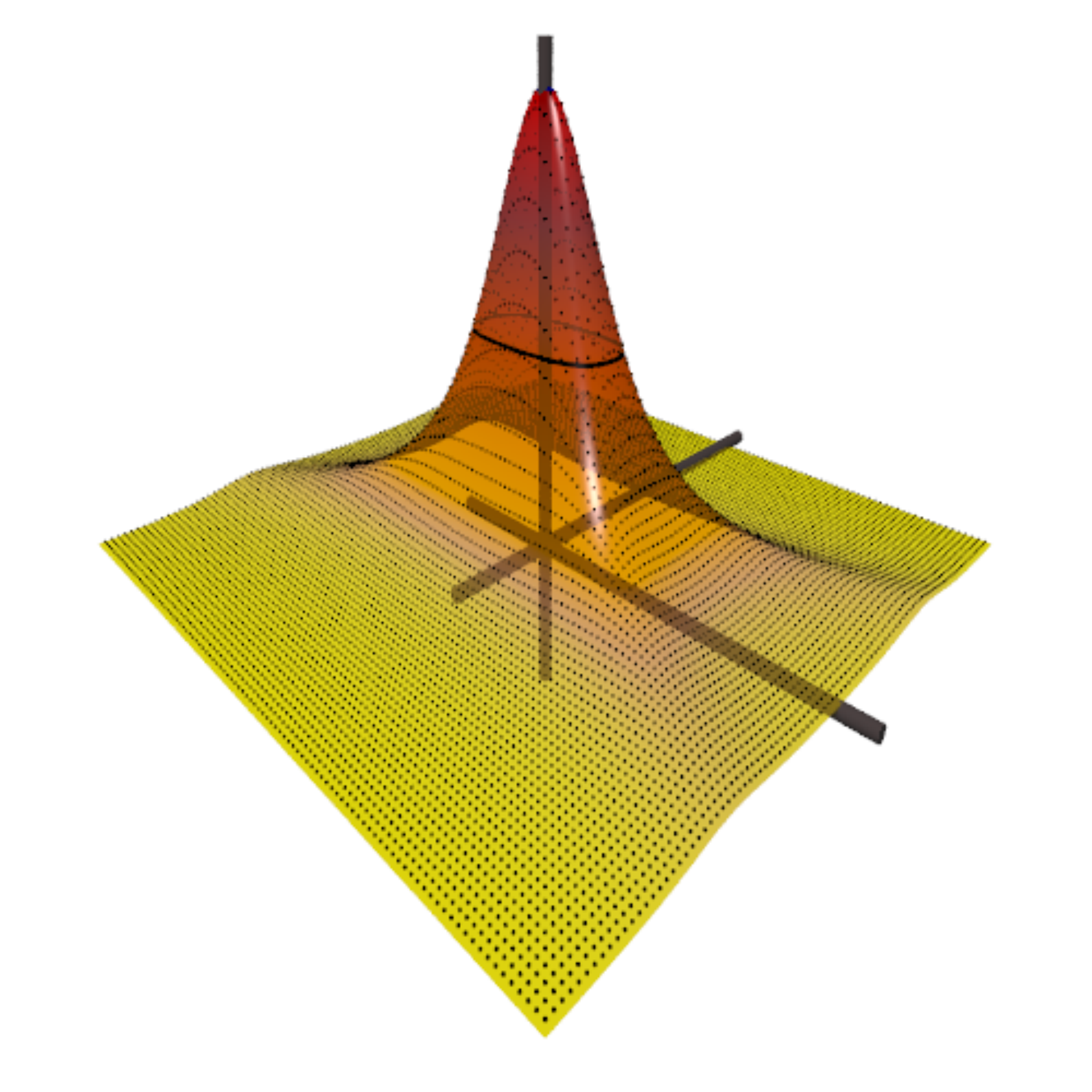}%
\end{picture}%
\setlength{\unitlength}{3947sp}%
\begingroup\makeatletter\ifx\SetFigFont\undefined%
\gdef\SetFigFont#1#2#3#4#5{%
  \reset@font\fontsize{#1}{#2pt}%
  \fontfamily{#3}\fontseries{#4}\fontshape{#5}%
  \selectfont}%
\fi\endgroup%
\begin{picture}(4509,5310)(2789,-6099)              
\put(6976,-4561){\makebox(0,0)[lb]{\smash{{\SetFigFont{84}{16.8}{\familydefault}{\mddefault}{\updefault}{\color[rgb]{0,0,0}$x_\varphi$}%
}}}}
\put(6300,-2911){\makebox(0,0)[lb]{\smash{{\SetFigFont{84}{16.8}{\familydefault}{\mddefault}{\updefault}{\color[rgb]{0,0,0}$x_\omega$}%
}}}}
\put(4500,-2511){\makebox(0,0)[lb]{\smash{{\SetFigFont{24}{16.8}{\familydefault}{\mddefault}{\updefault}{\color[rgb]{0,0,0}$\frac{1}{2\pi^2}$}%
}}}}
\put(4026,-1261){\makebox(0,0)[lb]{\smash{{\SetFigFont{84}{16.8}{\familydefault}{\mddefault}{\updefault}{\color[rgb]{0,0,0}$-\mathcal{Q}_\vartheta^{(edge)}f(\underline x)$}%
}}}}
\end{picture}%
}
}
\subfloat[$\vartheta=\pi/4$]{
     \resizebox{0.18\linewidth}{!}{
\begin{picture}(0,0)%
\includegraphics[width=0.7\textwidth]{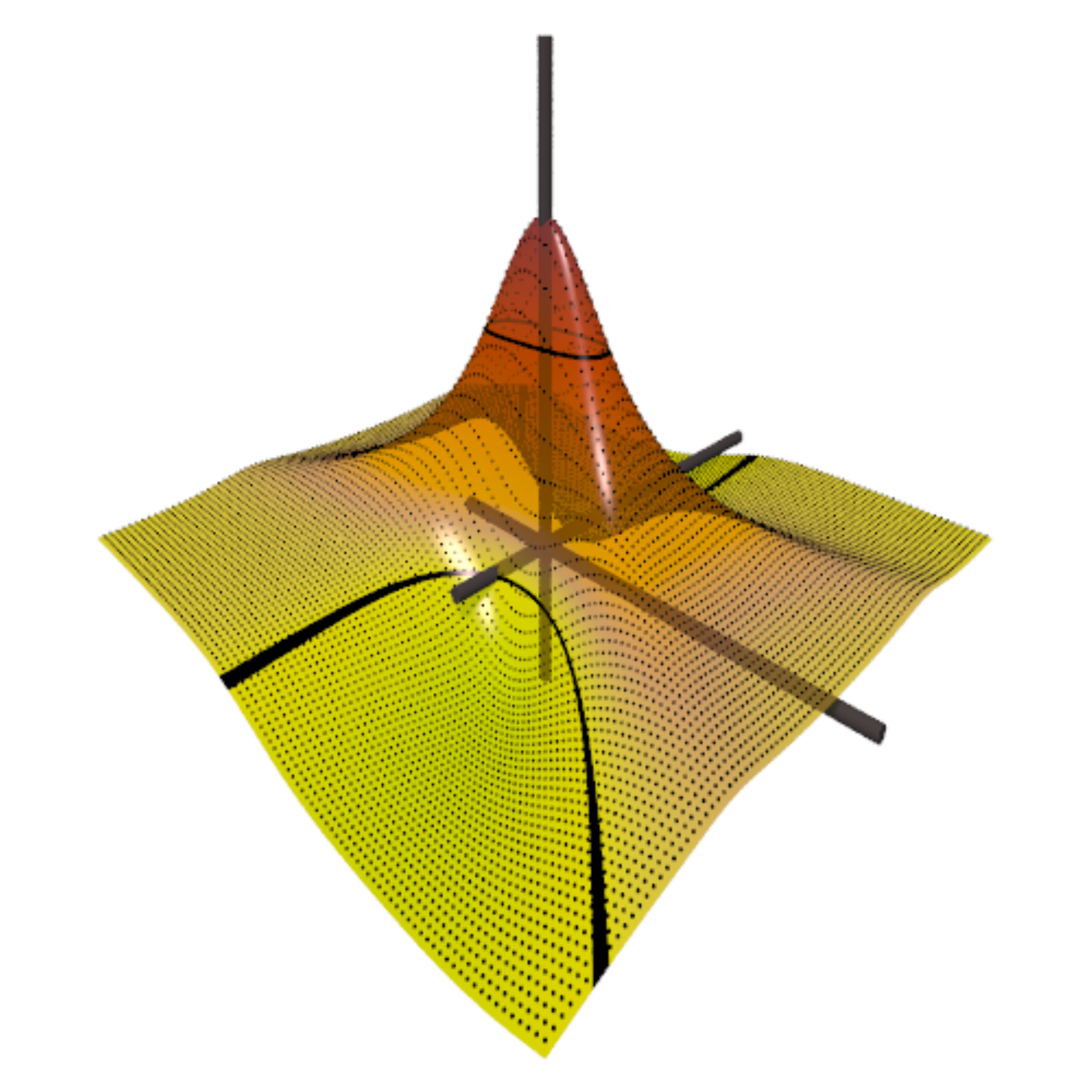}%
\end{picture}%
\setlength{\unitlength}{3947sp}%
\begingroup\makeatletter\ifx\SetFigFont\undefined%
\gdef\SetFigFont#1#2#3#4#5{%
  \reset@font\fontsize{#1}{#2pt}%
  \fontfamily{#3}\fontseries{#4}\fontshape{#5}%
  \selectfont}%
\fi\endgroup%
\begin{picture}(4509,5310)(2789,-6099)
\put(6976,-4561){\makebox(0,0)[lb]{\smash{{\SetFigFont{84}{16.8}{\familydefault}{\mddefault}{\updefault}{\color[rgb]{0,0,0}$x_\varphi$}%
}}}}
\put(6300,-2911){\makebox(0,0)[lb]{\smash{{\SetFigFont{84}{16.8}{\familydefault}{\mddefault}{\updefault}{\color[rgb]{0,0,0}$x_\omega$}%
}}}}
\put(4026,-1261){\makebox(0,0)[lb]{\smash{{\SetFigFont{84}{16.8}{\familydefault}{\mddefault}{\updefault}{\color[rgb]{0,0,0}$-\mathcal{Q}_\vartheta^{(edge)}f(\underline x)$}%
}}}}
\put(3686,-4401){\makebox(0,0)[lb]{\smash{{\SetFigFont{84}{16.8}{\familydefault}{\mddefault}{\updefault}{\color[rgb]{0,0,0}$0$}%
}}}}
\end{picture}%
}
}
\subfloat[$\vartheta=\pi/2$]{
     \resizebox{0.18\linewidth}{!}{
\begin{picture}(0,0)%
\includegraphics[width=0.7\textwidth]{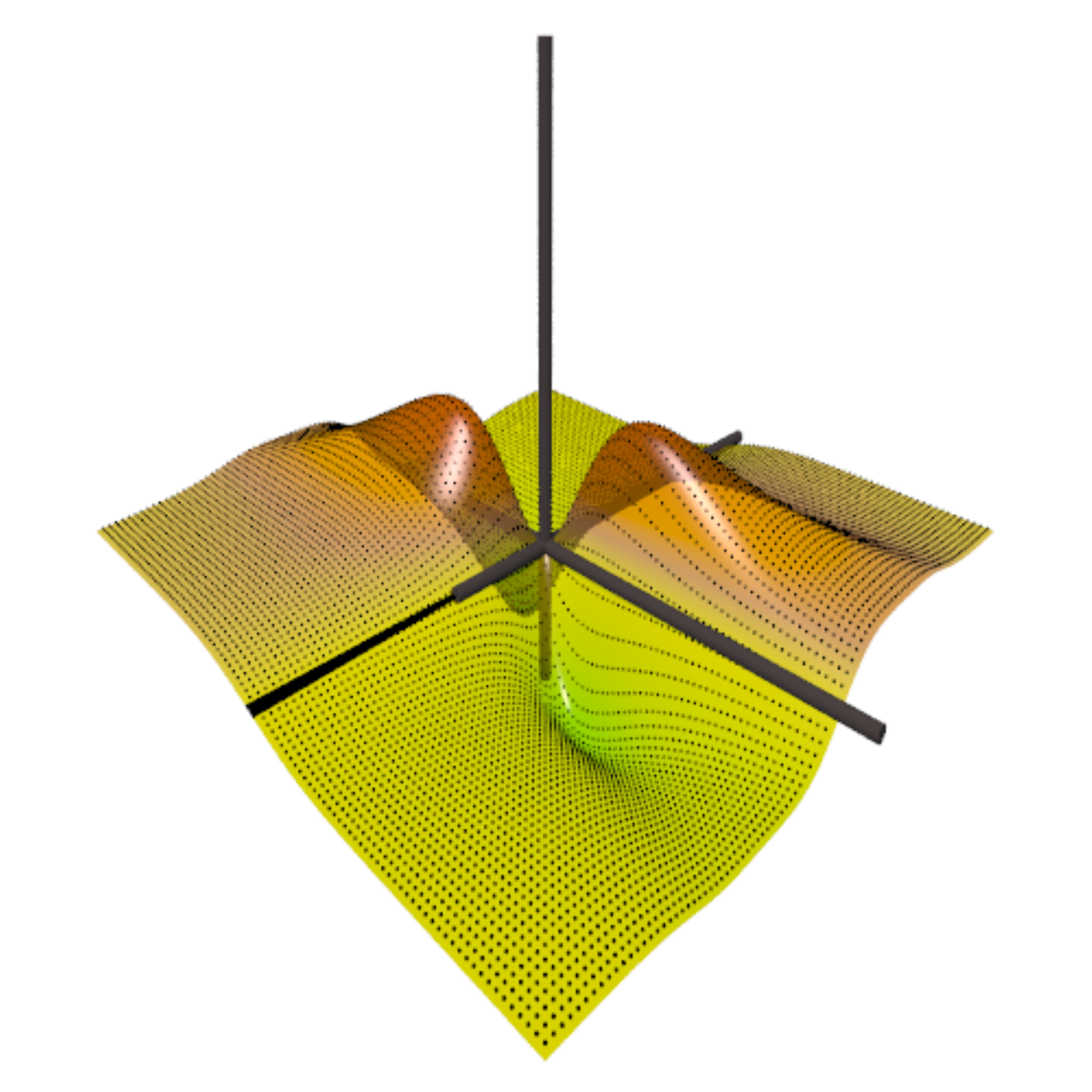}%
\end{picture}%
\setlength{\unitlength}{3947sp}%
\begingroup\makeatletter\ifx\SetFigFont\undefined%
\gdef\SetFigFont#1#2#3#4#5{%
  \reset@font\fontsize{#1}{#2pt}%
  \fontfamily{#3}\fontseries{#4}\fontshape{#5}%
  \selectfont}%
\fi\endgroup%
\begin{picture}(4509,5310)(2789,-6099)
\put(6976,-4561){\makebox(0,0)[lb]{\smash{{\SetFigFont{84}{16.8}{\familydefault}{\mddefault}{\updefault}{\color[rgb]{0,0,0}$x_\varphi$}%
}}}}
\put(6300,-2911){\makebox(0,0)[lb]{\smash{{\SetFigFont{84}{16.8}{\familydefault}{\mddefault}{\updefault}{\color[rgb]{0,0,0}$x_\omega$}%
}}}}
\put(4026,-1261){\makebox(0,0)[lb]{\smash{{\SetFigFont{84}{16.8}{\familydefault}{\mddefault}{\updefault}{\color[rgb]{0,0,0}$-\mathcal{Q}_\vartheta^{(edge)}f(\underline x)$}%
}}}}
\end{picture}%
}
     \label{subfig:Qedgetransformthetapi2}
}
\subfloat[$\vartheta=3\pi/4$]{
     \resizebox{0.18\linewidth}{!}{
\begin{picture}(0,0)%
\includegraphics[width=0.7\textwidth]{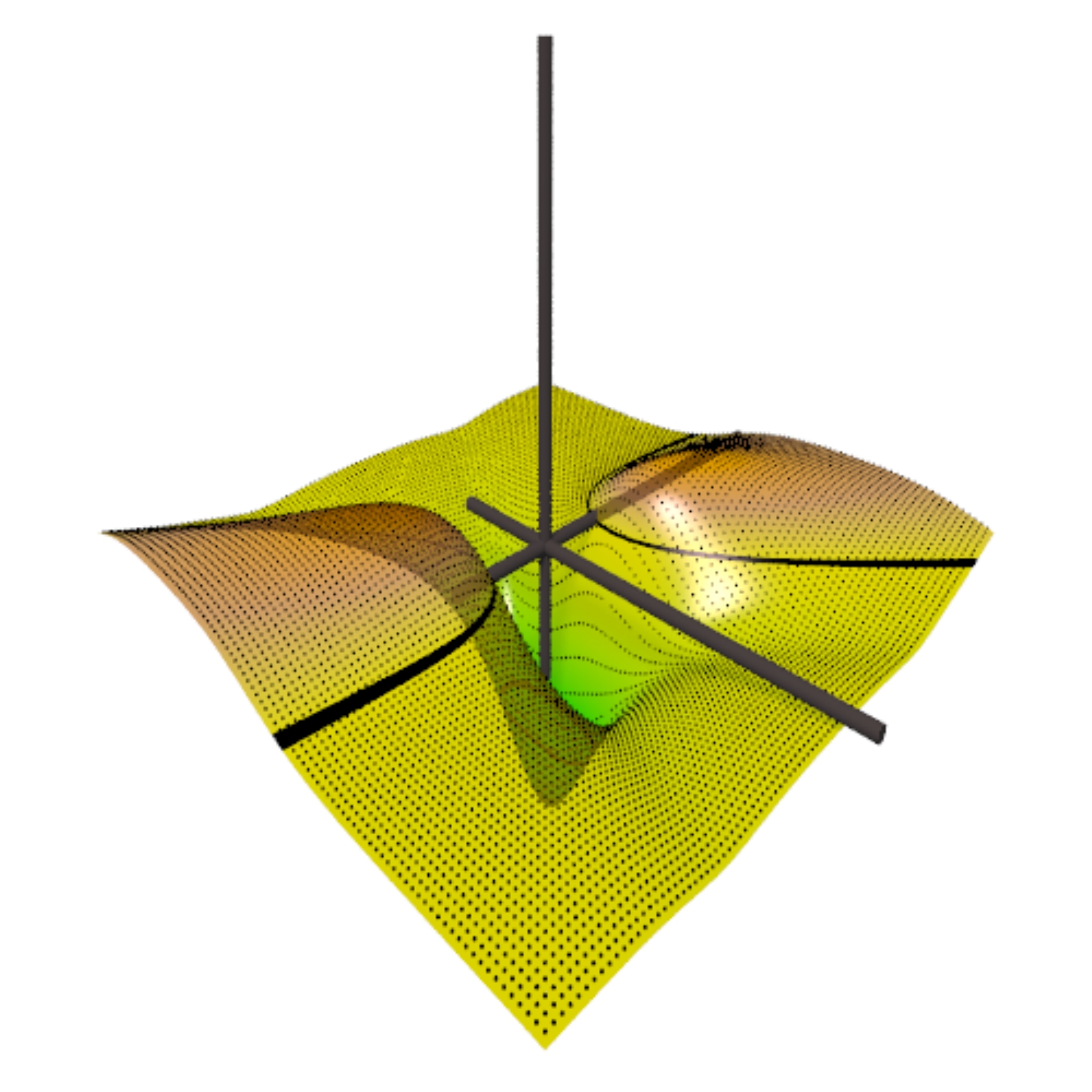}%
\end{picture}%
\setlength{\unitlength}{3947sp}%
\begingroup\makeatletter\ifx\SetFigFont\undefined%
\gdef\SetFigFont#1#2#3#4#5{%
  \reset@font\fontsize{#1}{#2pt}%
  \fontfamily{#3}\fontseries{#4}\fontshape{#5}%
  \selectfont}%
\fi\endgroup%
\begin{picture}(4509,5310)(2789,-6099)
\put(6976,-4561){\makebox(0,0)[lb]{\smash{{\SetFigFont{84}{16.8}{\familydefault}{\mddefault}{\updefault}{\color[rgb]{0,0,0}$x_\varphi$}%
}}}}
\put(6300,-2911){\makebox(0,0)[lb]{\smash{{\SetFigFont{84}{16.8}{\familydefault}{\mddefault}{\updefault}{\color[rgb]{0,0,0}$x_\omega$}%
}}}}
\put(4026,-1261){\makebox(0,0)[lb]{\smash{{\SetFigFont{84}{16.8}{\familydefault}{\mddefault}{\updefault}{\color[rgb]{0,0,0}$-\mathcal{Q}_\vartheta^{(edge)}f(\underline x)$}%
}}}}
\put(5600,-4641){\makebox(0,0)[lb]{\smash{{\SetFigFont{24}{16.8}{\familydefault}{\mddefault}{\updefault}{\color[rgb]{0,0,0}$-\frac{1}{2\pi^2}$}%
}}}}
\end{picture}%
}
}
\subfloat[$\vartheta=\pi$]{
     \resizebox{0.18\linewidth}{!}{
\begin{picture}(0,0)%
\includegraphics[width=0.7\textwidth]{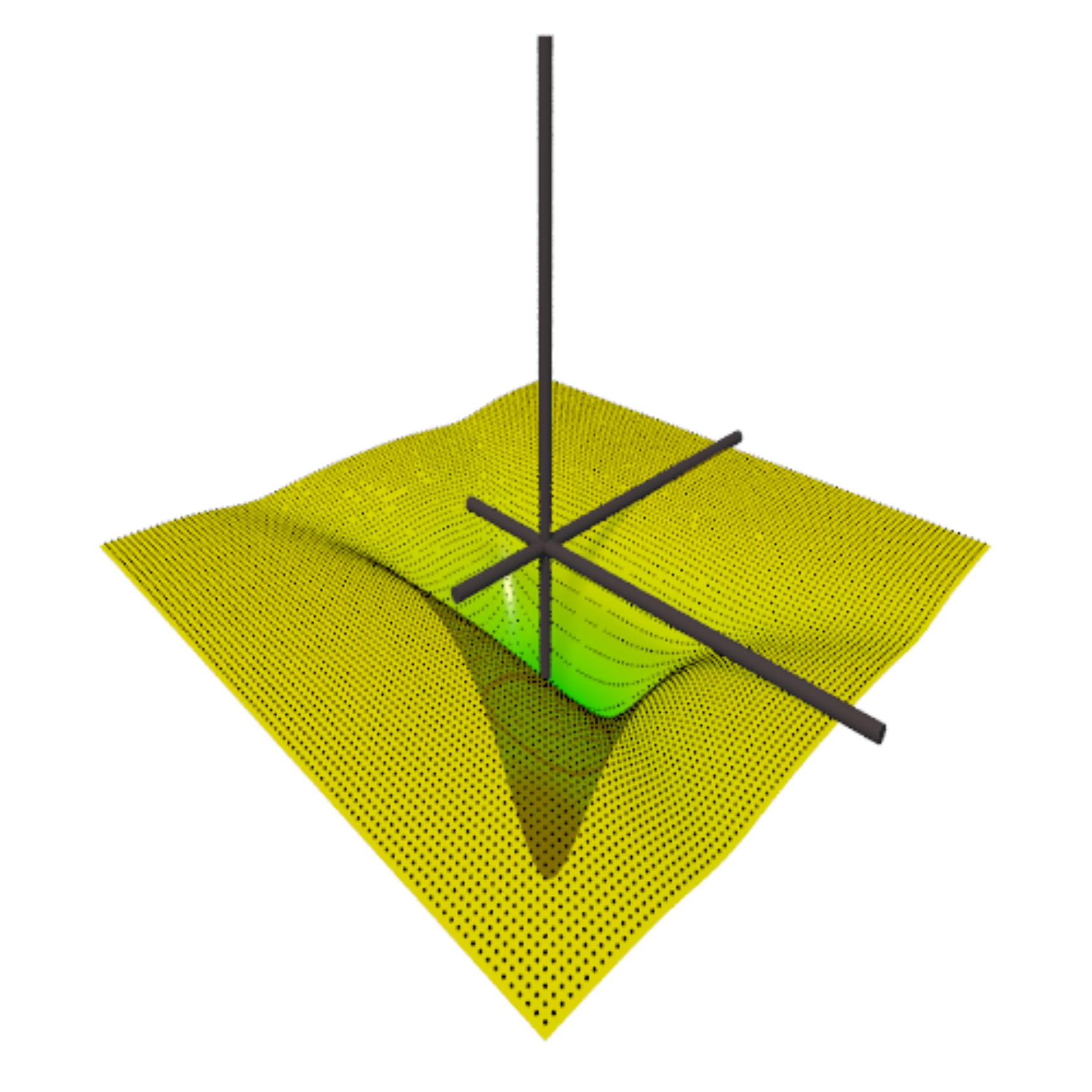}%
\end{picture}%
\setlength{\unitlength}{3947sp}%
\begingroup\makeatletter\ifx\SetFigFont\undefined%
\gdef\SetFigFont#1#2#3#4#5{%
  \reset@font\fontsize{#1}{#2pt}%
  \fontfamily{#3}\fontseries{#4}\fontshape{#5}%
  \selectfont}%
\fi\endgroup%
\begin{picture}(4509,5310)(2789,-6099)
\put(6976,-4561){\makebox(0,0)[lb]{\smash{{\SetFigFont{84}{16.8}{\familydefault}{\mddefault}{\updefault}{\color[rgb]{0,0,0}$x_\varphi$}%
}}}}
\put(6300,-2911){\makebox(0,0)[lb]{\smash{{\SetFigFont{84}{16.8}{\familydefault}{\mddefault}{\updefault}{\color[rgb]{0,0,0}$x_\omega$}%
}}}}
\put(4026,-1261){\makebox(0,0)[lb]{\smash{{\SetFigFont{84}{16.8}{\familydefault}{\mddefault}{\updefault}{\color[rgb]{0,0,0}$-\mathcal{Q}_\vartheta^{(edge)}f(\underline x)$}%
}}}}
\end{picture}%
}
}
\caption{(color online) Transformation behaviour of the test function $f(\underline x)$ 
as a function of the edge-to-edge map $-\mathcal{Q}^\text{(edge)}_\vartheta$
for different values of $\vartheta$. Function values are shown within the
range $[-5,5]\times [-5,5]$. Due to translational and scale invariance, it
represents the edge-to-edge transformation behaviour of a Dirac delta function under the
continuity assumption \eqref{eq:sharedRPassumption}.}
\label{fig:Qedgetransformviz}
\end{figure*}

The result of the analytical integration of the first two integrals, namely
$(-\mathcal{Q}_\vartheta^\text{(edge)}f)(\underline x)$ is shown in figure
\ref{fig:Qedgetransformviz} for selected edge orientations. We find that 
$(-\mathcal{Q}_\vartheta^\text{(edge)}f)(\underline x)$ is a rational
function which changes continuously
as a function of $\vartheta$, in contrast to the transformation behaviour of
$\Im G_0(\underline{x} + \imag 0^0)$.
Note that since $\mathcal{Q}_\vartheta^\text{(edge)}$ is scale-invariant,
Eq.~\eqref{eq:commutatorQthetaScaling}, the transformation behaviour of the
Dirac delta distribution is analogous to figure \ref{fig:Qedgetransformviz}.
Consequently, the transformed delta distribution on the edge
$\mathbb{R}+\imag 0^\vartheta$ is not a function but rather a distribution 
with a relatively complicated structure. Therefore, the limit $\delta \to 0$ in
\eqref{eq:Qsimplified} cannot be taken before the last two integrals from the
Poisson kernel $\mathcal{P}_{r,\vartheta}$ are computed.

In the special case $\vartheta = \pi/2$, figure
\eqref{subfig:Qedgetransformthetapi2},
 the asymptotic behaviour of the result decays $\propto \frac{1}{|\underline
x|}$ when $\underline x \to \infty$, in
contrast to the original test function behaviour 
$f(\underline x)\propto \frac{1}{|\underline x|^2}$. This is because the Hilbert transforms
are taken with respect to mutually orthogonal directions in $\mathbb{R}^2$,
here.

The angles between $0$ and $\pi/2$ interpolate smoothly between the extremal cases of
the unperturbed well-localized $f(\underline x)$ at $\vartheta=0$ and the long-range function at 
$\vartheta=\pi/2$. The solution at $\vartheta=\pi$ is again strongly
localized and equals $-f(\underline x)$. The behaviour in the interval $[\pi,
2\pi)$ is analogous due to symmetry reasons.

\paragraph{Implementation.} 
{
As mentioned above, the third integral
of the operator sequence \eqref{eq:Qsimplified} 
can still be computed analytically. However, each
integration of the sequence adds additional poles to
the resulting function, and more and more complex distinguishments have to be done in
order to decide whether a pole is on the upper or lower half-plane and
whether it contributes or not to certain residue sums.
}

Because translational and scale invariance do not seem to be as useful concepts as
applied to $\mathcal{P}_{r,\vartheta}$, not only the extra variable $r$
appears in the computation of the remaining expressions, but also the 
shift $\underline X$ and the scale $\lambda$ of the test function
\eqref{eq:numtestfuncQgeneral}. For the third integral, one can still determine the
poles and residues before doing the latter substitution with the computer
algebra system, however. 

At present, very lengthy expressions result for the last integrand. As a
consequence, the last integral was evaluated numerically for each matrix
element. The limit $\delta \to 0$ can only be taken numerically, depending on
the grid.
An algebraic determination of the poles of this expression is cumbersome, 
because high-order polynomials appear in the denominator of the
resulting expressions. Nevertheless, numerical computations indicate that the
limit $\delta\to 0$ yields well-defined functions after the fourth
integration. Once an algebraic expression is found, the expression for the
limit $\delta \to 0$ would be more simple than the intermediate terms.
As already stated in the beginning of this section, we compute the fourth integral 
with an adaptive numerical integration routine in practice. 
In the numerical MaxEnt implementation, one can adjust $\delta$ as a function of $\underline x$,
denoted by $\delta_{\underline x}$,
depending on how well a specific region of the edge should be resolved.

When defined according to the interacting branch cut geometry,
$\mathcal{Q}_{r,\vartheta}$ will be simply called $\mathcal{Q}$, in the
following.
Details on the numerical representation of $\mathcal{Q}$ are provided in
appendix \ref{app:numericsQ}.
A detailed description of implementation and setup of the maximum entropy method using
$\mathcal{Q}$ is given in appendix \ref{app:maxentmultiwedge}. 

\paragraph{Details of the MaxEnt procedure}
The quality of the results of the MaxEnt method using the
$\mathcal{Q}$-mapping critically relies on a careful a-posteriori
identification of the high-energy structure of the
{physical edge function}
 $\tilde A(x_\varphi,
x_\omega)$ along its singular directions. The information is 
incorporated into the default model as described in appendix 
\ref{app:maxentmultiwedge}. In brief, we first determine the
most probable lateral {width} of the default model \eqref{eq:defmodAtildevariedbroadness}. 
In the same fashion, a second step optimizes the low-energy 
structure of the default model, by a posteriori determining the most probable
low-energy {bandwidth}, i.e.~the quantity $\tilde \sigma_\text{def}$ in
equation \eqref{eq:xdepsigmadef} with the
highest posterior probability. In equation \eqref{eq:xdepsigmadef} we set 
the low-energy scale to $R=5\Gamma$. In tested examples, no strong dependence
of the inferred results on $R$ was observed. In future applications however, 
in order to increase accuracy, it may be advantegeous to also perform an 
optimization with respect to the posterior probability of $R$.

\begin{figure}
\centering
\includegraphics[width=\linewidth]{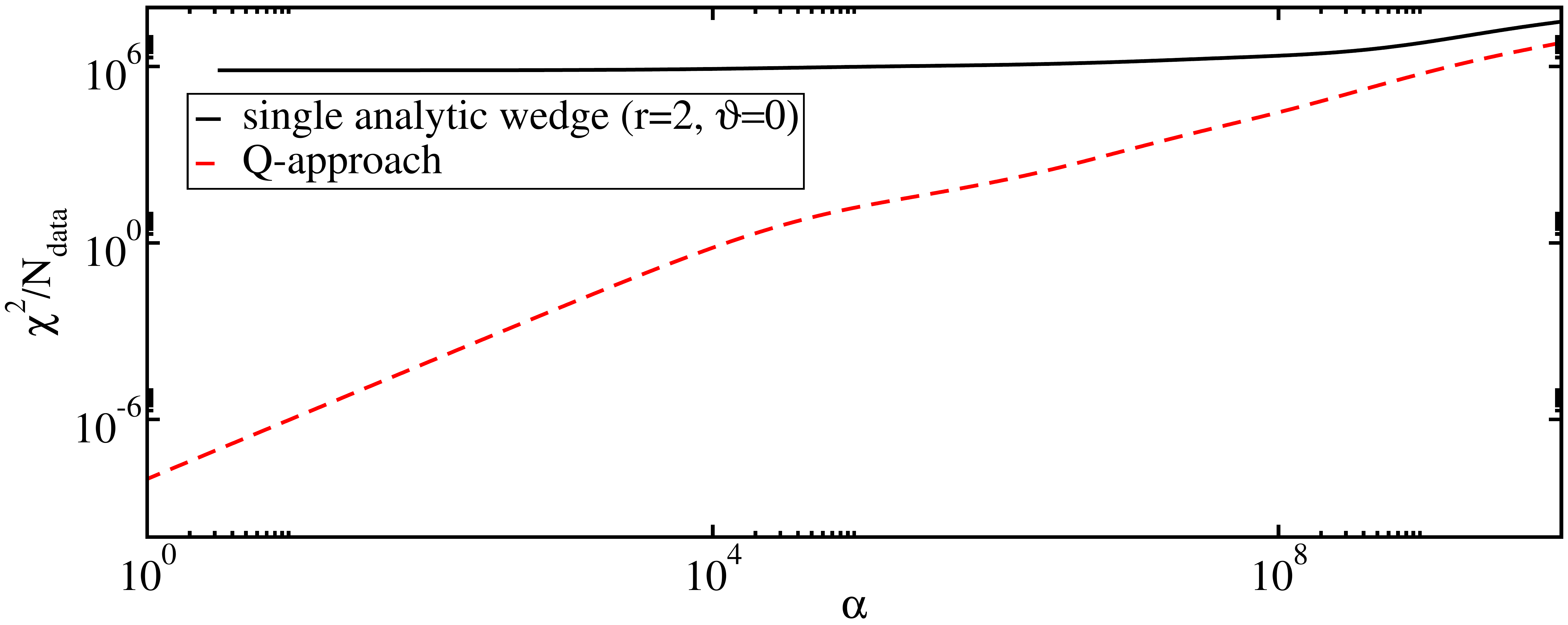}
\caption{(color online) Comparison of $\chi^2$ as a function of the MaxEnt regularization
parameter for single-wedge kernel $\mathcal{P}_{r,\vartheta}$ and
multi-wedge kernel $\mathcal{Q}$ at weak interaction $U=2\Gamma, \beta=5\Gamma^{-1},$ and
$e\Phi=\Gamma$ as
a function of the regularization parameter $\alpha$. For the same input set,
the single-wedge approach clearly fails to converge due to the presence of
higher-order branch cuts.}
\label{fig:chi2comparisonQvsSingleWedge}
\end{figure}
\begin{figure*}
\centering
\subfloat[a-posteriori identified best default model]{
   \includegraphics[width=0.45\textwidth]{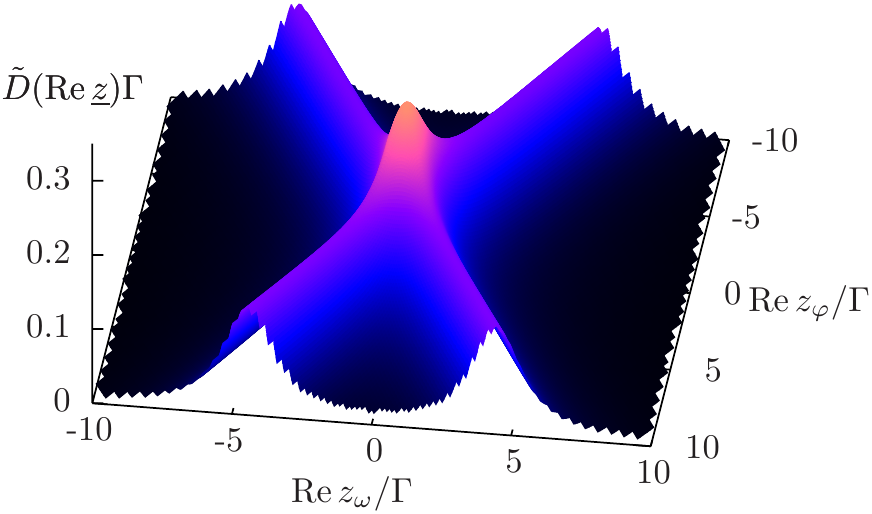}
   \label{subfig:mostprobD_U2Vb1beta5}
}
\subfloat[inferred $\tilde A(\underline x)$]{
   \includegraphics[width=0.45\textwidth]{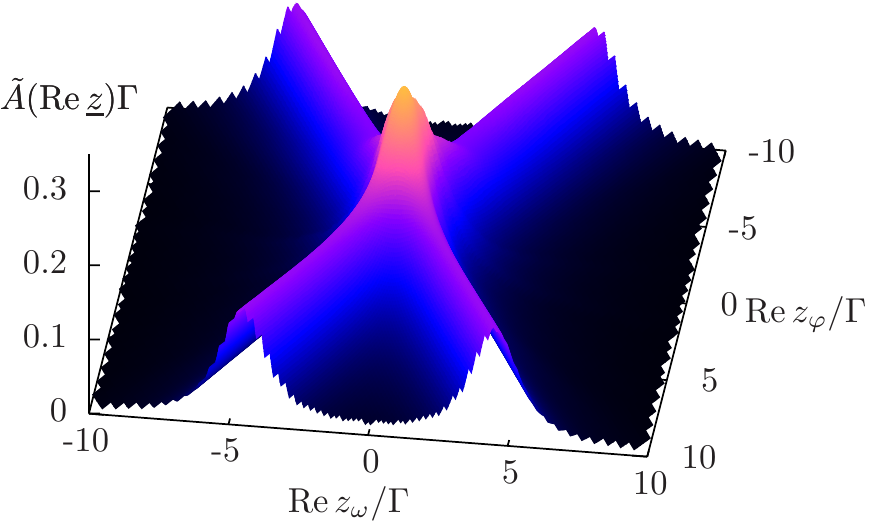}
   \label{subfig:Atilde_U2Vb1beta5}
}
\caption{(color online) Application of the MaxEnt procedure for the $\mathcal{Q}$-mapping to
the nonequilibrium weak-coupling case $U=2\Gamma$, $\beta=5\Gamma^{-1},
e\Phi=\Gamma$, with CT-QMC data as input. The default model has been
identified via its maximal posterior probability.}
\end{figure*}
Let us provide an example of how the method works for the weak-coupling case, i.e.\ for
parameters $U=2\Gamma$, $\beta=5\Gamma^{-1},
e\Phi=\Gamma$. The final results for
the spectral function were already presented in
Fig.~\ref{fig:AU2V1b5GenApproach}.
It turns out that the
applicability
of the $\mathcal{Q}$-approach is limited a posteriori by bad behavior of the
inferred spectral functions to input data with $\omega_n >
|\varphi_m/2|$ for $n\ge0$. This condition 
corresponds to not crossing the principal branch cut $\gamma = \pm 1$ in
figure
\ref{fig:branchcuts}, when coming from the retarded Green's function (edge
orientation $\vartheta=0$).
Apart from this restriction, there appears to be no further
problem with the approach. Consequently, at least for weak coupling, the central continuity assumption of the
$\mathcal{Q}$-approach is practically solely violated with respect to the
branch cuts $\gamma=\pm1$. 
{The violation already occurs at very
small values of the many-body interaction, $U\leq 2\Gamma$. However it
vanishes at $U=0$, since the continuity assumption is exact for $G_0$.
}
This observation is compatible with the observed strong violation of
the assumption within the Dyson series which was reported in section
\ref{subsec:rangesharedrp}.

We therefore can use all Matsubara data of our Monte-Carlo simulation subject to
$\omega_n>|\varphi_m/2|$. 
{For the inverse temperature $\beta=5\Gamma^{-1}$, these extend from $n=1$ to
$n=8$ for $\omega_n$ with $m=\pm1,\dots, \pm5$ for $\varphi_m$.}
As a first test, we show in Fig.~\ref{fig:chi2comparisonQvsSingleWedge} the
performance of the MaxEnt method for both, the single-wedge\cite{dirks} and the
multiple-wedge approach for the given data set. 
Because data from the comparably widely opened wedge $\omega_n > |\varphi_m / 2|$
are used in the single-wedge approach, it
implicitly assumes the interacting Green's function to be analytic for $\Im
z_\omega > |\Im z_\varphi / 2|$. Apparently, this wrong assumption makes it impossible 
to obtain a reasonable fit with a positive definite $\tilde A(\underline x)$.
Consequently, the $\chi^2$ value of the procedure does not drop below $10^6 \cdot
N_\text{data}$, and the MaxEnt fails to converge. In sharp
contrast, values of $\chi^2/N_\text{data} \approx 1$ may be reached with the
MaxEnt with
respect to $\mathcal{Q}$. Also, controls such as the MaxEnt error rescaling
merit do not indicate the presence of any abnormalities.

Thus, for the $\mathcal{Q}$-mapping, a well-behaved MaxEnt solution is obtained.
{
As further discussed in appendix \ref{app:maxentmultiwedge}, the quality of
the solution very much relies on appropriately including the prior knowledge
on singular directions of the Green's functions in $\mathbb{C}^2$ into the default model
of the edge function. For this purpose, within a set of smooth default models
with the correct singular directions as $\underline x \to \infty$, a most
probable one is identified within the Bayesian framework of the MaxEnt method.
The thus identified default model for the edge function is displayed in figure \ref{subfig:mostprobD_U2Vb1beta5}. 
}
Using this default
model, the well-behaved edge function $\tilde A(x_\varphi, x_\omega)$ 
shown in panel \ref{subfig:Atilde_U2Vb1beta5} is obtained.
 An overall moderate sharpening of
the edge function along the cross-like structure is observed as a result of
this final step of the Bayesian inference procedure.

{
With such appropriately optimized default models, the results presented in section \ref{sec:results} were obtained
from weak to intermediate coupling strengths.
Throughout, the only major data range constraint, $\omega_n > |\varphi_m/2|$
was found, which prohibits crossing the principal branch cut due to violations
of the continuity assumption. On occasion, for stronger correlation strengths, values
with small $\omega_n$ had to be discarded in order to obtain a converging
MaxEnt solution, i.e.~a solution which meets the continuity assumption constraints.
}
At the comparably small inverse temperature $\beta=5\Gamma^{-1}$ used,
calculations require only moderate computer resources, mainly due to the comparably 
small QMC data space of approximately
50 imaginary-time-theory data points. In general, the amount of data will
grow quadratically as a function of inverse temperature, due to the
simultaneous presence of Matsubara voltage and Matsubara frequency.
Additionally, at low
temperatures, sharp features in the spectral function and hence the 
{edge function} 
$\tilde A$ will have to be resolved, requiring an enhanced grid refinement.
Altogether, matrix sizes in the MaxEnt will increase substantially when the
temperature is decreased. In particular, the computational effort
for the generation of an appropriate kernel matrix (cf.~appendix
\ref{app:numericsQ}) grows dramatically
and the memory consumption of the MaxEnt itself
poses a
limitation at lower temperatures at the present stage of code development.
Additionally, it is well-known that the resolution of low-temperature
features with the MaxEnt method requires a careful Bayesian analysis based on
higher-temperature data, i.e.~an ``annealing procedure'', involving a sequence of 
QMC plus MaxEnt runs for a reasonably fine temperature grid \cite{mem, jarrell}.

\section{Perspective: unbiased $\mathcal{Q}$-approach}
\label{sec:unbiasedQ}
From a mathematical point of view, the underlying continuity assumption of 
the $\mathcal{Q}$-approach is only approximate, because in higher orders of 
perturbation theory, terms which do not characterize the full collection of
wedges, but rather just isolated wedges or subcollections of wedges, are generated. 
These terms are manifested in discontinuities of the real part of the Green's function at the
branch point $\Im \underline z = 0$.

In order to extend the $\mathcal{Q}$-approach to the full nonequilibrium
Kondo regime
$U\geq 2\pi\Gamma$, $e\Phi \sim T_K$, $\beta^{-1} \sim T_K$,
one has to take these contributions into account.
This requires the consideration of the
full analytic structure of the theory, i.e.~the full set of edge functions.

As a consequence, extra terms have to be
added to the representation of $G(z_\varphi, z_\omega)$ within the MaxEnt
procedure. Obvious candidates for such degrees of freedom are the residual
imaginary parts of edge functions
\begin{equation}
\tilde R_n (\underline x) := \Im G(\underline x + \imag
0^{\vartheta_n})
- \pi \cdot (\mathcal{Q}^\text{(edge)}_{\vartheta_n} \tilde A)(\underline x),
\end{equation}
for the edge of the $n$-th wedge with orientation $\vartheta_n$. Because the
$\mathcal{Q}$-mapping is exact at high energies $\|\underline x\|$,
the terms $\tilde R_n (\underline x)$ are essentially localized within a finite radius around
$0$. This range is expected to be of the order of magnitude of the energy
scales
$\Gamma$, $U$, $\epsilon_d$, and $e\Phi$. 

Regarding the inverse problem, for data in the $n$-th wedge, one has the
\emph{exact} representation
\begin{equation}
\begin{split}
\Im G(\imag\varphi_m, \imag\omega_n) = &
(\mathcal{P}_{r_n,\vartheta_n}\tilde R_n)(\imag\varphi_m,\imag\omega_n)  \\
&+ \pi\cdot(\mathcal{Q}\tilde A)(\imag\varphi_m,\imag\omega_n),
\end{split}
\label{eq:extendedQ}
\end{equation}
where $r_n$ is the opening ratio of the respective data wedge \footnote{For
the definitions see figure \ref{fig:wedgethetaparam}.}. The MaxEnt procedure 
must determine $\tilde R_n$ and $\tilde A$ simultaneously.
Practically, the terms $\tilde R_n$ would act as ``valves''
for the conceptual imperfection of the $\mathcal{Q}$-mapping within the
Bayesian information flow.

It is an interesting question if the formally infinitely many two-dimensional variable vectors
in practice lead to a dramatic increase in the fit space or not. Due to
locality of the terms $\tilde R_n (\underline x)$, the effort is probably less
than for the $\tilde A$ function which itself encodes many aspects of the analytic 
structure. Furthermore, the rather large Poisson kernel matrix elements at low energies 
will possibly lead to a comparably good MaxEnt performance in the determination of 
$\tilde R_n (\underline x)$, as long the opening ratio $\vartheta_n$ of the $n$-th
wedge is comparably large. 

Because the functions $\tilde R_n(\underline x)$ cannot be expected to be
positive, it is
necessary to introduce a shift to a positive function, such as for
the spectral functions of the static observables in paper 1. The terms 
$\tilde R_n(\underline x)$ are presumably most dominant for wedges next to the 
noninteracting Green's function's branch
cuts. A very careful Bayesian analysis, including an appropriate set of
choosable
default models constructed from a-priori information, is probably required
for a
successful application of the exact approach \eqref{eq:extendedQ}. It is also
possible that the perturbative structure of the theory reorganizes terms 
$\tilde R_n$ in subcollections of wedges which result in a more moderate MaxEnt 
problem than equation \eqref{eq:extendedQ}. In particular, the branch cut at
$\Im z_\varphi=0$ probably leads to a nonzero limit
$\lim_{n\to \infty} \tilde R_n$, where the limit $n\to\infty$ shall consider a
sequence of wedges with $\vartheta_{n\to\infty} = 0$ or
$\vartheta_{n\to\infty} = \pi$.

\section{Summary}
{
We systematically studied the mathematical structure of the dot-level Green's
function and the Bayesian inference of non-equilibrium spectral functions and
transport properties from effective-equilibrium quantum Monte-Carlo data
within the Matsubara-voltage theory. Furthermore, a continuity assumption on 
the analytic structure was introduced which strongly improved the numerics of 
the MaxEnt approach of an earlier publication.\cite{dirks}
}

{
Formal parts of the paper introduced the essential concepts
of the function theory of several complex variables and connected to the
respective mathematical literature. Using insights from perturbation theory, the Green's function was 
characterized axiomatically with regard to its function-theoretical structure.
As the fundamental domains of holomorphy, so-called wedges (tubular cones)
emerged. The Green's function is composed of sheets which are holomorphic on
the wedges enclosed by branch cuts. Within each wedge, the Matsubara data of
the Green's function uniquely map to a real-time limit on the so-called edge of the
wedge. For this purpose, an explicit integral representation was constructed.
However, depending on the considered wedge, the edge structure does not
necessarily have a direct physical interpretation.
}

{
In an earlier publication,\cite{dirks} we had been unable to compute reliable
\emph{non-equilibrium} spectral functions from a MaxEnt procedure based on integral
representations within wedges, due to rather strong assumptions on the
analytic structure and weak assumptions on the physical structure. The assumptions had 
limited us to a single wedge with a rather small opening ratio, on which the Green's function is not strictly 
analytic but which directly includes the physical limit procedure on its edge. While as 
compared to the present work the numerical effort of the MaxEnt procedure was rather low, 
due to the wedge structure and simple kernel structure, we had not 
been able to consider most available quantum Monte-Carlo data within the
MaxEnt procedure. The hereby implied loss of information from available
simulation data had not affected the equilibrium
spectra but the non-equilibrium spectra, due to the kernel structure.
}

{
In order to overcome these previous limitations, we introduced a continuity
assumption to the real-time structure of the Green's function, i.e.~its
structure at the branch point around which the edges of the wedges associated to the branches of the Green's
function are aligned. The assumption includes structures generated by the earlier fit approach 
introduced in Ref.~\onlinecite{prl07}, which was motivated by perturbation theory.
Mathematically, the assumption lead to a uniform description of data from all
branches of the Green's function and gave rise to a linear operator
$\mathcal{Q}$ which, while hard to implement, enhanced the MaxEnt procedure to a larger set of quantum
Monte-Carlo data.
}

{
We found that the continuity assumption appears to be valid for a very broad range of
data, up to intermediate coupling strengths, eventually yielding reasonable
non-equilibrium MaxEnt results for spectral function and transport
properties, which are dramatically improved as compared to the results of
Ref.~\onlinecite{dirks}. 
However, as the nonequilibrium Kondo regime is approached, we expect the continuity
assumption to break down eventually. For this parameter regime, the method
could be extended along the line discussed in section \ref{sec:unbiasedQ}.
}

\section{Acknowledgments}
The authors acknowledge useful discussions with J.~Freericks, F.B.~Anders,
S.~Schmitt, K.~Sch\"onhammer, and A.~Schiller.
AD acknowlegdes financial support by the DAAD through the PPP exchange program.
JH acknowledges the National Science Foundation with the
Grant number DMR-0907150.
MJ acknowledges the NSF LA-SiGMA cooperative agreement, EPS-100389.
AD and TP would also like to acknowledge
computer support by the HLRN, the GWDG and the GOEGRID initiative of the
University of G\"ottingen.
Parts of the implementation are based on the ALPS 1.3 library \cite{alps}.

\appendix
\section{Uniqueness of the Analytic Continuation of Dynamical Quantities}
\label{apx:uniqueness}
In the following, we would like to show that the continuation 
of Matsubara data $G(\imag\varphi_m, \imag \omega_n)$ to the multisheeted
holomorphic function $G(\underline z)$ is unique, i.e.~we will prove assumption 
3 in section \ref{subsec:holostructGF}, relation \eqref{eq:AssumptionIII}.
We will derive the uniqueness using the axiomatic statements 1, 2, and 3' of
section \ref{subsec:holostructGF}.
Since the proof will involve some elementary geometry, it will be accompanied
by several sketches.

We may focus our attention to a single wedge $T^C$ which is defined by subsequent branch cuts
from Eqs.~\eqref{eq:AssumptionI}. The data $G(\imag \varphi_m, \imag
\omega_n)$ which are located in the wedge are our starting point,
$(\varphi_m,\omega_n)^T\in C$.
Without loss of generality we can assume that we have entire lines of data,
$G(\imag\varphi_m,\imag \omega_I)$, $\omega_I\in \mathbb{R}$, because arbitrary 
continuous imaginary $\omega_I$ may be computed by Fourier transform in the 
$\varphi_m$th effective equilibrium theory, having again $(\varphi_m,\omega_I)^T\in C$.
Let us denote the effective equilibrium data range by
\begin{equation}
E_0 := \{\imag(\varphi_m,\omega_I)^T|m\in\mathbb{Z},\omega_I\in\mathbb{R}\}
  \cap T^C.
\end{equation}

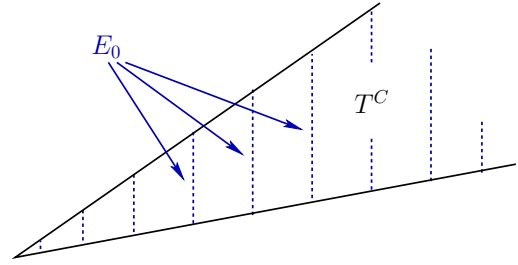
\begin{figure}
\begin{center}
\resizebox{0.9\linewidth}{!}{
\begin{picture}(0,0)%
\includegraphics{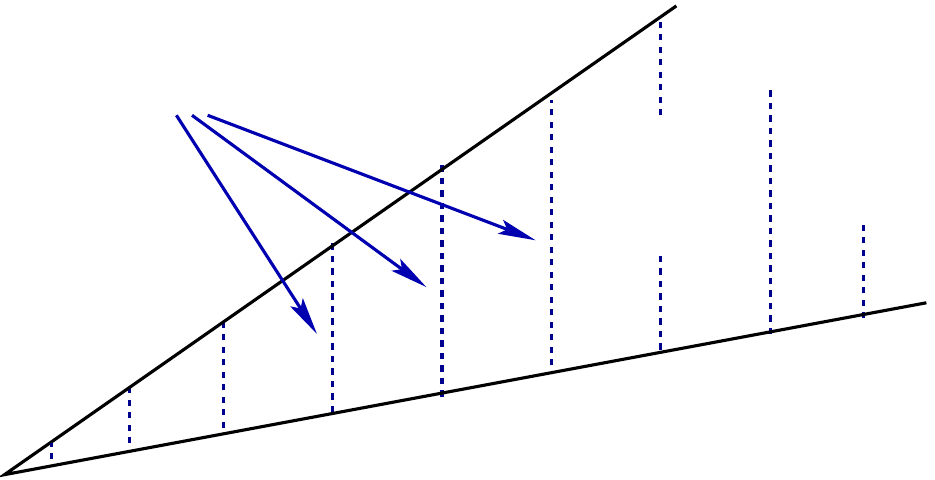}%
\end{picture}%
\setlength{\unitlength}{3947sp}%
\begingroup\makeatletter\ifx\SetFigFont\undefined%
\gdef\SetFigFont#1#2#3#4#5{%
  \reset@font\fontsize{#1}{#2pt}%
  \fontfamily{#3}\fontseries{#4}\fontshape{#5}%
  \selectfont}%
\fi\endgroup%
\begin{picture}(4469,2294)(2904,-3008)
\put(3601,-1186){\makebox(0,0)[lb]{\smash{{\SetFigFont{14}{16.8}{\familydefault}{\mddefault}{\updefault}{\color[rgb]{0,0,.69}$E_0$}%
}}}}
\put(5926,-1711){\makebox(0,0)[lb]{\smash{{\SetFigFont{14}{16.8}{\familydefault}{\mddefault}{\updefault}{\color[rgb]{0,0,0}$T^C$}%
}}}}
\end{picture}%
}
\end{center}
\caption{(color online) The wedge to be considered. The dash-dotted lines denote 
the data yielded by imaginary-time theory.}
\label{fig:wedge0}
\end{figure}

These lines of known data of the unknown function $G(\underline z)$ in the wedge 
$T^C$ are depicted in Fig.~\ref{fig:wedge0}. They constitute one-dimensional
lines in the four-dimensional wedge $T^C$ for which the function
$\left.G\right|_{T^C}$ shall be reconstructed.

We will, step by step, prove the uniqueness of the continuation of the
imaginary-time data by applying biholomorphic maps and the identity theorem
of complex analysis. The central idea will be to extend larger and
larger subsets for which a unique continuation is obtained.

\subsection{Reconstruction of edge values using \\ complex lines which are isomorphic to
$\mathbb{H}$}

\label{subsec:reconedge1}

Due to assumptions 1 and 2, we found that the Green's function
$\left.G\right|_{T^C}$ may be reconstructed from
their edge values, using Eqs.~\eqref{eq:VladimirovPoisson}
and \eqref{eq:VladimirovSchwarz}. Therefore, 
it suffices to show that we can reconstruct all edge
values of the function $G$ from the data
$\left.G\right|_{E_0}$.

We will first show that one may reconstruct a certain set of \emph{single
lines through zero}
on the edge. Each of these lines is defined by an angle $\vartheta$. All function
values on the line may be reconstructed if the angle $\vartheta$ is contained
by the cone $C$.

\begin{figure}
\begin{center}
\resizebox{0.9\linewidth}{!}{
\begin{picture}(0,0)%
\includegraphics{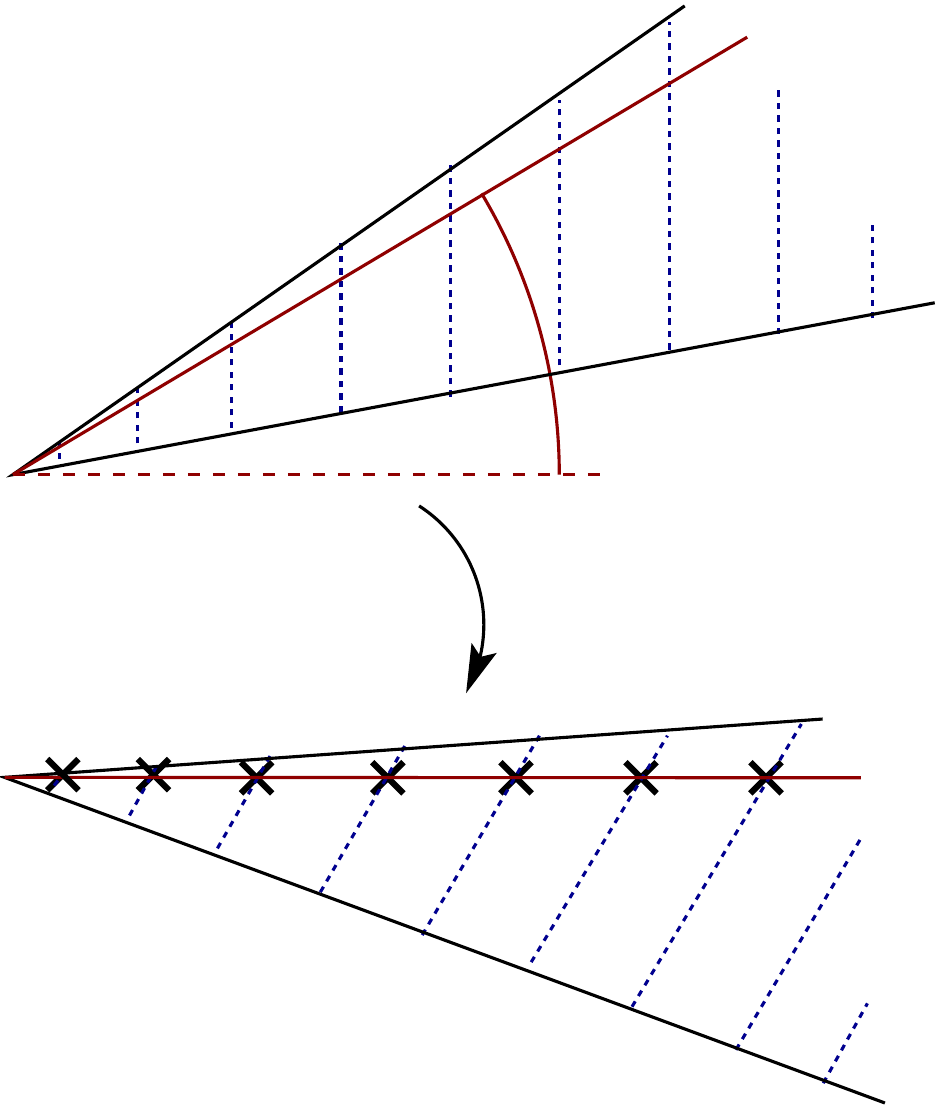}%
\end{picture}%
\setlength{\unitlength}{3947sp}%
\begingroup\makeatletter\ifx\SetFigFont\undefined%
\gdef\SetFigFont#1#2#3#4#5{%
  \reset@font\fontsize{#1}{#2pt}%
  \fontfamily{#3}\fontseries{#4}\fontshape{#5}%
  \selectfont}%
\fi\endgroup%
\begin{picture}(4509,5310)(2789,-6099)
\put(3976,-5611){\makebox(0,0)[lb]{\smash{{\SetFigFont{14}{16.8}{\familydefault}{\mddefault}{\updefault}{\color[rgb]{0,0,0}$R_{-\vartheta}T^C$}%
}}}}
\put(5251,-3661){\makebox(0,0)[lb]{\smash{{\SetFigFont{14}{16.8}{\familydefault}{\mddefault}{\updefault}{\color[rgb]{0,0,0}$R_{-\vartheta}$}%
}}}}
\put(6976,-4561){\makebox(0,0)[lb]{\smash{{\SetFigFont{14}{16.8}{\familydefault}{\mddefault}{\updefault}{\color[rgb]{.56,0,0}$\cong \mathbb{H}$}%
}}}}
\put(5626,-2911){\makebox(0,0)[lb]{\smash{{\SetFigFont{14}{16.8}{\familydefault}{\mddefault}{\updefault}{\color[rgb]{.56,0,0}$\vartheta$}%
}}}}
\put(4126,-1561){\makebox(0,0)[lb]{\smash{{\SetFigFont{14}{16.8}{\familydefault}{\mddefault}{\updefault}{\color[rgb]{0,0,0}$T^C$}%
}}}}
\end{picture}%
}
\end{center}
\caption{(color online) A single line contained by the cone $C$ may after biholomorphic
rotation $R_{-\vartheta}$ and subsequent complexification be interpreted as the 
upper half plane $\mathbb{H}$ of $\mathbb{C}$.}
\label{fig:wedge1}
\end{figure}
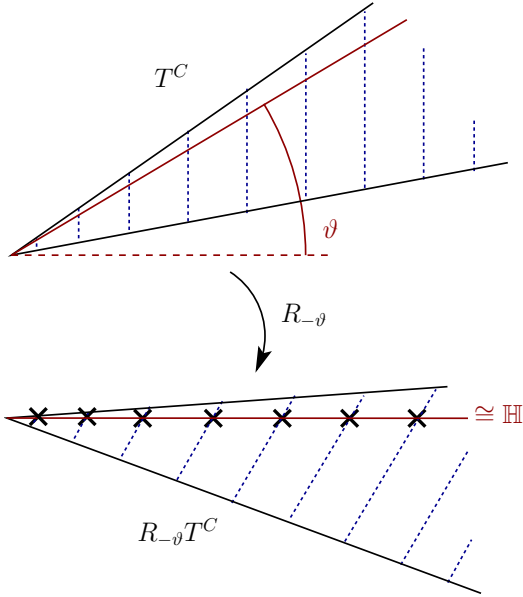

The proof of the latter statement is the following. Consider a single line in the cone $C$, given by
the angle $\vartheta$.
A biholomorphic rotation in the sense of section
\ref{subsubsec:anastructbiholomorphiceq} can then be applied in such a way
that the line is horizontal and may after complexification be interpreted as 
the upper half plane $\mathbb{H}$ of
$\mathbb{C}$, see Fig.~\ref{fig:wedge1}. The real line is then associated to
a horizontal line on the edge of $R_{-\vartheta}T^C$. 

The biholomorphic equivalent to the yet unknown function is now 
$\tilde G(\underline z) = G(R_{-\vartheta}^{-1}\cdot \underline z)$.
Note that the line $(0,\imag\lambda)^T$ in $R_{-\vartheta}T^C$, $\lambda >
0$,  contains
infinitely many known values of $\tilde G(\underline z)$. These are denoted
by the crosses in Fig.~\ref{fig:wedge1}. Extending to the upper half plane
$(0,\mathbb{H})$, one may apply the identity theorem of complex analysis for
reconstructing $\tilde G(\underline z)$ on the whole plane $(0,\mathbb{H})$.
In particular, the boundary values $\tilde G(0,\mathbb{R}+\imag 0^+)$ are recovered.
This proves the statement of this subsection.

\subsubsection{Identity Theorem}

Let us comment on the satisfaction of the assumptions of the identity theorem. Since $(0,\infty)$
is the accumulation point of the known data points on $(0,\mathbb{H})$, 
i.e.~of the ``series of crosses in Fig.~\ref{fig:wedge1}'', 
given by $(0,\mathbb{H}) \cap R_{-\vartheta}E_0$, we have to 
show that $\tilde G(0,1/z)$ may be extended to an analytic function at $z =0$.
Once this is possible, the function is uniquely determined by the set of
function values.

Combining assumptions 1 and 3', we know that 
$G(\zeta\underline x^{(0)})$ (in the sense of  assumption 3') behaves
like a conventional Green's function, because the singular case coincides
with a branch cut which is by construction not contained by the wedge.
Due to this rapid decay one may extend $\tilde G$ to the lower half plane
such that $\tilde G(0,z^*) = \tilde G(0,z)^*$ and is holomorphic at
$z=\infty$. This can be done explicitly using a spectral
representation with respect to the boundary values of $\Im \tilde G(0,z)$ on
the real axis. The spectral representation exists due to the $1/z$
asymptotics which lets the line integral contribution vanish on the
infinitely large semicircle attached to the real axis.
Note that this construction is also compatible with the symmetry relation
$G(-\imag\varphi_m,-\imag\omega_n) = G(\imag\varphi_m,\imag\omega_n)^*$.

As a consequence, the identity theorem is applicable for $\tilde G(0,z)$ at $z=\infty$ such as it is for
regular Matsubara Green's functions.

One may also think of $G(\zeta\underline x^{(0)})$, $\zeta\in\mathbb{H}$ as a meromorphic
function of $\zeta\in\mathbb{C}$, because it may due to boundedness and $1/\zeta$
asymptotics be approximated arbitrarily well by a meromorphic function, such
as in an infinite Pad\'e expansion. Since meromorphic 
functions on $\mathbb{C}$ are holomorphic on the Riemannian sphere, the identity theorem
holds at the accumulation point $\infty$.

\subsubsection{Resulting reconstruction of edge values}
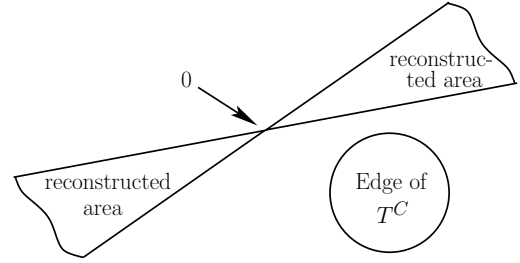
\begin{figure}
\begin{center}
\resizebox{0.9\linewidth}{!}{
\begin{picture}(0,0)%
\includegraphics{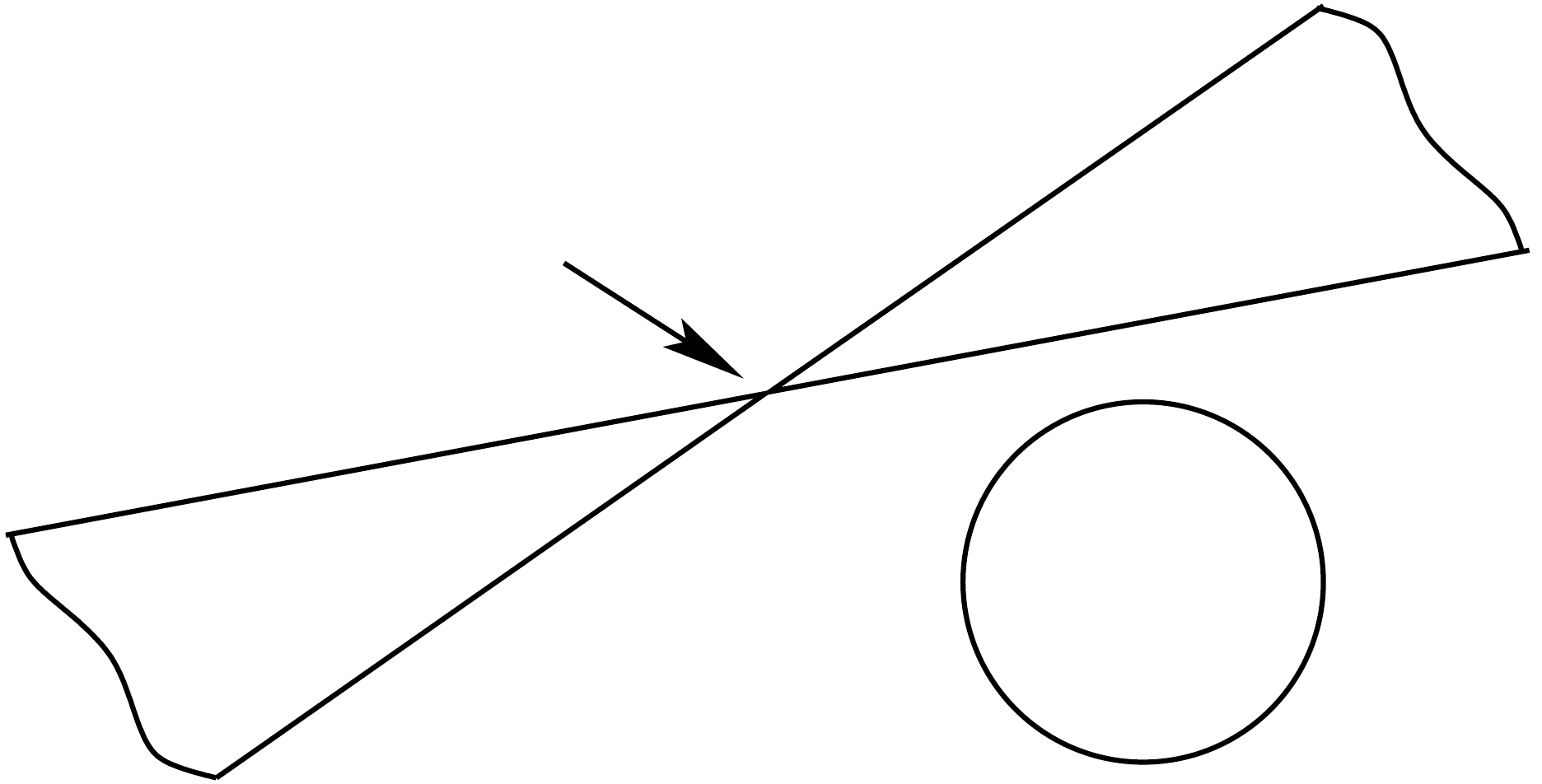}%
\end{picture}%
\setlength{\unitlength}{3947sp}%
\begingroup\makeatletter\ifx\SetFigFont\undefined%
\gdef\SetFigFont#1#2#3#4#5{%
  \reset@font\fontsize{#1}{#2pt}%
  \fontfamily{#3}\fontseries{#4}\fontshape{#5}%
  \selectfont}%
\fi\endgroup%
\begin{picture}(8968,4566)(1815,-4894)
\put(4801,-1861){\makebox(0,0)[lb]{\smash{{\SetFigFont{29}{34.8}{\familydefault}{\mddefault}{\updefault}{\color[rgb]{0,0,0}$0$}%
}}}}
\put(8251,-4261){\makebox(0,0)[lb]{\smash{{\SetFigFont{29}{34.8}{\familydefault}{\mddefault}{\updefault}{\color[rgb]{0,0,0}$T^C$}%
}}}}
\put(2401,-3661){\makebox(0,0)[lb]{\smash{{\SetFigFont{29}{34.8}{\familydefault}{\mddefault}{\updefault}{\color[rgb]{0,0,0}reconstructed}%
}}}}
\put(3076,-4111){\makebox(0,0)[lb]{\smash{{\SetFigFont{29}{34.8}{\familydefault}{\mddefault}{\updefault}{\color[rgb]{0,0,0}area}%
}}}}
\put(7876,-3661){\makebox(0,0)[lb]{\smash{{\SetFigFont{29}{34.8}{\familydefault}{\mddefault}{\updefault}{\color[rgb]{0,0,0}Edge of}%
}}}}
\put(8776,-1861){\makebox(0,0)[lb]{\smash{{\SetFigFont{29}{34.8}{\familydefault}{\mddefault}{\updefault}{\color[rgb]{0,0,0}ted area}%
}}}}
\put(8476,-1486){\makebox(0,0)[lb]{\smash{{\SetFigFont{29}{34.8}{\familydefault}{\mddefault}{\updefault}{\color[rgb]{0,0,0}reconstruc-}%
}}}}
\end{picture}%
}
\end{center}
\caption{Uniquely reconstructed range $\text{Edge}^{(\text{recon},0)}_{T^C}$ of 
$\left.G(\underline z)\right|_{T^C}$ on the edge of $T^C$ following from the
partial argument of subsection \ref{subsec:reconedge1}.
The wiggly lines in the boundary mean that the area extends to infinity.
}
\label{fig:edgerecon1}
\end{figure}

Sweeping through all possible angles $\vartheta$ which are contained by the
cone $C$, the uniquely reconstructed edge behaviour of 
$\left.G\right|_{T^C}$ is given by the area depicted in
Fig.~\ref{fig:edgerecon1}.

It is obviously given by
\begin{equation}
\text{Edge}^{(\text{recon,0})}_{T^C} := C \cup (-C).
\label{eq:edgereconarea1}
\end{equation}

\subsection{Extending the unique range to the entire edge}

In order to show that the function values of $\left.G\right|_{T^C}$ 
are also uniquely defined by $\left.G\right|_{E_0}$ for the complement of
$\text{Edge}^{(\text{recon,0})}_{T^C}$, the argument has to be extended in a similar
way. The trick is to consider yet another set of $\mathbb{H}$-isomorphic
subspaces and then apply the argument of the last section to a larger set of
data. 

\subsubsection{Extending the known data range within the wedge}
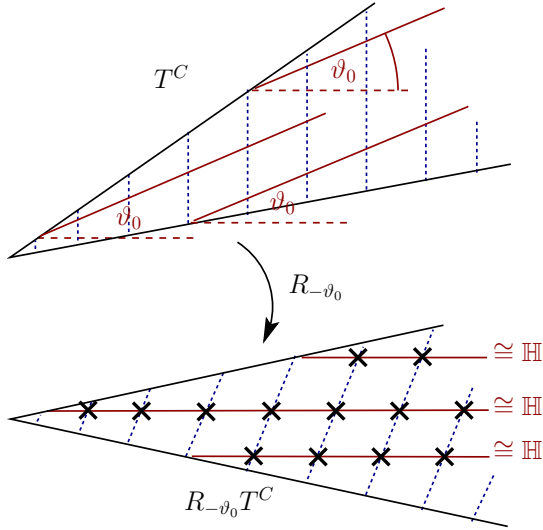
\begin{figure}
\begin{center}
\resizebox{0.9\linewidth}{!}{
\begin{picture}(0,0)%
\includegraphics{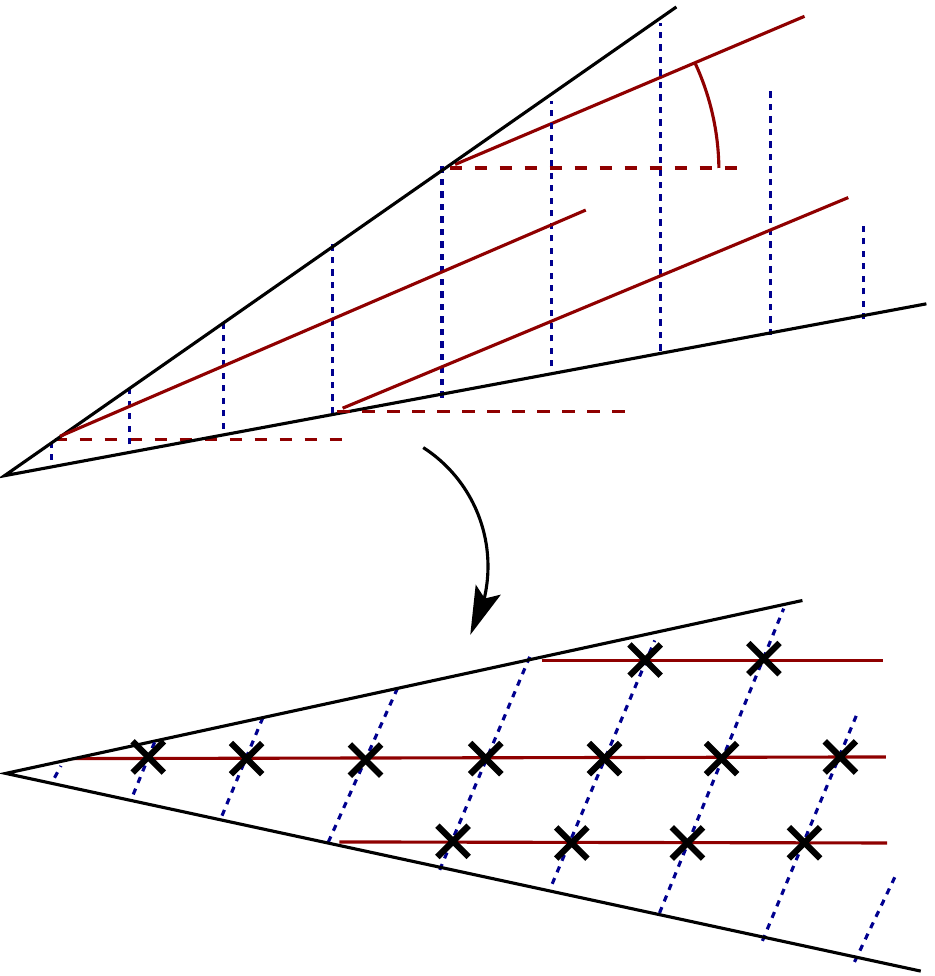}%
\end{picture}%
\setlength{\unitlength}{3947sp}%
\begingroup\makeatletter\ifx\SetFigFont\undefined%
\gdef\SetFigFont#1#2#3#4#5{%
  \reset@font\fontsize{#1}{#2pt}%
  \fontfamily{#3}\fontseries{#4}\fontshape{#5}%
  \selectfont}%
\fi\endgroup%
\begin{picture}(4469,4672)(2769,-5746)
\put(5626,-1756){\makebox(0,0)[lb]{\smash{{\SetFigFont{14}{16.8}{\familydefault}{\mddefault}{\updefault}{\color[rgb]{.56,0,0}$\vartheta_0$}%
}}}}
\put(4066,-1846){\makebox(0,0)[lb]{\smash{{\SetFigFont{14}{16.8}{\familydefault}{\mddefault}{\updefault}{\color[rgb]{0,0,0}$T^C$}%
}}}}
\put(5086,-2926){\makebox(0,0)[lb]{\smash{{\SetFigFont{14}{16.8}{\familydefault}{\mddefault}{\updefault}{\color[rgb]{.56,0,0}$\vartheta_0$}%
}}}}
\put(3729,-3061){\makebox(0,0)[lb]{\smash{{\SetFigFont{14}{16.8}{\familydefault}{\mddefault}{\updefault}{\color[rgb]{.56,0,0}$\vartheta_0$}%
}}}}
\put(7061,-4738){\makebox(0,0)[lb]{\smash{{\SetFigFont{14}{16.8}{\familydefault}{\mddefault}{\updefault}{\color[rgb]{.56,0,0}$\cong \mathbb{H}$}%
}}}}
\put(7061,-5128){\makebox(0,0)[lb]{\smash{{\SetFigFont{14}{16.8}{\familydefault}{\mddefault}{\updefault}{\color[rgb]{.56,0,0}$\cong \mathbb{H}$}%
}}}}
\put(7061,-4258){\makebox(0,0)[lb]{\smash{{\SetFigFont{14}{16.8}{\familydefault}{\mddefault}{\updefault}{\color[rgb]{.56,0,0}$\cong \mathbb{H}$}%
}}}}
\put(4331,-5578){\makebox(0,0)[lb]{\smash{{\SetFigFont{14}{16.8}{\familydefault}{\mddefault}{\updefault}{\color[rgb]{0,0,0}$R_{-\vartheta_0}T^C$}%
}}}}
\put(5251,-3661){\makebox(0,0)[lb]{\smash{{\SetFigFont{14}{16.8}{\familydefault}{\mddefault}{\updefault}{\color[rgb]{0,0,0}$R_{-\vartheta_0}$}%
}}}}
\end{picture}%
}
\end{center}
\caption{(color online) Enhancing the formal holomorphic reconstruction within the wedge.}
\label{fig:wedge2}
\end{figure}

The first step is depicted in Fig.~\ref{fig:wedge2}. In contrast to before,
we consider a constant angle $\vartheta_0$ and various lines which start at
different points on the boundary of the cone with the orientation $\vartheta_0$.

After biholomorphic rotation to the wedge $R_{-\vartheta_0}T^C$ we can again
complexify the lines 
\begin{equation}
\tilde{l}_{\underline{\tilde y}^{(0)}} :=\{\imag \underline{\tilde y}^{(0)} +
\imag (0,\lambda)^T, \lambda>0\};\,\,\underline{\tilde y}^{(0)} \in
\partial C
\end{equation}
to 
\begin{equation}
\tilde{L}_{\underline{\tilde y}^{(0)}} :=  \imag \underline{\tilde y}^{(0)} +
\begin{pmatrix}
0\\
\mathbb{H}
\end{pmatrix}
.
\end{equation}
The isomorphy of $\tilde{L}_{\underline{\tilde y}^{(0)}}$ to $\mathbb{H}$ and
assumption 3' again enable us to apply the identity theorem to the crossed
data in Fig.~\ref{fig:wedge2}, namely to the infinite sequence
$\left.(G\circ R_{\vartheta})\right|_{(R_{-\vartheta}E_0) \cap
\tilde{L}_{\underline{\tilde y}^{(0)}}}$. 

By this, the transformed Green's function 
$\tilde G = G\circ R_{\vartheta_0}$ is reconstructed for all
points of the set 
\begin{equation}
D := \bigcup_{\underline{\tilde y}^{(0)} \in \partial C}
     \tilde{L}_{\underline{\tilde y}^{(0)}}
= (R_{-\vartheta_0}T^C) \cap 
\begin{pmatrix}
\imag \mathbb{R} \\
\mathbb{C}
\end{pmatrix}.
\end{equation}

For simplicity, we may now just look at a subset of $D$, namely
\begin{equation}
\tilde{E}_1 := R_{-\vartheta_0}E_0 
+
\begin{pmatrix}
0\\
\mathbb{R}
\end{pmatrix}.
\end{equation}
It enables us to see that the values of $G$ are now known on the set
\begin{equation}
E_1 := E_0 
+
R_{\vartheta_0}
\begin{pmatrix}
0\\
\mathbb{R}
\end{pmatrix}.
\end{equation}

\subsubsection{Full Reconstruction of the Edge}
With the information from $E_1$ one may reinterpret the procedure associated
with Fig.~\ref{fig:wedge1} and described in section \ref{subsec:reconedge1}. 
The dashed lines of known data now contain an
\emph{additional real dimension} along the direction $R_{\vartheta_0}(0,1)^T$.

We can use each point $\lambda \cdot R_{\vartheta_0}(0,1)^T$ ($\lambda\in
\mathbb{R}$) of this new degree of freedom as an offset of the lines
used in section \eqref{subsec:reconedge1} and reapply the entire procedure. 
Using the resulting affine subspaces,
the Green's function may be reconstructed on further regions of the edge which
are affine to the one in Fig.~\ref{fig:edgerecon1}, namely
\begin{equation}
\begin{split}
\text{Edge}^{(\text{recon},\lambda)}_{T^C} := & \,
\text{Edge}^{(\text{recon},0)}_{T^C} \\
&+ \lambda \cdot R_{\vartheta_0}(0,1)^T.
\end{split}
\end{equation}

Applying the argument to all $\lambda \in \mathbb{R}$ reconstructs the entire
edge and hence the entire Green's function $\left.G\right|_{T^C}$, because
$\bigcup_{\lambda\in\mathbb{R}} \text{Edge}^{(\text{recon},\lambda)}_{T^C}
= \mathrm{Edge}_{T^C}$.

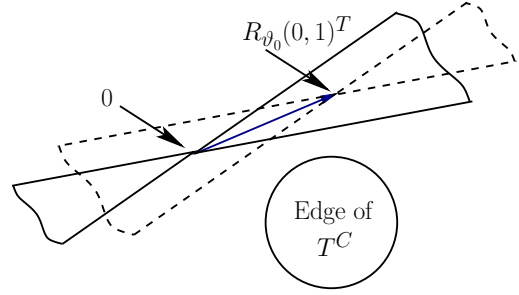
\begin{figure}
\begin{center}
\resizebox{0.9\linewidth}{!}{
\begin{picture}(0,0)%
\includegraphics{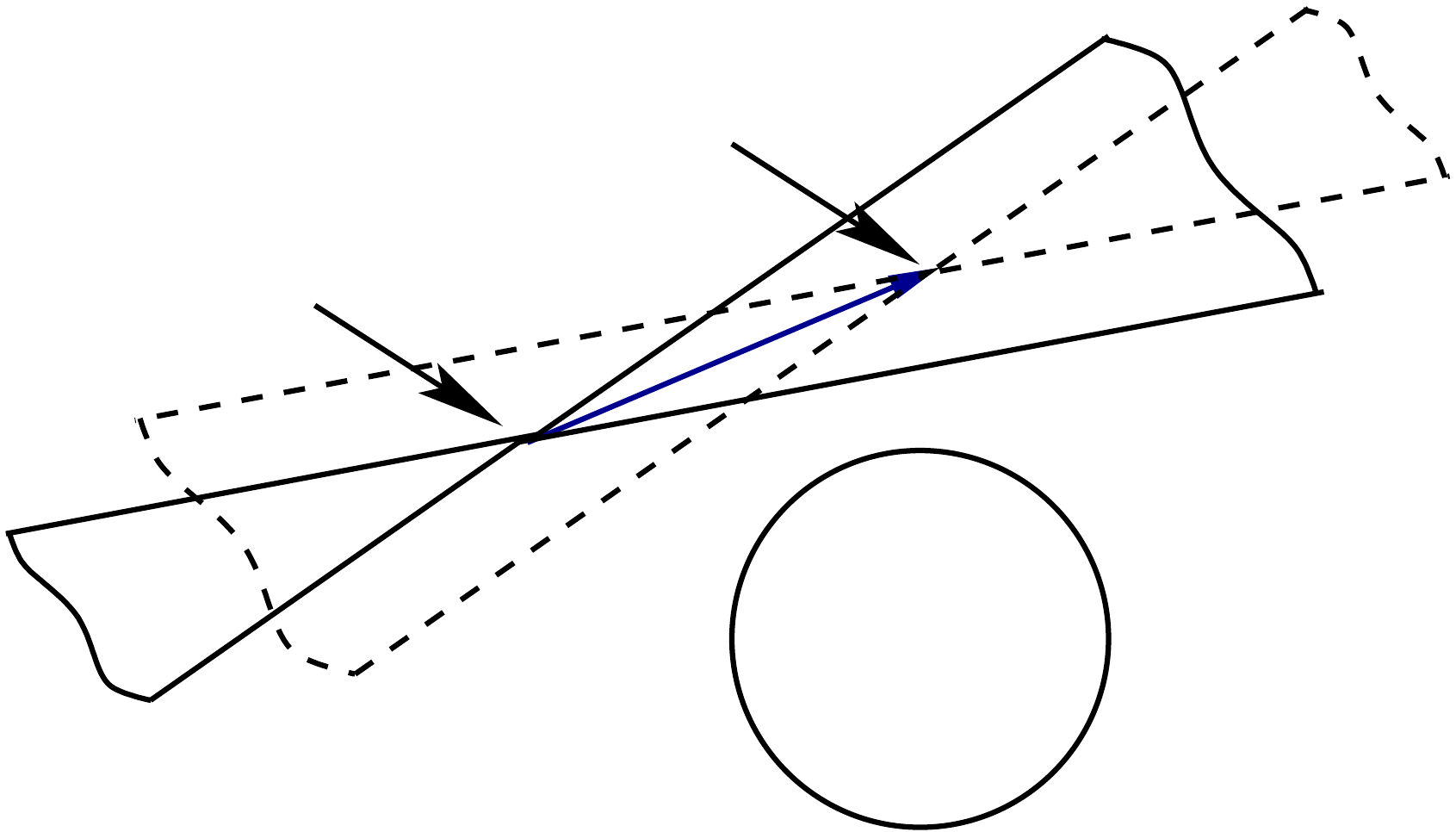}%
\end{picture}%
\setlength{\unitlength}{3947sp}%
\begingroup\makeatletter\ifx\SetFigFont\undefined%
\gdef\SetFigFont#1#2#3#4#5{%
  \reset@font\fontsize{#1}{#2pt}%
  \fontfamily{#3}\fontseries{#4}\fontshape{#5}%
  \selectfont}%
\fi\endgroup%
\begin{picture}(8114,4618)(3346,-4792)
\put(7051,-811){\makebox(0,0)[lb]{\smash{{\SetFigFont{29}{34.8}{\familydefault}{\mddefault}{\updefault}{\color[rgb]{0,0,0}$R_{\vartheta_0}(0,1)^T$}%
}}}}
\put(4801,-1861){\makebox(0,0)[lb]{\smash{{\SetFigFont{29}{34.8}{\familydefault}{\mddefault}{\updefault}{\color[rgb]{0,0,0}$0$}%
}}}}
\put(8251,-4261){\makebox(0,0)[lb]{\smash{{\SetFigFont{29}{34.8}{\familydefault}{\mddefault}{\updefault}{\color[rgb]{0,0,0}$T^C$}%
}}}}
\put(7876,-3661){\makebox(0,0)[lb]{\smash{{\SetFigFont{29}{34.8}{\familydefault}{\mddefault}{\updefault}{\color[rgb]{0,0,0}Edge of}%
}}}}
\end{picture}%
}
\end{center}
\caption{(color online) Reconstructing the Green's function on the complete edge.
The reconstructed area $\text{Edge}^{(\text{recon},0)}_{T^C}$ (see
Fig.~\ref{fig:edgerecon1} and Eq.~\eqref{eq:edgereconarea1}) may be extended by affine
transformations along the $R_{\vartheta_0}(0,1)^T$ direction using the
information from the set $E_1$. 
}
\label{fig:edgerecon2}
\end{figure}

This ``affine procedure'' is sketched in Fig.~\ref{fig:edgerecon2}.

\section{Empirical properties of the residual term}
\label{app:empiricalrelations}
The empirical observation that the quantum Monte Carlo data are continuous as
a function of $\imag\omega_n$, as one crosses higher-order branch cuts, yields the
following structure. Let $\vartheta_0$ be the orientation of the
corresponding branch cut. Then, due to the observed continuity, we have
\begin{equation}
\begin{split}
G(\imag \rho \sin (\vartheta_0-\delta),\imag \rho \cos (\vartheta_0-\delta))
= \\
=G(\imag \rho \sin (\vartheta_0+\delta),\imag \rho \cos
(\vartheta_0+\delta)),
\end{split}
\end{equation}
for any $\rho > 0$.

Using the identity theorem along the directions $\vartheta_0\pm\delta$, one
find that the relation
\begin{equation}
R^{(\vartheta_0-\delta,\vartheta_0+\delta)}(\rho \sin (\vartheta_0), \rho \cos
(\vartheta_0)) = 0
\end{equation}
holds for any $\rho > 0$.
Consequently, the continuity in Matsubara space induces a continuity relation
in the edge space. Similarly, one could construct relations for the
derivatives of $G$ which help constrain
$R^{(\vartheta,\vartheta')}(\underline x)$.

\section{Commutator relations of $\mathcal{Q}^\text{(edge)}_\vartheta$}
\label{sec:commrelQedge}

We will derive the commutator relations 
\eqref{eq:commutatorQthetaTranslation}, \eqref{eq:commutatorQthetaScaling},
and \eqref{eq:commutatorQthetaRotation}.

\subsection{Translational invariance}
Let us consider the action of $\mathcal{Q}^\text{(edge)}_\vartheta=
\mathcal{R}_\vartheta \mathcal{H}\mathcal{R}_\vartheta^{-1}\mathcal{H}$ on a
function $\tilde A(x,y)$ which is translated by the operator
$\mathcal{T}_{\underline X}$, where we set $\underline{X} = (X,Y)^T$.
We also write $\underline x:= (x,y)^T$.

As a first step, we apply the Hilbert transform,
\begin{equation}
\begin{split}
(\mathcal{H}\mathcal{T}_{\underline X} \tilde A)(\underline x) =& 
\frac{1}{\pi} 
\mathcal{P}\!\!\!\!\!\!\int\!\!\Dfrtl{\tilde y}
\frac{\tilde A(x-X, \tilde y - Y)}
{y-\tilde y}\\
=&
\frac{1}{\pi} 
\mathcal{P}\!\!\!\!\!\!\int\!\!\Dfrtl{\tilde y}
\frac{\tilde A(x-X, \tilde y)}
{(y-Y)-\tilde y}.\\
\end{split}
\end{equation}
Using the short-hand notation $c:=\cos \vartheta$, $s:=\sin\vartheta$, we
then have
\begin{equation}
(\mathcal{R}_\vartheta^{-1}\mathcal{H}\mathcal{T}_{\underline X} \tilde A)
(\underline x)
=
\frac{1}{\pi} 
\mathcal{P}\!\!\!\!\!\!\int\!\!\Dfrtl{\tilde y}
\frac{\tilde A(cx+sy-X, \tilde y)}
{(-sx+cy-Y)-\tilde y}.
\end{equation}
In the next two steps one obtains
\begin{widetext}
\begin{equation}
(\mathcal{Q}^\text{(edge)}_\vartheta \mathcal{T}_{\underline X} \tilde A)
(\underline x) = 
\frac{1}{\pi^2}
\mathcal{P}\!\!\!\!\!\!\int\!\!\Dfrtl{\tilde{\tilde y}}
\frac{1}{\underbrace{sx+cy-\tilde{\tilde y}}_{\to \eqref{eq:eqnsystsymmI}}}
\mathcal{P}\!\!\!\!\!\!\int\!\!\Dfrtl{\tilde y}
\frac
{
\tilde A (\overbrace{(cx-sy)c+s\tilde{\tilde y} - X}^{\to \eqref{eq:eqnsystsymmII}}, \tilde y)
}
{
\underbrace{-(cx-sy)s + c\tilde{\tilde y} - Y}_{\to \eqref{eq:eqnsystsymmIII}} - \tilde y
}.
\label{eq:QTAtilde}
\end{equation}
By substituting $\underline X = 0$ and then applying $\mathcal{T}_{\underline
X}$ from the left, one finds
\begin{equation}
(\mathcal{T}_{\underline X} \mathcal{Q}^\text{(edge)}_\vartheta \tilde A)
(\underline x) =
\frac{1}{\pi^2}
\mathcal{P}\!\!\!\!\!\!\int\!\!\Dfrtl{\tilde{\tilde y}}
\frac{1}{s(x-X)+c(y-Y)-\tilde{\tilde y}}
\mathcal{P}\!\!\!\!\!\!\int\!\!\Dfrtl{\tilde y}
\frac
{
\tilde A ((c(x-X)-s(y-Y))c+s\tilde{\tilde y}, \tilde y)
}
{
-(c(x-X)-s(y-Y))s + c\tilde{\tilde y} - \tilde y
}.
\label{eq:TQAtilde}
\end{equation}
In order to verify that the expressions \eqref{eq:QTAtilde} and
\eqref{eq:TQAtilde} are, in fact, equal, we consider the following system of
linear equations:
\begin{eqnarray}
\label{eq:eqnsystsymmI}
sx + cy - \tilde {\tilde y} &=& s x_0 + c y_0 - y^* \\
\label{eq:eqnsystsymmII}
(cx-sy)c + s\tilde{\tilde y} -X &=& (cx_0  -sy_0 ) c + sy^*  \\
\label{eq:eqnsystsymmIII}
-(cx - sy)s + c \tilde {\tilde y} -Y &=& - (cx_0  - sy_0 ) s + y^* c,
\end{eqnarray}
in matrix form:
\begin{equation}
\underbrace{
\begin{pmatrix}
-1 & s & c \\
s & c^2 & -sc \\
c & -sc & s^2
\end{pmatrix}
}_{=:M} \cdot
\begin{pmatrix}
\tilde {\tilde y} \\
x \\
y
\end{pmatrix}
-
\begin{pmatrix}
0 \\
X \\
Y
\end{pmatrix}
=
\underbrace{
\begin{pmatrix}
-1 & s & c \\
s & c^2 & -sc \\
c & -sc & s^2
\end{pmatrix}
}_{=:M} \cdot
\begin{pmatrix}
y^* \\
x_0 \\
y_0
\end{pmatrix}.
\end{equation}

The equations correspond to the idea of substituting the terms in equation
\eqref{eq:QTAtilde} as denoted there in such a way that $y^*$ is the new
integration variable, $\int \Dfrtl{\tilde{\tilde y}} \to \kappa \int \Dfrtl
{y^*}$, where $\kappa$ is some regular prefactor from the integral transformation,
and such that the external variables $x$ and $y$ are replaced by $x_0$ and $y_0$.
We will see that the resulting form will exactly be
\eqref{eq:TQAtilde}:

The system \eqref{eq:eqnsystsymmI}-\eqref{eq:eqnsystsymmIII} can be solved by
inversion of the matrix $M$: One has 
\begin{equation}
M^{-1} = 
\begin{pmatrix}
0 & s & c \\
s & 1 & 0 \\
c & 0 & 1 
\end{pmatrix},
\end{equation}
$\det M = -1$, and consequently the well-defined solution
\begin{eqnarray}
y^* &=& \tilde{\tilde y} - (sX + cY), \\
x_0 &=& x -X, \\
y_0 &=& y-Y.
\end{eqnarray}
The integral transformation constant $\kappa = 1$, and performing the
substitutions \eqref{eq:eqnsystsymmI}-\eqref{eq:eqnsystsymmIII} in 
\eqref{eq:QTAtilde} yields \eqref{eq:TQAtilde}, when $y^*$ is again renamed
$\tilde{\tilde y}$.

Therefore, $[\mathcal{Q}^\text{(edge)}_\vartheta,\mathcal{T}_{\underline X}] =0$.

\subsection{Scale invariance}
Similarly, we show the scale invariance \eqref{eq:commutatorQthetaScaling}. 
We have 
\begin{eqnarray}
(\mathcal{Q}^\text{(edge)}_\vartheta \Lambda_\lambda \tilde A) (\underline x)
&=&
\frac{\lambda^2}{\pi^2}
\mathcal{P}\!\!\!\!\!\!\int\!\!\Dfrtl{\tilde{\tilde y}}
\frac{1}{sx + cy - \tilde{\tilde y}}
\mathcal{P}\!\!\!\!\!\!\int\!\!\Dfrtl{\tilde y}
\frac{
\tilde A(\lambda [(cx-sy)c + s \tilde {\tilde y}], \lambda \tilde y)
}
{
-(cx - sy)s + c \tilde{\tilde y} - \tilde y
} \\
&=& 
\frac{1}{\pi^2}
\mathcal{P}\!\!\!\!\!\!\int\!\!\Dfrtl{\tilde{\tilde y}}
\frac{1}{sx + cy - \tilde{\tilde y}/ \lambda}
\mathcal{P}\!\!\!\!\!\!\int\!\!\Dfrtl{\tilde y}
\frac{
\tilde A((c\lambda x-s\lambda y)c + s \tilde {\tilde y}, \tilde y)
}
{
-(cx - sy)s + c \tilde{\tilde y}/\lambda - \tilde y/\lambda
} \\ 
&=& 
\frac{\lambda^2}{\pi^2}
\mathcal{P}\!\!\!\!\!\!\int\!\!\Dfrtl{\tilde{\tilde y}}
\frac{1}{s\lambda x + c\lambda y - \tilde{\tilde y}}
\mathcal{P}\!\!\!\!\!\!\int\!\!\Dfrtl{\tilde y}
\frac{
\tilde A((c\lambda x-s\lambda y)c + s \tilde {\tilde y}, \tilde y)
}
{
-(c\lambda x - s\lambda y)s + c \tilde{\tilde y} - \tilde y
} \\
&=&
(\Lambda_\lambda \mathcal{Q}^\text{(edge)}_\vartheta \tilde A) (\underline
x).
\end{eqnarray}

\subsection{Absence of rotational invariance}
We provide a simple example for which $[\mathcal{Q}^\text{(edge)}_\vartheta,
\mathcal{R}_{\vartheta'}] \neq 0$.

We consider the bare $\tilde A_0$, equation \eqref{eq:nonintAtilde}.
Setting $\vartheta=\vartheta'=\pi/2$, we find 

\begin{eqnarray}
(\mathcal{Q}^\text{(edge)}_{\pi/2} \tilde A_0)(\underline x) & =&
\sum_{\alpha=\pm1}
\frac{\alpha\Gamma_\alpha / \pi} 
{
(x_\omega -\alpha (x_\varphi - \Phi) / 2 - \varepsilon_d)^2 + 
\Gamma^2
},\text{ and} \\
(\mathcal{R}_{\pi/2}\mathcal{Q}^\text{(edge)}_{\pi/2} \tilde A_0)(\underline
x) &=&
\sum_{\alpha=\pm1}
\frac{\alpha\Gamma_\alpha / \pi} 
{
(x_\varphi -\alpha (-x_\omega - \Phi) / 2 - \varepsilon_d)^2 + 
\Gamma^2
}.
\end{eqnarray}

On the other hand,
\begin{eqnarray}
(\mathcal{R}_{\pi/2}\tilde A_0)(\underline x) &=& 
\sum_{\alpha=\pm1}
\frac{\Gamma_\alpha / \pi}
{
(x_\varphi -\alpha (-x_\omega - \Phi) / 2 - \varepsilon_d)^2 + 
\Gamma^2
},\text{ and} \\
(\mathcal{Q}^\text{(edge)}_{\pi/2}\mathcal{R}_{\pi/2} \tilde A_0)(\underline
x) &=&
\sum_{\alpha=\pm1}
\frac{-\alpha\Gamma_\alpha / \pi}
{
(x_\varphi -\alpha (-x_\omega - \Phi) / 2 - \varepsilon_d)^2 + 
\Gamma^2
}.
\end{eqnarray}
Hence, $\mathcal{Q}^\text{(edge)}_{\pi/2}\mathcal{R}_{\pi/2} \tilde A_0 =
-\mathcal{R}_{\pi/2}\mathcal{Q}^\text{(edge)}_{\pi/2} \tilde A_0$, and
therefore
$[\mathcal{Q}^\text{(edge)}_{\pi/2},\mathcal{R}_{\pi/2}]\neq 0$.

\section{Numerical representation of the multi-wedge map $\mathcal{Q}$}
\label{app:numericsQ}
In this appendix, a recipe for the numerical computation of the quadruple 
integral $\mathcal{Q}$ is given. The application to the test function
\eqref{eq:numtestfuncQgeneral},
$f_{\underline X, \delta}$, for a function value at
$(\imag\varphi_m,\imag\omega_n)$ is to be computed.
The first three integrals can be computed analytically by use of a computer
algebra system. A numerical quadrature method can be used for the
approximation of the remaining integral.

\subsection{Analytic computation of first three integrals}
Using the translational and scale invariance of the edge-to-edge contribution
$\mathcal{Q}^\text{(edge)}_\vartheta$, only the action of
$\mathcal{Q}^\text{(edge)}_\vartheta$ on our test function
\begin{equation}
f(x,y) = \frac{1}{\pi^2} \frac{1}{(y-x/2)^2+1} \cdot \frac{1}{(y+x/2)^2+1}
\end{equation}
 has to be computed, yielding the
results shown in figure \ref{fig:Qedgetransformviz}. For brevity we set 
$y=x_\omega$ and $x=x_\varphi$.
The first principal integral can be eliminated by straightforward application 
of the residue theorem:
\begin{equation}
\begin{split}
(\mathcal{H}f)(x,\tilde y) =& \mathcal{P}\!\!\!\!\!\!\int \Dfrtl y 
  \frac{\pi^{-1}}{\tilde y - y} f(x,y) \\
=& -
\frac{8}{{
\pi }^{2}}\,{\frac { \left( {x}^{2}-12-4\,{{\it \tilde y}}^{2} \right) {\it \tilde y
}}{ \left( 16\,{{\it \tilde y}}^{4}+32\,{{\it \tilde y}}^{2}-8\,{{\it \tilde y}
}^{2}{x}^{2}+16+8\,{x}^{2}+{x}^{4} \right)  \left( {x}^{2}+4 \right) }}.
\end{split}
\end{equation}
As a next step, introducing the short-hand notation $s=\sin \vartheta$ and
$c=\cos \vartheta$, imposing the rotation operator $\mathcal{R}_\vartheta^{-1}$, one
obtains
\begin{equation}
(\mathcal{R}_\vartheta^{-1}\mathcal{H}f)(k,l) = \mathrm{subs}\,(x \to kc+ls, \tilde y \to lc-ks;
(\mathcal{H}f)(x,\tilde y)).
\end{equation}
Here, ``$\mathrm{subs}$'' denotes the operation of a variable substitution.
In order to apply the second Hilbert transform, it is necessary to determine
the poles of the corresponding integrand
\begin{equation}
\begin{split}
g(k, \tilde{\tilde y}; l) := & \frac{\pi^{-1}}{\tilde{\tilde y} - l}
 \cdot (\mathcal{R}_\vartheta^{-1}\mathcal{H}f)(k,l) \\
= & 8\,{\frac { \left( {k}^{2}{c}^{2}+10\,kcls+{l}^{2}{s}^{2}-12-4\,{l}^{2
}{c}^{2}-4\,{k}^{2}{s}^{2} \right)  \left( lc-ks \right) }{{\pi }^{3}
 \left( {k}^{2}{c}^{2}+2\,kcls+{l}^{2}{s}^{2}+4 \right)  }}
\cdot
\\ 
&\cdot
\big( 4
\,{l}^{2}{c}^{2}-6\,kcls+4\,{k}^{2}{s}^{2}+4+4\,k{c}^{2}l
-\\&\qquad-4\,{k}^{2}cs
+4\,{l}^{2}sc-4\,l{s}^{2}k+{k}^{2}{c}^{2}+{l}^{2}{s}^{2} \big)^{-1}\cdot
\\ &\cdot
{  
 \left( -\tilde {\tilde y}+l \right) 
}^{-1}
\end{split}
\end{equation}
with respect to the integration variable $l$.
One finds that the function has the following seven poles in the complex
plane:
\begin{eqnarray}
l_1 &:=& \tilde{\tilde y}, \\
l_{2,3} &:=& \frac{-kc + 2ks\pm 2\imag}{s}, \\
l_{4,5} &:=& \frac{-kc + 2ks \pm 2\imag}{s+2c}, \\
l_{6,7} &:=& \frac{kc + 2ks \pm 2\imag}{-s+2c}.
\end{eqnarray}
The Hilbert transform can now be evaluated through the residue sum
\begin{equation}
(\mathcal{H}\mathcal{R}_\vartheta^{-1}\mathcal{H}f)(k,\tilde{\tilde y})
= \sum_{n=1}^7 \Re \left[2 \pi \imag \cdot \mathrm{Res}_{l = l_n} g(k,
\tilde{\tilde y}; l)\right]
   \cdot \theta (\Im l_n).
\label{eq:ressumHRHf}
\end{equation}
The Heaviside function $\theta(x)$ ensures that only poles from the upper
half plane are taken into account for the evaluation of the contour integral
which corresponds to the principal value integral. An explicit evaluation of the
residue sum \eqref{eq:ressumHRHf} can be accomplished with a computer algebra
system. A dramatic increase in complexity is 
coming along with the constraints $\Im l_n > 0$ which depend on the
wedge orientation angle $\vartheta$. In fact, six separate cases emerge as a function of
$\vartheta$. They can be parametrized by three overlapping different cases, 
namely (A) terms which are proportional to $\sign (s)$, (B) proportional to
$\sign (2c+s)$, and (C) terms which are proportional to $\sign (2c-s)$,
discriminating between the different signs of $\Im l_n$.
The term
\begin{equation}
(\mathcal{R}_\vartheta\mathcal{H}\mathcal{R}_\vartheta^{-1}\mathcal{H}f)(x,y) = 
\mathrm{subs}\,(
k\to xc-ys,
\tilde{\tilde y} \to yc + xs
;
(\mathcal{H}\mathcal{R}_\vartheta^{-1}\mathcal{H}f)(k,\tilde{\tilde y})
)
\end{equation}
is then best reorganized into rational functions as coefficients of the sign
functions, 
\begin{equation}
\begin{split}
(\mathcal{Q}^\text{(edge)}_\vartheta f)(x,y) = 
(\mathcal{R}_\vartheta\mathcal{H}\mathcal{R}_\vartheta^{-1}\mathcal{H}f)(x,y)
= &
A(x,y) \cdot \sign (s) + \\
& +B(x,y) \cdot \sign (2c+s) + \\
&+C(x,y) \cdot \sign
(2c-s).
\end{split}
\label{QedgefkoeffABCrepresentation}
\end{equation}
This explicit split is necessary for the study of the interplay of
the rational functions $A(x,y)$, $B(x,y)$, and $C(x,y)$ in a
computer algebra system.

\subsubsection{Rational coefficients of the transformed edge test function}
The rational functions
\begin{eqnarray}
A(x,y) &=& \frac{A_\text{enum}(x,y)}{A_\text{denom}(x,y)}, \\
B(x,y) &=& \frac{B_\text{enum}(x,y)}{B_\text{denom}(x,y)}, \\
C(x,y) &=& \frac{C_\text{enum}(x,y)}{C_\text{denom}(x,y)}.
\end{eqnarray}
have the following polynomials as enumerators and denominators:
\begin{equation}
\begin{split}
A_\text{enum}(x,y) =&
-8\, \Bigg(  \left( -1+{x}^{2}+1/4\,{y}^{2}-{\frac {41}{16}}\,xy
 \right) {c}^{4}-{\frac {13}{16}}\, \left( {\frac {12}{13}}\,xy+{x}^{2
}+{\frac {12}{13}}-{\frac {28}{13}}\,{y}^{2} \right) s{c}^{3} \\
&\quad + \left( 
1+{\frac {13}{8}}\,xy-1/2\,{y}^{2} \right) {c}^{2}+1/16\,s \left( -4+{
x}^{2}-4\,xy-12\,{y}^{2} \right) c \\
&\quad -1/16\,y \left( -4\,y+x \right) 
 \Bigg),
\end{split}
\end{equation}
\begin{equation}
A_\text{denom}(x,y) =
 {\pi }^{2} \left( {x}^{2}+4-4\,xy+4\,{y}^{2} \right) \cdot  K(x,y),
\end{equation}
\begin{equation}
\begin{split}
B_\text{enum}(x,y) = &
-8\, \Bigg(  \left( -1+{x}^{2}+1/4\,{y}^{2}+{\frac {41}{16}}\,xy
 \right) {c}^{4}+ \\
&\quad +{\frac {13}{16}}\,s \left( -{\frac {12}{13}}\,xy+{x}^
{2}+{\frac {12}{13}}-{\frac {28}{13}}\,{y}^{2} \right) {c}^{3}+
 \left( 1-{\frac {13}{8}}\,xy-1/2\,{y}^{2} \right) {c}^{2} \\
&\quad -1/16\,s
 \left( -4+{x}^{2}+4\,xy-12\,{y}^{2} \right) c+1/16\,y \left( 4\,y+x
 \right)  \Bigg), 
\end{split}
\end{equation}
\begin{equation}
B_\text{denom}(x,y) = 
 \left( {x}^{2}+4\,xy+4\,{y}^{2}+4 \right){
\pi }^{2} \cdot K(x,y),
\end{equation}
\begin{equation}
\begin{split}
C_\text{enum}(x,y) = &
-3\, \left( -1/3\,{y}^{2}-3+{x}^{2} \right) yx{c}^{4}-s
 \left(  \left( -3\,{x}^{2}+4 \right) {y}^{2}-5\,{x}^{2}+4+{x}^{4}
 \right) {c}^{3} \\
&  +3\, \left( -10/3+{x}^{2}-2/3\,{y}^{2} \right) yx{c}^{
2} \\
& + \left(  \left( -3\,{x}^{2}+4 \right) {y}^{2}-{x}^{2}+4 \right) sc+
yx \left( {y}^{2}+1 \right), 
\end{split}
\end{equation}
and
\begin{equation}
C_\text{denom}(x,y) =
 \left( {x}^{2}+4 \right) {\pi }^{2} \cdot K(x,y). 
\end{equation}
For the denominators, we introduced the shared polynomial
\begin{equation}
\begin{split}
K(x,y) = &
  \left( {y}^{4}+ \left( -3-6\,{x}^{2} \right) {y}^{2}+3\,{x}^{
2}-4+{x}^{4} \right) {c}^{4} \\
& \quad-4\, \left( 3/2+{x}^{2}-{y}^{2} \right) xs
y{c}^{3}+ \left( -2\,{y}^{4}+ \left( 2+6\,{x}^{2} \right) {y}^{2}+4+{x
}^{2} \right) {c}^{2} \\
& \quad -4\, \left( 1/2+{y}^{2} \right) xsyc+{y}^{2}+{y}^
{4}.
\end{split}
\end{equation}
\end{widetext}
\subsubsection{Composition of the rational coefficients}
\label{app:subsec:compositionrationalcoefficients}
In contrast to the actual edge functions
$\mathcal{Q}_\vartheta^\text{(edge)}f$, the coefficients 
$A(x,y)$, $B(x,y)$, and $C(x,y)$ are comparably ill-behaved. For example, one
obtains
\begin{equation}
\left.B(x,y)\right|_{\vartheta=0} = -8\,{\frac {x+y}{{\pi }^{2} \left(
{x}^{2}+4 \right)  \left( {x}^{2}+
4\,xy+4\,{y}^{2}+4 \right) }}\cdot \frac{1}{x}.
\end{equation}
Apparently, the function diverges for $x\to 0$. 
The functions $A(x,y)$ and $C(x,y)$ are similarly structured. However, the 
actual edge functions, as
displayed in figure \ref{fig:Qedgetransformviz}, have no such singularities,
but do rather represent smooth deformations of the test function when
the wedge orientation angle $\vartheta$ is tuned. As a
consequence, the real singularities are cancelled as the rational functions 
are added up in equation \eqref{QedgefkoeffABCrepresentation}. For a 
consistent further evaluation of the action of $\mathcal{Q}$ on the test 
function it is thus necessary to study each of the full combinations
\eqref{QedgefkoeffABCrepresentation} separately. There are six possible
combinations, namely the sectors
\begin{eqnarray}
(a)& 0 &\leq\vartheta \leq \arctan 2, \label{eq:contribsectora} \\
(b)& \arctan 2 &\leq \vartheta \leq \pi -\arctan 2, \label{eq:contribsectorb} \\
(c)& \pi - \arctan 2 &\leq \vartheta \leq \pi, \label{eq:contribsectorc} \\
(d)& \pi             &\leq \vartheta \leq \pi + \arctan 2, \label{eq:contribsectord} \\
(e)& \pi + \arctan 2 &\leq \vartheta \leq 2\pi - \arctan 2, \label{eq:contribsectore} \\
(f)& 2\pi - \arctan 2 &\leq \vartheta \leq 2\pi. \label{eq:contribsectorf} 
\end{eqnarray}

For example, the expression for the sector $(a)$ reads
\begin{equation}
\left.(\mathcal{Q}^\text{(edge)}_\vartheta f)\right|_{(a)}(x,y)
=
A(x,y) + B(x,y) + C(x,y).
\end{equation}

\subsubsection{Contraction with the Poisson kernel}
As a next step, one of the integrals introduced by Vladimirov's Poisson kernel 
\eqref{eq:PoissonOpFiniteTheta} will be evaluated
analytically. For this, the pole structure of both, the Poisson kernel
\eqref{eq:poissonkernelrotatedepsdomain}, and
the edge-transformed test function have to be analyzed.
\paragraph{Pole structure of edge-transformed test functions}
In fact, we are interested in the pole structure of the scaled and then translated
edge-transformed test functions $(\mathcal{T}_{\underline X} \Lambda_{1/\delta} 
\mathcal{Q}^\text{(edge)}_\vartheta f)(x',y')$, from equation
\eqref{eq:Qsimplified}. The poles and also the residues of these functions
are however easily calculated from the poles and residues of
$(\mathcal{Q}^\text{(edge)}_\vartheta f)(x,y)$. This is the crucial advantage
of translational and scale invariance of
$\mathcal{Q}^\text{(edge)}_\vartheta$.

For example, in sector $(a)$, the following poles of
$(\mathcal{Q}^\text{(edge)}_\vartheta f)(x,y)$ with respect to $x$ are
obtained: $x_{1,2} = \pm2\imag$, $x_{3,4} = s y \pm 2 \imag$, 
$x_{5,6} = -2y \pm 2 \imag$,
$x_{7,8} = 2y \pm 2 \imag$. The corresponding residues are 
$r_{1,2} = \frac{1}{2\pi^2 y}$, $r_{3,4} = - \frac{1}{2\pi^2 y}$, 
$r_{5,6} = \frac{1}{4 \pi^2} \cdot \frac{-1 \mp \imag y} {(y^2 +1) y}$,
$r_{7,8} = \frac{1}{4 \pi^2} \cdot \frac{1 \mp \imag y} {(y^2 +1) y}$.
The resulting residues of $(\mathcal{T}_{\underline X} \Lambda_{1/\delta} 
\mathcal{Q}^\text{(edge)}_\vartheta f)(x',y')$ are then given by
\begin{equation}
r_i' = \frac{1}{\delta} \cdot \mathrm{subs}\,\left(y\to \frac{y'-Y}{\delta};
r_i(y)\right),
\end{equation}
where the center of mass of the test function is $\underline X = (X,Y)$.
They are associated to the poles of $(\mathcal{T}_{\underline X} \Lambda_{1/\delta} 
\mathcal{Q}^\text{(edge)}_\vartheta f)(x',y')$, which are similarly given by
\begin{equation}
x_i' = X + \delta \cdot \mathrm{subs}\,\left( y \to \frac{y' - Y}{\delta}; x_i \right).
\end{equation}
In order to obtain the true residues with respect to the $x'$-contraction
with the Poisson kernel, one only has to evaluate the Poisson kernel
at the poles \eqref{eq:poissonkernelrotatedepsdomain} and multiply $r_i'$ with the value.

\paragraph{Pole structure of Poisson kernel}
The pole structure of Vladimirov's Poisson kernel is rather straightforward
to compute, however rather lengthy expressions result for the poles. Similar
to above, the residues have to be multiplied by the function values of the
edge-transformed test function $(\mathcal{T}_{\underline X}
\Lambda_{1/\delta}  \mathcal{Q}^\text{(edge)}_\vartheta f)(x',y')$ at the pole of the Poisson kernel.
Poles of the rotated Poisson kernel $\mathcal{P}_{r,\vartheta}(\varphi_m, \omega_n; x', y')$ with 
respect to $x'$ are
\begin{eqnarray}
x_{1,2}' &=& 
         \frac{s-c\epsilon} {s\epsilon + c}y' \pm
     \imag \frac{\eta_1 + \varepsilon \eta_2}{s\varepsilon + c}, \\
x_{3,4}' &=&
          \frac{c\epsilon + s}{c - s \epsilon} y' \pm \imag
           \frac{\eta_1-\epsilon \eta_2}{s\epsilon - c}. 
\end{eqnarray}
We introduced the short-hand notations $\eta_1 = \varphi_m c - \omega_n s$, 
$\eta_2 = \varphi_m s + \omega_n c$. The associated residues of
$\mathcal{P}_{r,\vartheta}$ are easily determined.

\paragraph{Residue sum for the $x'$ integral}
An algebraic expression for the residue sum corresponding to the $x'$ integral can be generated symbolically
by evaluating  
\begin{widetext}
\begin{equation}
\begin{split}
I_3(\varphi_m, \omega_n; y') =
&
\int \Dfrtl{x'} 
\underbrace{\mathcal{P}_{r,\vartheta}(\varphi_m, \omega_n; x',y')}_\text{``Poisson''}
\cdot
\underbrace{(\mathcal{T}_{\underline X}
\Lambda_{1/\delta}  \mathcal{Q}^\text{(edge)}_\vartheta f)(x',y')
}_\text{``edge''} = \\
=& \sum_{x_i'}^\text{``edge''}
\Re \left[
2\pi\imag \cdot
\mathcal{P}_{r,\vartheta}(\varphi_m, \omega_n; x_i',y')
\cdot r_i'
\right] \cdot \theta (\Im x_i') \,\, + \\
& + \sum_{x_i'}^\text{``Poisson''}
\Re \Big[ 2\pi \imag \cdot
(\mathcal{T}_{\underline X}
\Lambda_{1/\delta}  \mathcal{Q}^\text{(edge)}_\vartheta f)(x_i',y')
\cdot  \\
& \quad\qquad\qquad\qquad \cdot \mathrm{Res}_{x'=x_i'}\mathcal{P}_{r,\vartheta}(\varphi_m, \omega_n; x',y')
\Big] \cdot \theta (\Im x_i').
\end{split}
\label{eq:residuesumthirdintegral}
\end{equation}
\end{widetext}
 However, it yields rather lengthy formulae, because the integrations 
with respect to $x'$ have to be done separately, for each of the sectors \eqref{eq:contribsectora} to
\eqref{eq:contribsectorf}. The growth in complexity is also due to amount of
parameters which increased dramatically by introducing the Poisson kernel and
inserting the translations $\mathcal{T}_{\underline X}$ and scaling
$\Lambda_{1/\delta}$ of the edge functions. 
It is possible to export the expressions resulting from
\eqref{eq:residuesumthirdintegral} from the computer algebra system to a file of Fortran code which is 500
kilobytes in size. Similar to the procedure described in section
\ref{app:subsec:compositionrationalcoefficients} for the second integral, the
expression \eqref{eq:residuesumthirdintegral} includes removable
discontinuities.
 
\subsection{Numerical quadrature of the fourth integral}
Due to the vast complexity of expression \eqref{eq:residuesumthirdintegral},
the last remaining integral
\begin{equation}
I_4 (\varphi_m,\omega_n) = \int \Dfrtl{y'} I_3(\varphi_m,\omega_n; y')
\end{equation} 
is evaluated numerically, making use of the exported Fortran code. Adaptive
integration routines from the 
GNU Scientific Library are imposed.\cite{RefManGSL} 
Because the resulting matrix elements give rise to an inverse problem,
it is compulsory to achieve a high integration accuracy.
By definition, the integrand is most distinguishly structured in the area $y' \approx Y$, 
on a scale $\delta$. Special attention has to be drawn to the appropriate integration of
this range.

The high-frequency tails $(-\infty,-R]$ and $[R,\infty)$ need to be
integrated out separately, where $R$ is the integration range of the
conventional quadrature. For some parameter values of $\delta$, $\underline X$, etc.,
problems with the convergence of these high-energy integrals may occur, due
to floating point precision.
Choosing a finite interval extending to $\pm \max(10^6, |X|\cdot 10^3,
|Y|\cdot 10^3)$ is then usually sufficient for numerically satisfactory data.

\begin{figure}
\centering
\includegraphics[width=\linewidth]{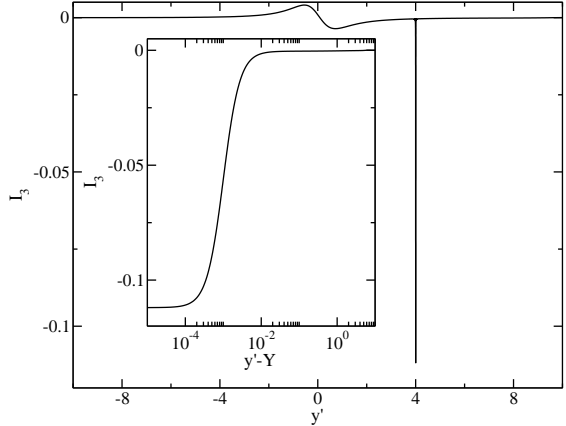}
\caption{Example for $I_3$ as a function of $y'$, using $\varphi_m = 2.0$, $\omega_n=2.0$,
with test function location $X=0.0$, $Y=4.0$, and test function width $\delta=10^{-3}$.
The interacting branch cut geometry is used for the determination of $r$ and
$\vartheta$.}
\label{fig:I3asfuncofyprime}
\end{figure}

A typical shape of the integrand $I_3$ is shown in figure
\ref{fig:I3asfuncofyprime}. The structure at $y' \approx Y$ is not
necessarily $\delta$-shaped, but depending on the values of $\varphi_m$ and
$\omega_n$ it may rather look like the Hilbert transform of such. 
The integral may be computed at each point $(X, Y)$ of the $\tilde A$ 
discretization lattice for all values of the simulation data
$(\imag\varphi_m,\imag\omega_n)$ on a computer cluster. In practice,
the computation of the matrix elements has to be done only once for each 
temperature $\beta$, regardless of the bias voltage. This is because an 
adjustment of the $\tilde A$ grid is not necessary in the latter case.

In future applications, one should aim at symbolically programming the
residue sum of the fourth integral and then take the limit
$\delta\to 0$ analytically.

\section{MaxEnt implementation for data from multiple wedges}
\label{app:maxentmultiwedge}
In this appendix chapter, the implementation of the MaxEnt algorithm for the
$\mathcal{Q}$-mapping is described. Details on the computation of the
numerical representation of $\mathcal{Q}$ were provided in appendix
\ref{app:numericsQ}.

The local test function width $\delta_{\underline x}$ for the map
$\mathcal{Q}_{r,\vartheta}$ can be adjusted to the local grid
resolution when the function $\tilde A (\underline x)$ is discretized. The
inverse problem for the inference of spectral properties using assumption
\eqref{eq:sharedRPassumption} is, by construction
\begin{equation}
\Im G(\imag \varphi_m,\imag \omega_n) = (\mathcal{Q}_{r,\vartheta} \tilde
A)(\imag \varphi_m, \imag \omega_n).
\end{equation}
The values $r,\vartheta$ are those which specify the $T^{C_{r,\vartheta}}$
branch of $G$ the vector $(\imag \varphi_m,\imag \omega_n)^T$ is located in,
as defined by point 1 of section \ref{subsec:holostructGF}.
The spectral function of the dot electrons can then again be gained by
evaluating along the physical line,
\begin{equation}
A(\omega) = \tilde A(\Phi,\omega),
\end{equation}
of the inferred $(\vartheta=0)$-edge function. See reference \onlinecite{dirks} for
details.

\subsubsection{Discretization of $\tilde A(x_\varphi,x_\omega)$} 
The single-wedge MaxEnt-based analytic continuation problem proposed in
Ref.~\onlinecite{dirks} only required a rather straightforward discretization of
the function $\tilde A(x_\varphi, x_\omega)$.
In contrast, for the multiwedge mapping, the 
discretization of the edge function has to pay tribute to the strong
intertwining of edge structure and branch cut structure which is revealed by
$\mathcal{Q}^\text{(edge)}$ (cf.~paragraph \ref{paragraph:propQedge}).

Especially, it turns out that the limiting behaviour along the singular directions of $\tilde A$
has to be captured numerically. In terms of the multiwedge approach, the
singular directions dominate the mathematical structure. In our experience, 
also the \emph{lateral} structure of $\tilde
A(\underline x)$ along the singular directions has to be resolved.
We constructed a grid as follows. Let $\tilde{x}^{(i)}_1$ and $\tilde
x^{(j)}_2$ be two
variables which are discretized on $i$-th and $j$-th logarithmic grid points
around zero, respectively. Then the grid
$
\underline{x}^{(i,j)} = 
\frac{\sqrt 5}{20}
\begin{pmatrix}
8 & 10 \\
-4       & 5
\end{pmatrix}
\underline{\tilde x}^{(i,j)}
$
yields an appropriate discretization of the edge, because the given matrix maps the double-cone
$\mathbb{R}^+\times\mathbb{R}^+ \cup \mathbb{R}^-\times\mathbb{R}^-$ and its complement to the wedges defined by the 
singular directions.
The numerical test function width $\delta_{\underline{x}}$ can then be adjusted to the
local grid resolution.

Also the high-energy structure of the Green's function has to be taken into
account explicitly, because along the singular directions it does not decay.
In practice it seems to be important to have a very large logarithmically
discretized fit region, for which in practice a $x_\varphi/\Gamma$ region of at most $[-800,
800]$ is subject to modifications by the MaxEnt algorithm and a
$x_\omega/\Gamma$
region of at most $[-400,400]$. The singular-direction contributions
beyond this range also prove not to be negligible, in a test with the bare
Green's function (see also the $G_0$ benchmark below). In order to take
them into account, their contribution up to very large energies ($x_\varphi
\approx 10^5\Gamma$) is computed assuming a $G_0$-like structure along the 
directions, positioning adequately weighted $\delta$-spikes along it and 
substracting the corresponding contributions from the raw
data, as done for the negative-spectral-function contributions of static 
observables in the first paper.

\subsubsection{Kernel structure}
The kernel $\mathcal{Q}_{r,\vartheta}$ may exhibit rather sharp structures
with respect to the $\tilde A$ function space. 
In particular, this may be the case in regions where the to-be-determined $\tilde A$ is expected to 
be very smooth and physically noninteresting. Consequently, for these
regions, the MaxEnt discretization grid would be chosen rather coarse-grained. 
These potentially disturbing structures can already be seen from the formal
structure of $\mathcal{P}_{r,\vartheta}$ which features strong 
anisotropies. 
The convolution with the (transformed) test functions $\mathcal{Q}^\text{(edge)}_\vartheta
f_{\underline X, \delta}$ is in general no
cure for this problem, because $\mathcal{Q}^\text{(edge)}_\vartheta
f_{\underline X, \delta}$ is even more sharply structured, on the scale
$\delta$, which is of the order of the discretiziation scale (see figure
\ref{fig:Qedgetransformviz}). 

\begin{figure}
\centering
\subfloat[$m=1$, $n=0$ ($\beta\Gamma=5.0$)]{\includegraphics[width=0.8\linewidth]{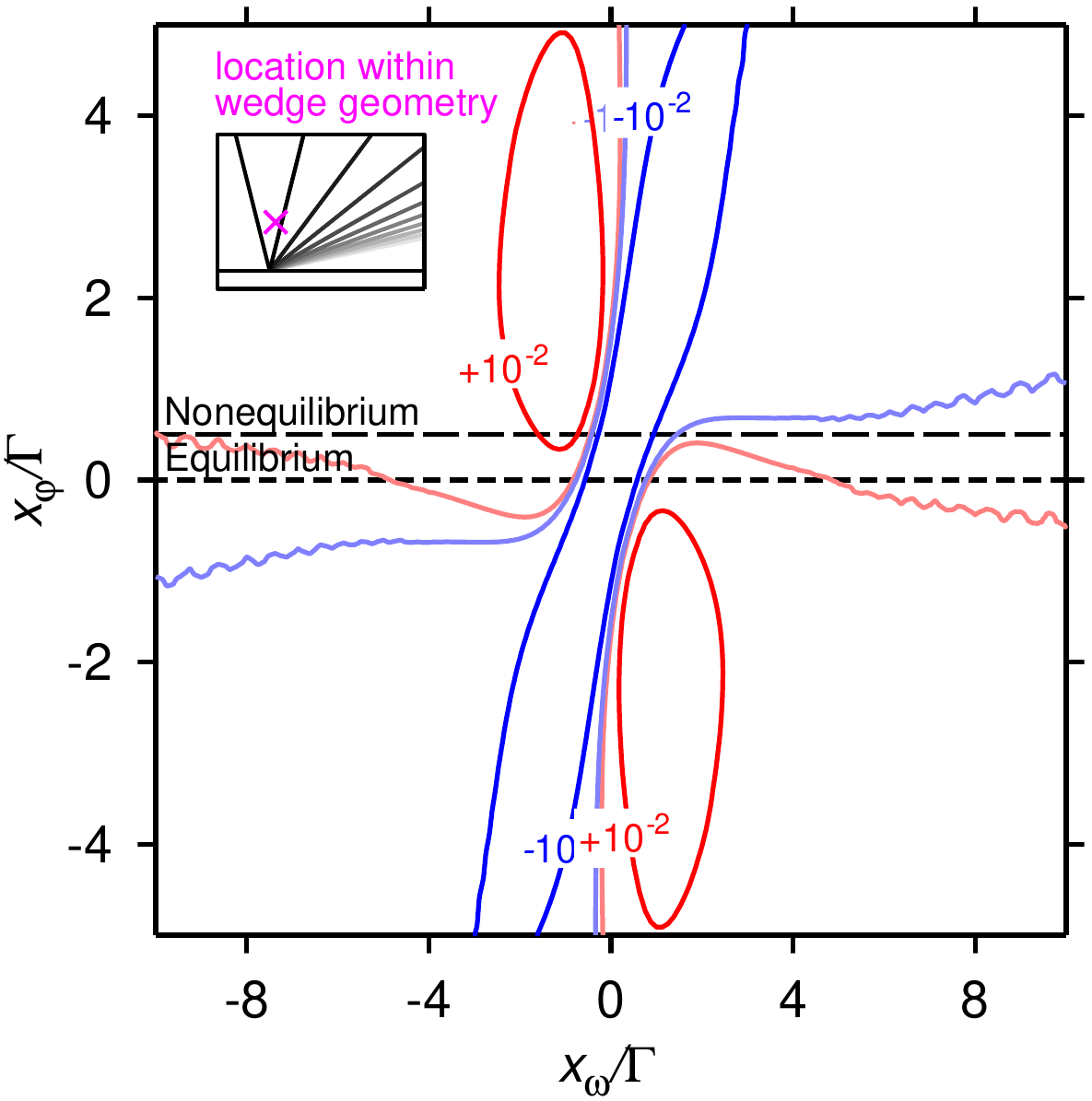}}
\\
\subfloat[$m=1$, $n=3$ ($\beta\Gamma=5.0$)]{\includegraphics[width=0.8\linewidth]{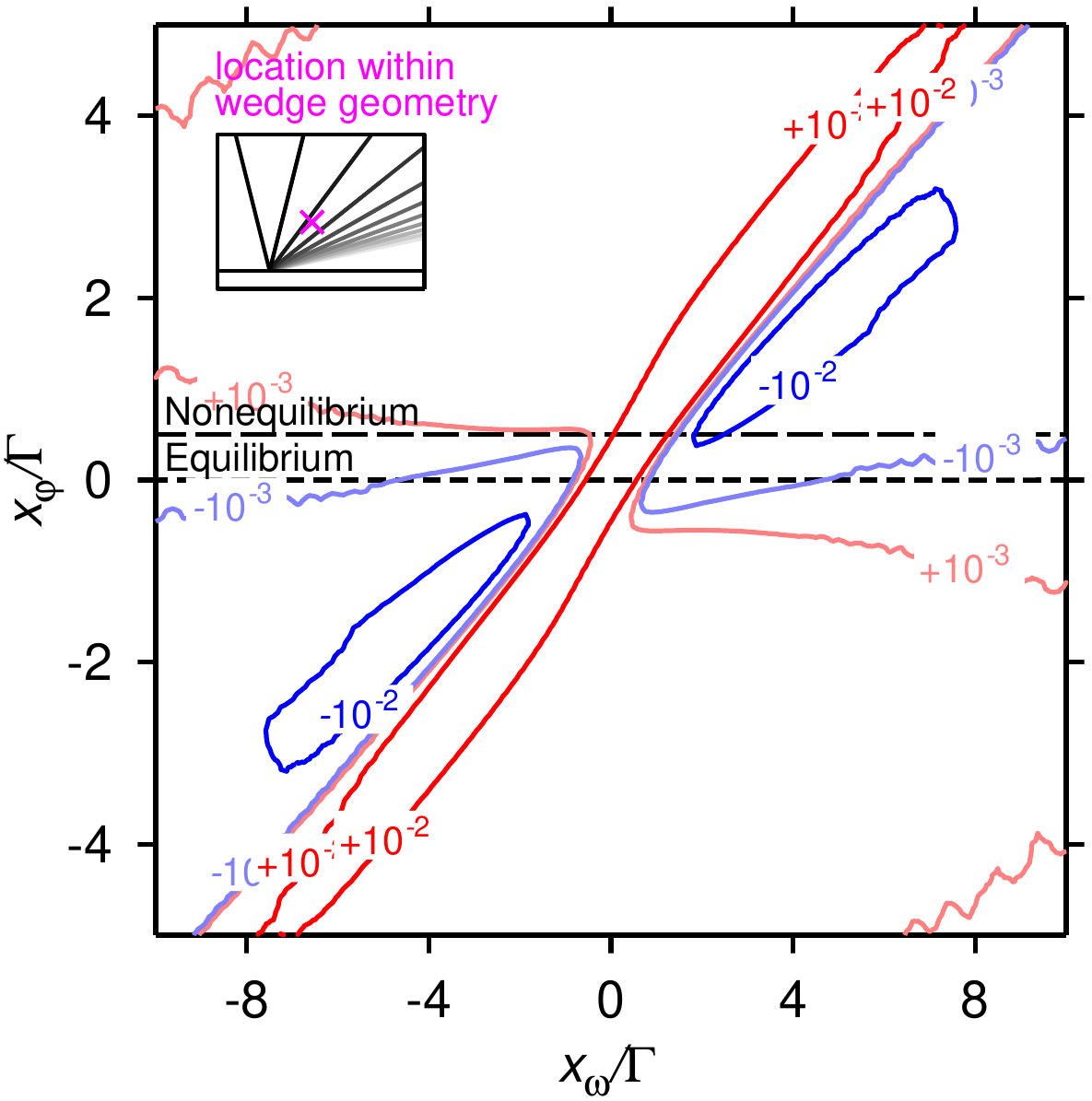}}
\caption{(color online) Cut through $\mathcal{Q}_{r,\vartheta}$ in $\tilde A$-space
for different pairs of $\imag\varphi_m$ and $\imag\omega_n$, at $\beta\Gamma=5.0$.
The wedge opening ratios $r$ and wedge orientations $\vartheta$ (figure
\ref{fig:wedgethetaparam}) are chosen
according to the interacting branch-cut geometry.
 The ``nonequilibrium''
line represents the location of the dot-electron spectral function for a system with 
source-drain voltage $e\Phi=0.5\Gamma$. Wiggly structures at higher energies result
from the increasingly coarse-grained $\tilde A$ grid.}
\label{fig:genkerplot}
\end{figure}

In order to discuss this in more detail, some
matrix elements of $\mathcal{Q}_{r,\vartheta}$ are plotted in figure 
\ref{fig:genkerplot}. 
The orientation of the considered data point in Matsubara space defines the
orientation of the structure which emerges in the kernel with respect to
the $(x_\varphi, x_\omega)$ coordinates of $\tilde A$-space.
A major qualitative difference to the structure of the single-wedge kernels $\mathcal{P}_{r,\vartheta}$ is the
emergence of distinguished negative regions. They are
generated by the combinations of Hilbert transforms within the edge-to-edge map 
$\mathcal{Q}^\text{(edge)}_\vartheta$. As such, they are a direct consequence
of the branch cuts.
The negative and positive regions spread over a comparably wide range and
will compete in the process of Bayesian inference, in
which several $(\imag\varphi_m,\imag\omega_n)$ pairs and differently
overlapping combinations of positive/negative regions are involved. The wide range
of the regions appears to result from superimposing the $1/x$ tails of
$\mathcal{Q}^\text{(edge)}_\vartheta
f_{\underline X, \delta}$ which are dominant for
$\vartheta\approx \pi/2$ and  $\vartheta\approx 3\pi/2$ and absent for
$\vartheta\approx 0$, as well as $\vartheta \approx \pi$.
Note that since the continuity assumption \eqref{eq:sharedRPassumption} becomes 
exact for larger energies, this feature can be expected to be contained in the kernel 
of an optimal continuation theory of Green's functions within the Matsubara 
voltage formalism. 

The kernel structure moreover indicates that due to the leverage of
the single-wedge constraint (as it applied to the MaxEnt calculations in
Ref.~\onlinecite{dirks}), 
the nonequilibrium spectral function could now well 
be resolved. In the following, the interacting branch cut geometry will
always be used for the operator $\mathcal{Q}_{r,\vartheta}$. For brevity, the
accordingly defined operator will be shortly written as $\mathcal{Q}$,
since $r$ and $\vartheta$ are now well-determined.

\subsubsection{Non-interacting Green's function as benchmark}
The fundamental assumption of this chapter, equation
\eqref{eq:sharedRPassumption}, is exact for $G_0$ (equation
\eqref{eq:ReG0Edge}). As a 
consequence, we use the noninteracting Green's function as a benchmark
for our multiple-wedge numerical analytic continuation procedure, already 
assuming the interacting branch cut geometry for the construction of
$\mathcal{Q}_{r,\vartheta}$, which is certainly also valid for $G_0$. 
\emph{At present, from a numerical point of view, the method is comprised of two technically 
challenging consecutive steps:} 
First, the kernel and its high-energy convolution with the Green's function 
have to be evaluated numerically up to a certain precision. Second, an 
appropriate default model has to be defined and the MaxEnt must converge 
to a good estimate in a controlled way.

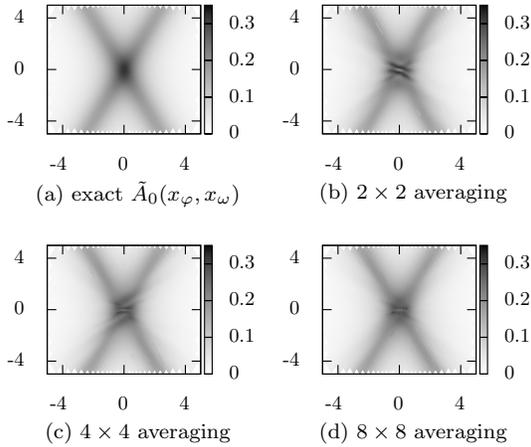
\begin{figure}
\centering
\subfloat[exact $\tilde
A_0(x_\varphi,x_\omega)$]{\resizebox{0.47\linewidth}{!}{\begin{minipage}{4.2cm}
\setlength{\unitlength}{0.0500bp}
\begin{picture}(-100,0)
\includegraphics{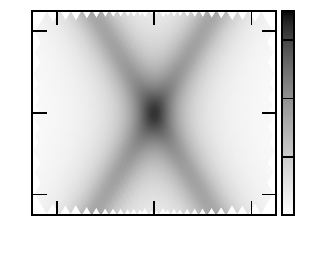}
\end{picture}
\begin{picture}(1870.00,1474.00)
      \put(327,-64){\makebox(0,0){\strut{}-4}}%
      \put(888,-64){\makebox(0,0){\strut{} 0}}%
      \put(1449,-64){\makebox(0,0){\strut{} 4}}%
      \put(16,339){\makebox(0,0)[r]{\strut{}-4}}%
      \put(16,810){\makebox(0,0)[r]{\strut{} 0}}%
      \put(16,1281){\makebox(0,0)[r]{\strut{} 4}}%
      \put(1826,221){\makebox(0,0)[l]{\strut{} 0}}%
      \put(1826,557){\makebox(0,0)[l]{\strut{} 0.1}}%
      \put(1826,893){\makebox(0,0)[l]{\strut{} 0.2}}%
      \put(1826,1229){\makebox(0,0)[l]{\strut{} 0.3}}
\end{picture}
\vspace{0.2cm}\end{minipage} } \label{fig:genkernelavgingexact}}
\subfloat[$2\times 2$ averaging]{\resizebox{0.47\linewidth}{!}{\begin{minipage}{4.2cm}
\setlength{\unitlength}{0.0500bp}
\begin{picture}(-100,0)
\includegraphics{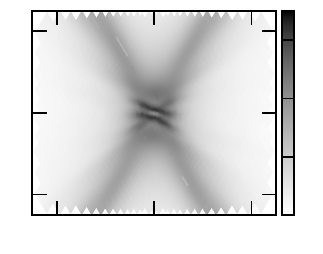}
\end{picture}
\begin{picture}(1870.00,1474.00)
      \put(327,-64){\makebox(0,0){\strut{}-4}}%
      \put(888,-64){\makebox(0,0){\strut{} 0}}%
      \put(1449,-64){\makebox(0,0){\strut{} 4}}%
      \put(16,339){\makebox(0,0)[r]{\strut{}-4}}%
      \put(16,810){\makebox(0,0)[r]{\strut{} 0}}%
      \put(16,1281){\makebox(0,0)[r]{\strut{} 4}}%
      \put(1826,221){\makebox(0,0)[l]{\strut{} 0}}%
      \put(1826,557){\makebox(0,0)[l]{\strut{} 0.1}}%
      \put(1826,893){\makebox(0,0)[l]{\strut{} 0.2}}%
      \put(1826,1229){\makebox(0,0)[l]{\strut{} 0.3}}
\end{picture}
\vspace{0.2cm}\end{minipage} } }
\\
\subfloat[$4\times 4$ averaging]{\resizebox{0.47\linewidth}{!}{\begin{minipage}{4.2cm}
\setlength{\unitlength}{0.0500bp}
\begin{picture}(-100,0)
\includegraphics{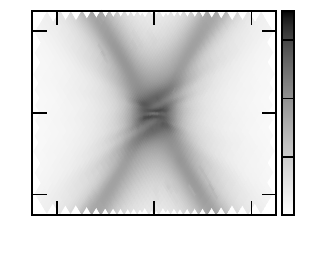}
\end{picture}
\begin{picture}(1870.00,1474.00)
      \put(327,-64){\makebox(0,0){\strut{}-4}}%
      \put(888,-64){\makebox(0,0){\strut{} 0}}%
      \put(1449,-64){\makebox(0,0){\strut{} 4}}%
      \put(16,339){\makebox(0,0)[r]{\strut{}-4}}%
      \put(16,810){\makebox(0,0)[r]{\strut{} 0}}%
      \put(16,1281){\makebox(0,0)[r]{\strut{} 4}}%
      \put(1826,221){\makebox(0,0)[l]{\strut{} 0}}%
      \put(1826,557){\makebox(0,0)[l]{\strut{} 0.1}}%
      \put(1826,893){\makebox(0,0)[l]{\strut{} 0.2}}%
      \put(1826,1229){\makebox(0,0)[l]{\strut{} 0.3}}
\end{picture}
\vspace{0.2cm}\end{minipage} } }
\subfloat[$8\times 8$ averaging]{\resizebox{0.47\linewidth}{!}{\begin{minipage}{4.2cm}
\setlength{\unitlength}{0.0500bp}
\begin{picture}(-100,0)
\includegraphics{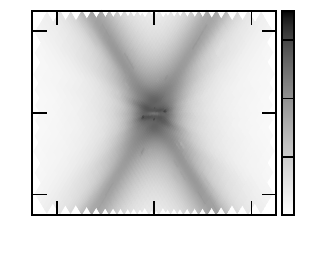}
\end{picture}
\begin{picture}(1870.00,1474.00)
      \put(327,-64){\makebox(0,0){\strut{}-4}}%
      \put(888,-64){\makebox(0,0){\strut{} 0}}%
      \put(1449,-64){\makebox(0,0){\strut{} 4}}%
      \put(16,339){\makebox(0,0)[r]{\strut{}-4}}%
      \put(16,810){\makebox(0,0)[r]{\strut{} 0}}%
      \put(16,1281){\makebox(0,0)[r]{\strut{} 4}}%
      \put(1826,221){\makebox(0,0)[l]{\strut{} 0}}%
      \put(1826,557){\makebox(0,0)[l]{\strut{} 0.1}}%
      \put(1826,893){\makebox(0,0)[l]{\strut{} 0.2}}%
      \put(1826,1229){\makebox(0,0)[l]{\strut{} 0.3}}
\end{picture}
\vspace{0.2cm}\end{minipage} } \label{fig:genkernelavging8x8} }
\caption{Successive improvement of kernel quality by averaging out the local
kernel structure within the local $\tilde A$ grid resolution. Data are shown
for $U=0$, $\Phi=0$, $\beta\Gamma=5$, $n=0, \dots, 9$, $m=-3,\dots, 3$ and a
realistic mock diagonal covariance matrix $C=\diag (\frac{10^{-13}}{\Gamma^{2}})$.
Abscissa denotes $x_\varphi/\Gamma$, ordinate denotes $x_\omega/\Gamma$,
grayscale denotes $\tilde A(x_\varphi,x_\omega)$ in units of $\Gamma^{-1}$.}
\label{fig:genkernelavging}
\end{figure}
In order to test the performance of the \emph{first step}, we can take the exact 
solution as default model and run the MaxEnt with the discretized kernel. 
By construction, due to the design of Bryan's algorithm \cite{bryanmem}, MaxEnt changes of the 
$\tilde A(x_\varphi, x_\omega)$ function will directly correspond to the 
numerical errors in the computation of the kernel matrix elements: evidence
for changes of the exact solution is taken from the exact data due to
numerical imperfections in the kernel.
Without integrating out the sharp
structures of the kernel properly for these regions, serious artifacts are obtained
even for larger test function broadnesses $\delta_{\underline x}$.
This can be seen in the ``historic'' MaxEnt data shown figure \ref{fig:genkernelavging}. Here,
$\delta_{\underline x}$ is chosen adaptively with respect to the local kernel
resolution, namely $\delta_{\underline{x}} = 0.3\cdot (\text{local kernel grid resolution})$.
The MaxEnt is able to modify the $\tilde A$ function on a large grid varying over the ranges 
$x_\varphi\in [-800,800]$, $x_\omega\in [-400,400]$.
As the local kernel resolution is increased, averaging out its
structure within the $\tilde A$ grid, an increasingly appropriate 
discretization of the kernel is obtained. In the computations shown in 
figure \ref{fig:genkernelavging}, realistic covariance weights for the 
imaginary-time data were assumed. If
numerical errors $\gtrapprox\sqrt{10^{-13}}$ were included into the realization of the kernel,
there would probably be stronger deviations from $\tilde A_0(x_\varphi,x_\omega)$ than observed.
For some single points the $\delta_{\underline x}$ is so small that the
adaptive quadrature of the fourth integral in $\mathcal{Q}$ does
not converge. This can be seen best in figure 
\ref{fig:genkernelavging8x8}, because here the kernel discretization grid is
eight times finer than the $\tilde A(x_\varphi, x_\omega)$ discretization grid.
Similarly, in the nonequilibrium situation, $\Phi\neq 0$, the function $\tilde A_0$ is not significantly altered 
by the $8\times 8$-averaging kernel. This was tested explicitly also for large bias voltages, such as $e\Phi=\Gamma$.

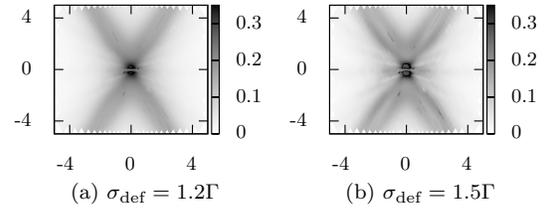
\begin{figure}
\centering
\subfloat[$\sigma_\text{def}=1.2\Gamma$]{\resizebox{0.47\linewidth}{!}{\begin{minipage}{4.2cm}
\setlength{\unitlength}{0.0500bp}
\begin{picture}(-100,0)
\includegraphics{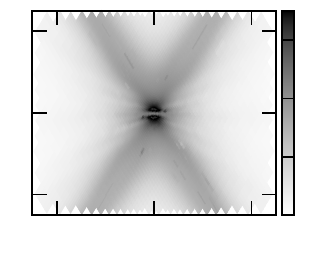}
\end{picture}
\begin{picture}(1870.00,1474.00)
      \put(327,-64){\makebox(0,0){\strut{}-4}}%
      \put(888,-64){\makebox(0,0){\strut{} 0}}%
      \put(1449,-64){\makebox(0,0){\strut{} 4}}%
      \put(16,339){\makebox(0,0)[r]{\strut{}-4}}%
      \put(16,810){\makebox(0,0)[r]{\strut{} 0}}%
      \put(16,1281){\makebox(0,0)[r]{\strut{} 4}}%
      \put(1826,221){\makebox(0,0)[l]{\strut{} 0}}%
      \put(1826,557){\makebox(0,0)[l]{\strut{} 0.1}}%
      \put(1826,893){\makebox(0,0)[l]{\strut{} 0.2}}%
      \put(1826,1229){\makebox(0,0)[l]{\strut{} 0.3}}
\end{picture}
\vspace{0.2cm}\end{minipage} } }
\subfloat[$\sigma_\text{def}=1.5\Gamma$]{\resizebox{0.47\linewidth}{!}{\begin{minipage}{4.2cm}
\setlength{\unitlength}{0.0500bp}
\begin{picture}(-100,0)
\includegraphics{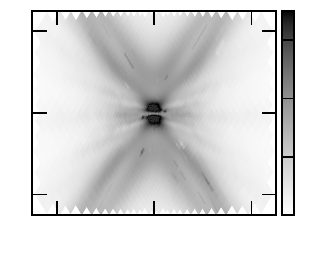}
\end{picture}
\begin{picture}(1870.00,1474.00)
      \put(327,-64){\makebox(0,0){\strut{}-4}}%
      \put(888,-64){\makebox(0,0){\strut{} 0}}%
      \put(1449,-64){\makebox(0,0){\strut{} 4}}%
      \put(16,339){\makebox(0,0)[r]{\strut{}-4}}%
      \put(16,810){\makebox(0,0)[r]{\strut{} 0}}%
      \put(16,1281){\makebox(0,0)[r]{\strut{} 4}}%
      \put(1826,221){\makebox(0,0)[l]{\strut{} 0}}%
      \put(1826,557){\makebox(0,0)[l]{\strut{} 0.1}}%
      \put(1826,893){\makebox(0,0)[l]{\strut{} 0.2}}%
      \put(1826,1229){\makebox(0,0)[l]{\strut{} 0.3}}
\end{picture}
\vspace{0.2cm}\end{minipage} } \label{fig:genkernelvarydefwidthsigma1.5} }
\caption{Sensitivity to the default model, using the same parameters and scales as in figure \ref{fig:genkernelavging8x8}.}
\label{fig:genkernelvarydefwidth}
\end{figure}
The performance of the \emph{second step} can be tested by using a kernel realization which succeeded in the first step
and then performing runs with a modified default model.
Because the noninteracting $\tilde A_0 (x_\varphi,x_\omega)$ function has the correct singular
behaviour as $\underline x \to \infty$ we investigate the dependence of the
MaxEnt results on the following default models:
\begin{equation}
\tilde D_{\sigma_\text{def}}(x_\varphi, x_\omega)
= \frac{1}{2\pi}
\sum_{\alpha =\pm 1} \frac{\sigma_\text{def}}
                          {(x_\omega - \frac{\alpha}{2}(x_\varphi-\Phi))^2 + \sigma_\text{def}^2}.
\label{eq:defmodAtildevariedbroadness}
\end{equation}
As compared to $\tilde A_0(x_\varphi,x_\omega)$, the width of the Lorentzians is varied.
Using the best-quality kernel, i.e.~$8\times 8$-averaging (see figure
\ref{fig:genkernelavging8x8}), increasing the default-model width quickly results in spurious features 
in the low- to intermediate-energy region, even though $\mathcal{Q}$ represents an exact relation between data and $\tilde A$ 
and the numerical representation of  $\mathcal{Q}$ is sufficiently accurate.
Away from the low-energy region also for $\sigma=1.5\Gamma$ a good agreement with $\tilde A_0$ is obtained, i.e.~a sharpened 
structure along the cross-shaped directions with an approximately correct amplitude (as compared to figure 
\ref{fig:genkernelavgingexact}). The strong sensitivity of especially the low-energy range on the 
default model may be interpreted as a result of the subtle interplay of positive and negative regions of high-amplitude 
kernel matrix  elements for different $(\imag\varphi_m,\imag\omega_n)$. The structure of the matrix elements was discussed 
above and plotted in figure \ref{fig:genkerplot}. 
As shown in figure \ref{fig:genkernelvarydefwidth} a problem often encountered for not well-chosen default models is 
apparently an increase of spectral weight in the low-energy region $|\underline x| \approx 0$, which exceeds the color scale 
used in the plots by up to a factor of three, even for moderate deviations of $\sigma_\text{def}$ from $\Gamma$. This is 
unfortunate, because not only for spectral functions unphysically high
values may be deduced, but also the overall weight of the spectral function is too large. However, since the kernel 
$\mathcal{Q}$ imposes an exact relation on $G_0$ and is resolved well enough,
this unfortunate aspect is identified
as a pure MaxEnt (``\emph{second step}'') artifact. As such it is no conceptual problem of the $\mathcal{Q}$-approach and
 can in principle be removed by developing a more sophisticated MaxEnt algorithm which imposes the physical constraints as prior 
information. In fact, this issue can be significantly reduced by a careful
but straightforward analysis of the posterior probabilities within a set of
smooth default models.

From our data we can conclude that default models with the shape \eqref{eq:defmodAtildevariedbroadness} are apparently not of much use 
for functions whose high-energy behaviour along the singular directions is a Lorentzian with width $\Gamma$. Once the high-energy structure
is known to be such, an interesting experiment is to flatten out the low- to intermediate-energy structure of the default model, by 
imposing an $\underline x$-dependent 
\begin{equation}
\sigma_\text{def}(x_\varphi) = \Gamma + (\tilde\sigma_\text{def}-\Gamma)\frac{R^2}{x_\varphi^2 + R^2},
\label{eq:xdepsigmadef}
\end{equation}
 where $R$ is the flattening radius and $\tilde\sigma_\text{def} \gg \Gamma$
is a strong flattening of the default model's low-energy region. For $G_0$ it turns out that the resulting 
MaxEnt solution is practically identical to the $\sigma_\text{def}=\Gamma$
solution. Consequently, the ``\emph{second step}''-artifact for $G_0$ of 
overshooting low-energy values (Fig.~\ref{fig:genkernelvarydefwidth}) can just be cured
by imposing the correct high-energy limit.
The low-energy artifact is thus caused by missing a-priori information about the high-energy structure.
This appears to be another manifestation of the fact that the kernel
$\mathcal{Q}$ puts a large range of energy scales in relation to each other. 
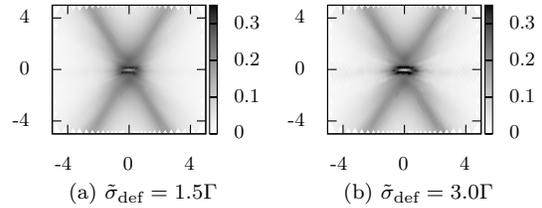
\begin{figure}
{
\centering
\subfloat[$\tilde \sigma_\text{def}=1.5\Gamma$]{\resizebox{0.47\linewidth}{!}{\begin{minipage}{4.2cm}
\setlength{\unitlength}{0.0500bp}
\begin{picture}(-100,0)
\includegraphics{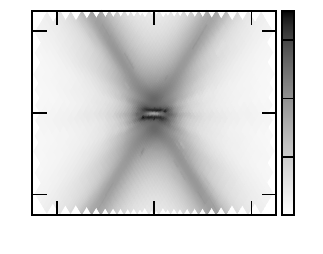}
\end{picture}
\begin{picture}(1870.00,1474.00)
      \put(327,-64){\makebox(0,0){\strut{}-4}}%
      \put(888,-64){\makebox(0,0){\strut{} 0}}%
      \put(1449,-64){\makebox(0,0){\strut{} 4}}%
      \put(16,339){\makebox(0,0)[r]{\strut{}-4}}%
      \put(16,810){\makebox(0,0)[r]{\strut{} 0}}%
      \put(16,1281){\makebox(0,0)[r]{\strut{} 4}}%
      \put(1826,221){\makebox(0,0)[l]{\strut{} 0}}%
      \put(1826,557){\makebox(0,0)[l]{\strut{} 0.1}}%
      \put(1826,893){\makebox(0,0)[l]{\strut{} 0.2}}%
      \put(1826,1229){\makebox(0,0)[l]{\strut{} 0.3}}
\end{picture}
\vspace{0.2cm}\end{minipage} } }
\subfloat[$\tilde \sigma_\text{def}=3.0\Gamma$]{\resizebox{0.47\linewidth}{!}{\begin{minipage}{4.2cm}
\setlength{\unitlength}{0.0500bp}\begin{picture}(-100,0)
\includegraphics{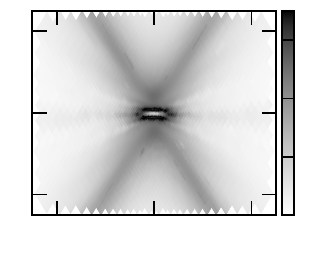}
\end{picture}
\begin{picture}(1870.00,1474.00)
      \put(327,-64){\makebox(0,0){\strut{}-4}}%
      \put(888,-64){\makebox(0,0){\strut{} 0}}%
      \put(1449,-64){\makebox(0,0){\strut{} 4}}%
      \put(16,339){\makebox(0,0)[r]{\strut{}-4}}%
      \put(16,810){\makebox(0,0)[r]{\strut{} 0}}%
      \put(16,1281){\makebox(0,0)[r]{\strut{} 4}}%
      \put(1826,221){\makebox(0,0)[l]{\strut{} 0}}%
      \put(1826,557){\makebox(0,0)[l]{\strut{} 0.1}}%
      \put(1826,893){\makebox(0,0)[l]{\strut{} 0.2}}%
      \put(1826,1229){\makebox(0,0)[l]{\strut{} 0.3}}
\end{picture}
\vspace{0.2cm}\end{minipage} } }
}
\caption{MaxEnt results for flat low-energy default models \eqref{eq:xdepsigmadef}, $R=5$. 
As compared to Fig.~\ref{fig:genkernelvarydefwidthsigma1.5}, the quality of low-energy data is increased significantly, due to the correct
high-energy behaviour of the default model. Scales are as in figure \ref{fig:genkernelavging8x8}.
}
\label{fig:genkernelvarydefwidth}
\end{figure}

\subsection{Application to the interacting model}
Switching on a finite Coulomb interaction, one has to be aware of the fact that the $\mathcal{Q}$-mapping can no longer be 
expected to be fully exact. However, a special case of the assumption, 
namely the fitting ansatz in Ref.~\onlinecite{prl07}, is found to yield reasonable results which agree with other methods up to a certain extent \cite{han10}. 
Therefore, it seems worthwhile to investigate how far one can go with the controlled MaxEnt approach to the inversion of the 
$\mathcal{Q}$-mapping \footnote{To be more precise, for technical
reasons (conservation of the spectral function normalization for \emph{any} causal
selfenergy) the fits in Refs.~\onlinecite{prl07, han10} were performed with
respect to the selfenergy, not with respect to the Green's function. Because the $\mathcal{Q}$-mapping is also exact for ansatz functions
of the fits presented in Ref.~\onlinecite{prl07}, the fits are essentially special cases of the present work, though. The first-order Pad\'e approximant fits presented
in Ref.~\onlinecite{han10} represent a different kind of generalization.}.

\subsubsection{Lateral structure along singular directions}
\begin{figure}
\centering
\subfloat[nontrivial physics at $x_\omega\approx \pm x_\varphi/2$]
{
 \resizebox{0.45\linewidth}{!}{
\begin{picture}(0,0)%
\includegraphics{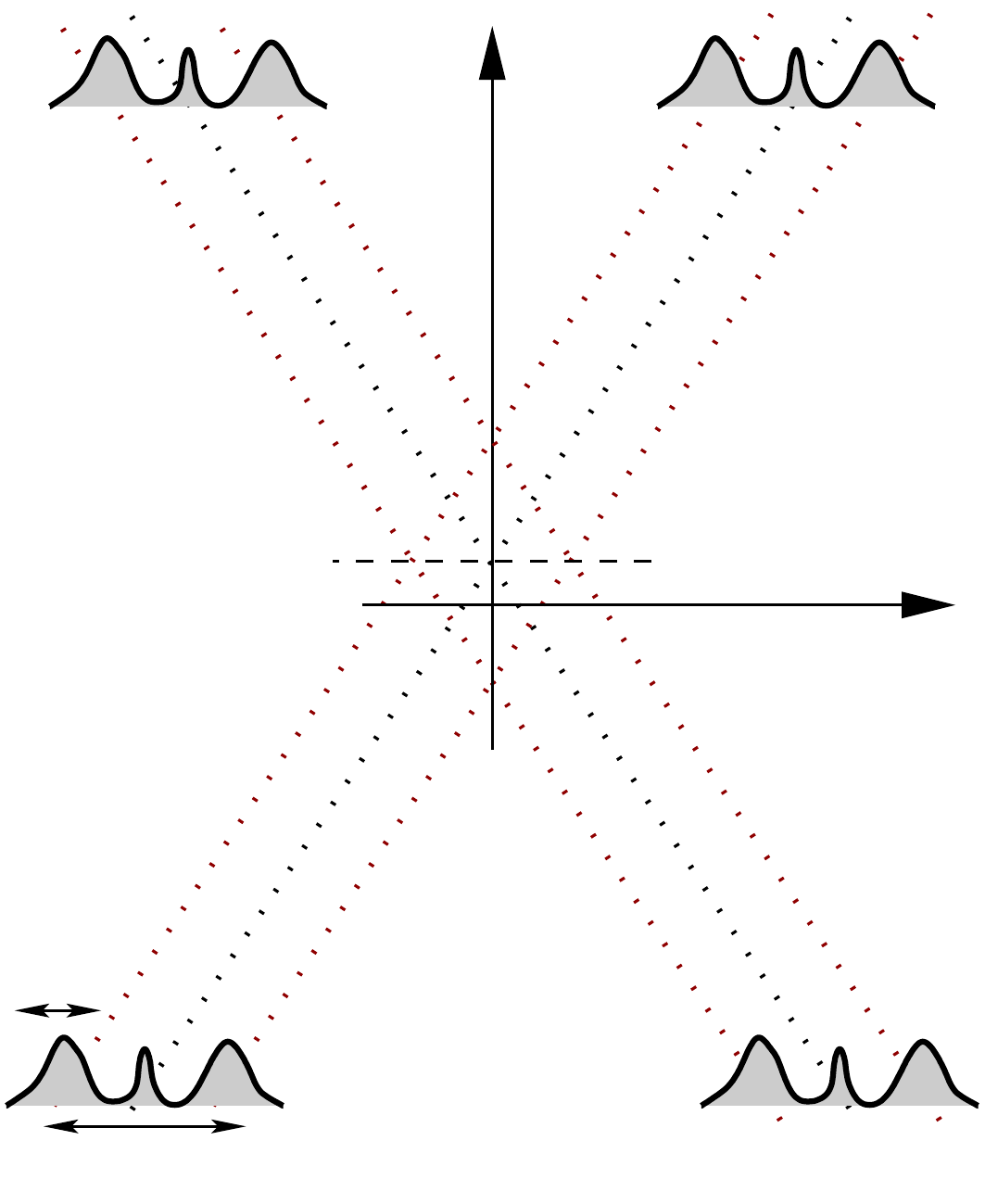}%
\end{picture}%
\setlength{\unitlength}{3947sp}%
\begingroup\makeatletter\ifx\SetFigFont\undefined%
\gdef\SetFigFont#1#2#3#4#5{%
  \reset@font\fontsize{#1}{#2pt}%
  \fontfamily{#3}\fontseries{#4}\fontshape{#5}%
  \selectfont}%
\fi\endgroup%
\begin{picture}(5100,6228)(3676,-5636)
\put(3751,-4561){\makebox(0,0)[lb]{\smash{{\SetFigFont{20}{24.0}{\familydefault}{\mddefault}{\updefault}{\color[rgb]{0,0,0}$\sim$\-$\Gamma$}%
}}}}
\put(6301,314){\makebox(0,0)[lb]{\smash{{\SetFigFont{20}{24.0}{\familydefault}{\mddefault}{\updefault}{\color[rgb]{0,0,0}$x_\varphi$}%
}}}}
\put(8251,-2386){\makebox(0,0)[lb]{\smash{{\SetFigFont{20}{24.0}{\familydefault}{\mddefault}{\updefault}{\color[rgb]{0,0,0}$x_\omega$}%
}}}}
\put(7051,-2386){\makebox(0,0)[lb]{\smash{{\SetFigFont{20}{24.0}{\familydefault}{\mddefault}{\updefault}{\color[rgb]{0,0,0}$A(\omega)$}%
}}}}
\put(4276,-5536){\makebox(0,0)[lb]{\smash{{\SetFigFont{20}{24.0}{\familydefault}{\mddefault}{\updefault}{\color[rgb]{0,0,0}$U$}%
}}}}
\end{picture}%
 }
}
\subfloat[``asymptotic freedom'']{
 \resizebox{0.39\linewidth}{!}{
\begin{picture}(0,0)%
\includegraphics{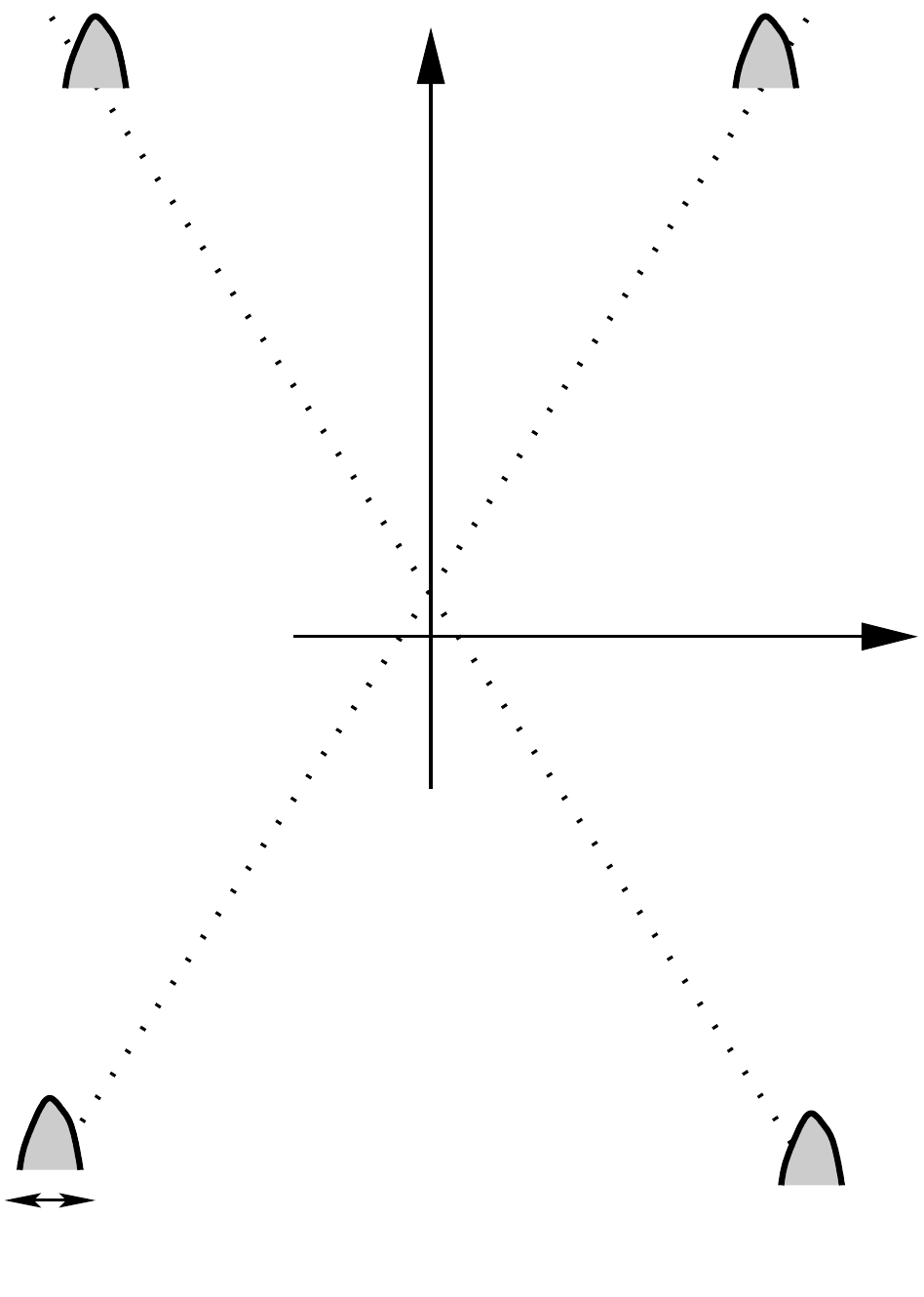}%
\end{picture}%
\setlength{\unitlength}{3947sp}%
\begingroup\makeatletter\ifx\SetFigFont\undefined%
\gdef\SetFigFont#1#2#3#4#5{%
  \reset@font\fontsize{#1}{#2pt}%
  \fontfamily{#3}\fontseries{#4}\fontshape{#5}%
  \selectfont}%
\fi\endgroup%
\begin{picture}(4544,6367)(4104,-5786)
\put(6301,314){\makebox(0,0)[lb]{\smash{{\SetFigFont{20}{24.0}{\familydefault}{\mddefault}{\updefault}{\color[rgb]{0,0,0}$x_\varphi$}%
}}}}
\put(8251,-2386){\makebox(0,0)[lb]{\smash{{\SetFigFont{20}{24.0}{\familydefault}{\mddefault}{\updefault}{\color[rgb]{0,0,0}$x_\omega$}%
}}}}
\put(4276,-5686){\makebox(0,0)[lb]{\smash{{\SetFigFont{20}{24.0}{\familydefault}{\mddefault}{\updefault}{\color[rgb]{0,0,0}$\Gamma$}%
}}}}
\end{picture}%
  \par~\par~ 
 }
}
\caption{At high energies, one might (a) expect the lateral structure of $\tilde A(x_\varphi, x_\omega)$ to be 
composed of two Hubbard peaks and possibly a quasi-particle resonance which combine to the physical 
spectrum $A(\omega)$ at the intersection point.
In the complementary scenario (b), the function $\tilde A$ would not differ from the
noninteracting one at high energies.}
\label{fig:lateralstruchighenergy}
\end{figure}
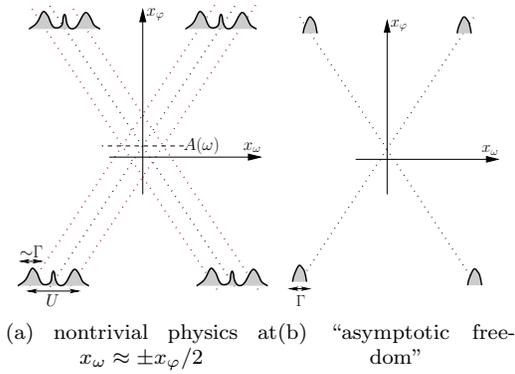

As shown in the preceding section, the a-posteriori determination of a most
adequate approximate a-priori picture of the high-energy structure is crucial for the
success of the MaxEnt procedure. 
As sketched in figure \ref{fig:lateralstruchighenergy}, at finite $U$, one may, for example, expect the 
lateral structure be an unphysical copy of a spectral function, i.e.~two Hubbard peaks
with possibly an additional peak associated to a quasi-particle resonance. Such a structure would extend over a range $\approx U$. However, the two parallel
Hubbard peaks can be expected to approximately have a Lorentzian structure of width $\approx \Gamma$ and would generate a type of branch cut in the 
$\mathcal{Q}$-mapping which is equivalent to the one in $G_0$. In the strongly correlated regime, Hubbard satellites may
be broadened up to a width of $2\Gamma$, due to many-body correlations
\cite{glossop}.

It is \emph{a priori} uncertain to which extent either of the intuitive
pictures in figure \ref{fig:lateralstruchighenergy} is correct. 
However, one of the conceptual strengths of the $\mathcal{Q}$-mapping is the precise rendering of the high-energy structure 
of the imaginary-voltage theory (cf.~section \ref{subsec:strucresterm}). One can expect that only a characteristic width of the lateral structure along the singular directions is 
needed in order to model the correct high-energy contribution to the amplitude of the discontinuity of $G(z_\varphi,z_\omega)$ at the low-to-intermediate energy
portions of the branch cuts. Based on this, we can investigate the 
posterior probability $\Pr(\sigma_\text{def}|\bar G)$ for default models 
\eqref{eq:defmodAtildevariedbroadness} as a function of their width $\sigma_\text{def}$. 

The thus determined most probable $\sigma_\text{def}$ then serves as
an effective description of the high-energy structure $\tilde A(x_\varphi, x_\omega)$ 
for the actual computations. However, as input data from the QMC simulations, only low- to intermediate-energy data are available. Therefore, the posterior probability 
probe with respect to default models \eqref{eq:defmodAtildevariedbroadness}
has to be interpreted with care.
\begin{figure}
\centering
\includegraphics[width=\linewidth]{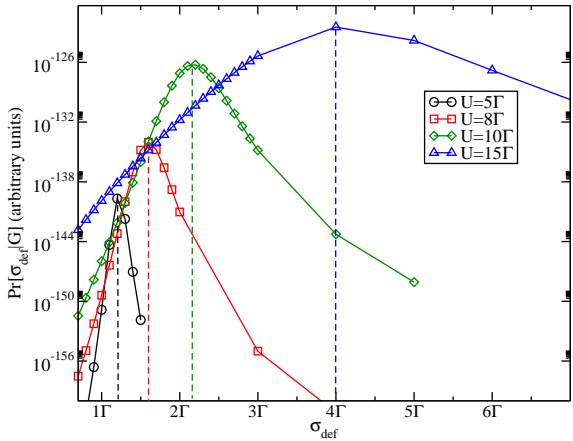}
\caption{(color online) Posterior probability of the default model
\eqref{eq:defmodAtildevariedbroadness} at $\beta\Gamma = 5$, $e\Phi=\Gamma$
for several interactions strengths. The result is found to be essentially
independent of the bias voltage. The kernel validated in figure 
\ref{fig:genkernelavging8x8} has been used.}
\label{fig:pprobsHI}
\end{figure}
In figure \ref{fig:pprobsHI}, posterior probabilities for different
interaction strengths are displayed. Due to the width being significantly larger than
$2\Gamma$ for $U=15\Gamma$ it is
obvious that the lateral width cannot solely be interpreted as a signature
of the Hubbard bands. Merely, the overall Lorentzian broadness of the spectral function
seems to be obtained.
Based on our data, neither of the scenarios of figure \ref{fig:lateralstruchighenergy}
can be preferred. However, based on our experience, the most probable
high-energy structure also yields reasonable results in the case of
comparably strong interactions. Thus, in the practical computations, first
the most probable default model is identified. As a next step, the actual
spectral functions are estimated.

\end{document}